\documentclass[pdflatex,sn-mathphys-num,iicol]{sn-jnl}

\usepackage{graphicx}%
\usepackage{multirow}%
\usepackage{amsmath,amssymb,amsfonts}%
\usepackage{amsthm}%
\usepackage{mathrsfs}%
\usepackage{mathtools}
\usepackage[title]{appendix}%
\usepackage{xcolor}%
\usepackage{textcomp}%
\usepackage{manyfoot}%
\usepackage{booktabs}%
\usepackage{algorithm}%
\usepackage{algorithmicx}%
\usepackage{algpseudocode}%
\usepackage{tabularx}
\usepackage{booktabs}
\usepackage{array}
\usepackage{soul}
\usepackage{url}        
\usepackage{hyperref}   

\usepackage{longtable, xcolor, tabularx}
\usepackage[table]{xcolor}
\usepackage{ltablex}   
\keepXColumns
\usepackage{array}
\usepackage{booktabs}
\usepackage[table]{xcolor}

\usepackage{empheq}
\usepackage{ragged2e}
\newcolumntype{Y}[1]{>{\raggedright\arraybackslash}p{#1}} 
\newcolumntype{Z}{>{\raggedright\arraybackslash}X}       

\usepackage{subcaption}    

\usepackage{dblfloatfix} 

\usepackage{listings}
\newcommand{\mcode}[1]{%
  \lstinline[language=Matlab, basicstyle=\ttfamily\normalsize]{#1}%
}

\usepackage{xcolor}

\definecolor{codeblue}{rgb}{0,0,0.6}
\definecolor{codegreen}{rgb}{0,0.5,0}
\definecolor{codegray}{rgb}{0.5,0.5,0.5}
\definecolor{codepurple}{rgb}{0.58,0,0.82}
\definecolor{backcolour}{rgb}{0.97,0.97,0.97}

\usepackage{listings}
\usepackage{upquote}

\lstdefinestyle{matlabstyle}{
    language=Matlab,
    backgroundcolor=\color{backcolour},
    basicstyle=\footnotesize\ttfamily,
    keywordstyle=\color{codeblue}\bfseries,
    commentstyle=\color{codegreen}\itshape,
    stringstyle=\color{codepurple},
    numberstyle=\tiny\color{codegray},
    numbers=left,
    numbersep=6pt,
    frame=single,
    rulecolor=\color{black},
    breaklines=true,
    showstringspaces=false,
    tabsize=2
}

\usepackage{siunitx}
\renewcommand{\arraystretch}{1.2}
\definecolor{lightgray}{gray}{0.93}

\newcolumntype{L}[1]{>{\raggedright\arraybackslash}p{#1}}
\newcolumntype{C}[1]{>{\centering\arraybackslash}p{#1}} 
\newcolumntype{Y}{>{\raggedright\arraybackslash}X}
\newcommand{\bx}{{\mathbf{x}}}

\newcommand{\bp}{{\mathbf{p}}}

\newcommand{\bK}{{\mathbf{K}}}

\newcommand{\TS}{\mathcal{T}}
\newcommand{\DS}{\mathcal{D}}
\newcommand{\bk}{{\mathbf{k}}}

\newcommand{\R}{{\mathbb{R}}}

\newcommand{\bd}{{\mathbf{d}}}

\DeclareMathOperator{\sign}{sign}

\newcommand{\minimize}{\mathop{ \text{minimize}}}

\usepackage{textcomp}
	
\usepackage{bm}

\newcommand{\blambda}{\bm{\lambda}}

\usepackage{mathrsfs}

\newcommand{\AmirAdded}[1]{\textcolor{teal}{#1}}

\theoremstyle{thmstyleone}%
\theoremstyle{thmstyletwo}%

\theoremstyle{thmstylethree}%

\let\orcidlogo\relax
\usepackage{orcidlink}

\begin{document}


\title[STORX: An Open-Source Object-Oriented Framework for Shape and Topology Optimization in MATLAB]{STORX: An Open-Source Object-Oriented Framework for Shape and Topology Optimization in MATLAB}


\author*[1]{
\fnm{Amir M.} \sur{Mirzendehdel}
\,\orcidlink{0000-0002-4407-1877}%
}
\email{amirzend@ku.edu}
\author[2]{\fnm{Krishnan} \sur{Suresh} \,\orcidlink{0000-0002-9688-9697}}\email{ksuresh@wisc.edu}

\affil*[1]{\orgdiv{Aerospace Engineering Department}, \orgname{University of Kansas},  \city{Lawrence},  \state{KS}, \country{USA}}

\affil[2]{\orgdiv{Mechanical Engineering Department}, \orgname{UW-Madison}, \city{Madison}, \state{WI}, \country{USA}}


\abstract{This paper presents \textsc{STORX}: Shape and Topology Optimization for Research and Experimentation, an open-source MATLAB-based educational framework for learning and teaching computational design optimization. 
Unlike existing educational codes, which are typically built around a single formulation, \textsc{STORX} is, to the best of our knowledge, the first open-source MATLAB framework to unify parametric shape, level-set shape, and multiple families of topology optimization. 
All modules in \textsc{STORX} follow a consistent object-oriented structure and integrate visualization, sensitivity analysis, and finite element routines, enabling users to explore the continuum between shape and topology optimization in a transparent and reproducible manner. The code is designed to complement graduate-level coursework and independent research by emphasizing modularity and extensibility through a clear separation of intent. Core software interfaces are defined via abstract base classes, enabling new objective functionals and design/manufacturing constraints to be implemented by adding derived classes without modifying the core code. The paper also describes the software architecture and demonstrates how the framework maps mathematical formulations directly to executable code through a series of illustrative problems.}


\maketitle

\section{Introduction}\label{sec:intro}

Shape and topology optimization (SO/TO) are often taught and implemented as separate algorithmic families, even though many of their computational steps are shared. In shape optimization, the design is typically represented through geometric parameters, boundary motion, or mesh deformation \cite{pironneau2005optimal,sokolowski1992introduction,haftka2012elements}. In topology optimization, the design may be represented through element densities, level-set fields, or topological sensitivities that permit more fundamental changes in material layout and connectivity \cite{bendsoe1988generating,allaire2005structural,novotny2007topological}. These representations differ substantially in what they can express and how they are updated, but they rely on a common design-optimization pipeline: solve the governing state problem, evaluate objectives and constraints, compute sensitivities or shape derivatives, regularize the design space when needed, and perform a constrained design update. For students and new researchers, this shared structure is often obscured because each method is introduced with its own notation, data structures, discretization choices, and implementation style.

\begin{table*}[b]
\caption{Overview of educational and reference code bases in TO.
A more detailed catalog is provided in ~\ref{app:reviewTable}.}
\label{tab:to_roadmap}
\centering
\small
\renewcommand{\arraystretch}{1.15}
\begin{tabular}{p{2.7cm} p{6cm} p{4cm} p{1.5cm}}
\hline
\textbf{Topic} & \textbf{Value proposition} & \textbf{Code bases} & \textbf{Language}\\
\hline
2D compliance & Transparent educational baseline & \texttt{top99}~\cite{Sigmund2001}, \texttt{top88}~\cite{Andreassen2011}, \texttt{PolyTop}~\cite{talischi2012polytop} & MATLAB\\
3D + scaling & Larger models / speed & \texttt{top3d}~\cite{LiuTovar2014}, \texttt{TOP3D\_XL}~\cite{wang2025efficient}, \texttt{TopOpt\_PETSc}~\cite{AagePETSc2014} & MATLAB, C++\\
Level-set family & Crisp boundaries / geometry control & \texttt{top\_levelset}~\cite{challis2010discrete}, \texttt{MMC188}~\cite{zhang2016new}, \texttt{TOPRBF}~\cite{Wei2018} & MATLAB\\
Stress constraints & Strength-aware design & \texttt{minVpnorm\_adpt}~\cite{amir2021efficient}, \texttt{stress\_minimize}~\cite{deng2022efficient}, \texttt{PolyStress}~\cite{giraldo2021polystress} & MATLAB\\
Fluids & Flow control / Stokes-Navier-Stokes & \texttt{topFlow}~\cite{Alexandersen2022}, \texttt{stokes-topology}~\cite{DolfinAdjointStokesTO} & MATLAB, Python\\
Multimaterial & Multiple phases and functions & \texttt{multi\_top}~\cite{tavakoli2014alternating}, \texttt{Multimaterial2d/3d}~\cite{zheng2024efficient} & MATLAB\\
ML-assisted & Learned parameterizations / surrogates & \texttt{TOuNN}~\cite{TOuNN2021}, \texttt{deep-topopt}~\cite{DeepTopoptGitHub} & Python\\
\hline
\end{tabular}
\end{table*}

Educational optimization codes have been essential for making individual methods transparent and reproducible. However, many are intentionally written around a single formulation, such as a compact density-based compliance-minimization code or a specific level-set implementation. This ``script-per-method'' style is effective for teaching one algorithm at a time, but it makes it difficult to see which parts of the solver are method-specific and which parts are reusable across SO/TO. It also limits systematic comparison and slows extension to new objectives, constraints, geometries, and coupled physics. A framework that preserves the clarity of teaching codes while exposing shared abstractions across design representations can therefore help bridge the gap between textbook formulations and extensible research software.

\begin{figure*}[!h]
  \centering 
  \includegraphics[width=\linewidth]{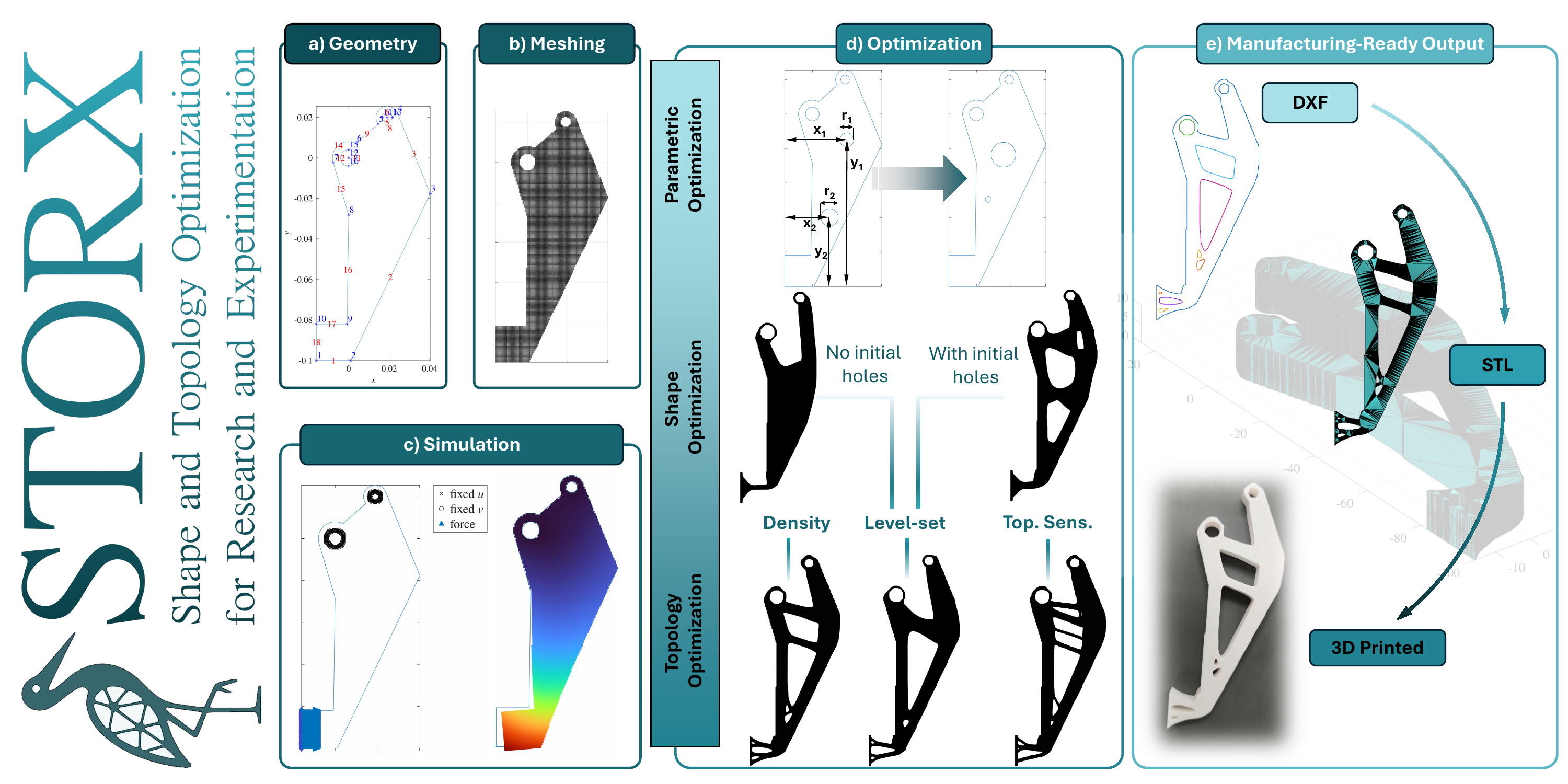}
  \caption{STORX overview. (a) The initial design is specified using a boundary representation (B-Rep). (b) The geometry is discretized. (c) The state problem is solved, e.g., via finite element analysis (FEA). (d) The design is updated through parametric shape, level-set shape, or topology optimization. (e) The resulting geometry is exported to DXF and then to a watertight STL for downstream manufacturing (e.g., 3D printing).}
  \label{fig:storx_overview}
\end{figure*}

\subsection{Related Work}

Early classroom examples emphasized the essential mechanics of material redistribution via simple density updates and removal strategies \cite{xie1997basic,martin2004topology}. MATLAB implementations subsequently made the SIMP methodology broadly accessible, notably through \texttt{top99} and its successors, which prioritized clarity, filtering, and robust behavior on canonical compliance benchmarks \cite{Sigmund2001,Andreassen2011,ferrari2020new}.

As the field matured, educational implementations expanded beyond a single representation while preserving the same pedagogical objective: \textit{making the modeling and algorithmic choices explicit}. 

Representative teaching-oriented codes have explored level-set formulations \cite{challis2010discrete,Wei2018}, geometry projection \cite{SmithNorato2020,coniglio2019generalized}, moving morphable components \cite{zhang2016new}, polygonal and isogeometric discretizations \cite{talischi2012polytop,gao2021igatop}, and selected extensions to nonlinearity and dynamics \cite{chen2019213,giraldo2021polydyna}. 

Since the teaching-code literature is dominated by TO, Table~\ref{tab:to_roadmap} is intentionally TO-centered, while shape optimization is discussed primarily where it shares the same analysis-sensitivity-update abstractions or appears through boundary-evolution and geometry-projection formulations.
Table~\ref{tab:to_roadmap} summarizes this progression using representative code bases; Appendix Table~\ref{tab:educational_tutorials} provides the full catalog. 
For a comprehensive review of educational articles on SO/TO published through 2021, see \cite{wang2021comprehensive}.

Despite this breadth, many widely used resources remain centered on two-dimensional structural compliance and favor minimal MATLAB scripts with vectorization and filtering to keep runtimes low and the algorithmic structure transparent \cite{Andreassen2011,talischi2012polytop,ferrari2020new}. Recent teaching-oriented efforts have also addressed a broader set of objectives and physics, including stress, buckling, compliant mechanisms, multiscale design, fluids, and photonics, while remaining intentionally compact and editable \cite{giraldo2021polystress,ferrari2021topology,huang2023matlab,gao2019concurrent,Alexandersen2022,christiansen2021compact}; extensions toward larger three-dimensional problems and faster solvers (e.g., matrix-free strategies) exist but are still less common in classroom-facing materials \cite{LiuTovar2014,wang2025efficient}.

In parallel, Python and Julia ecosystems have begun to provide more structured educational entries that move beyond single scripts toward modern software practices and reproducible workflows \cite{FEniCSSIMPDemo,FEniCSTopOptRepo,DolfinAdjointStokesTO,DTUPythonCodes,TopOptjlDocs,TopOptjlGitHub,TopOptJLPkg,AagePETSc2014,DeepTopoptGitHub}. 
However, across languages and representations, recurring simplifications limit reuse for more general research experimentation: many examples assume box-like geometries and regular grids, and many implementations are organized as monolithic scripts, which is ideal for transparency but makes it difficult to systematically swap geometry models, solvers, objectives, constraints (including manufacturing constraints), and coupled-physics workflows \cite{Sigmund2001,Andreassen2011,talischi2012polytop,coniglio2019generalized,SmithNorato2020}.

While prior educational codes typically support a single formulation or extend a formulation to new objectives and physics, none, to our knowledge, unifies parametric shape, level-set shape, and multiple TO families under a shared solver, sensitivity, and manufacturing-constraint framework. This has serious pedagogical implications, formulation-specific implementation hides the common theme and precludes reuse of data structures and algorithms.

This paper presents \textit{\textsc{STORX}: Shape and Topology Optimization for Research and Experimentation}, an open-source MATLAB-based educational framework for learning, teaching, and prototyping computational design optimization. As illustrated in Fig. \ref{fig:storx_overview}, \textsc{STORX} supports parametric and level-set shape optimization, as well as topology optimization methods spanning density-based, level-set, and topological-sensitivity-driven approaches (including evolutionary and Pareto-tracing strategies). All modules follow a consistent object-oriented structure and integrate finite element analysis (FEA), sensitivity analysis, and visualization, enabling users to study the continuum between shape and topology optimization in a transparent and reproducible workflow. The paper's central methodological contribution and a key element of \textsc{STORX} is its abstract-class architecture.

\begin{figure*}[!h]
  \centering 
  \includegraphics[width=\linewidth]{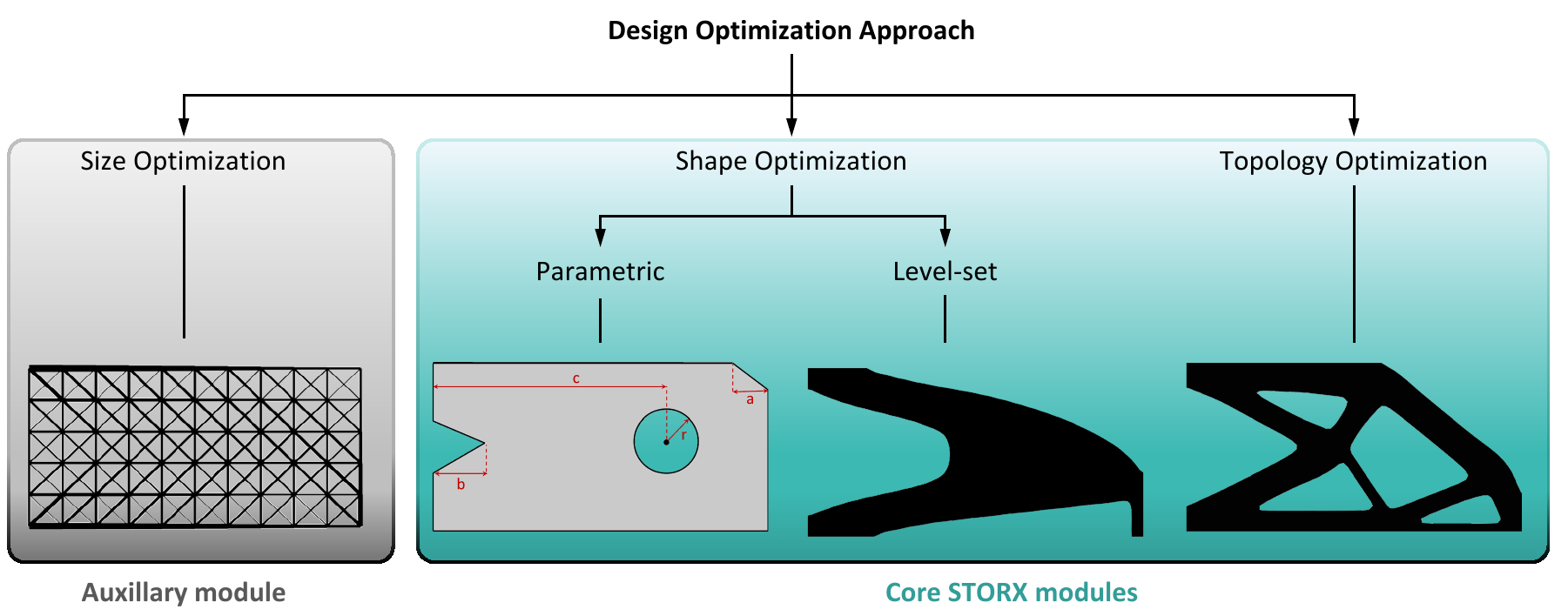}
  \caption{Design optimization approaches for size, shape, and topology optimization. \textsc{STORX} focuses on continuum shape and topology optimization; a truss size-optimization example is provided as an auxiliary educational module.}

  \label{fig:DesOpt_Approaches}
\end{figure*}

\subsection{Contributions \& Outline}

The central design goal of \textsc{STORX} is modularity through \textit{separation of intent}. Core software interfaces are defined via abstract base classes, enabling new objective functionals, constraints (including manufacturing-oriented constraints), and problem-specific components to be implemented as derived classes without modifying the core code. The framework is designed to support general planar geometries rather than being restricted to box-like domains, and it decouples boundary-condition specification from the underlying discretization so that loads and supports can be prescribed consistently across different meshing and representation choices.

The main contributions of this work are summarized as follows:

\begin{enumerate}
\item A \textit{unified computational platform} that enables controlled comparisons between shape, level-set, and topology optimization methods under consistent physical and numerical conditions.
\item Support for \textit{general 2D geometries}, allowing users to move beyond standard rectangular domains to complex, boundary-defined problems.
\item A \textit{highly modular architecture} that decouples solvers from problem definitions, facilitating the rapid implementation of new objective functionals and constraints.
\item A \textit{vectorized implementation} for maximum computational efficiency, featuring optional explicit implementations of key routines to verify and understand the underlying algorithms.
\item A \textit{flexible prototyping environment} for testing novel optimization algorithms, material models, objectives, and constraints  without changing the underlying solver infrastructure.
\end{enumerate}

The paper describes this architecture (Section \ref{sec:overview}) and demonstrates how standard mathematical formulations map directly to executable implementations through a series of benchmark problems (Sections \ref{sec:repAnalysis}-\ref{sec:TopOpt}). It further presents extensions that incorporate additional physics, objective functions, manufacturing constraints, and external solver wrappers (Section \ref{sec:advTopics}), with the aim of accelerating hands-on understanding and supporting rapid experimentation.

\section{Framework Overview}\label{sec:overview}

\begin{figure*}[!h]
  \centering 
  \includegraphics[width=0.8\linewidth]{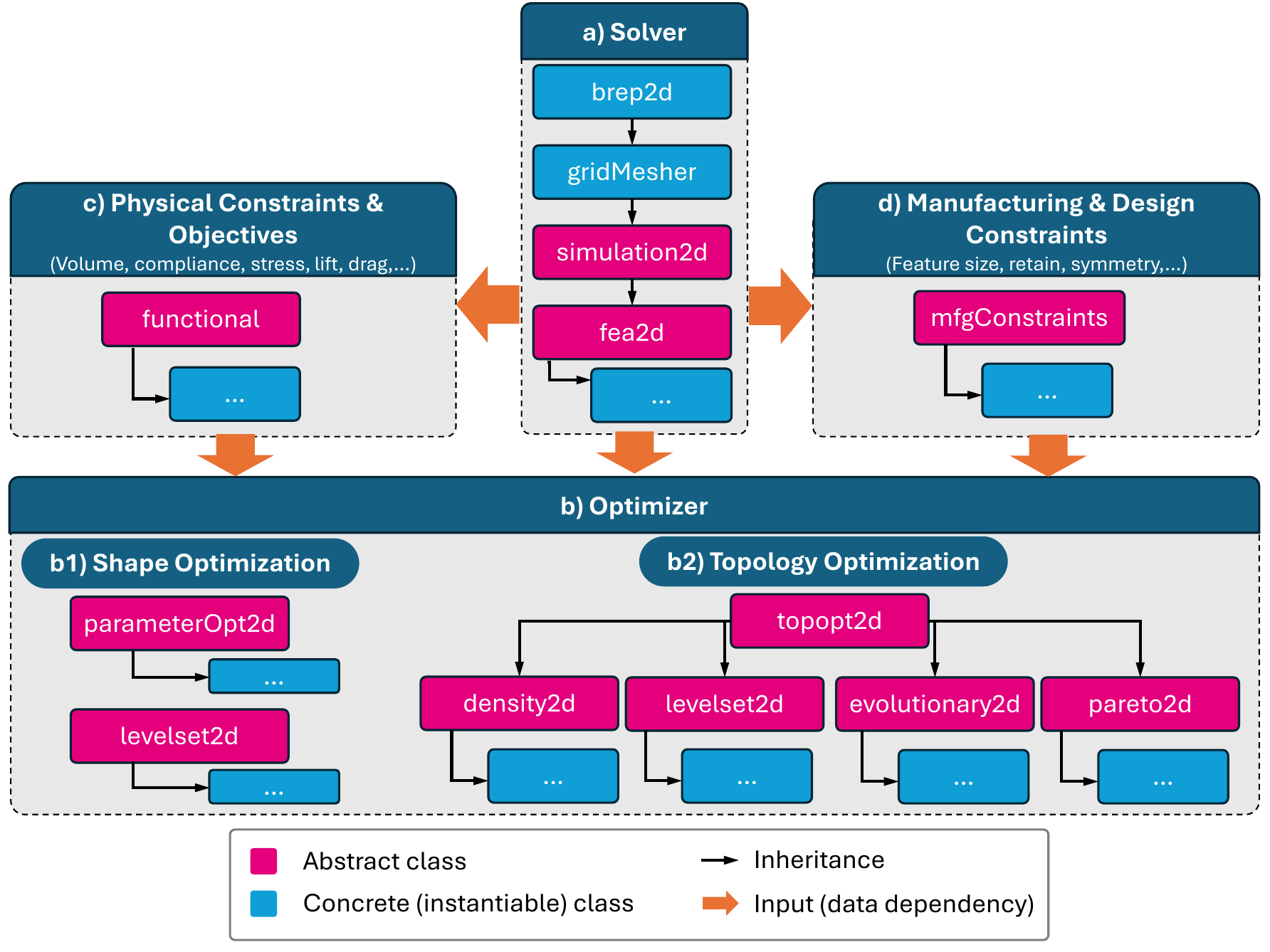}
  \caption{Class diagrams for STORX.}
  \label{fig:storx_class_diagram}
\end{figure*}

Figure~\ref{fig:DesOpt_Approaches} illustrates how \textsc{STORX} fits in the broader context of design optimization, spanning size, shape, and topology optimization. The present work focuses on continuum structural optimization in two dimensions, including density, level-set, and topological sensitivity methods. Size optimization appears in Fig.~\ref{fig:DesOpt_Approaches} for completeness; a compact truss size optimization example is provided in the repository as an auxiliary educational module, but it is not part of the core framework interfaces.

In the remainder of this section, we will summarize the object-oriented structure used across modules and describe the principal classes and abstract interfaces that connect geometry modeling, FEA, sensitivity evaluation, and optimization updates. These interfaces are designed to make the common SO/TO workflow explicit and reusable while allowing new objectives, constraints, and solution strategies to be incorporated through derived classes with minimal changes to the core code.

The core functionalities of the software are implemented in MATLAB using object-oriented programming (OOP). This approach is chosen to help students better understand the underlying concepts while enhancing the organization of the code. By employing OOP, the software facilitates better management of code structure, defining the application programming interfaces (API),  dependencies, and the classification of SO/TO methods, making it easier for users to navigate and extend the software.

The goal has been to keep the code accessible and easy to explain in an educational setting while remaining practical to run without significant computational overhead. This balance allows the code to serve as both a learning tool and a functional implementation of optimization methods for research.

There are four major components as illustrated in Fig.~\ref{fig:storx_class_diagram}: (a) a \textit{solver} for the state equation, which plays a central role in the framework and requires the specification of the geometry, boundary conditions (B.C.), meshing, and FEA solver; the resulting instantiated FEA object is directly passed to other modules as an input; (b) an \textit{optimizer} that updates the design, which takes a solver object as input; (c) \textit{physical constraints and objective} modules that evaluate the required functionals and compute their gradients with respect to the active design variables; and (d) \textit{manufacturing and design constraints} modules that act on the design variables and sensitivity fields to promote manufacturable solutions. 

With these four components in mind, we next describe how they are realized in \textsc{STORX} through a small set of core classes and abstract interfaces. The intent is twofold: (1) to make the end-to-end workflow in Fig.~\ref{fig:DesOpt_Approaches} explicit in code, geometry $\rightarrow$ discretization $\rightarrow$ state solve $\rightarrow$ sensitivity evaluation $\rightarrow$ design update, and (2) to provide stable software contracts so that new physics, objectives, constraints, and optimization strategies can be introduced by extending derived classes rather than rewriting core routines. 

\section{Representations and Analysis} \label{sec:repAnalysis}

\subsection{Boundary Representation} \label{sec:brep}

In computer-aided design, we typically start by defining the boundary of the design domain, including its functional surfaces where boundary conditions are applied. This boundary is often represented through the Boundary Representation (B-Rep) model. Let us focus on 2D B-Reps, which are collections of points (or vertices) and line\AmirAdded{/arc} segments (or edges) that outline the boundary.

To ensure that the B-Rep model is valid, it must satisfy the following conditions:

\begin{enumerate} 
\item Each vertex must be connected to exactly two other vertices (unless they are virtual edges).
\item Edges must only intersect at vertices.
\end{enumerate}
	
	Figure \ref{fig_invalidbrepfiles2d} shows some invalid 2d B-Reps:
	
	\begin{figure} [t]
		\centering
		\begin{subfigure}[t]{0.24\linewidth}
			\centering
			\includegraphics[width=0.9\linewidth]{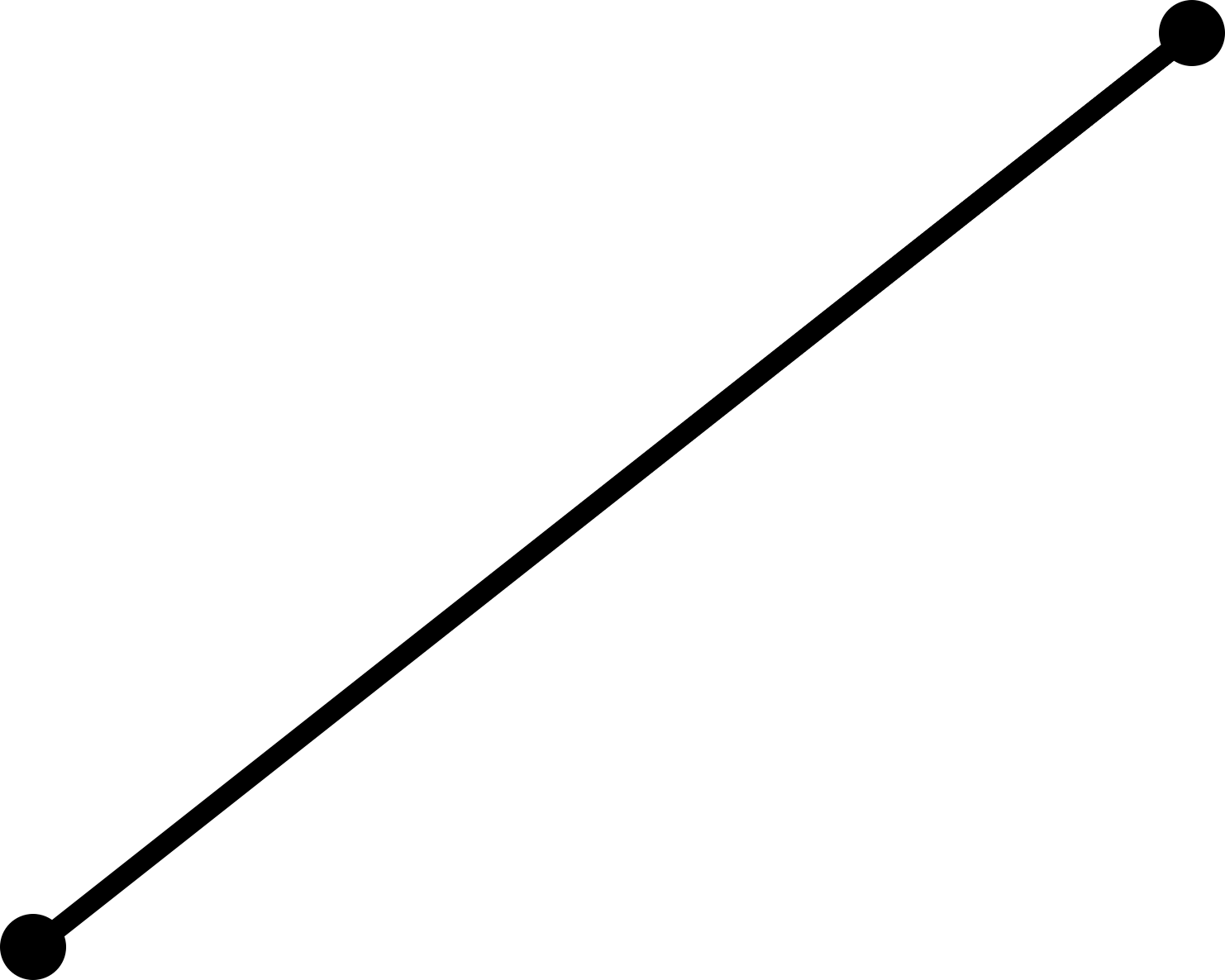}%
			\caption{}
		\end{subfigure}
		\begin{subfigure}[t]{0.24\linewidth}
			\centering
			\includegraphics[width=0.9\linewidth]{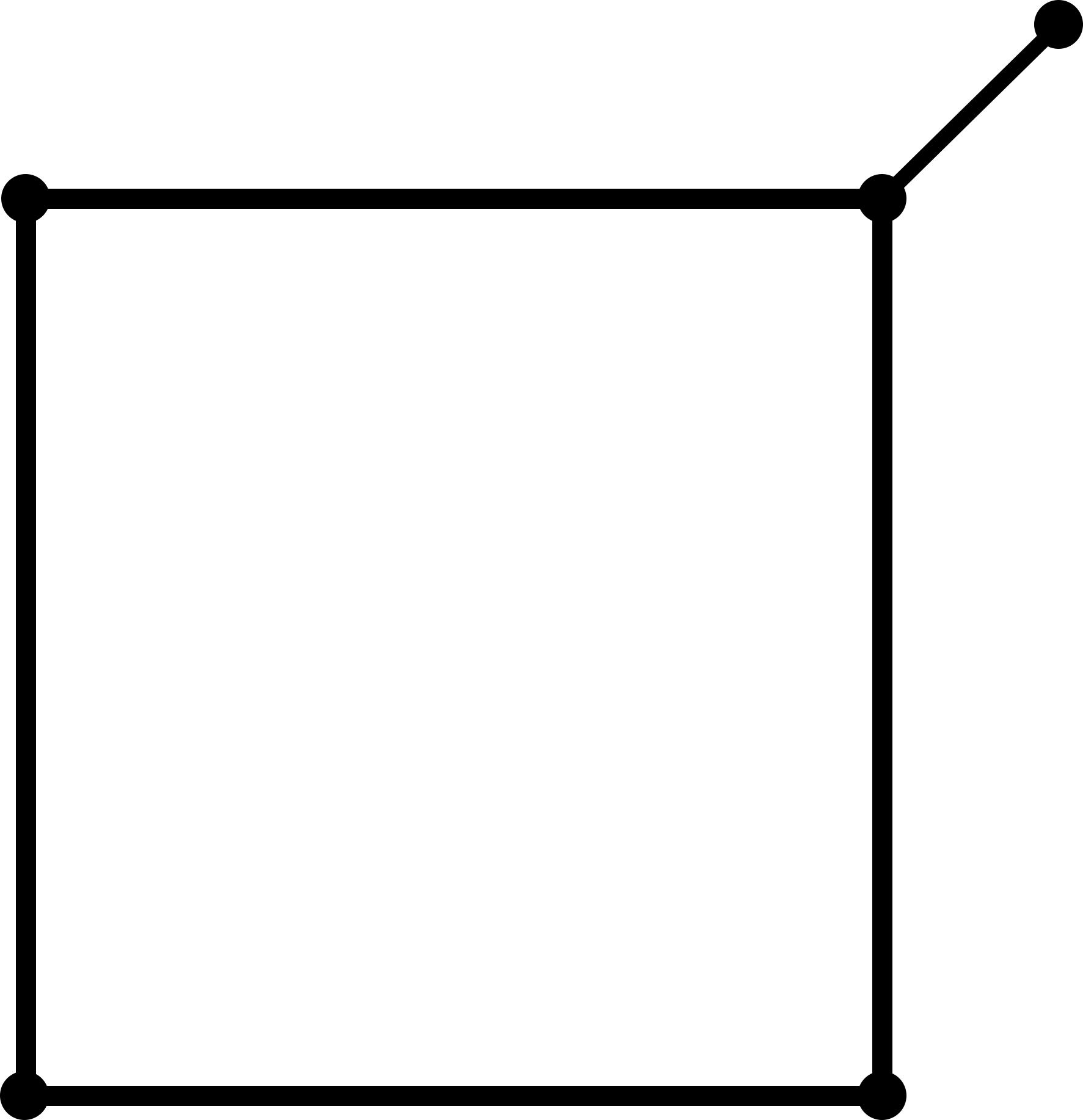}%
			\caption{}
		\end{subfigure}
	\begin{subfigure}[t]{0.24\linewidth}
	\centering
	\includegraphics[width=0.9\linewidth]{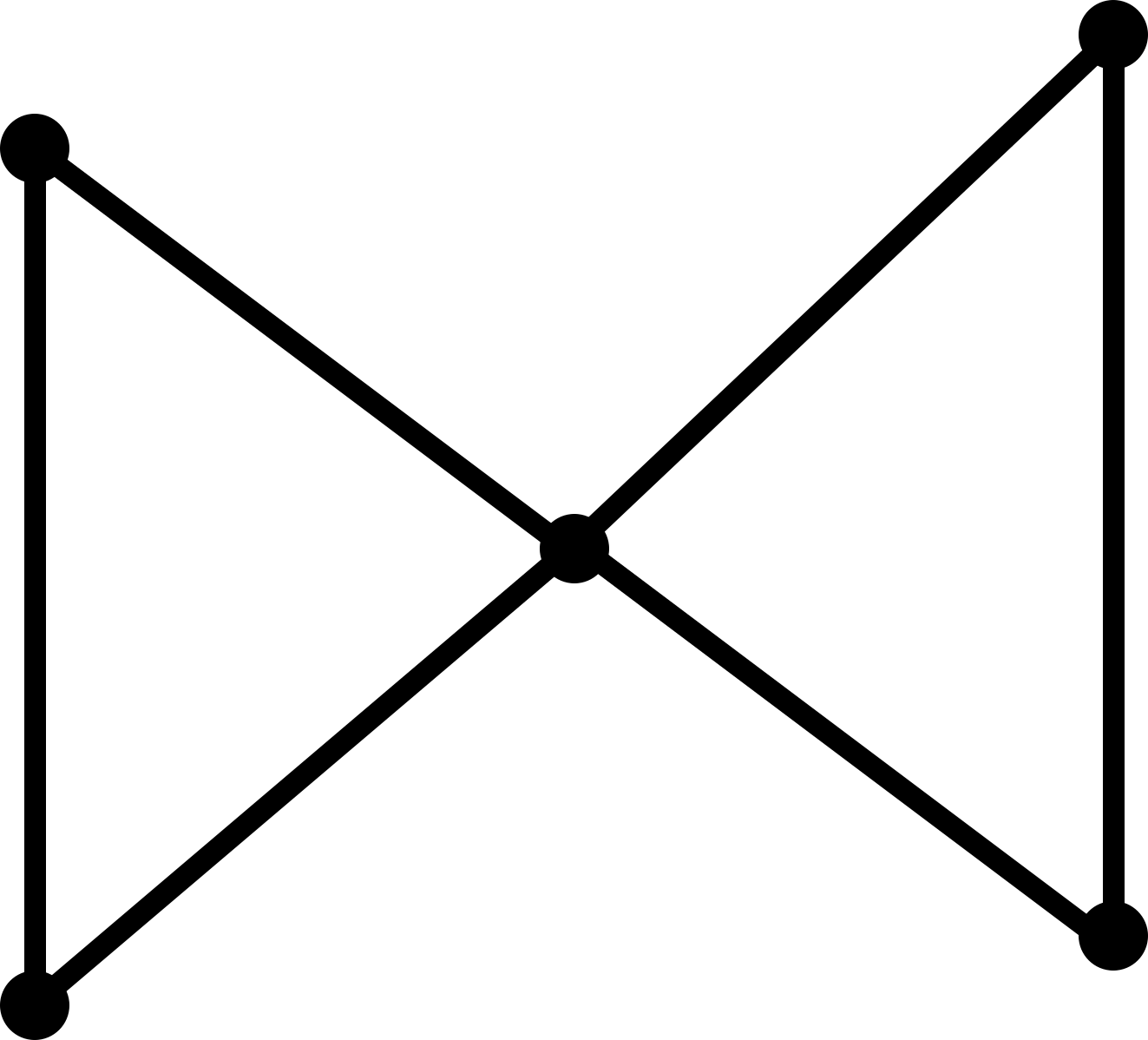}%
	\caption{}
\end{subfigure}
\begin{subfigure}[t]{0.24\linewidth}
\centering
\includegraphics[width=0.9\linewidth]{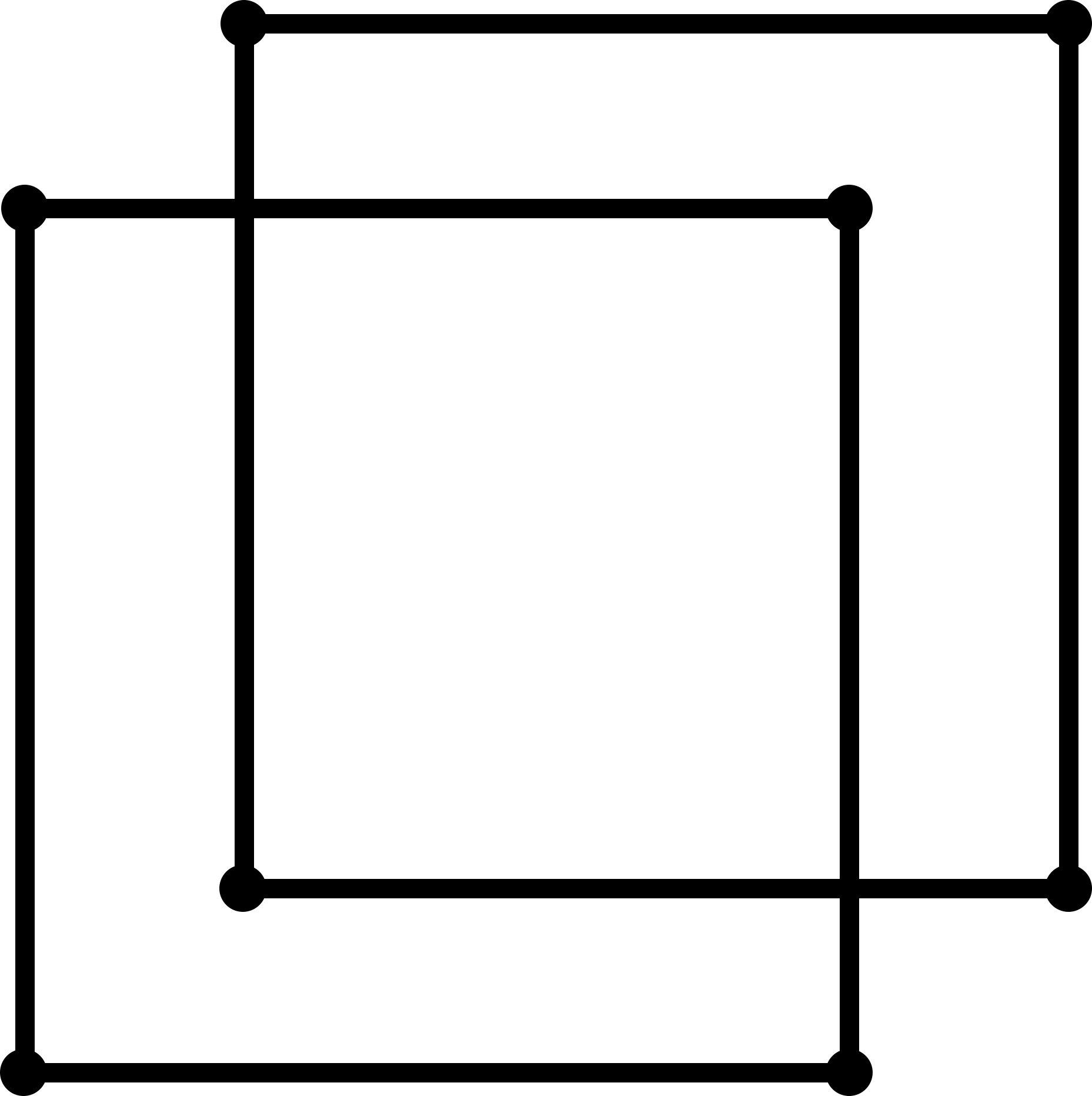}%
\caption{}
\end{subfigure}
		\caption{Examples of invalid B-Rep models in 2D. } \label{fig_invalidbrepfiles2d}
	\end{figure}
	
On the other hand, Fig. \ref{fig_validbrepfiles2d} illustrates some examples of valid B-Rep models:
	\begin{figure} [t]
		\centering
		\begin{subfigure}[t]{0.3\linewidth}
			\centering
			\includegraphics[width=0.8\linewidth]{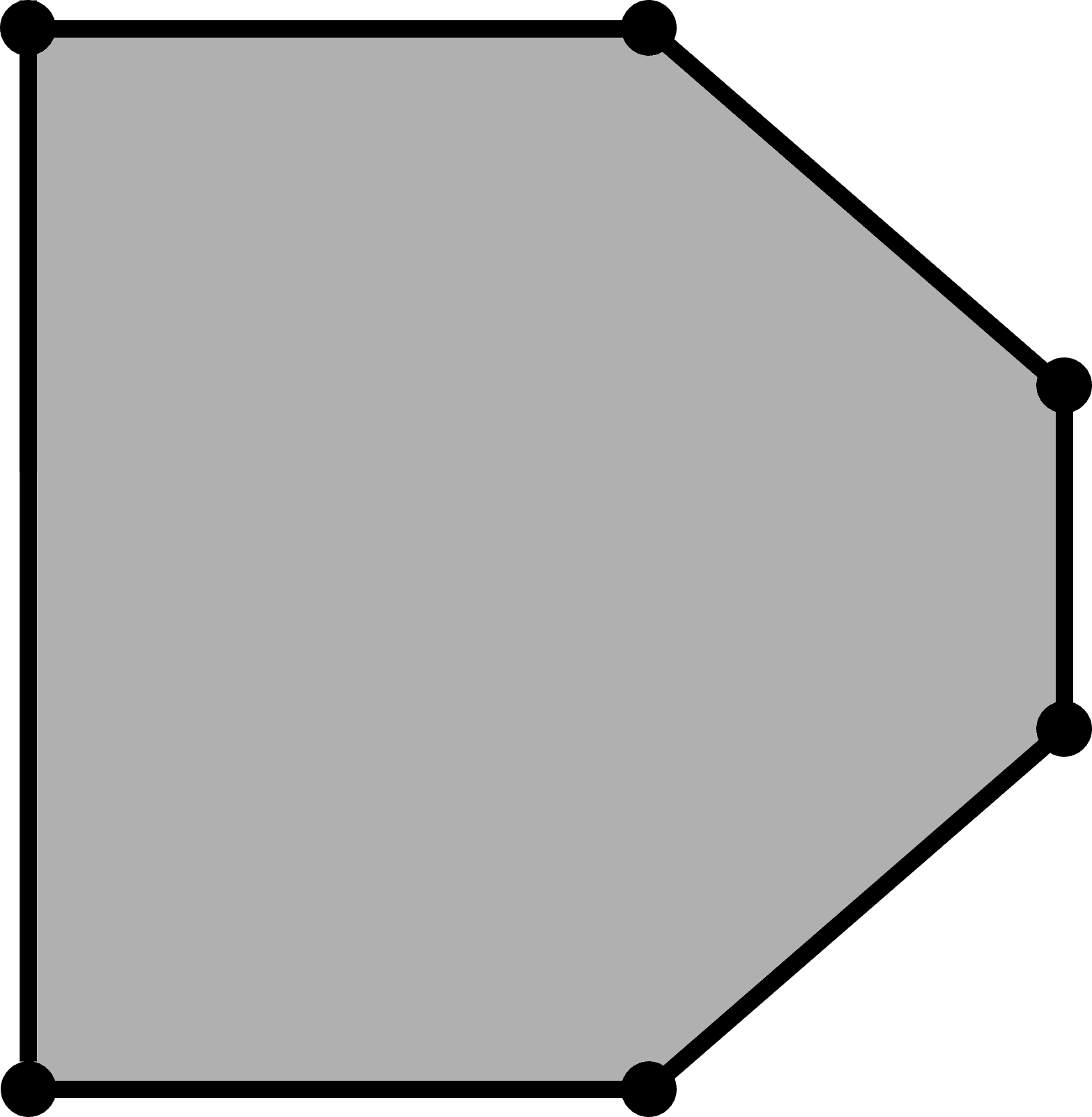}%
			\caption{}
		\end{subfigure}
		\begin{subfigure}[t]{0.3\linewidth}
			\centering
			\includegraphics[width=0.8\linewidth]{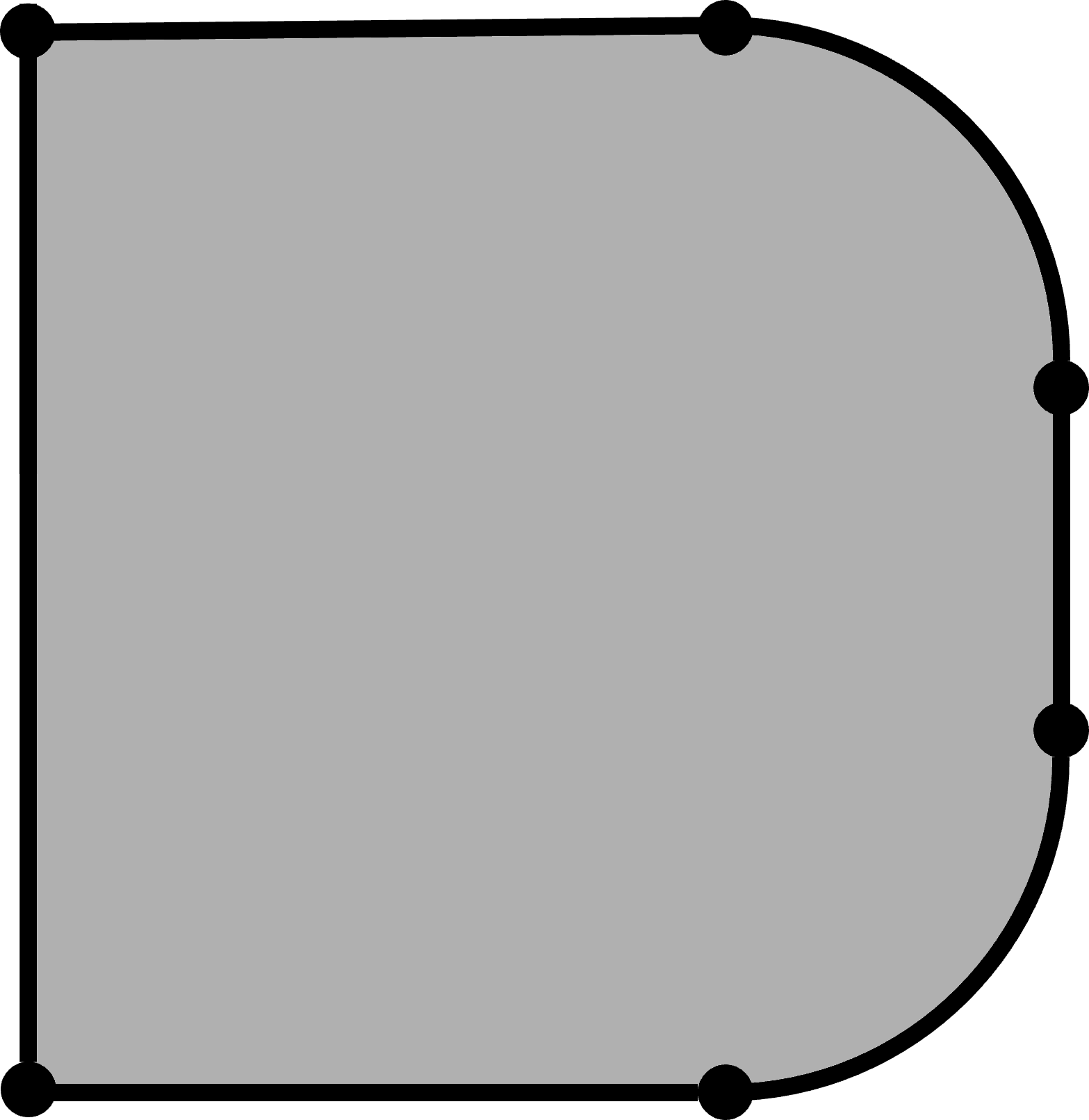}%
			\caption{}
		\end{subfigure}
		\begin{subfigure}[t]{0.3\linewidth}
			\centering
			\includegraphics[width=0.8\linewidth]{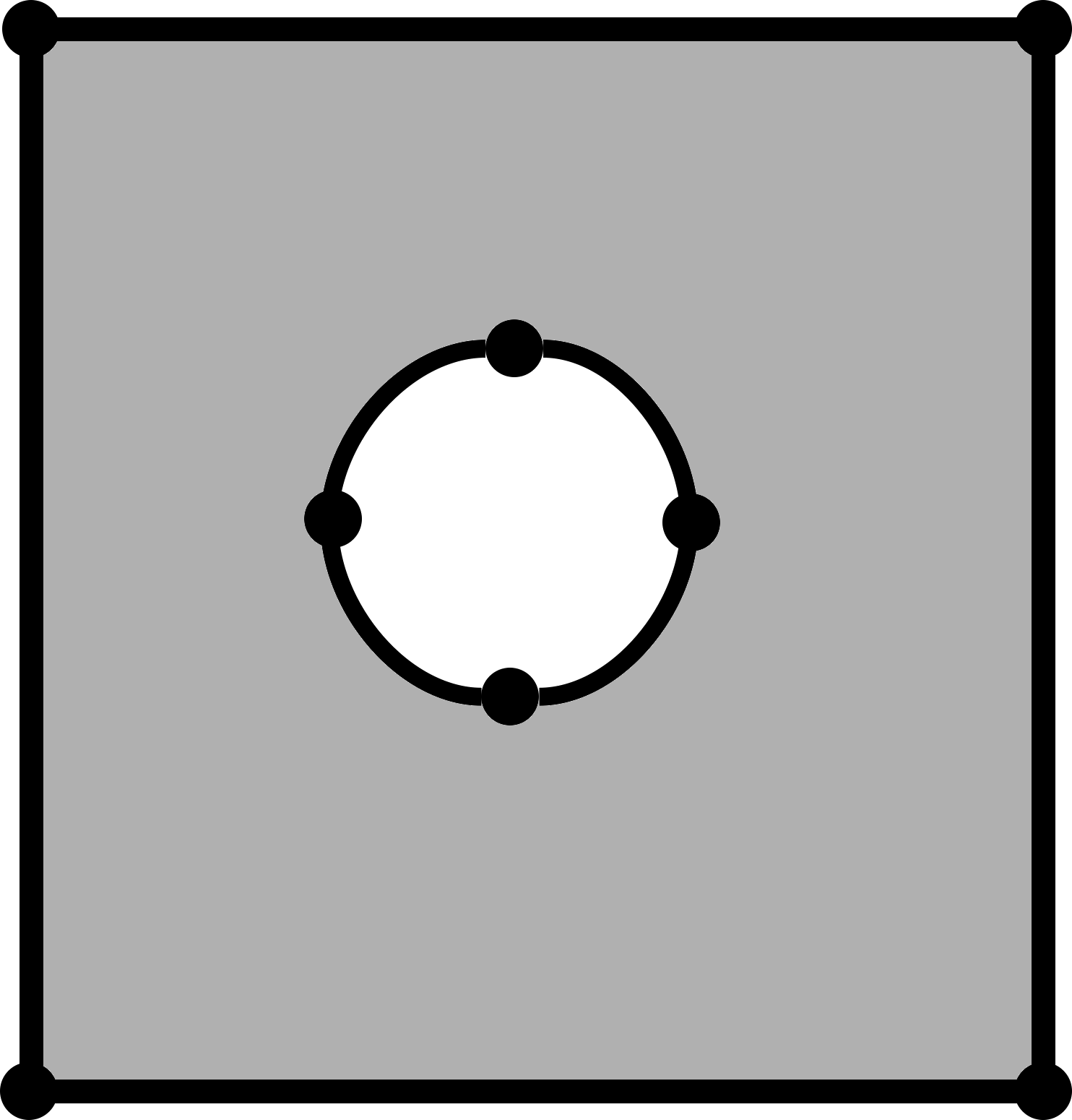}%
			\caption{}
		\end{subfigure}
		\caption{Examples of valid B-Rep models in 2D. } \label{fig_validbrepfiles2d}
	\end{figure}

\begin{figure}[t]
\centering\includegraphics[width=0.45\linewidth]{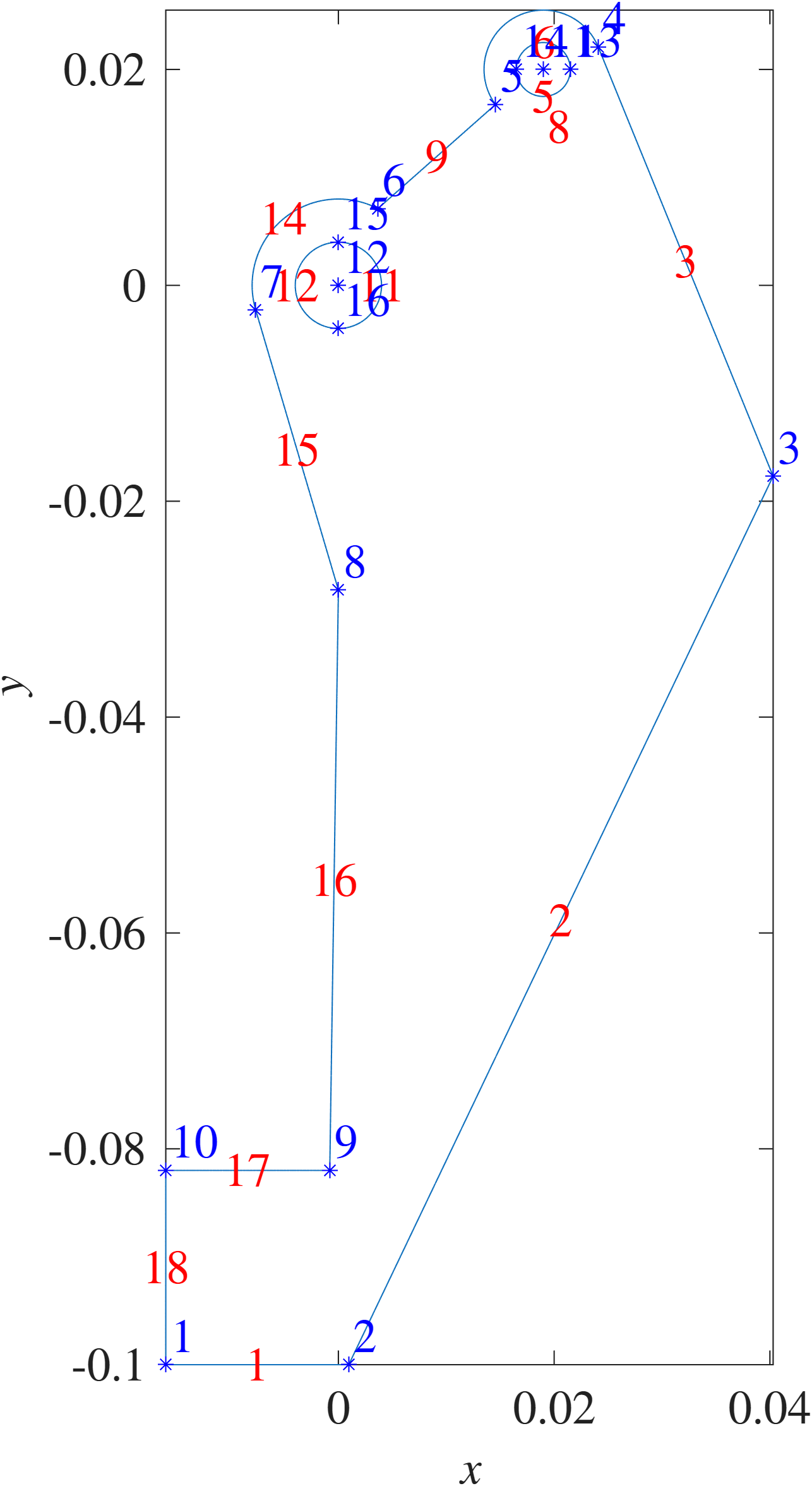}
	\caption{Gripper B-Rep with node and edge labels.}
	\label{fig_gripperBrep}
\end{figure}

In STORX, we represent 2D B-Rep models using two data structures, one for the position of vertices and another for the edges that capture how the vertices are connected to each other. 
To define vertex data, we first declare the number of vertices, followed by vertex coordinates. To define edge data, we declare the number of edges, followed by four properties of each segment,  namely, 1) type, 2) starting vertex, 3) end vertex, and 4) in case of arc segments arc center vertex and direction of the arc.
For instance, as a non-trivial B-Rep example, consider the gripper geometry shown in Fig. \ref{fig_gripperBrep} that is created using the following code snippet: 

\begin{lstlisting}
    brep = 'GripperComplex.brep'; % geometry
    geom = brep2d(brep); % class instantiation
    geom.plotGeometryWithLabels(); % plot 
\end{lstlisting}

Further explanation and an examples are provided in \ref{app:brep}.

\subsection{Finite Element Analysis}\label{sec:fea}
\begin{figure*}[t]
\centering\includegraphics[width=0.9\linewidth]{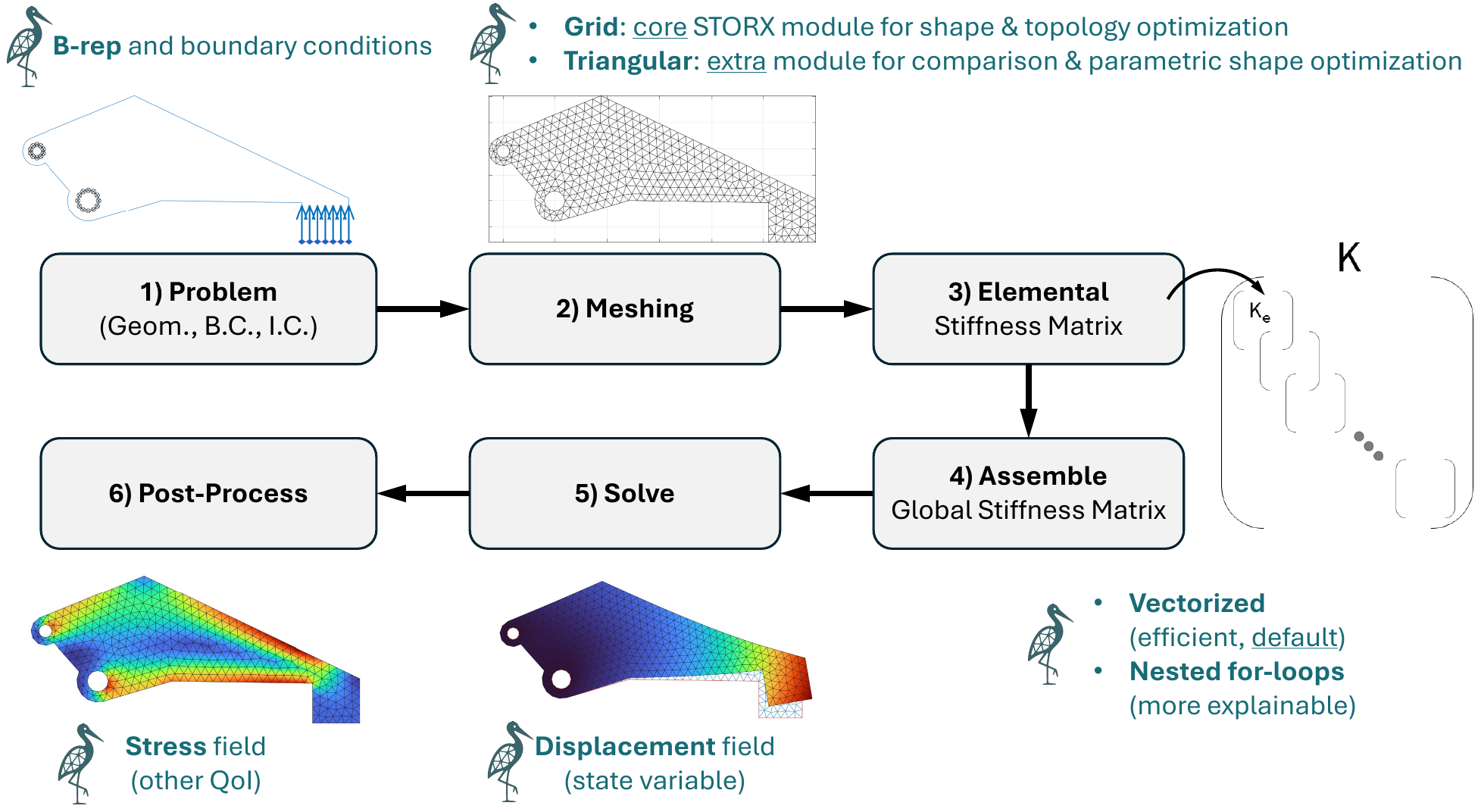}
	\caption{Finite element analysis workflow in STORX.}
	\label{fig_FEAWorkFlow}
\end{figure*}

\begin{figure}[t]
\centering\includegraphics[width=0.5\linewidth]{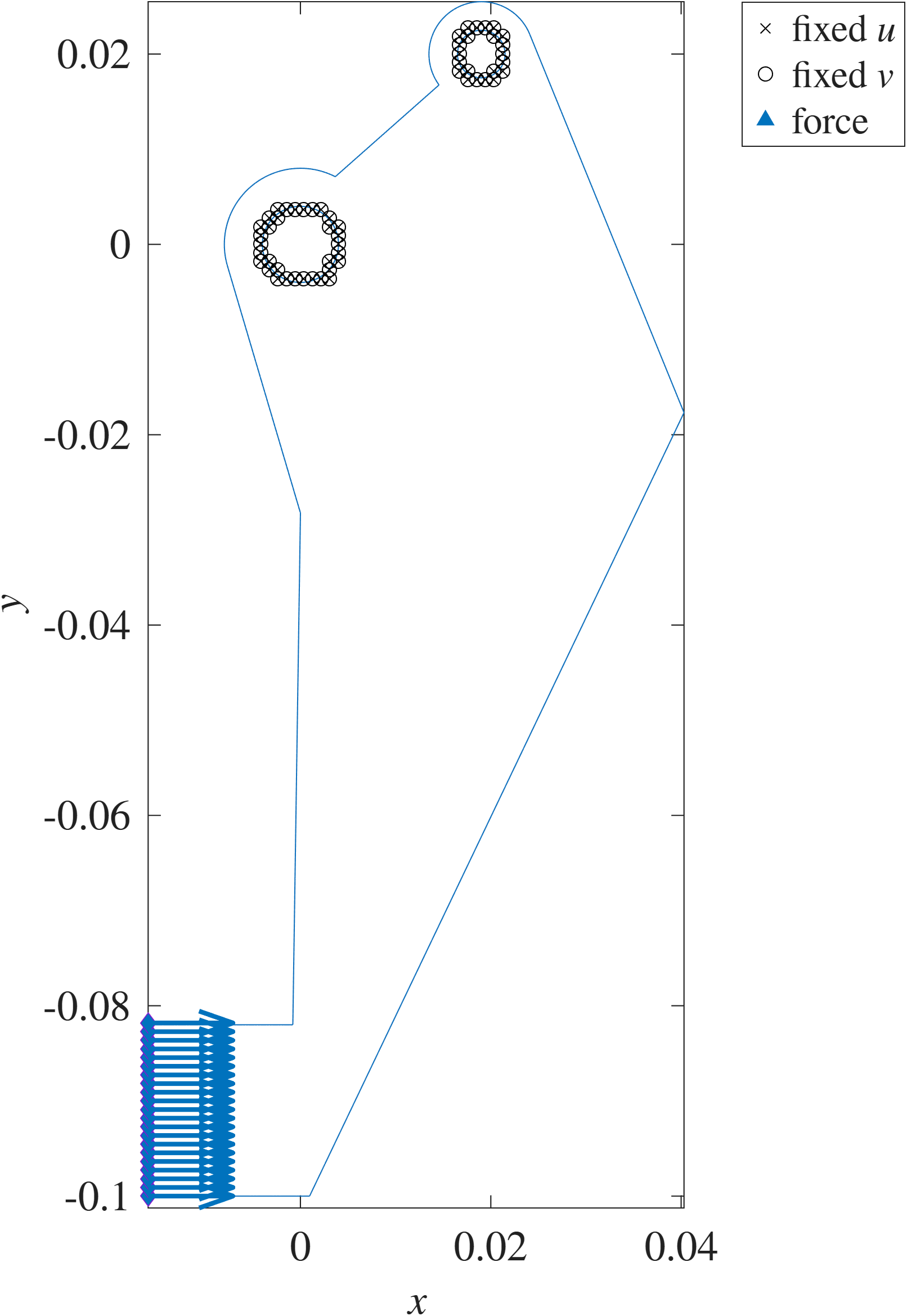}
	\caption{Gripper boundary condition (dimensions in m).}
	\label{fig_gripperBC}
\end{figure}

Finite element analysis is a powerful numerical method for solving partial differential equations (PDEs) that govern physical phenomena, and it plays a central role in SO/TO. The FEA workflow, as illustrated in Fig. \ref{fig_FEAWorkFlow} consists of the following steps:

\begin{enumerate}
	\item One starts with a physical problem (i.e., PDE) defined over a geometry; here a structural problem is illustrated.
\item The geometry is discretized into finite elements through a meshing procedure. The core STORX framework uses structured grid meshing for SO/TO. In addition, a \mcode{triMesher} utility and an extra module under \mcode{extra/triFEA/} are provided. These can be used for comparison purposes (e.g., for a more accurate stress computation) and for parametric shape optimization, which will be discussed in Section~\ref{sec:paramSO}.

\item A piecewise, typically polynomial, approximation is then assumed for the field variables (e.g., displacement fields in structural problems) over each element in the mesh.

\item Using this assumed field representation, an element stiffness matrix is constructed for each element. These element matrices are subsequently assembled into the global stiffness matrix. In STORX, this step can be performed in a vectorized manner, which is computationally efficient for structured grids, or via a non-vectorized implementation using nested loops, which is more explicit and pedagogically transparent.

\item The resulting system of algebraic equations is solved to obtain the unknown approximate field variables.

\item Finally, post-processing is performed to compute derived quantities of interest, such as stresses, strains, compliance or other response measures.
\end{enumerate}

Figure~\ref{fig_gripperBC} illustrates the boundary conditions for the gripper example. The structure is clamped from the circular cutout edges, while a horizontal load of $10~\mathrm{N}$ is applied on the bottom edge. The material is modeled as linear elastic with Young’s modulus $E = 2~\mathrm{GPa}$, Poisson’s ratio $\nu = 0.35$. The domain is discretized using approximately $4000$ elements. 

The problem can be defined as follows, 

\begin{lstlisting}
%% Problem Definition
brep = 'GripperComplex.brep'; % geometry
numElements = 4000; % mesh
material.E = 2e9; material.nu = 0.35; % material
force = 10; % N
\end{lstlisting}

Let us compare the results obtained for the grid-based FEA used throughout STORX, against a  triangular finite element discretization. 
The triangular FEA predicts a maximum displacement of $1.00\times10^{-6}\; \text{m}$ and a maximum von Mises stress of $6.87\; \text{MPa}$, whereas the grid-based FEA yields a maximum displacement 
of $1.05\times10^{-6}\; \text{m}$ and a maximum von Mises stress of $6.31\; \text{MPa}$. The close agreement between the two solutions indicates that the grid-based formulation provides sufficient accuracy for stiffness and stress prediction in this example, with discrepancies on the order of a few percent.

\begin{figure} [t]
		\centering
		\begin{subfigure}[t]{0.4\linewidth}
			\centering
			\includegraphics[width=\linewidth]{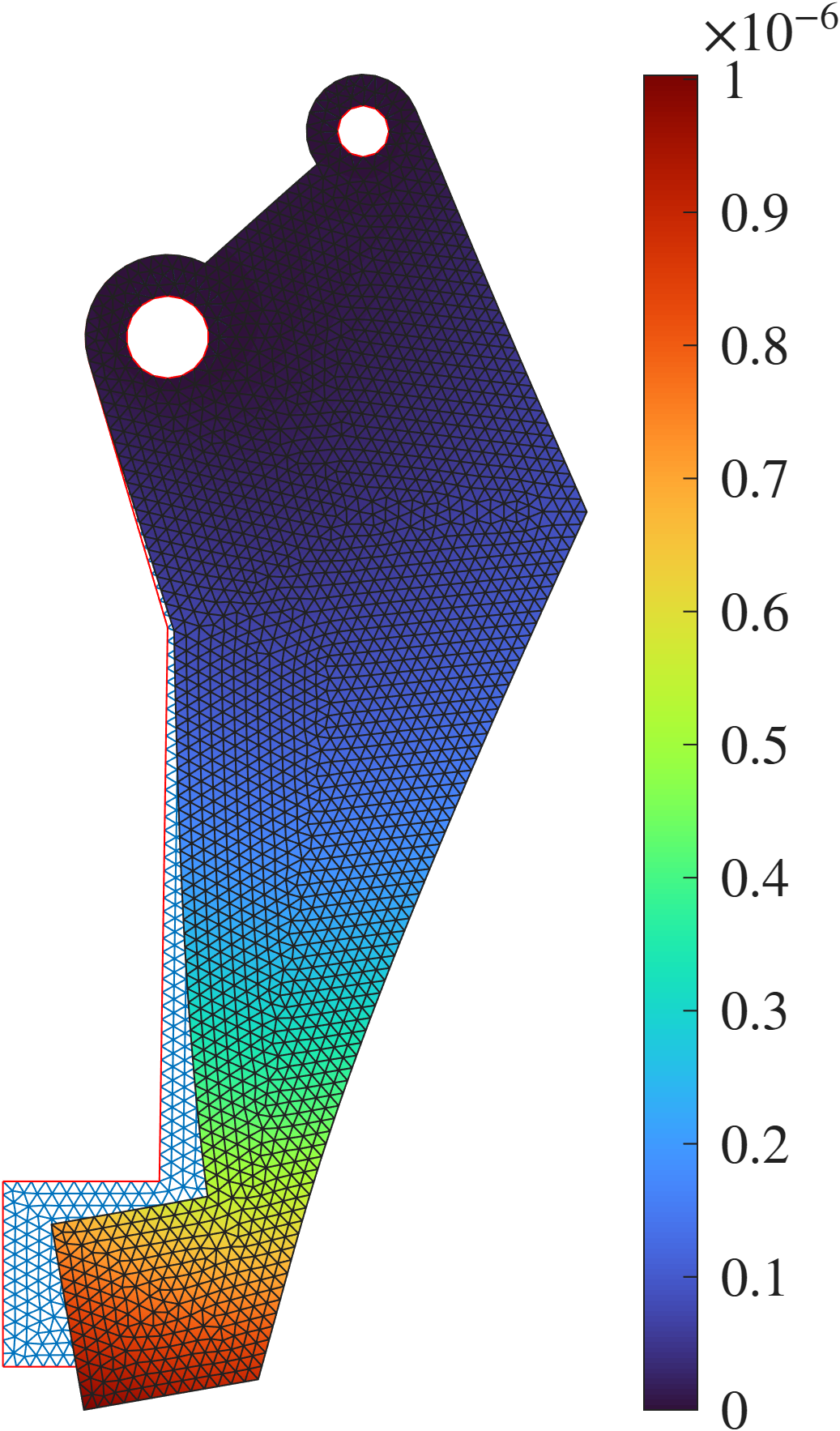}%
			\caption{}
		\end{subfigure}
		\begin{subfigure}[t]{0.4\linewidth}
			\centering
			\includegraphics[width=\linewidth]{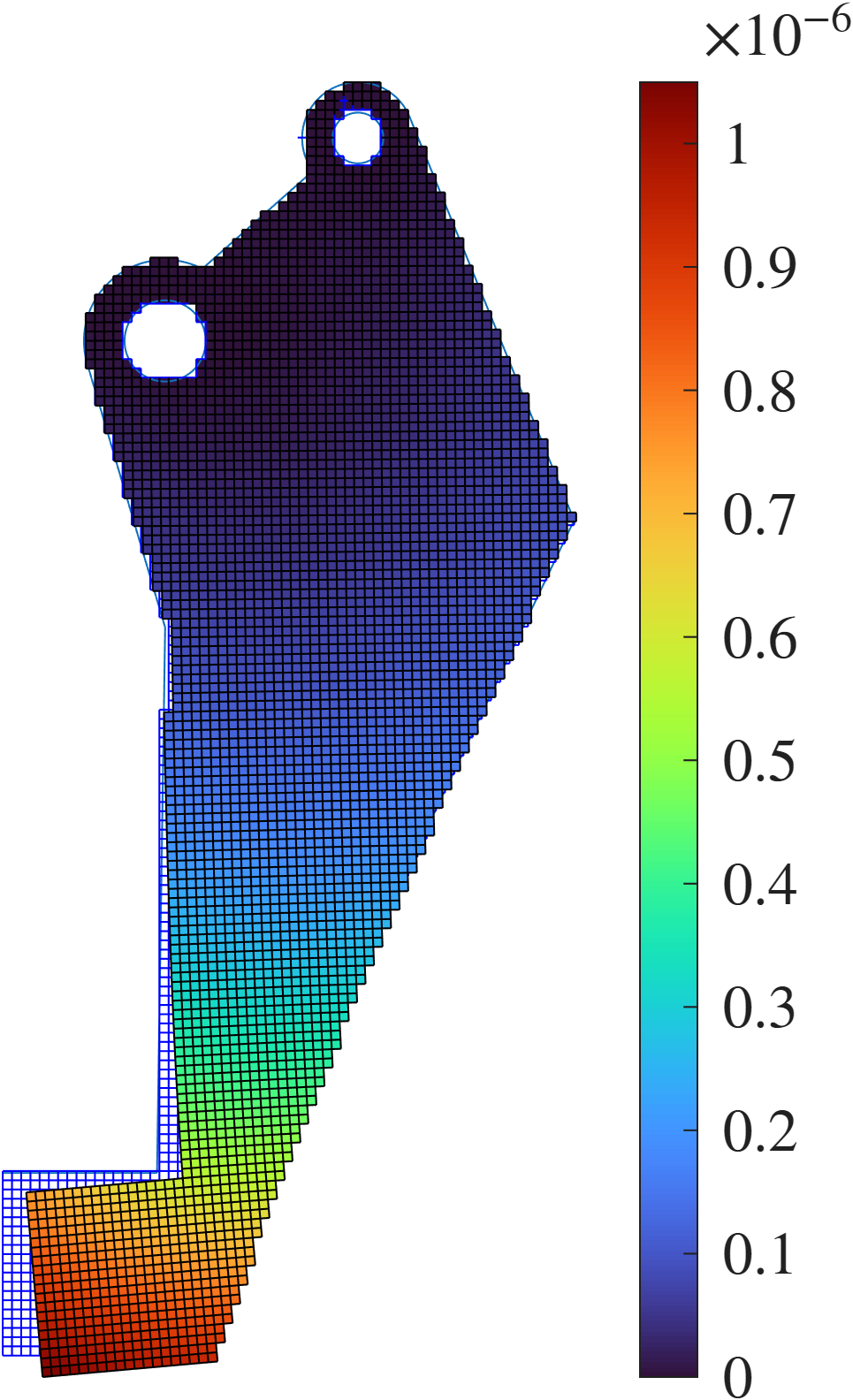}%
			\caption{}
		\end{subfigure}
		\caption{Gripper deformation: triangular versus grid meshes.} \label{fig_triGridFEA}
	\end{figure}

Figure~\ref{fig_triGridFEA} compares the deformation fields obtained from the triangular and grid-based discretizations. Both approaches capture the same global deformation mode and load transfer path, with only minor local differences near geometric features and boundaries. 

The \mcode{fea2d_elasticity} can be instantiated and the boundary conditions can be applied as: 
\begin{lstlisting}
%% Construct FEA Solver
solver = fea2d_elasticity(brep,numElements,material,vectorize); % call superclass
solver = solver.fixEdge([5,6,11,12]);
solver = solver.applyXForceOnEdge(18,force);
\end{lstlisting}
Subsequently, we perform pre-processing (assembly of the stiffness matrix), solve, and post-processing as follows: 
\begin{lstlisting}
%% Assemble, Solve, and Post-process
solver = solver.preProcess(); % assmble K and f
solver = solver.solve(); % d = K\f
solver = solver.postProcess(); % def., stress
%% Plot Results
solver.plotDeformation(); 
solver.plotVonMisesStress();
\end{lstlisting}

Unless otherwise stated, linear elasticity is assumed to be the governing model. Linear elasticity describes the equilibrium of solids under the assumptions of small strains and small displacements $\mathbf{d}$ computed by solving the following algebraic equilibrium system:

\begin{equation}
\label{eq:discrete_residual_form_elasticity}
{R}_{el}(\mathbf{d}) \coloneqq \mathbf{K}_{el}\mathbf{d}-\mathbf{f}_{el}=\mathbf{0},
\end{equation}

where $\mathbf{K}_{el}$ and $\mathbf{f}_{el}$ are the elasticity stiffness matrix and external force vectors, respectively.
Equation \eqref{eq:discrete_residual_form_elasticity} is the state equation and is imposed as an equality constraint in SO/TO.
\section{Parametric Shape Optimization}\label{sec:paramSO}

In parametric shape optimization, the domain geometry is controlled by a low-dimensional
vector of design parameters $\mathbf{p} \in \mathbb{R}^n$ (e.g., hole radius, chamfer
depth, notch position), and the optimizer searches for values that minimize a structural
objective subject to geometric constraints.  The framework supports both triangular
(\mcode{triFEA2d\_elasticity}) and structured-grid (\mcode{fea2d\_elasticity}) meshes;
triangular meshes are strongly recommended because they conform to arbitrary boundaries
and enable the semi-analytic gradient described in Section~\ref{sec:paramSO:FD}.  All
examples in this chapter use triangular meshes.

\subsection{Common setup}
Every optimizer in the framework is constructed from two user-supplied function handles:
\mcode{brepHandle}, which maps a parameter vector to a B-Rep struct, and
\mcode{solverHandle}, which builds and returns a pre-configured FEA object from that
B-Rep.  Design parameters are packaged in a struct with fields \mcode{.value} (initial
values), \mcode{.lb}, and \mcode{.ub}.  The objective (here \mcode{'compliance'})
and a geometric constraint (area or perimeter, equality or inequality) are specified at
construction.  After optimization, the initial and final solver objects are available
through \mcode{m\_solverInitial} and \mcode{m\_solverFinal} for postprocessing.

 \begin{figure}[t]
	\centering
		\includegraphics[width=\linewidth]{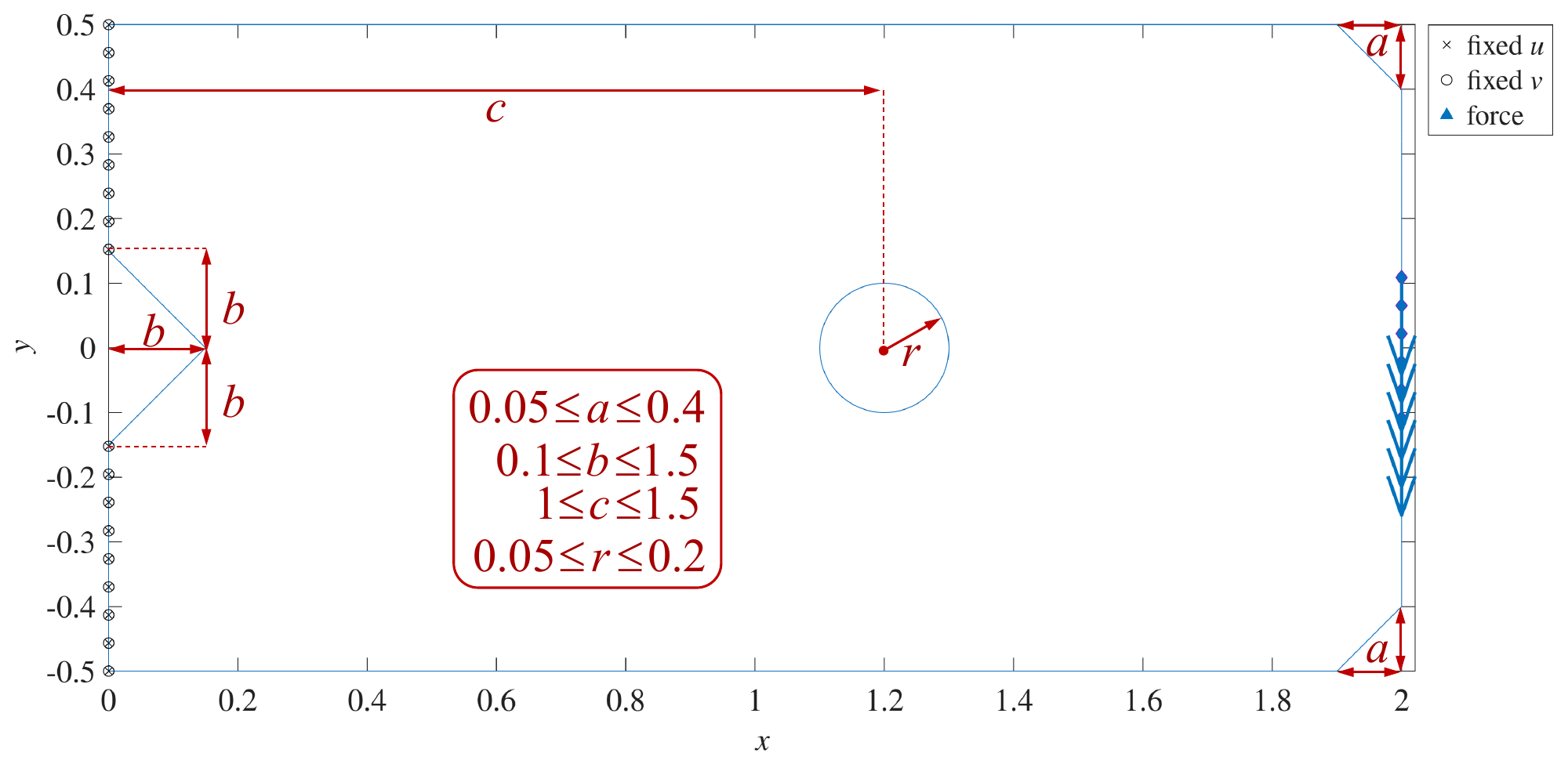}
	\caption{Parametric shape optimization of cantilever beam with circular hole.}
	\label{fig_paramSObenchmarkExample}
\end{figure}

Consider the SO problem of Fig.  \ref{fig_paramSObenchmarkExample} with four shape parameters $\bp = [a,b,c,r]^\top$. We can formally pose the compliance minimization SO problem as: Starting from any design with initial shape parameters $\bp_0 = [a_0,b_0,c_0,r_0]^\top$, find the shape parameters such that the compliance $C$ is minimized while area remains less than an upper-bound $A_{\text{max}}$: 

\begin{subequations} \label{SO_minCompliance}
	\begin{align}
		\minimize\limits_{\bp = [a,b,c,r]^\top}  \quad & C (\bd;\bp) \label{seq_SO_obj}\\  
		 \textrm{s.t.} \quad &\mathbf{K}_{el} (\bp)\mathbf{d}-\mathbf{f}_{el}=\mathbf{0} \label{seq_SO_fea}\\
         &  A - A_{\text{max}}  \le 0\label{seq_SO_g}\\	
		&  a_{min} \le a \le a_{max}\\
		&	b_{min} \le b \le b_{max}\\
		&	c_{min} \le c \le c_{max}\\
         &   r_{min} \le r \le r_{max}
	\end{align}
\end{subequations}

The code snippet below shows the complete setup for a cantilever-with-hole example
parameterized by four shape variables: corner chamfer~$a$, edge notch~$b$, horizontal
position~$c$, and hole radius~$r$.

\begin{lstlisting}
%% Design parameters
params0.value = [0.2  0.15  1.2  0.1];  % [a, b, c, r]
params0.lb    = [0.05 0.05  1.0  0.05];
params0.ub    = [0.4  0.4   1.5  0.2 ];

objective        = 'compliance';
constraints.area = 1.8;          % area <= 1.8
constraints.type = 'ineq';

brepHandle   = @createGeom;     % params -> B-Rep
solverHandle = @createProblem;  % B-Rep   -> FEA object

function fem = createProblem(brep)
    material.E = 100e9; material.nu = 0.3; material.rho = 1;
    fem = triFEA2d_elasticity(brep, 1000, material);
    fem = fem.fixEdge([2, 15]);
    fem = fem.applyYForceOnEdge(11, -1e5);
    fem = fem.preProcess();
end

function geom = createGeom(params)
    a = params(1); b = params(2); c = params(3); r = params(4);
    geom.vertices = [b 0; 0 -b; 0 -H/2; c -H/2; c -r; c r; ...
                     c 0; L-a -H/2; L -H/2+a; L -h/2; ...
                     L h/2; L H/2-a; L-a H/2; 0 H/2; 0 b]';
    geom.segments = [1 1 2 0; 1 2 3 0; ... 
                     2 5 6 7; 2 6 5 7; ...]'; % line and arc descriptors
end
\end{lstlisting}

All four subclasses expose the same postprocessing interface once \mcode{optimize()} returns:
\begin{lstlisting}[language=Matlab]
parOpt.m_solverInitial.plotGeometry(1, 0, 'Initial Geometry');
parOpt.m_solverFinal.plotDeformation();
parOpt.m_solverFinal.plotVonMisesStress();
\end{lstlisting}

\subsection{Finite Difference Method}\label{sec:paramSO:FD}

\mcode{parameterOpt2d\_FD} drives \mcode{fmincon} for gradient-based local search.
Internally, design parameters are normalized as $\mathbf{x} = \mathbf{p}/\mathbf{p}_0$
so that the initial iterate is $\mathbf{x}_0 = \mathbf{1}$ regardless of physical units,
which improves the conditioning seen by the optimizer.  The mesh type determines which
gradient strategy is used.

\subsubsection{Direct finite difference}
For structured-grid solvers, \mcode{fmincon} approximates gradients by perturbing each
normalized parameter by \mcode{finiteDifferenceStepSize}.  Each gradient evaluation
therefore costs $n$ additional FEA solves, so the total count per major iteration is
$1 + n$.

\subsubsection{Semi-analytic method}
\begin{figure} [!h]
	\centering
	\begin{subfigure}[t]{0.75\linewidth}
		\centering
		\includegraphics[width=\linewidth]{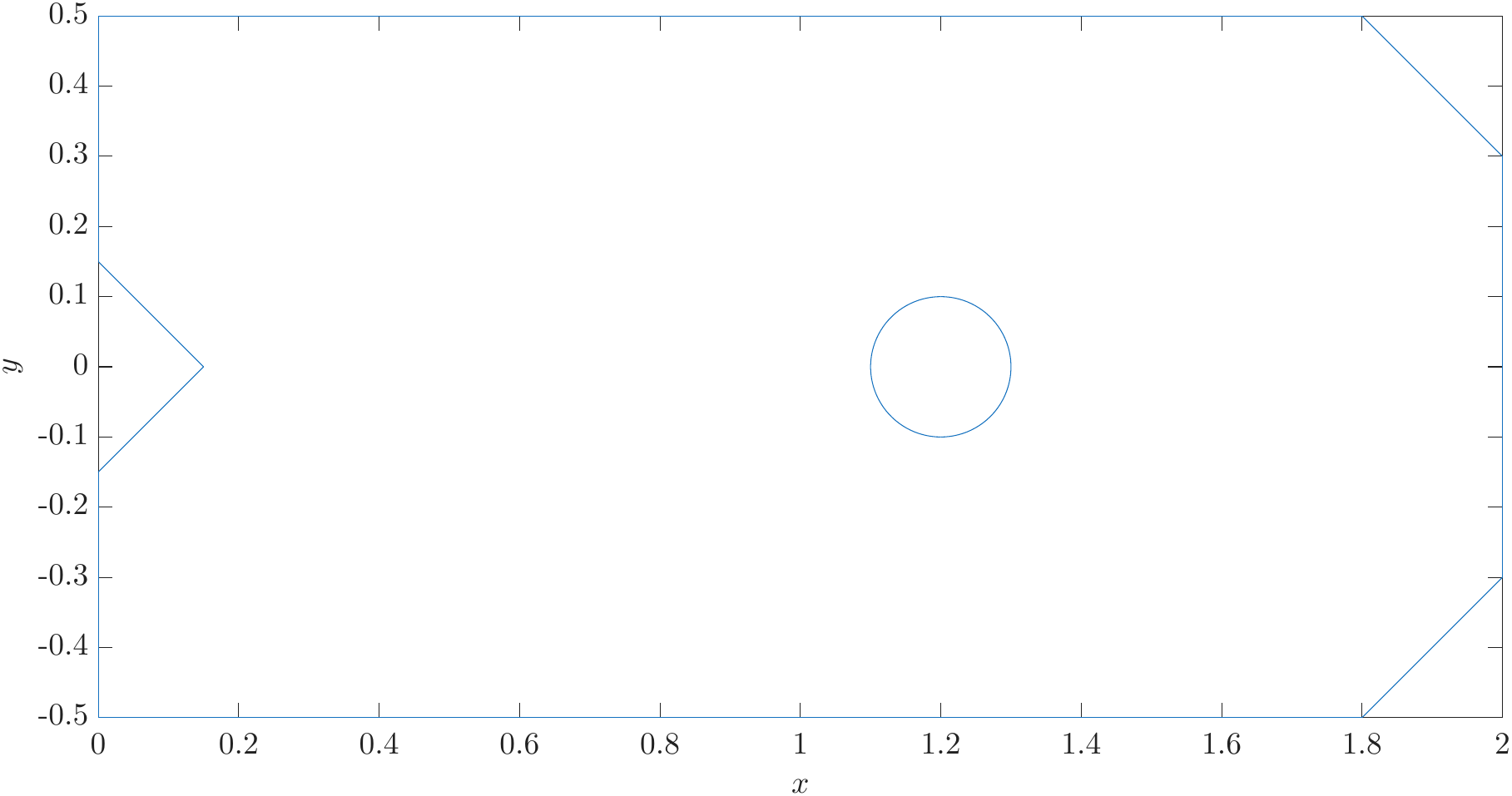}%
		\caption{Initial shape}
	\end{subfigure}
	\begin{subfigure}[t]{0.75\linewidth}
		\centering
		\includegraphics[width=\linewidth]{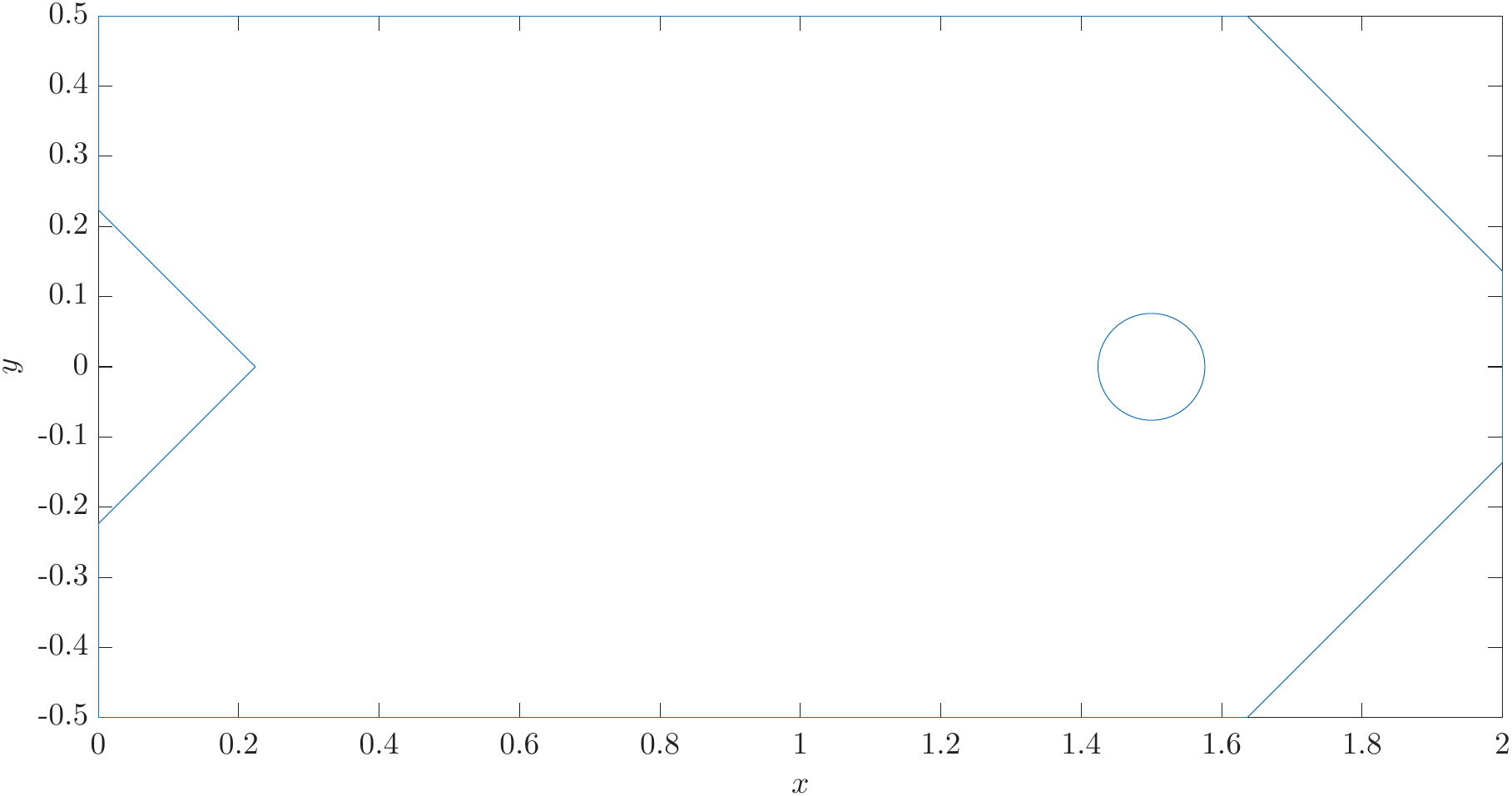}%
		\caption{Optimized shape}
	\end{subfigure}

	\begin{subfigure}[t]{0.78\linewidth}
		\centering
		\includegraphics[width=\linewidth]{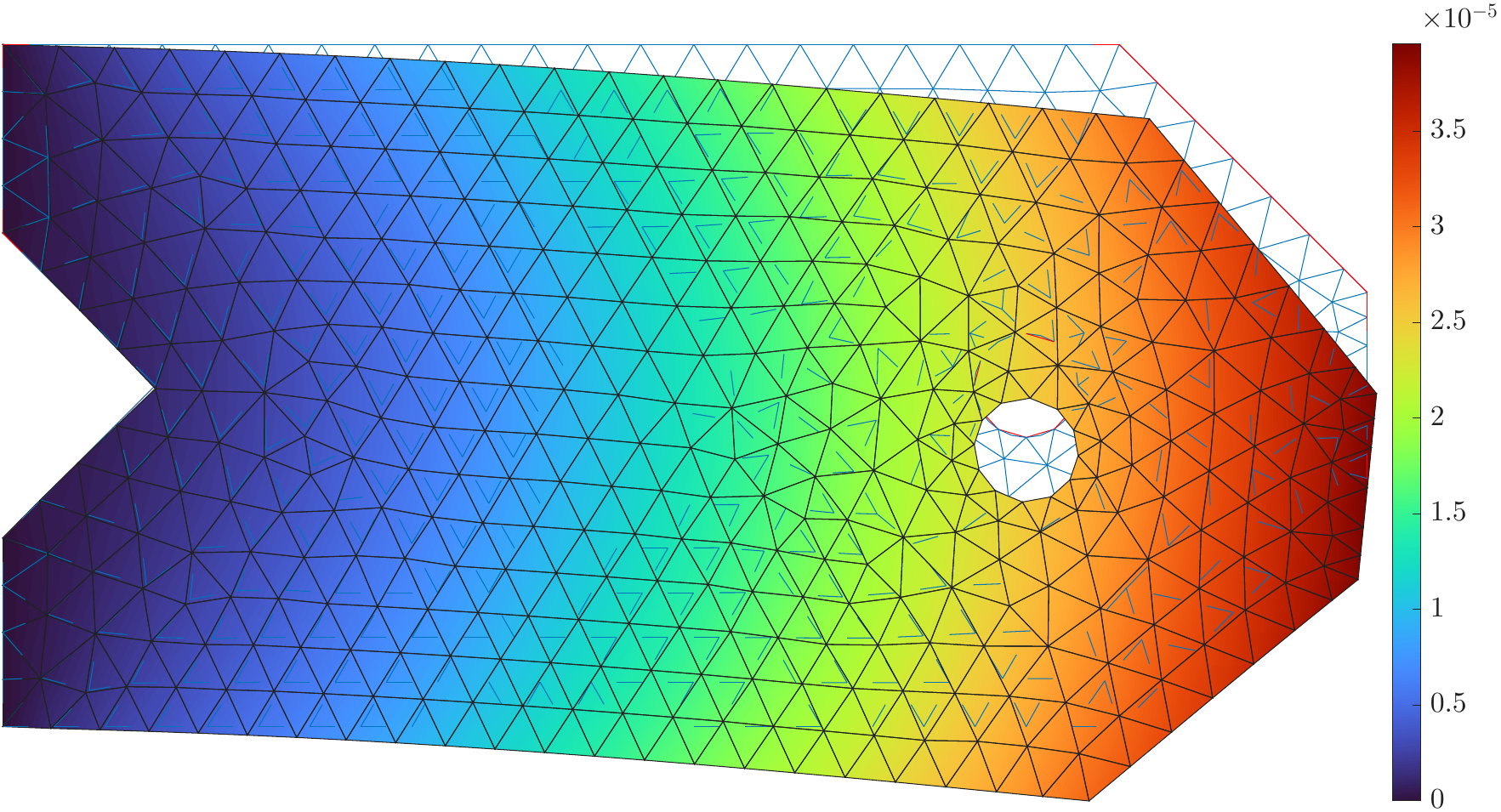}%
		\caption{Final deformation}
	\end{subfigure}
    \begin{subfigure}[t]{0.9\linewidth}
		\centering
		\includegraphics[width=\linewidth]{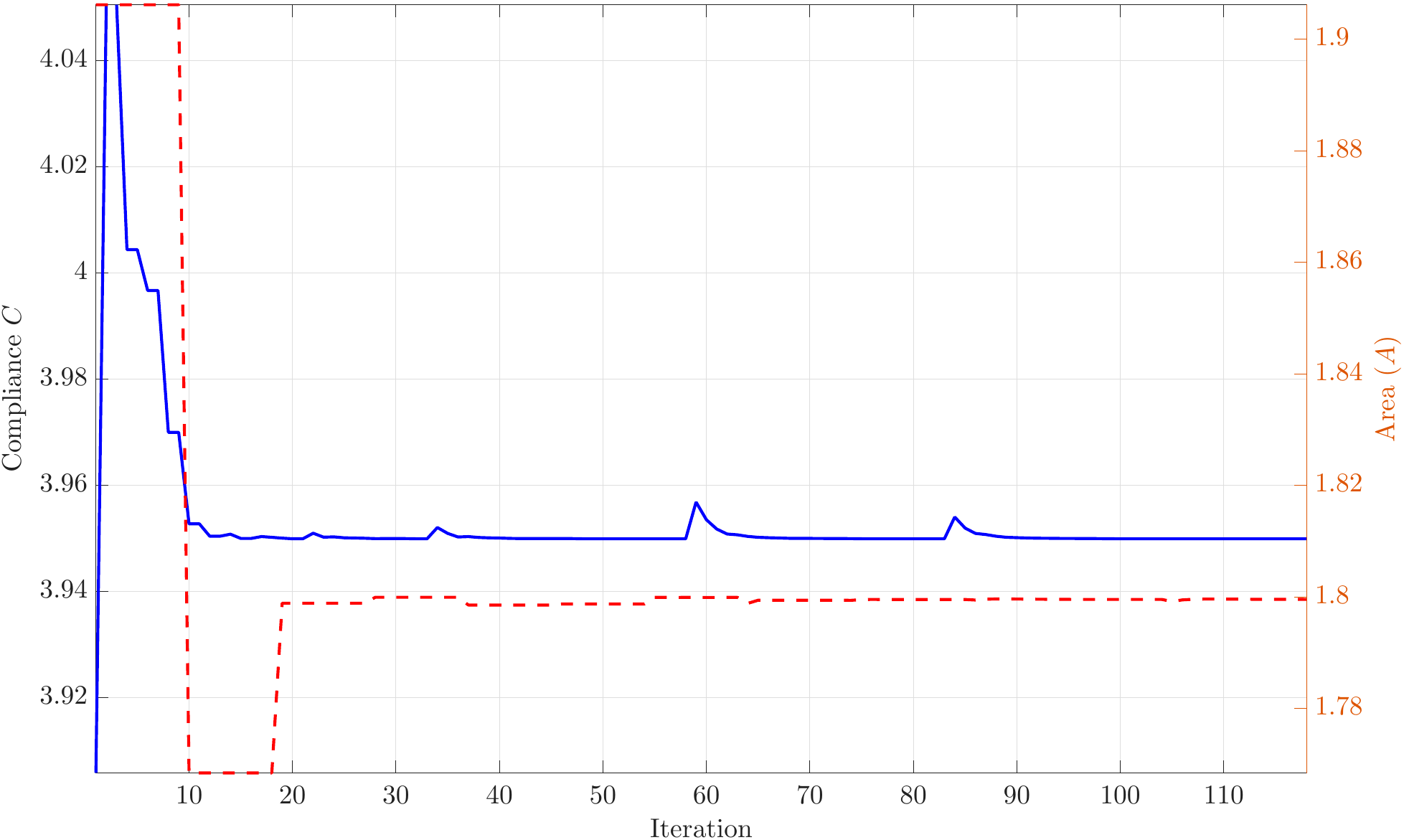}%
		\caption{Convergence}
	\end{subfigure}

	\caption{SO results using semi-analytic FD method with triangular mesh.} \label{fig_SO_FD}
\end{figure}

Triangular-mesh solvers activate a semi-analytic path in which the gradient is computed
analytically using finite-difference stiffness perturbations, bypassing full re-analysis
for each parameter. At major iteration \(k\), after the baseline solve
\[
  \mathbf{K}^{(k)} \mathbf{u}^{(k)} = \mathbf{f},
\]
each perturbed B-Rep morphs the existing mesh via \mcode{snapNodesToBRep} by projecting
nodes onto the perturbed boundary without re-meshing, and only the perturbed stiffness
matrix \(\mathbf{K}^{(k)}_{i,\mathrm{pert}}\) is reassembled. The compliance sensitivity
with respect to \(p_i\) is then obtained from the adjoint-like relation
\begin{align}
  C'_i &= \mathbf{f}^{\!\top} \mathbf{u}'_i, \\
  \mathbf{K}^{(k)} \mathbf{u}'_i &= -\mathbf{K}'_i \mathbf{u}^{(k)}, \\
  \mathbf{K}'_i &= \frac{\mathbf{K}^{(k)}_{i,\mathrm{pert}} - \mathbf{K}^{(k)}}{\Delta p_i}.
\end{align}
The factorization of \(\mathbf{K}^{(k)}\) is reused across all \(n\) sensitivity solves,
so the dominant cost per major iteration is one full FEA solve plus \(n\) stiffness
assemblies, a substantial reduction relative to the direct finite-difference approach.
The gradient is supplied to \mcode{fmincon} via
\mcode{SpecifyObjectiveGradient = true}, so no internal finite-difference objective
perturbations are performed.


\begin{lstlisting}
tol      = 1e-6;
stepSize = 1e-6;
parOpt = parameterOpt2d_FD(brepHandle, solverHandle, params0, ...
    objective, constraints, tol, stepSize, false);
parOpt = parOpt.optimize();
\end{lstlisting}

Starting from the initial parameter vector $\bp_0=(0.20,0.15,1.20,0.10)$, shown in Fig. \ref{fig_SO_FD}a, the finite-difference-based optimization converged after 118 iterations and 1070 FEA runs to $\bp_\text{FD}=(0.363,0.224,1.500,0.076)$, shown in Fig. \ref{fig_SO_FD}b. The optimized design reduced the area from $A_0=1.906~\mathrm{m}^2$ to $A_\text{FD}=1.7995~\mathrm{m}^2$, while the compliance increased slightly from $C_0=3.91~\mathrm{N.m}$ to $C_\text{FD}=3.95~\mathrm{N.m}$. The deformation field of the optimized design is shown in Fig. \ref{fig_SO_FD}c, with a maximum deflection of approximately $0.04~\mathrm{mm}$. The convergence histories in Fig. \ref{fig_SO_FD}d show an initial rapid reduction in area, driven by the infeasibility of the starting design with respect to the area constraint. This early shape change leads to a temporary increase in compliance, after which the optimizer gradually recovers stiffness while maintaining feasibility until termination.

\subsection{Global Search}\label{sec:paramSO:GS}
 \begin{figure}[t]
	\centering
		\includegraphics[width=\linewidth]{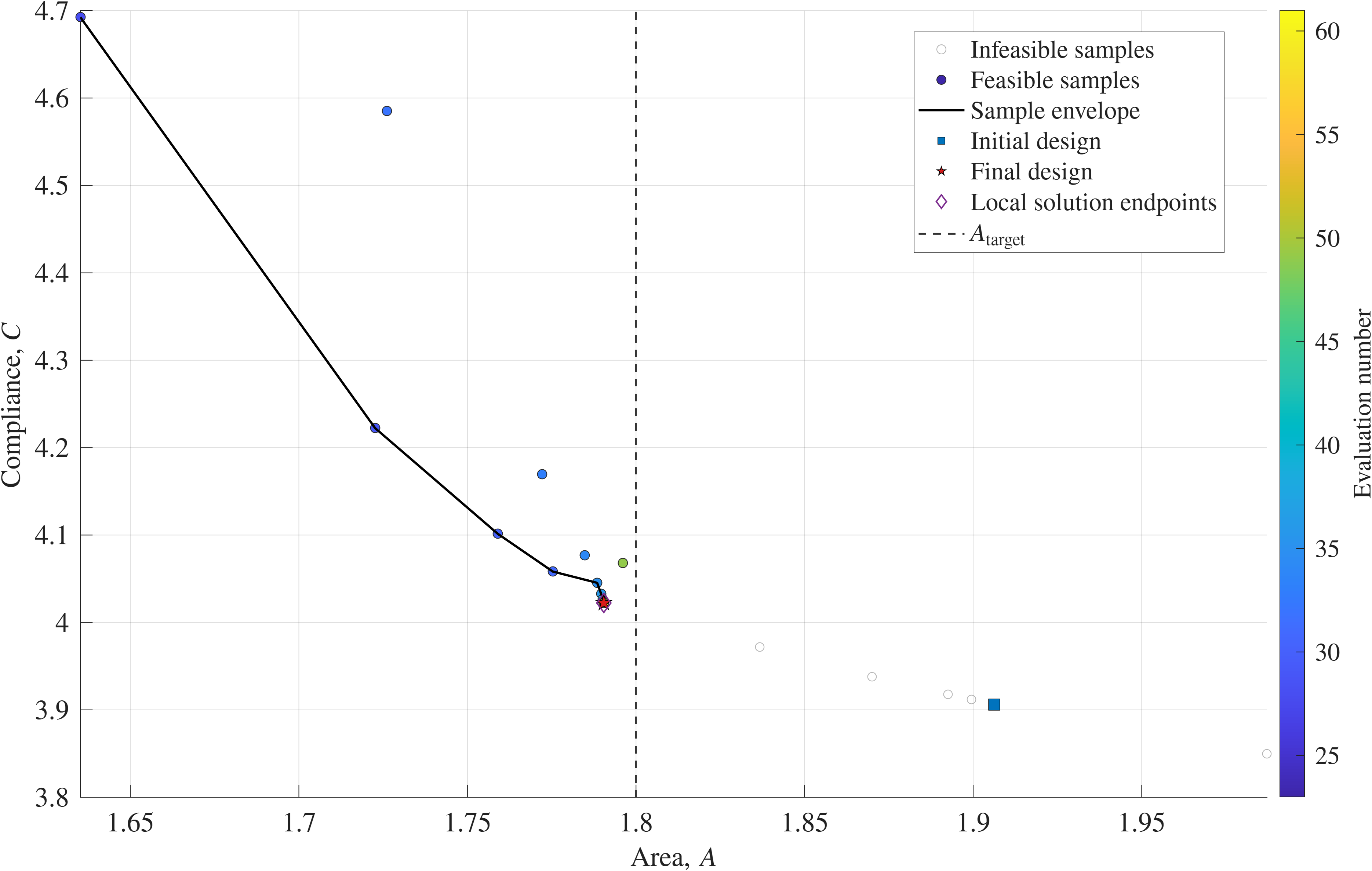}
	\caption{GS sample space.}
	\label{fig_SampleSpace_GS}
\end{figure}

\mcode{parameterOpt2d\_GS} wraps MATLAB's \mcode{GlobalSearch} algorithm, which
scatters trial points across the feasible domain and launches \mcode{fmincon} from each
promising start.  This strategy mitigates convergence to poor local minima at the cost of
many more FEA evaluations (up to $300 \times (1 + n)$ by default).  All feasible
solutions discovered during the run are retained in \mcode{m\_feasibleExploredSolutions}
and visualized in objective-constraint space via \mcode{plotParetoSpace}.

\begin{lstlisting}[language=Matlab]
parOpt = parameterOpt2d_GS(brepHandle, solverHandle, params0, ...
    objective, constraints, tol, stepSize, false);
parOpt = parOpt.optimize();
\end{lstlisting}

The optimization converged after 305 finite element analyses to the final parameter vector
$\bp_\text{GS} = (0.310,\;0.226,\;1.248,\;0.141)$.
The optimized design reduced the area to $A_\text{GS} = 1.79~\mathrm{m}^2$, while increasing the compliance to $C_\text{GS} = 4.022~\mathrm{N.m}$.

Figure \ref{fig_SampleSpace_GS} illustrates the GS sample space, in which 62 design variations were evaluated and 39 satisfied the feasibility criteria.

\subsection{Multi-Start}\label{sec:paramSO:MS}

\mcode{parameterOpt2d\_MS} uses MATLAB's \mcode{MultiStart} to launch a user-specified
number of independent local \mcode{fmincon} runs from randomly distributed starting
points.  Compared with \mcode{GlobalSearch}, the number of starts is directly controlled
(default: 5), making the evaluation budget more predictable.

\begin{lstlisting}[language=Matlab]
parOpt = parameterOpt2d_MS(brepHandle, solverHandle, params0, ...
    objective, constraints, tol, stepSize, false);
% parOpt.setNumberOfMultiStartLocalProblems(10);  % optional
parOpt = parOpt.optimize();
\end{lstlisting}

The optimization converged after 2545 FEA runs to the final parameter vector
$\bp^{}_\text{MS} =
(0.327,\;0.138,\;1.466,\;0.155)$.
The optimized design reduced the area to $A_\text{MS} = 1.798~\mathrm{m}^2$, while increasing the compliance to $C_\text{MS} = 4.001~\mathrm{N.m}$.

\begin{figure}[t]
	\centering
		\includegraphics[width=\linewidth]{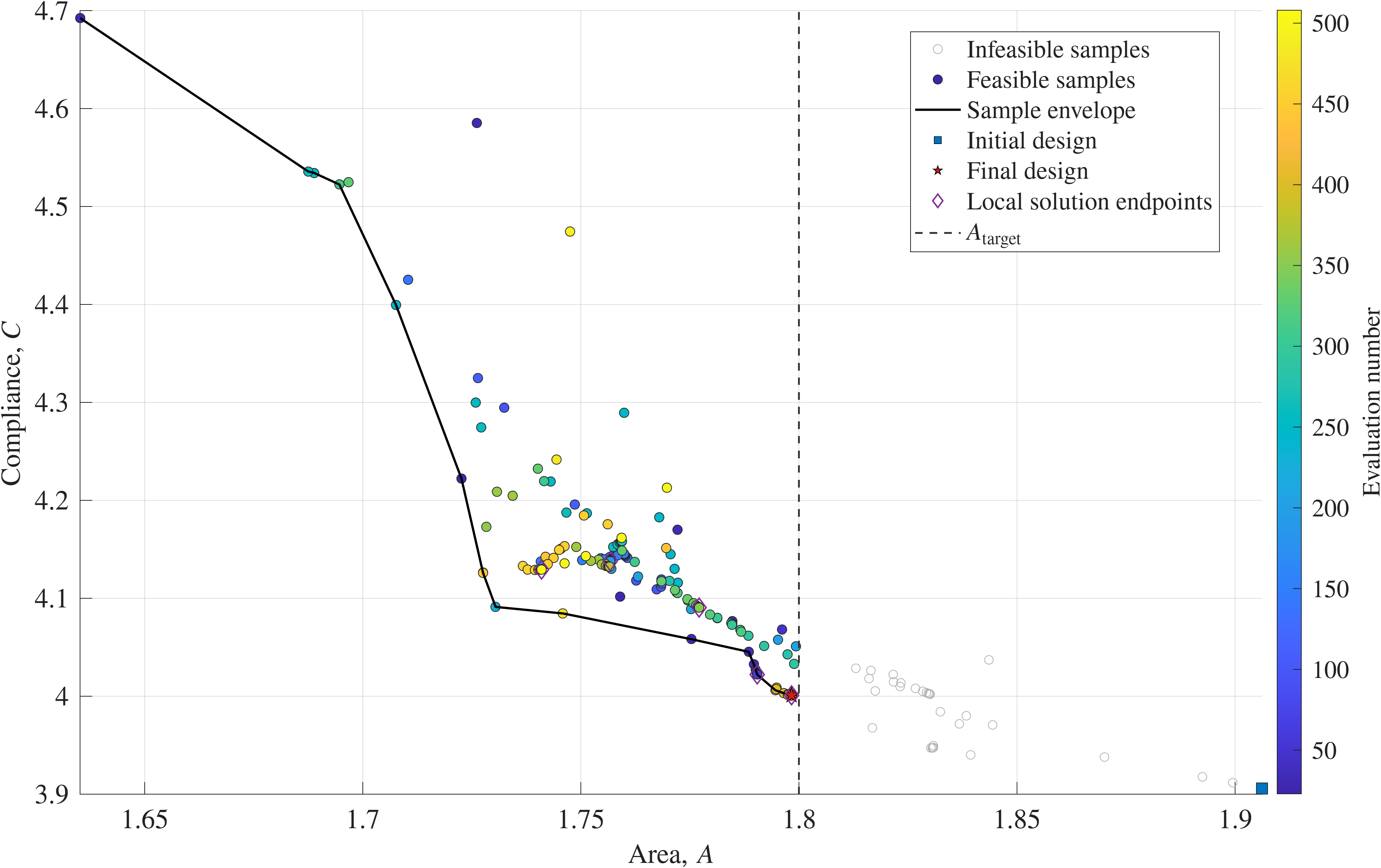}
	\caption{MS sample space.}
	\label{fig_SampleSpace_MS}
\end{figure}

Figure \ref{fig_SampleSpace_MS} illustrates the MS sample space, in which 5 local solvers were run, 508 design variations were evaluated, and 412 satisfied the feasibility criteria. 

\subsection{Random Search}\label{sec:paramSO:RS}

 \begin{figure}[t]
	\centering
		\includegraphics[width=\linewidth]{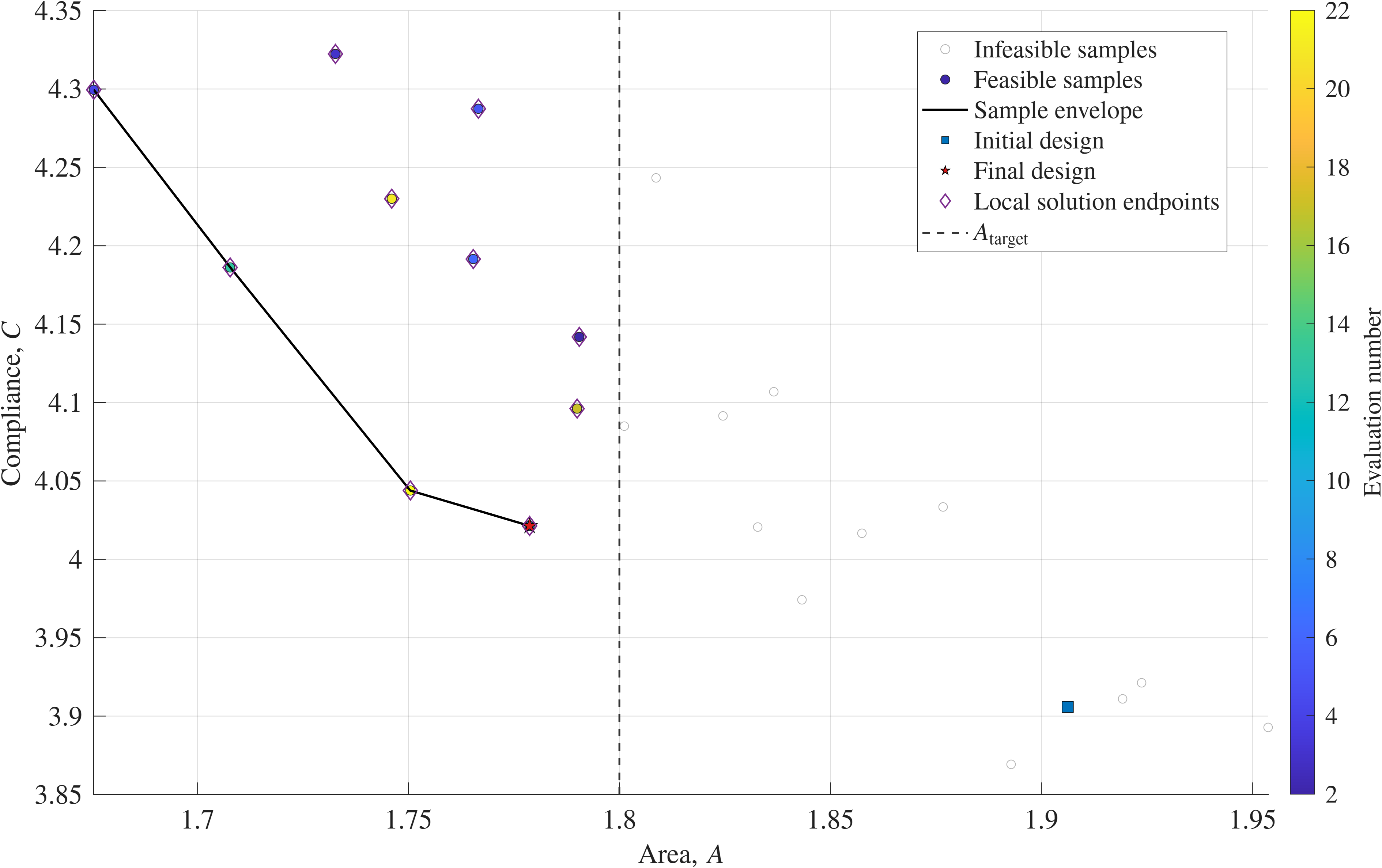}
	\caption{RS sample space.}
	\label{fig_SampleSpace_RS}
\end{figure}

\mcode{parameterOpt2d\_RS} samples the parameter space uniformly at random, evaluates
the objective at each feasible point, and retains the best.  It is entirely
derivative-free, requires no \mcode{terminationTolerance} or step-size arguments, and
terminates when a target number of feasible samples is reached or an evaluation budget is
exhausted.

\begin{lstlisting}[language=Matlab]
parOpt = parameterOpt2d_RS(brepHandle, solverHandle, params0, ...
    objective, constraints, false);
% parOpt.setNumberOfRandomSearchSamples(20, 200);  % nFeasible, nBudget
parOpt = parOpt.optimize();
\end{lstlisting}

Random search is a useful baseline and suffers from no local-minimum pathologies, but its
convergence degrades rapidly with $n$ (curse of dimensionality), so it is best reserved
for problems with very few parameters or as a warm-start strategy.

The optimization converged after 22 finite element analyses to the final parameter vector
$\bp_\text{RS} =
(0.359,\;0.150,\;1.337,\;0.150)$.
The optimized design reduced the area to $A_\text{RS} = 1.779~\mathrm{m}^2$, while increasing the compliance to $C_\text{RS} = 4.021~\mathrm{N.m}$.

Figure \ref{fig_SampleSpace_RS} illustrates the RS sample space, in which 22 design variations were evaluated and 10 satisfied the feasibility criteria. 
\begin{figure*}[t]
	\centering\includegraphics[width=1.0\linewidth]{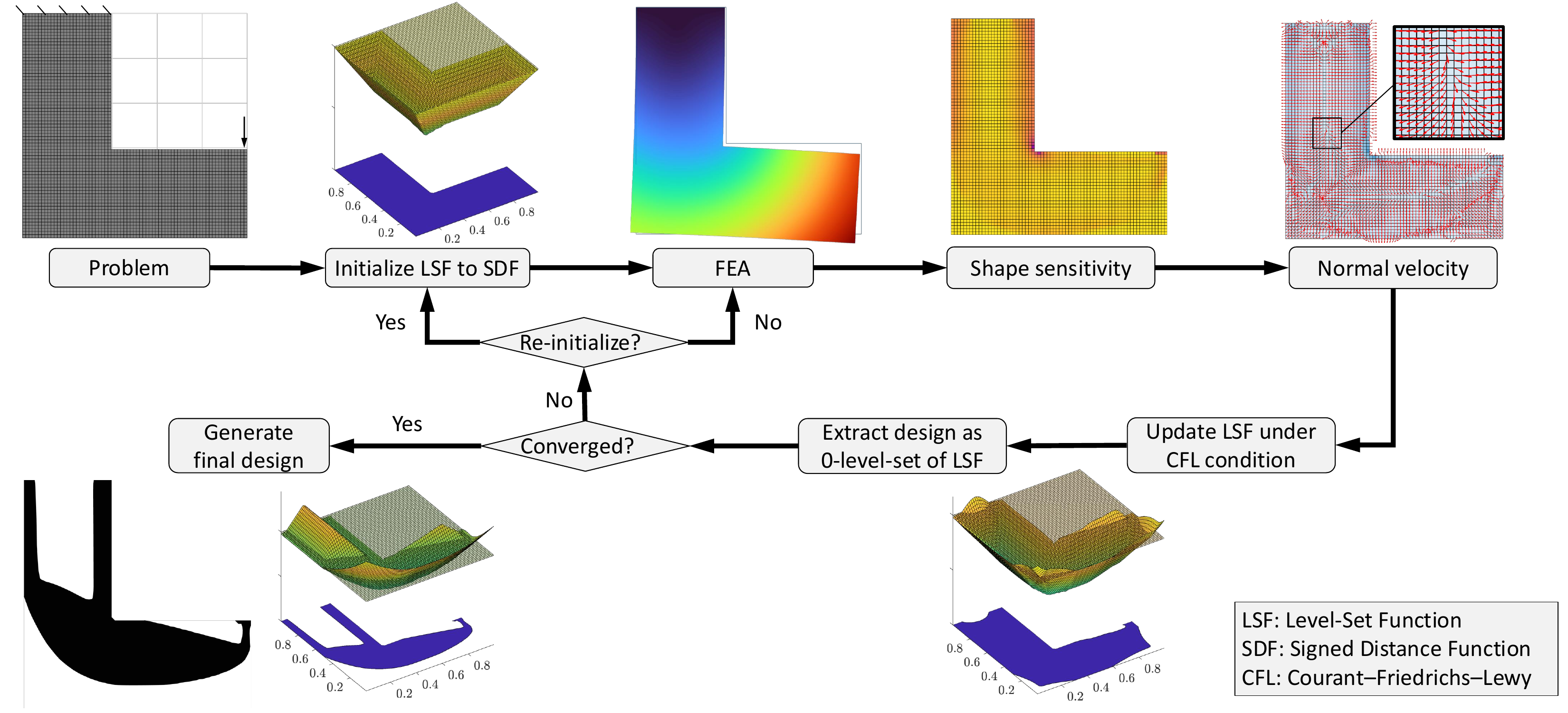}
	\caption{Shape optimization via LSSO using standard Hamilton-Jacobi equation.}
	\label{fig_LSworkflow_SO}
\end{figure*}
\section{Level-Set Shape Optimization}\label{sec:LSSO}

In this section, we will consider level-set shape optimization (LSSO), which enables the creation of optimized free-form shapes while maintaining a constant design topology, meaning that no new holes or cavities are introduced in the design. 
Figure \ref{fig_LSworkflow_SO} illustrates the shape optimization workflow via LSSO.
For a comprehensive overview of level-set methods for TO, the interested readers are referred to \cite{van2013level}.

\subsection{Level-Set Function}
Given a well-defined geometry $\Omega \in \R^d$, the level-set function (LSF) is defined as an \textit{implicit} function $\psi(\bx)$ such that:
\begin{equation} \label{eq_LSfunc}
	\psi(\bx)  \begin{cases} 
		< 0 &\bx \in \Omega\\
		= 0 & \bx \in \partial \Omega\\
		> 0 & \bx \notin \Omega \cup \partial \Omega 
	\end{cases} 
\end{equation}

\begin{figure} [t]
	\centering
	\begin{subfigure}[t]{0.68\linewidth}
		\centering
		\includegraphics[width=0.9\linewidth]{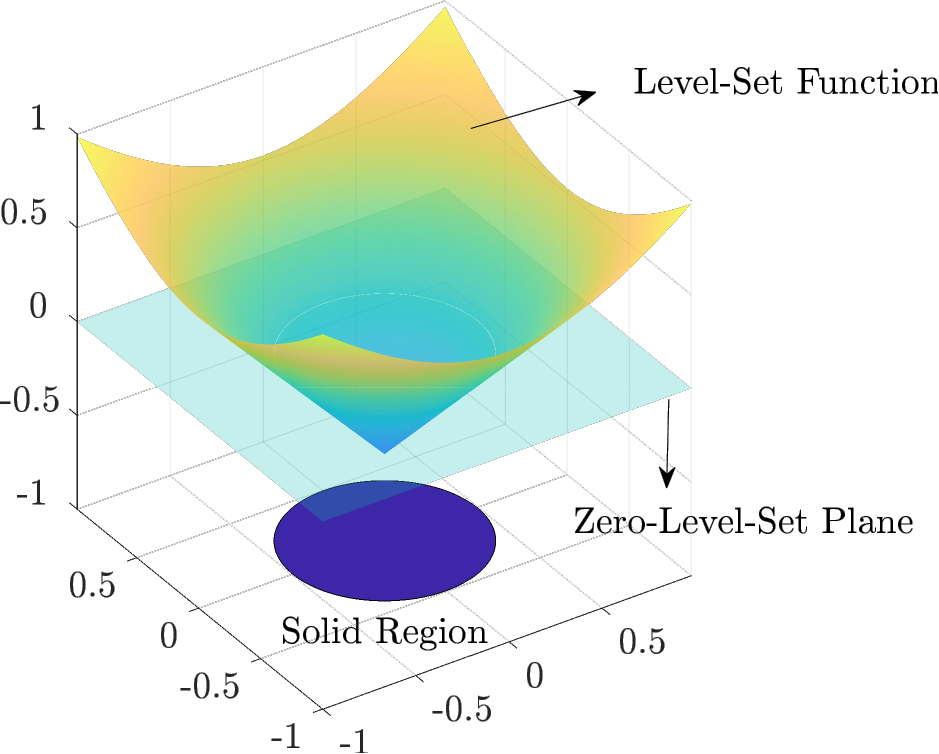}%
		\caption{Level-set function of a disc.}
	\end{subfigure}
	\begin{subfigure}[t]{0.3\linewidth}
	
		\centering
			\raisebox{1cm}{\includegraphics[width=\linewidth]{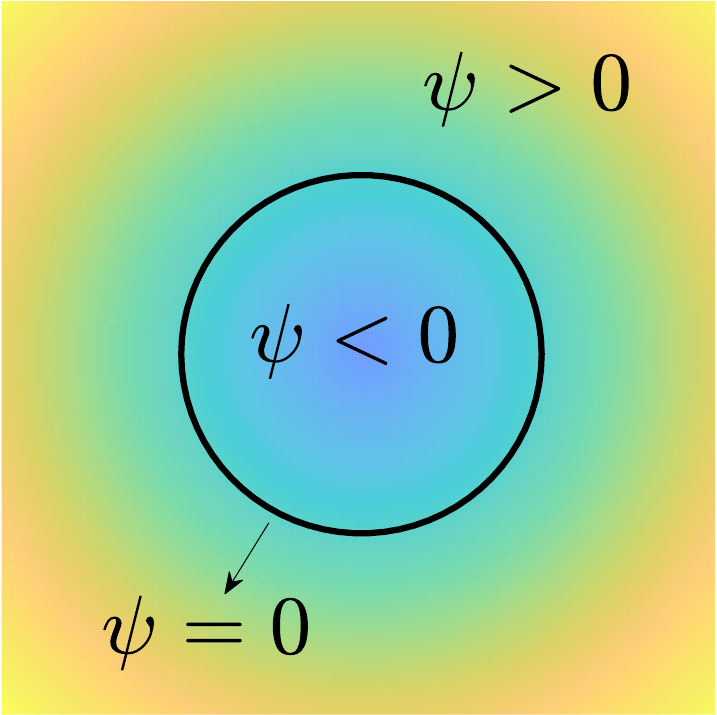}}
		\caption{Top view.}
	\end{subfigure}
	\caption{LSF of a disc; the 0-contour of the LSF separates inside from outside.} \label{fig_levelsetDomains}
\end{figure}

This discrete parameterization of \(\psi\) is critical for computational approaches, as it allows the level-set function to be represented and manipulated in numerical simulations. The grid resolution and parameterization directly impact the accuracy and computational cost of the simulation. Figure \ref{fig_levelsetDomains} illustrates the continuous LSF, represented as interpolated values from the discrete grid points. 
The evolution of the LSF is commonly posed as a Hamilton-Jacobi equation (HJE). It captures how the level-set evolves as the boundary changes and is widely used in studying wavefronts in fluid mechanics and optics \cite{van2013level,sigmund2013topology}. 

\subsection{Shape Optimization using HJE}
To compute the interface velocity, we adopt the reduced formulation in which the displacement field \(u(\psi)\) is defined implicitly by the linear elasticity equilibrium equations ${R}_{el}(\mathbf{d}) = \mathbf{0}$ (solved by FEA at each iteration). We therefore augment only the design constraint and write the shape optimization problem as
\begin{equation} \label{eq_SOproblem}
\begin{aligned}
\minimize_{\psi}\quad & \varphi(\psi) \\
\text{s.t.}\quad & |\Omega(\psi)|-V^*\le 0,\\
&{R}_{el}(\mathbf{d}) \coloneqq \mathbf{K}_{el}\mathbf{d}-\mathbf{f}_{el}=\mathbf{0}
\end{aligned}
\end{equation}
For this problem, we assume that the state equation is solved and already satisfied at each iteration. The Problem \eqref{eq_SOproblem} can be solved using the augmented Lagrangian as described in \cite{challis2010discrete}.

\subsection{Re-initialization}

\begin{figure} [t]
	\centering
	\begin{subfigure}[t]{0.45\linewidth}
		\centering
		\includegraphics[width=\linewidth]{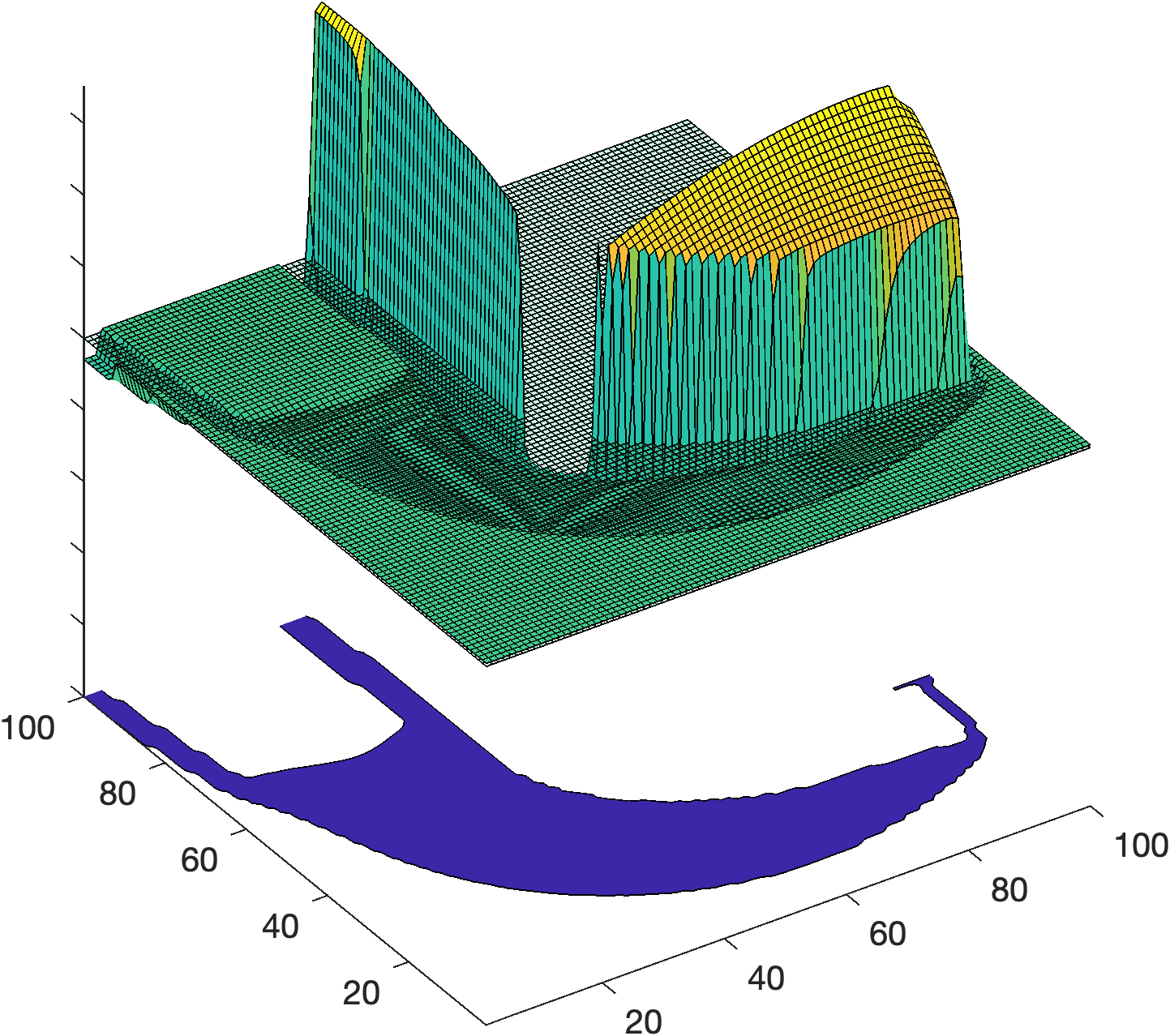}%
		\caption{LSF without reinitialization.}
	\end{subfigure}
	\begin{subfigure}[t]{0.45\linewidth}
		\centering
		\includegraphics[width=\linewidth]{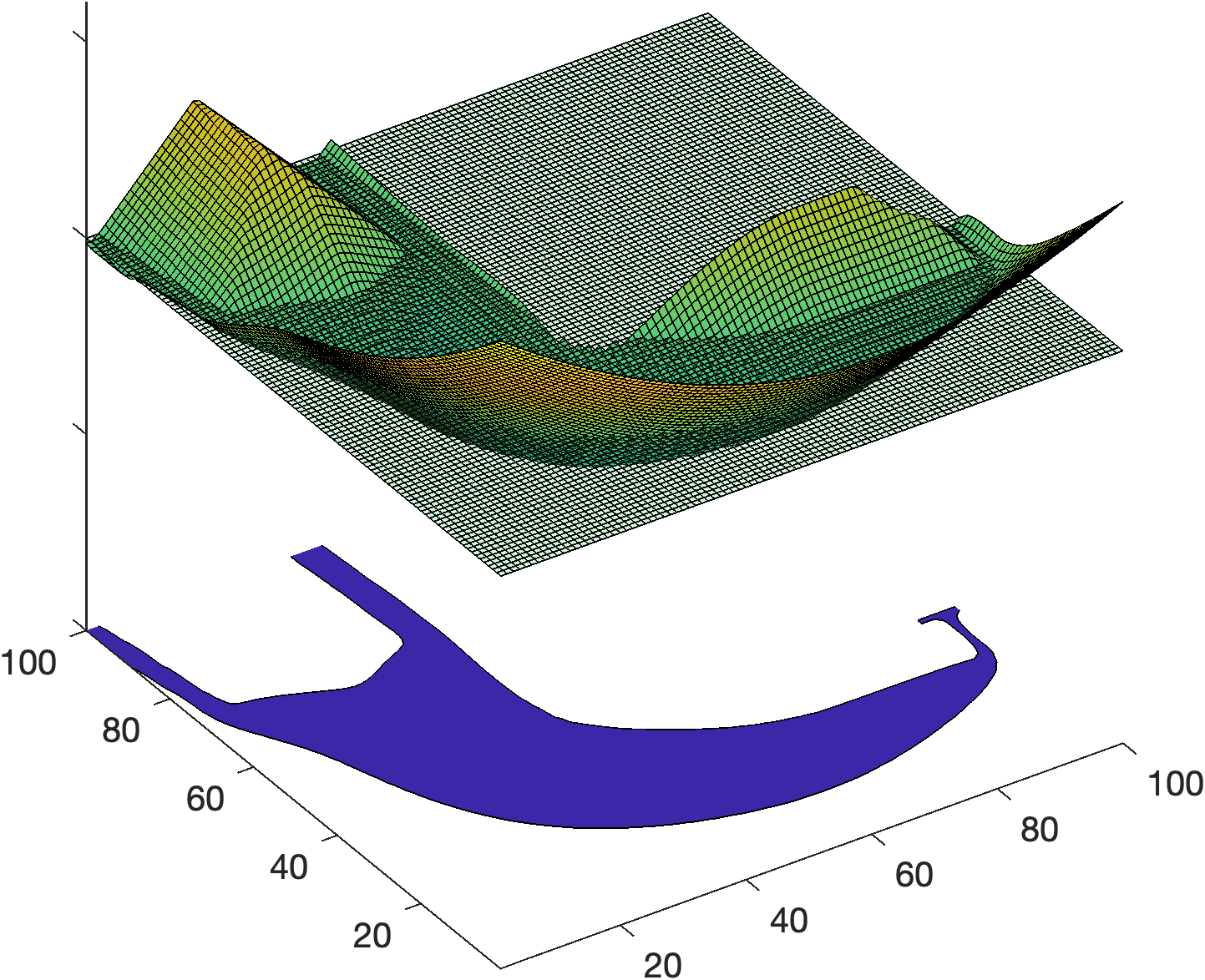}%
		\caption{LSF with reinitialization to SDF.}
	\end{subfigure}
	\caption{An intermediate LSF with and without reinitialization.} \label{fig_levelsetReinit}
\end{figure}

 \begin{figure}[!h]
	\centering
	\begin{subfigure}[b]{\linewidth}
		\centering
		\includegraphics[width=0.7\linewidth]{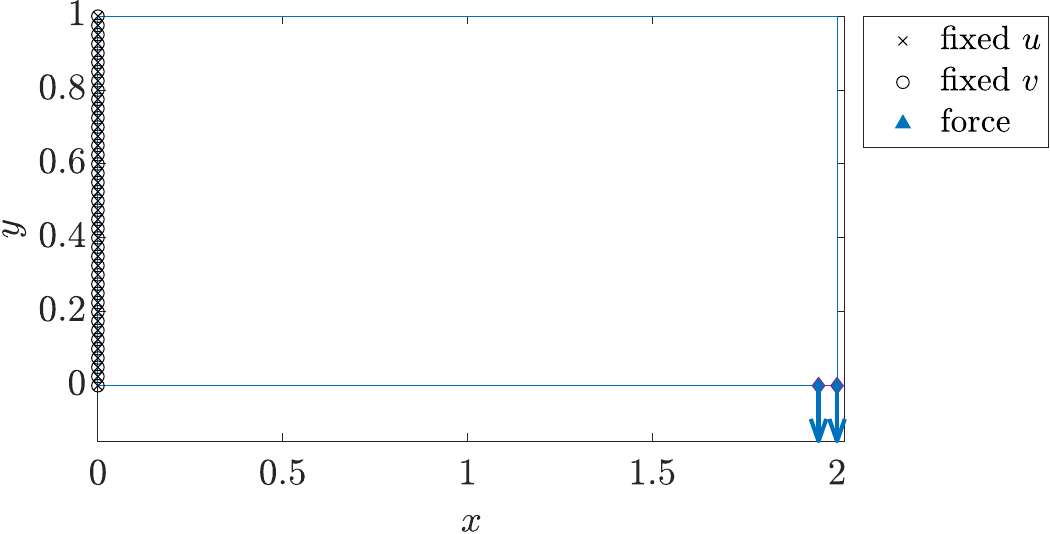}
		\caption{}
        \label{fig:cantileverBeamBC}
	\end{subfigure}
    
	\begin{subfigure}[b]{\linewidth}
		\centering
		\includegraphics[width=0.7\linewidth]{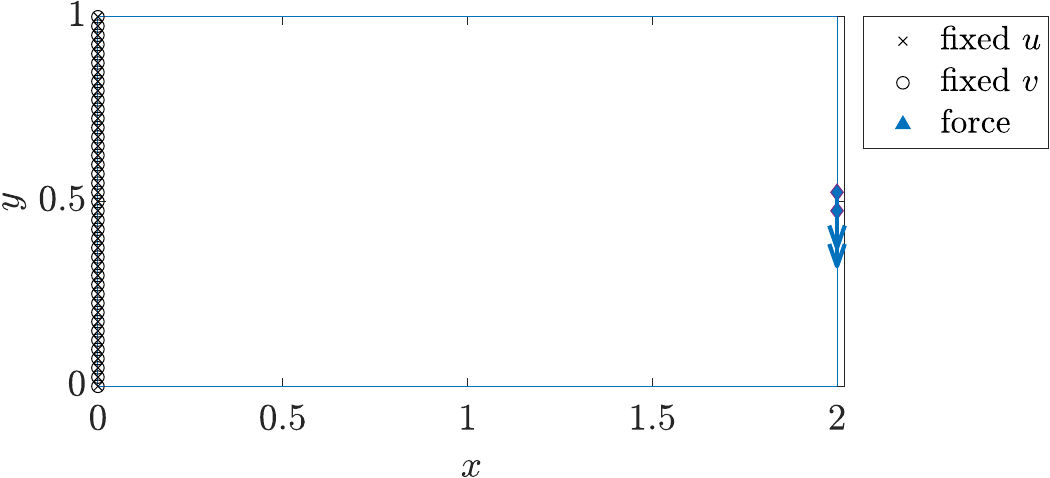}
		\caption{}
        \label{fig_BC_BeamMid}
	\end{subfigure}

	\begin{subfigure}[b]{\linewidth}
		\centering
		\includegraphics[width=0.7\linewidth]{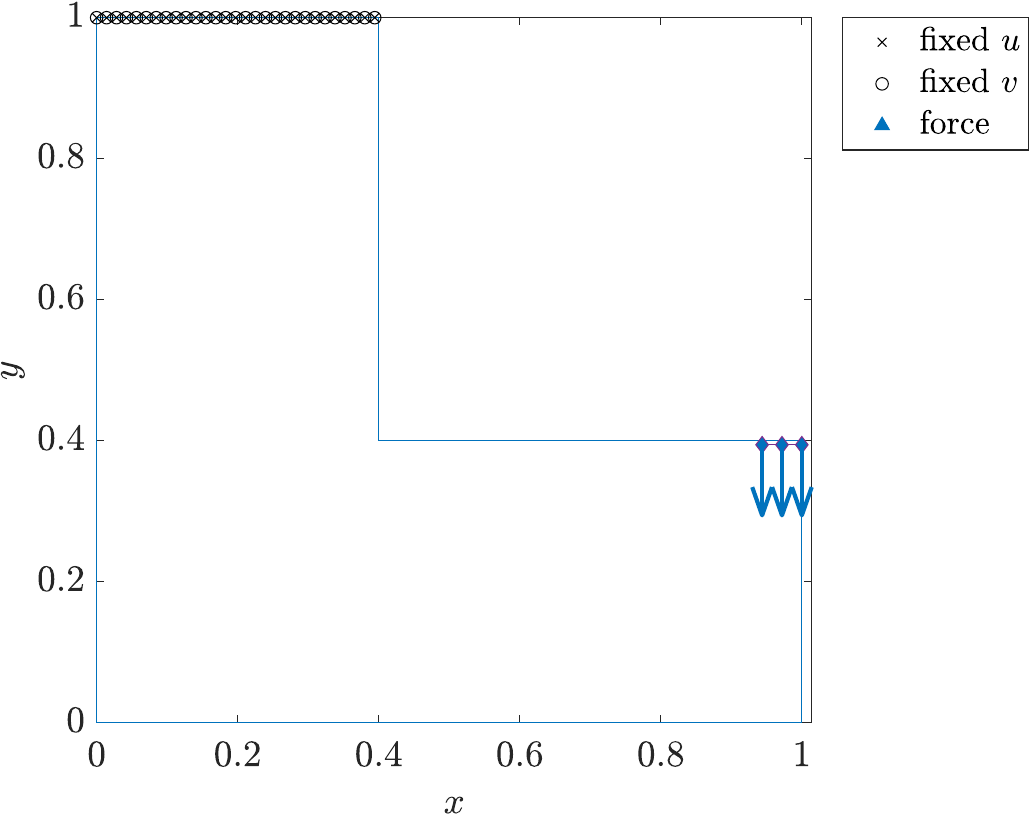}
		\caption{}
	\end{subfigure}

	\begin{subfigure}[b]{\linewidth}
		\centering
		\includegraphics[width=0.7\linewidth]{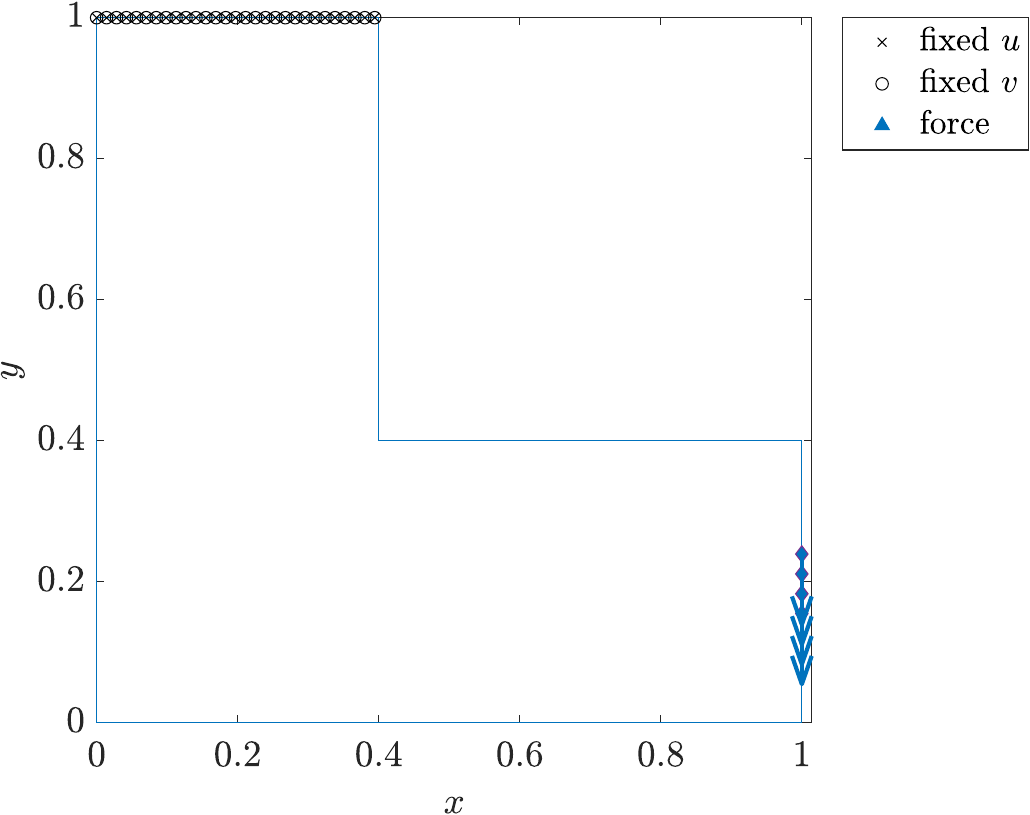}
		\caption{}
	\end{subfigure}

	\begin{subfigure}[b]{\linewidth}
		\centering
		\includegraphics[width=0.7\linewidth]{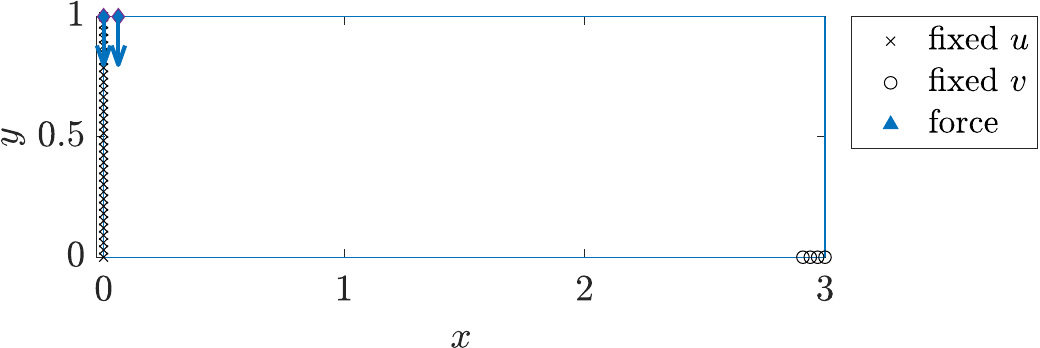}
		\caption{}
	\end{subfigure}
    
	\caption{Boundary conditions for SO/TO benchmark examples.}
	\label{fig_benchmarkExamples}
\end{figure}

 \begin{figure}[!h]
	\centering
	\begin{subfigure}[b]{0.45\linewidth}
		\centering
		\includegraphics[width=0.9\linewidth]{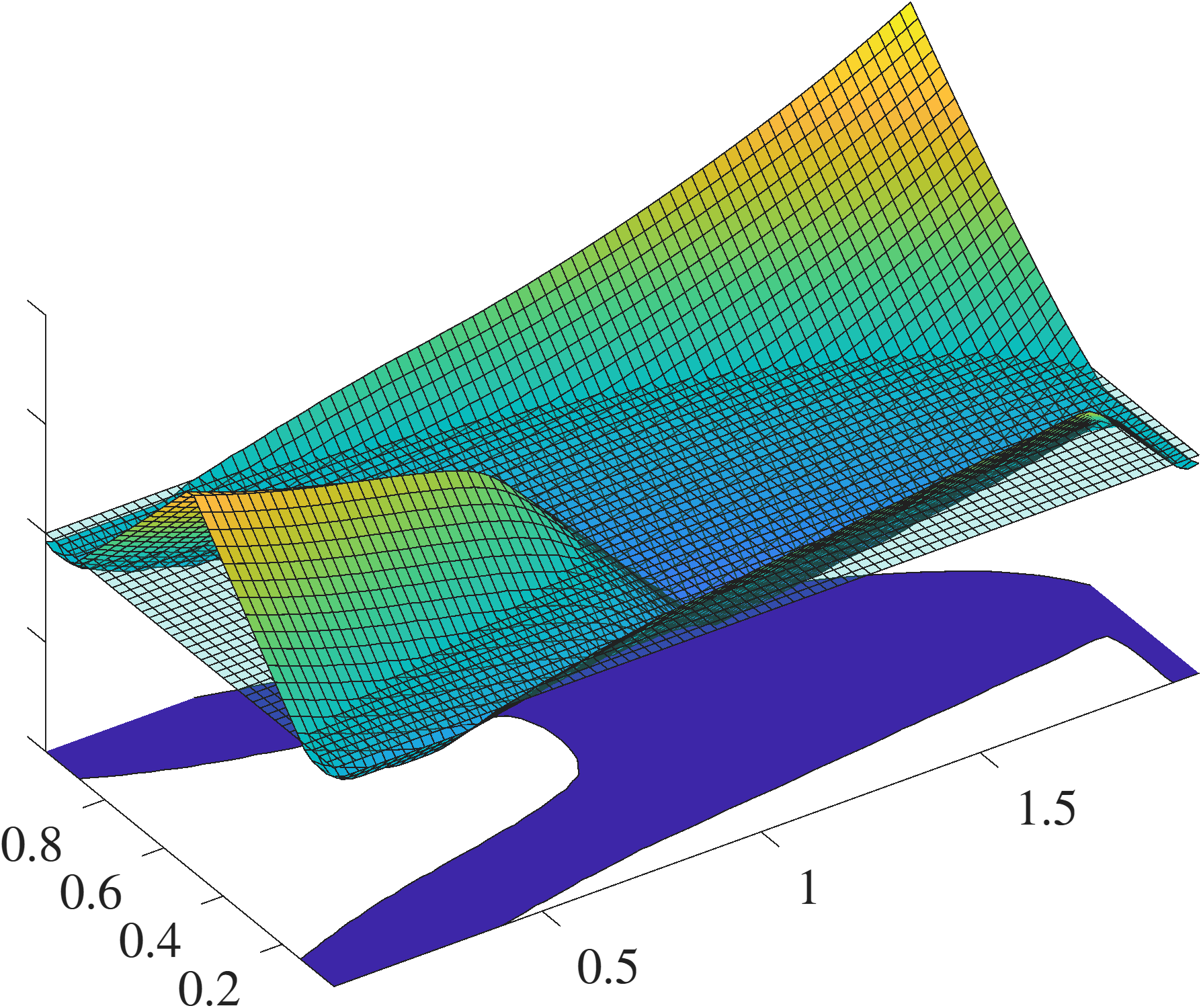}
		\caption{}
	\end{subfigure}
	\begin{subfigure}[b]{0.45\linewidth}
		\centering
		\includegraphics[width=0.9\linewidth]{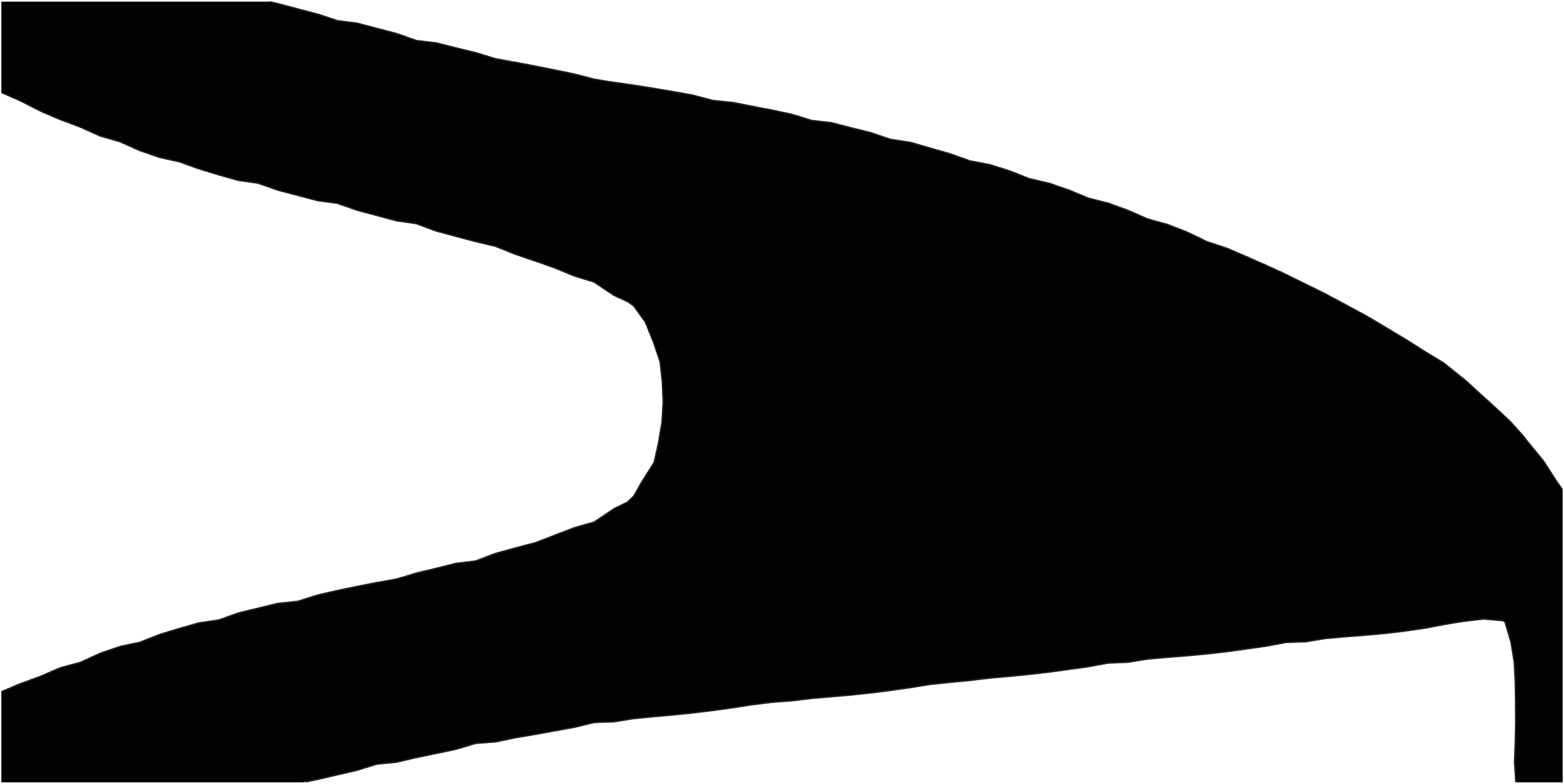}
		\caption{}
	\end{subfigure}

    \begin{subfigure}[b]{0.45\linewidth}
		\centering
		\includegraphics[width=0.9\linewidth]{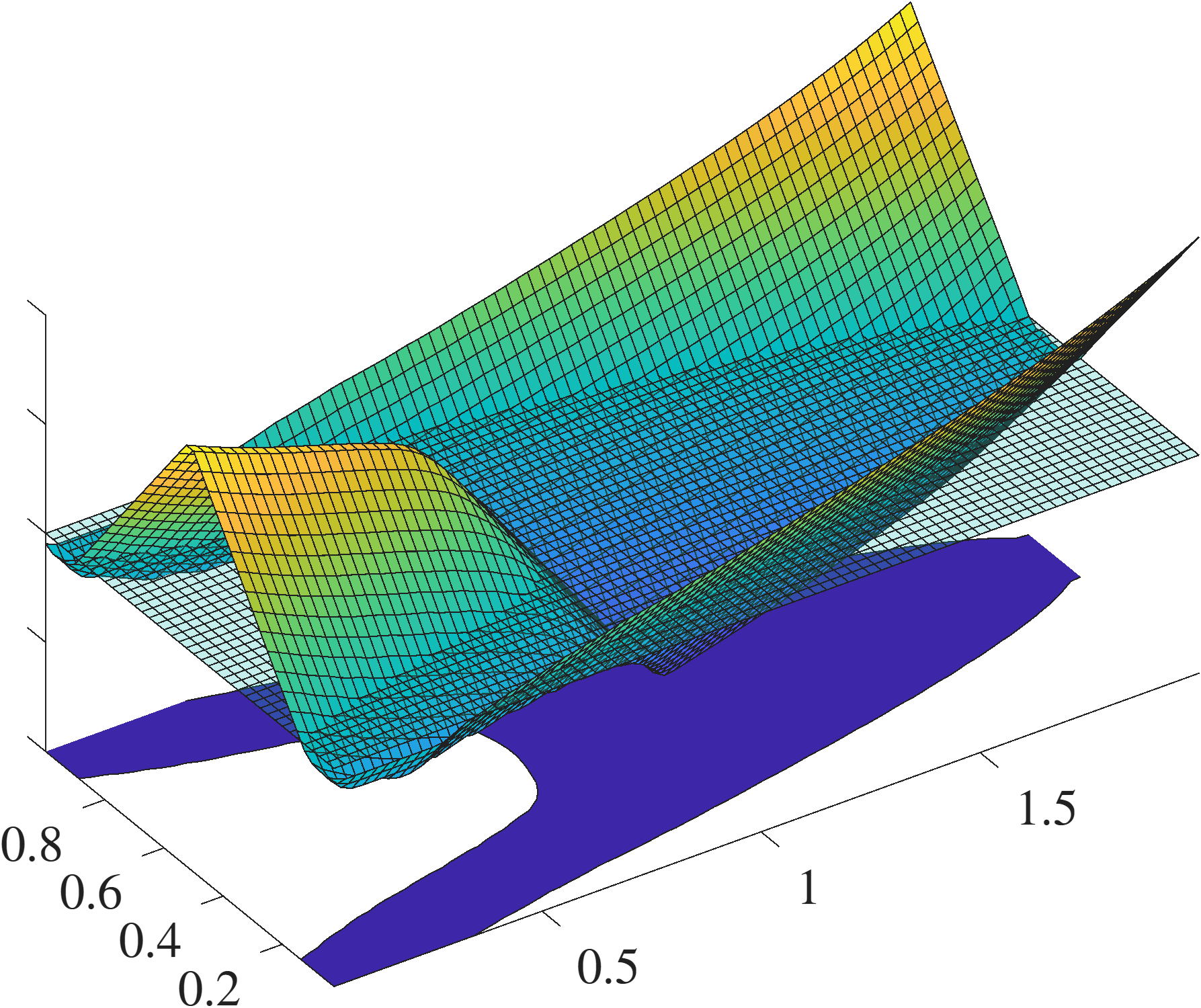}
		\caption{}
	\end{subfigure}
	\begin{subfigure}[b]{0.45\linewidth}
		\centering
		\includegraphics[width=0.9\linewidth]{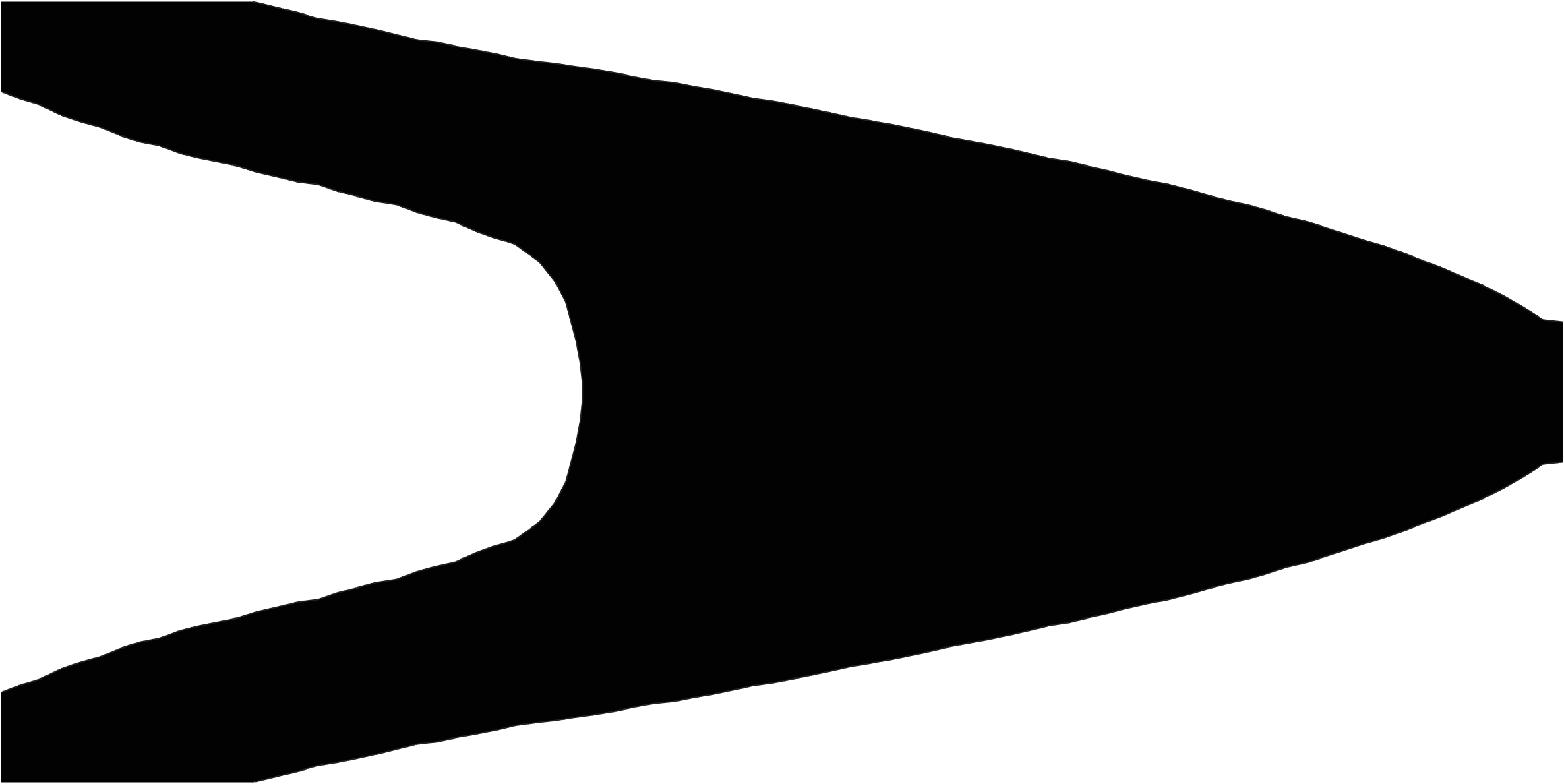}
		\caption{}
	\end{subfigure}

    \begin{subfigure}[b]{0.45\linewidth}
		\centering
		\includegraphics[width=0.9\linewidth]{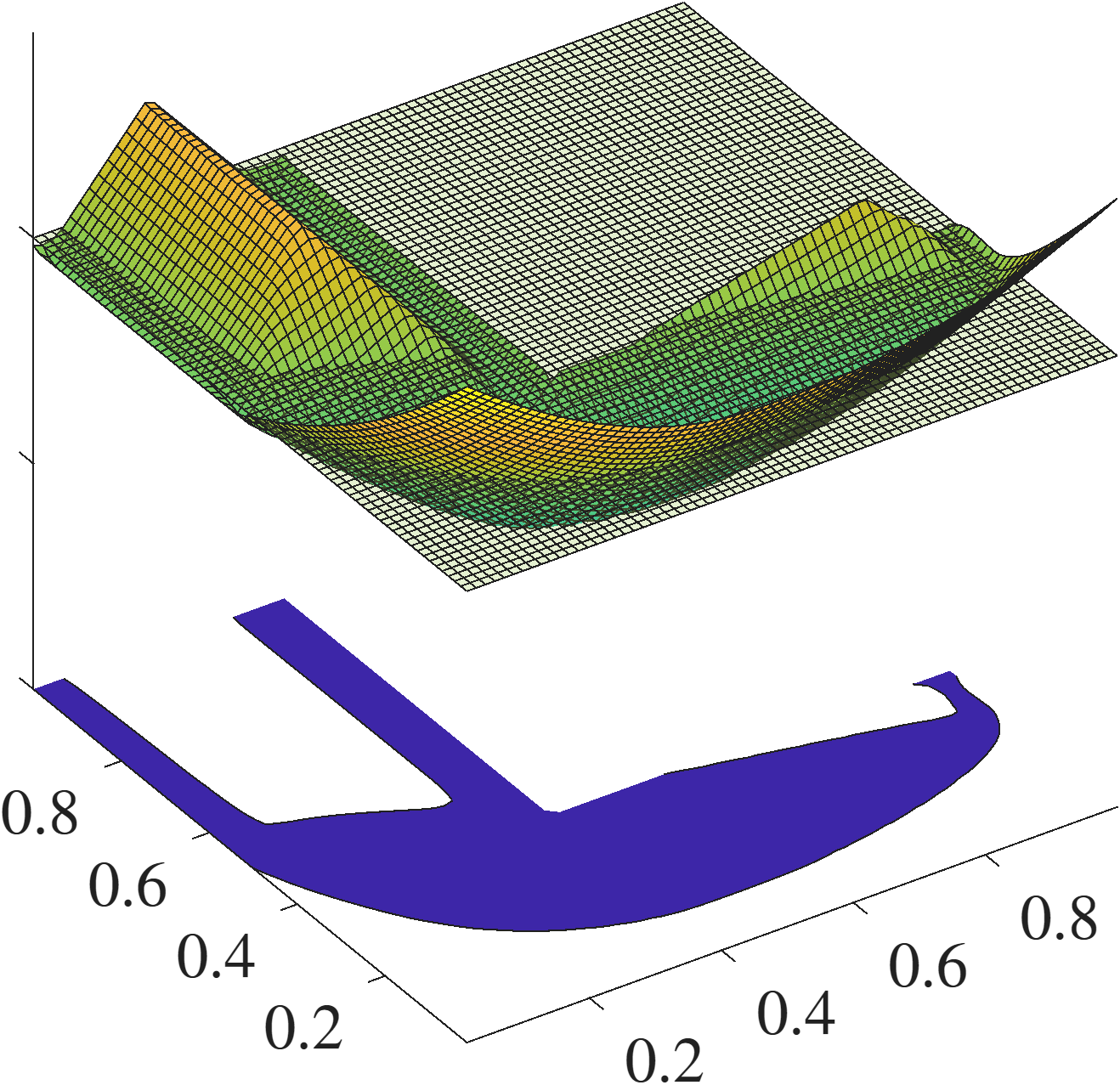}
		\caption{}
	\end{subfigure}
	\begin{subfigure}[b]{0.45\linewidth}
		\centering
		\includegraphics[width=0.9\linewidth]{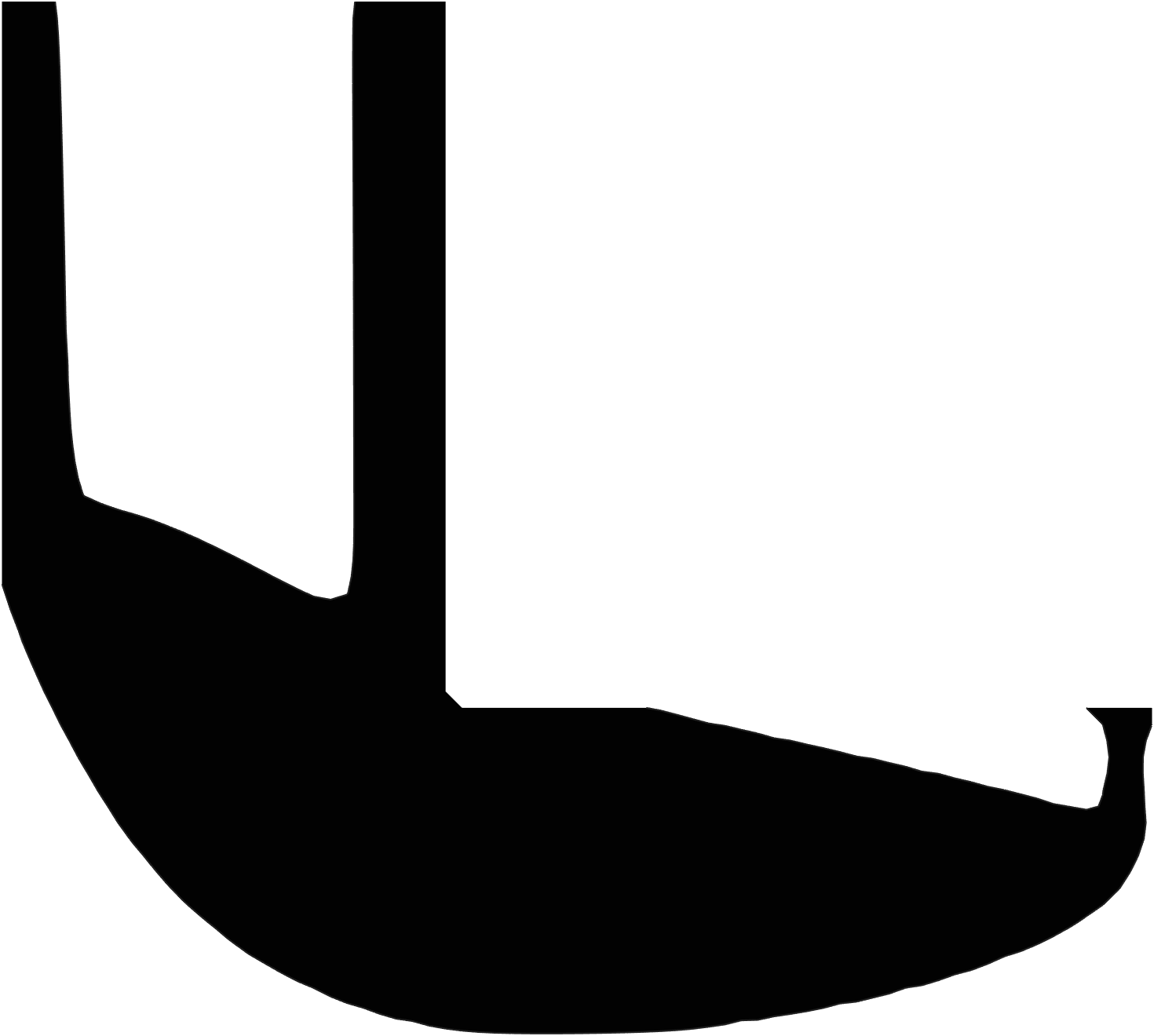}
		\caption{}
	\end{subfigure}

    \begin{subfigure}[b]{0.45\linewidth}
		\centering
		\includegraphics[width=0.9\linewidth]{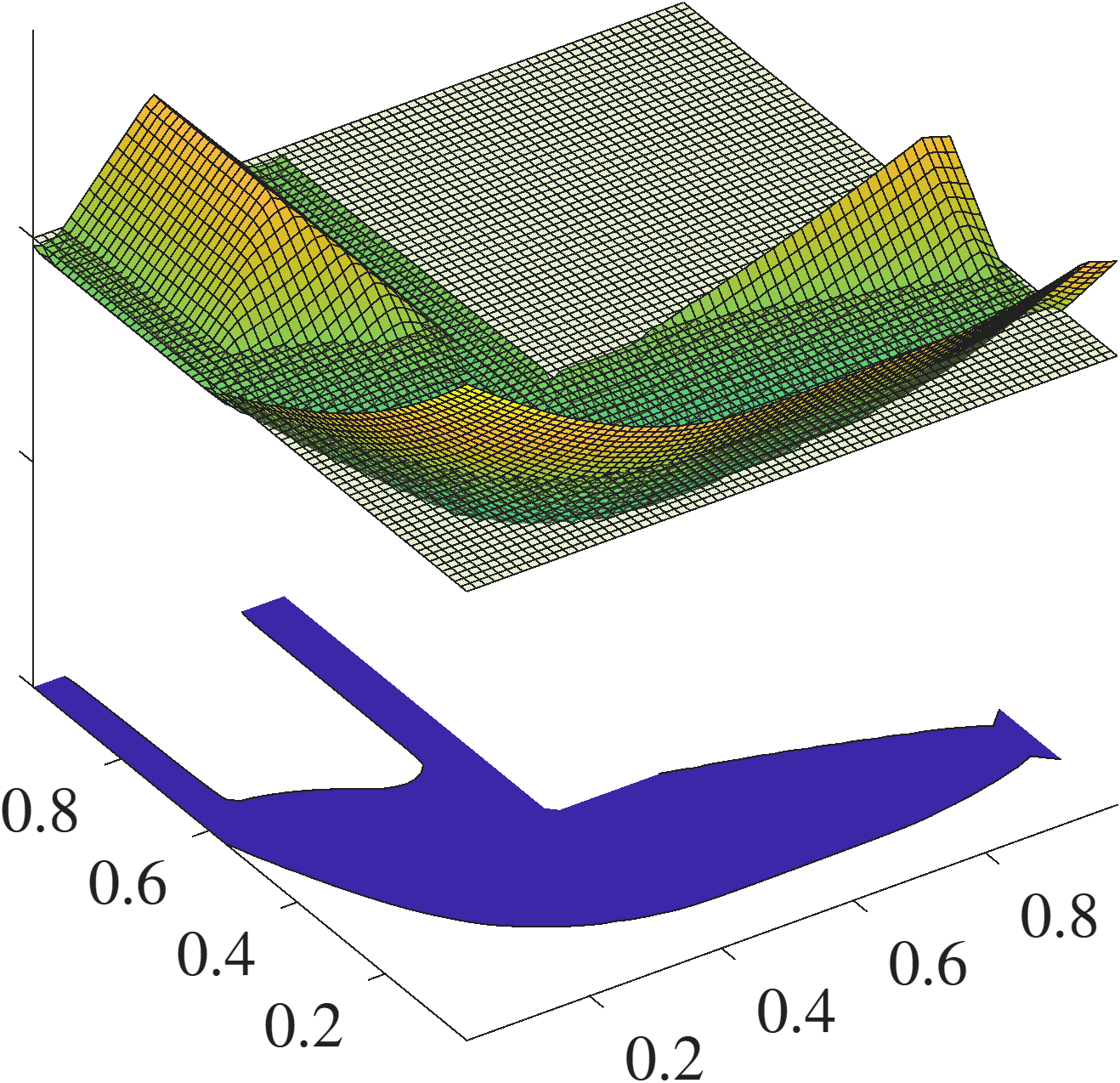}
		\caption{}
	\end{subfigure}
	\begin{subfigure}[b]{0.45\linewidth}
		\centering
		\includegraphics[width=0.9\linewidth]{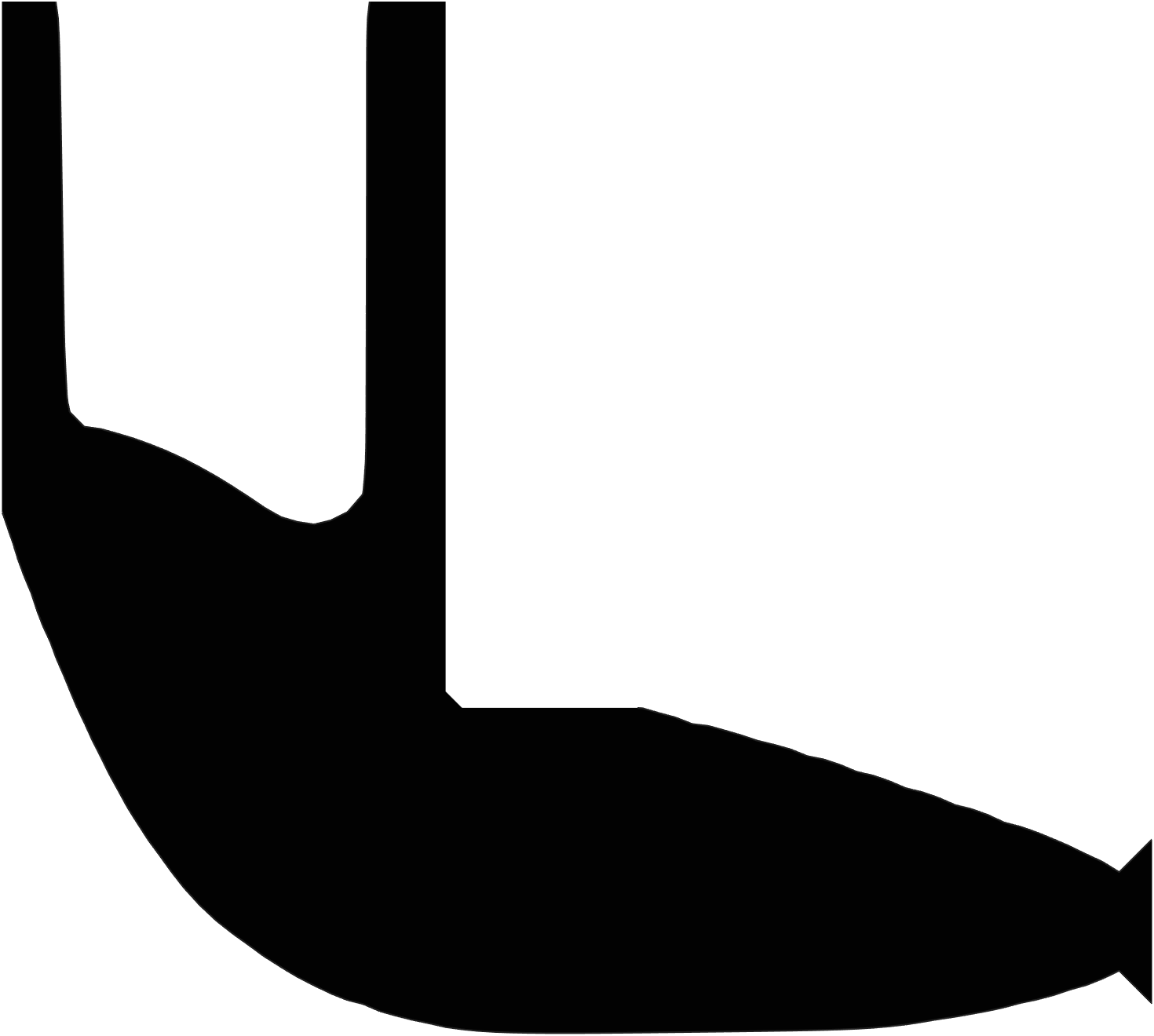}
		\caption{}
	\end{subfigure}

    \begin{subfigure}[b]{0.45\linewidth}
		\centering
		\includegraphics[width=0.9\linewidth]{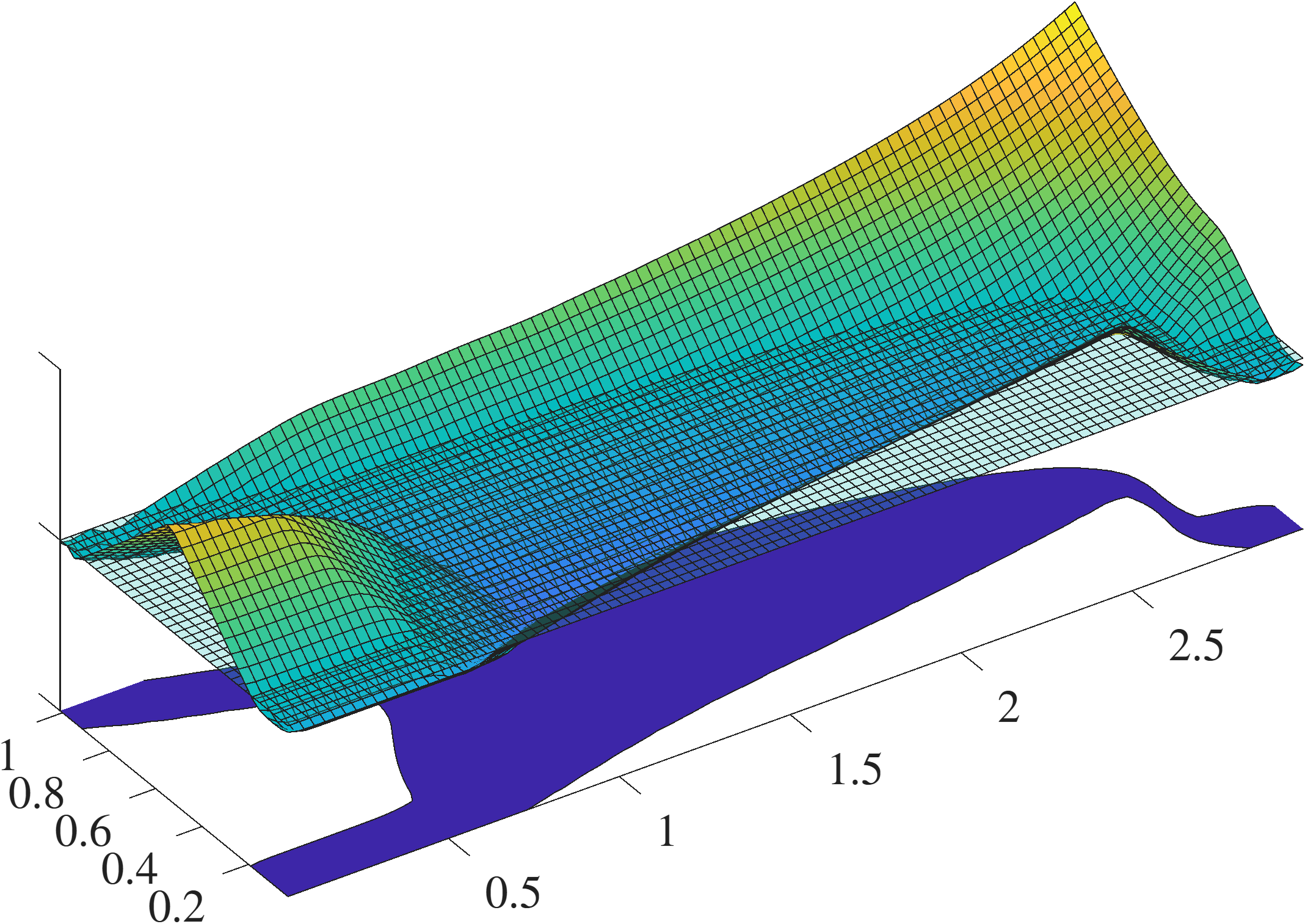}
		\caption{}
	\end{subfigure}
	\begin{subfigure}[b]{0.45\linewidth}
		\centering
		\includegraphics[width=0.9\linewidth]{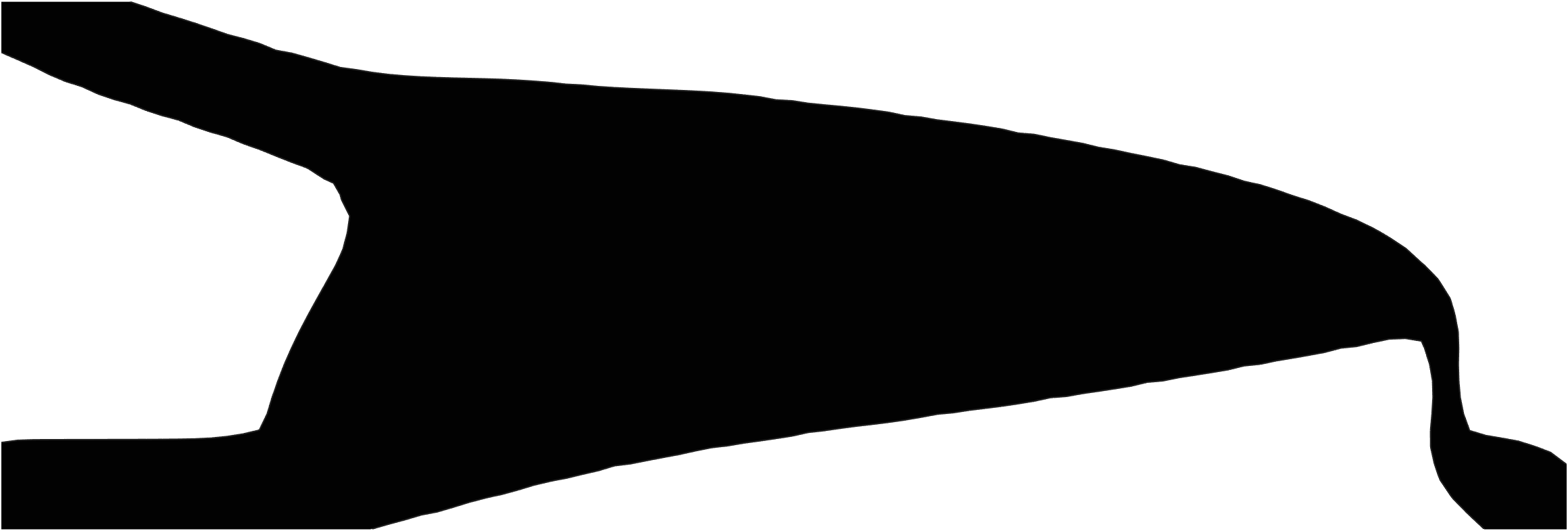}
		\caption{}
	\end{subfigure}

	\caption{Level-set SO benchmark examples.}
	\label{fig_example_LSSO}
\end{figure}

Throughout the optimization process, the LSF might become too steep or too flat, which can cause convergence issues. To ensure numerical accuracy, the level-set function is re-initialized to the SDF ($\lvert \nabla\psi \rvert \approx 1$) without changing its zero-level-set every few iterations by solving:
\begin{equation}\label{eq_SDF}
	\dfrac{\partial \psi}{\partial t} + \overline{\sign}(\psi_0)\left( \lvert \nabla\psi \rvert - 1\right)= 0
\end{equation}
where $\psi_0$ is the level-set before reinitialization and $\overline{\sign}$ is a smooth approximation of the sign function:
\begin{equation}\label{eq_smoothSign}
	\overline{\sign}(\psi) = \dfrac{\psi}{\sqrt{\psi^2 + \lvert \nabla\psi \rvert ^2 \epsilon^2}}
\end{equation}
where $\epsilon$ is a small value, typically the size of each finite element. Fig. \ref{fig_levelsetReinit} illustrates the impact of re-initialization. Observe in Fig. \ref{fig_levelsetReinit}a that the LSF exhibits a steep increase whereas the re-initialized LSF in Fig. \ref{fig_levelsetReinit}b exhibits a more gradual change across the domain. 

\subsection{Benchmark Examples} \label{sec:benchmarkExamples}
The SO/TO benchmark examples are illustrated in Fig. \ref{fig_benchmarkExamples}. For all benchmarks, we assume linear elasticity with \(E=100~\mathrm{GPa}\) and \(\nu=0.3\), and apply a downward load of \(100~\mathrm{kN}\) at the location indicated in Fig.~\ref{fig_benchmarkExamples}. 

\noindent \textbf{General parameters.} For the cantilever beam example, the FEA and SO classes are selected first, together with the general solver and optimization settings.  A uniform grid is enforced because it is required by the HJE update scheme, while vectorized FEA is enabled to improve computational efficiency. Since no material interpolation is needed for this benchmark, the interpolation and penalty parameters are kept inactive:
\begin{lstlisting}
%% Solvers
feaClass = @fea2d_elasticity;
shapeoptClass = @standardHJ2d_elasticity;
%% General Parameters
vectorize = true; % vectorized FEA
uniformGrid = 1; % needed for the Hamilton-Jacobi solver
exportGIF = false; % .gif of optimization 
%% Optimizer Parameters
interpolation = 'none';
penaltyStruct = struct('min',1,'max',1,'inc',0);
maxNumIters = 500;
\end{lstlisting}

\noindent \textbf{Problem definition.} The benchmark geometry is defined through the \mcode{'CantileverBeam.brep'} file and discretized using 3,200 active quadrilateral elements over the solid domain. The material properties and boundary conditions are specified for a single loading scenario. In this example, edge $\#5$ of the B-Rep is fixed, and a downward force of $100~\mathrm{kN}$ is applied on edge $\#2$ (Fig.~\ref{fig:cantileverBeamBC}):
\begin{lstlisting}
%% Problem Definition
brep = 'CantileverBeam.brep'; % geometry
numElements = 3200; % mesh
material.E = 100e9; material.nu = 0.3; 
numScenarios = 1;
%% Construct FEA Solver
solver = feaClass(brep,numElements,material,vectorize,numScenarios, ...
    interpolation,penaltyStruct,uniformGrid); % call superclass
solver = solver.fixEdge(5);
solver = solver.applyYForceOnEdge(2,-1e5);
solver = solver.preProcess(); % FEA pre-process
\end{lstlisting}

\noindent \textbf{Objective.} The optimization objective is structural compliance, with sensitivities evaluated for the HJE-based boundary evolution:
\begin{lstlisting}
%% Objective Functional
objective = standardHJComplianceElasticity(solver);
\end{lstlisting}

\noindent \textbf{Constraints.} The constraints are defined as a cell array, allowing multiple constraint objects to be included if needed. Each constraint is implemented as a class derived from the \mcode{functional} base class. In this example, only a single \mcode{volume} functional is used to impose a volume constraint, constructed by passing the solver object and the target volume fraction of 0.5:
\begin{lstlisting}
%% Constraints Functionals
volumeFraction = 0.5;
constraints  = {volume(solver, volumeFraction)};
\end{lstlisting}

\noindent \textbf{Manufacturing Constraints.} In addition to the global volume constraint, a manufacturing constraint is introduced to enforce a minimum feature size. Manufacturing constraints are also defined as a cell array, allowing multiple design and manufacturing restrictions to be included if needed. Each constraint is implemented as a class derived from the \mcode{mfgConstraints} base class. In this example, a single \mcode{minimumFeatureSize_conv} constraint is used to regularize the evolving boundary and suppress unrealistically thin members through a convolution-based filter similar to \cite{challis2010discrete}. The first argument, \mcode{solver}, provides the constraint with the mesh and active-design-domain information needed to filter only valid elements and avoid spreading sensitivities outside the geometry. The optional second argument, \mcode{R}, controls the filter radius in element units; for example, \mcode{R = 1} gives a \mcode{3 x 3} kernel, while larger values produce stronger smoothing and a larger minimum feature size. If \mcode{R} is omitted, the default value \mcode{R = 1} is used:
\begin{lstlisting}
% manufacturing constraints
mfgConstraints = {minimumFeatureSize_conv(solver)}; 
% or, with a user-specified filter radius:
% mfgConstraints = {minimumFeatureSize_conv(solver, R)};
\end{lstlisting}

\noindent \textbf{Optimize.} The shape optimizer is then instantiated by combining the solver, objective, constraints, and manufacturing restrictions. The initialization parameters controlling the number of seeded holes are set to zero in this example, so the optimization starts directly from the reference geometry:
\begin{lstlisting}
%% Construct Optimizer
nHolesX = 0; nHolesY = 0; r0 = 0;
shapeopt = shapeoptClass(solver, ...
    objective,constraints,mfgConstraints, ...
    nHolesX,nHolesY,r0, ...
    maxNumIters,exportGIF);
%% Optimize
shapeopt = shapeopt.optimize();
\end{lstlisting}

\noindent \textbf{Results.} Finally, the optimized design and its corresponding mechanical response are post-processed. The boundary conditions, displacement field, von Mises stress, and principal stress distributions are plotted to assess both the structural behavior and the quality of the final design:
\begin{lstlisting}
%% Plotting
shapeopt.m_solver.plotDeformation();
shapeopt.m_solver.plotVonMisesStress();
shapeopt.m_solver.plotPrincipalStress();
\end{lstlisting}

\begin{table}[t]
\centering
\caption{Final performance metrics for optimized designs obtained with LSSO fro the illustrative examples: compliance \(C\), maximum displacement \(\delta_{\max}\), and maximum von Mises stress \(\sigma_{\mathrm{vm},\max}\).}
\label{tab:LSSO_example_metrics}

\small
\renewcommand{\arraystretch}{1.15}
\begin{tabular}{
>{\centering\arraybackslash}p{2.5cm}
>{\centering\arraybackslash}p{1cm}
>{\centering\arraybackslash}p{1.5cm}
>{\centering\arraybackslash}p{1cm}}
\toprule
\textbf{Example} &
{$C$} &
{$\delta_{\max}$} &
{$\sigma_{\text{vm},\max}$} \\
&
{(N.m)} &
{(m)} &
{(MPa)} \\
\midrule

{Cantilever Beam (bottom load)}            & {7.85} & {9.37e-05} & {10.5} \\ \addlinespace
{Cantilever Beam (middle load)} & {12.5} & {9.99e-05} & {4.93} \\
\addlinespace
{L-bracket (top load)}     & {25.8} & {2.76e-04} & {37.8} \\
\addlinespace
{L-bracket (mid load)}     & {25.7} & {2.89e-04} & {54.3} \\
\addlinespace
{MBB (symmetry)}     & {22.7} & {2.48e-04} & {6.18} \\
\bottomrule
\end{tabular}
\end{table}

Figure~\ref{fig_example_LSSO} summarizes the LSSO results for the illustrative examples, showing the optimized level-set field (left), the corresponding extracted geometry (middle), and the objective history over iterations (right). Table~\ref{tab:LSSO_example_metrics} reports the final compliance \(C\), maximum displacement \(\delta_{\max}\), and maximum von Mises stress \(\sigma_{\mathrm{vm},\max}\) for each optimized design. 

\section{Topology Optimization} \label{sec:TopOpt}

The field of TO \cite{eschenauer2001topology,rozvany2009critical,sigmund2013topology,deaton2014survey,wu2021topology} was formally established in 1988 when Martin P. Bends{\o}e and Noboru Kikuchi introduced the homogenization method \cite{bendsoe1988generating}, enabling optimal material distribution within a fixed domain by modeling materials as composites with effective properties. Shortly after, Bends{\o}e proposed the Solid Isotropic Material with Penalization (SIMP) method \cite{bendsoe1989optimal}, which introduced element-wise pseudo-densities and a penalization scheme to promote discrete solid-void designs. Further refinements by Ole Sigmund improved numerical robustness and computational efficiency, leading SIMP to become the dominant practical approach in TO \cite{sigmund200199,bendsoe1999material}.

In parallel, alternative formulations emerged. Xie and Steven (1993) \cite{xie_simple_1993} proposed Evolutionary Structural Optimization (ESO), an intuitive material-removal strategy based on stress or energy criteria. Level set methods were adapted to TO in the early 2000s by Allaire, Jouve, and Toader, enabled smooth boundary evolution and natural topological changes through implicit interface representation \cite{allaire2004topology,allaire2005structural}. Around the same period, topological sensitivity analysis provided rigorous tools for hole nucleation \cite{feijoo2005topological,novotny2007topological,burger2004incorporating}.

Since the 2000s, TO has expanded to multi-physics applications including thermal \cite{sigmund1997design,deaton2016stress,dbouk2017review}, fluid \cite{borrvall2003topology,lin2015topology,AlexandersenReview2020}, acoustic \cite{dilgen2019topology,duhring2008acoustic,yi2016comprehensive}, and electromagnetic \cite{sigmund2003systematic,lucchini2022topology} problems, alongside the incorporation of manufacturing constraints such as minimum feature size \cite{zhou2015minimum,sigmund2007morphology} and process-specific limitations \cite{liu2018current,mirzendehdel2016support,mirzendehdel2018strength,mirzendehdel2020topology,mirzendehdel2022topology}. The rise of additive manufacturing in the 2010s further enabled fabrication of complex optimized geometries, while advances in parallel computing, GPUs \cite{suresh2013efficient,challis2014high,herrero2021multi}, and more recently machine learning \cite{iyer2024pato,chandrasekhar2023frc,shin2023topology,padhy2025pilltop,chandrasekhar2021tounn,regenwetter2022deep,herrmann2024neural,woldseth2022use} have supported large-scale, high-fidelity, and accelerated optimization workflows.
\subsection{Density-based Topology Optimization}\label{sec:densityTO}

\begin{figure*}[!h]
	\centering\includegraphics[width=1.0\linewidth]{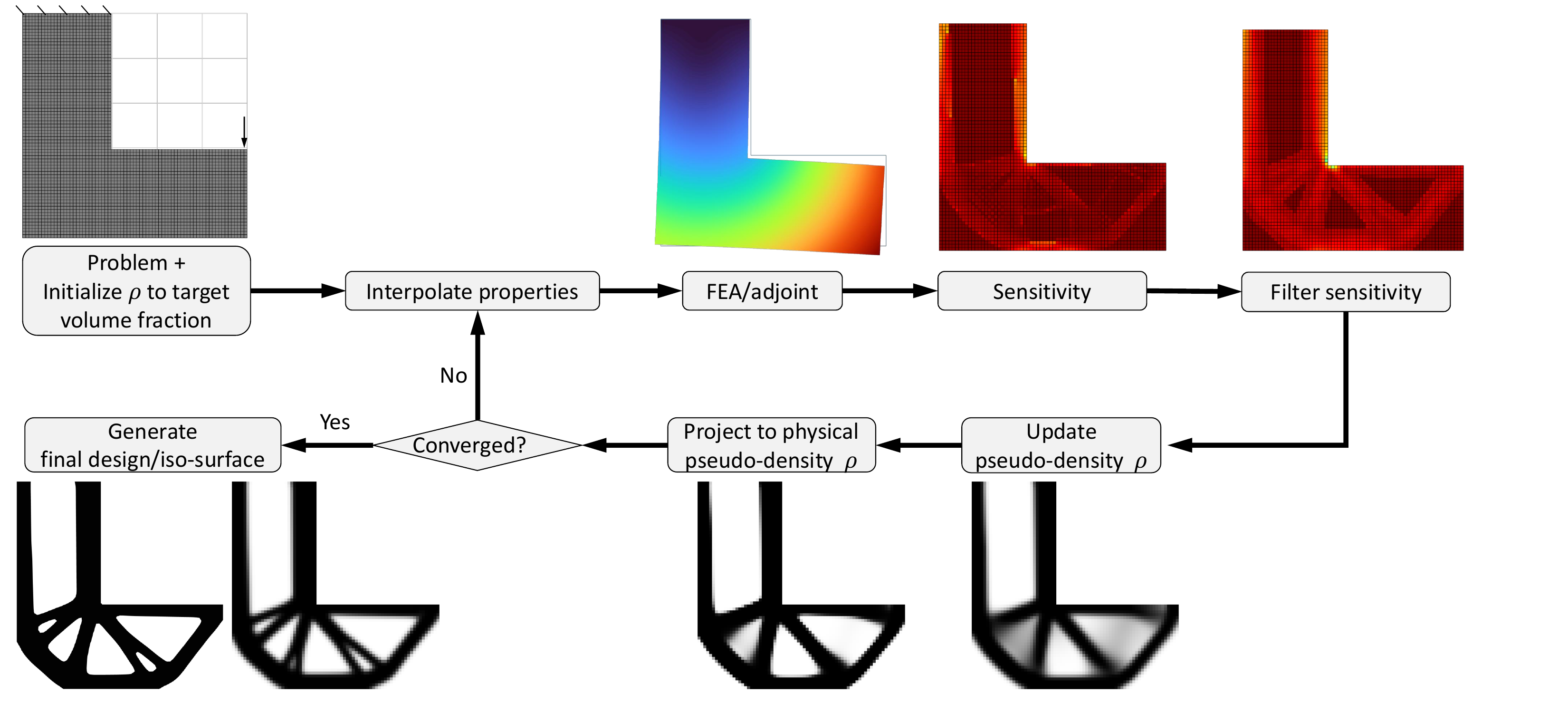}
	\caption{Density-based TO workflow.}
	\label{fig_SIMPworkflow}
\end{figure*}

Today, the density-based TO method stands as the most popular and extensively studied approach in the field. Its development over the past few decades reflects a collaborative effort to balance theoretical rigor with practical applicability, enabling the design of innovative structures that maximize performance and efficiency. The method's evolution underscores the significant impact of homogenization theory on modern engineering design and the ongoing advancements that continue to expand its capabilities.
By formulating the optimization problem based on effective constitutive properties, we are essentially \textit{relaxing} the condition that any point in a valid design must be either solid or void, which would require Integer Programming. In other words, 1) we expand the design space of feasible shapes to at least converge to a local optimum, 2) by replacing a discrete variable with a continuous one, we can gradient-based optimization algorithms, and 3) we can evaluate the macroscopic response through FEA more efficiently.

The general TO problem may be posed as:
\begin{subequations} \label{eq_simpTOProblem}
	\begin{align}
		\minimize\limits_{\boldsymbol{\rho}} \quad & \varphi (\bd;\boldsymbol{\rho}) \label{seq_simp_obj}\\  
		 \textrm{s.t.} \quad & g \coloneqq \sum\limits_{e}\rho_e v_e -  V^* \le 0\label{seq_simp_g}\\	
		&{R}_{el}(\mathbf{d;\boldsymbol{\rho}}) \coloneqq \mathbf{K}_{el} (\boldsymbol{\rho})\mathbf{d}-\mathbf{f}_{el}=\mathbf{0} \label{seq_simp_fea}\\
		& 0<\rho_{min} \le \rho_e \le 1; \quad \forall e \label{seq_simp_rho}
	\end{align}
\end{subequations}

 where $\varphi$ is the objective function to be minimized, \eqref{seq_simp_g} imposes a volume constraint, \eqref{seq_simp_fea} is the \textit{state equation} that governs the displacement $\bd$, \eqref{seq_simp_rho} imposes a \textit{box} or \textit{bound} constraint  on  $\rho_e$, and $\bK$ is the global stiffness matrix obtained through finite element assembly:  
\begin{equation} \label{eq_assemblyK}
	\bK = \sum_{assemble}\bk_e(\rho_e)
\end{equation}

The main concept behind this method is that each finite element is associated with a fictitious pseudo-density variable $\rho_e$, where $ 0 < \rho_{min} \le \rho_e \le 1$,  i.e., $\rho_e$ are the design variables; the design variables will be denoted by the set  $\boldsymbol{\rho}$. We subsequently construct the Lagrangian, and take the derivatives w.r.t. state variable and solve the adjoint problem, and take the derivative w.r.t. psuedo-densities to find the sensitivity. 
In this section, we will mainly focus on minimizing compliance:  
\begin{equation} \label{eq_simpCompliance}
\varphi := C = \bd^\top \bK\bd =\sum\limits_e \rho_e^p \bd_e^\top \bK_0 \bd_e
\end{equation} 
for which the compliance sensitivity is:
\begin{equation} \label{eq_simpComplianceSens}
D_{\rho} C = -\bd^\top \bK' \bd
\end{equation} 

The term $\bK'$ depends on the interpolation scheme.
In STORX, we have two options:
\begin{enumerate}
    \item Solid Isotropic Material with Penalization (SIMP): \mcode{interpolation = 'simp';}
    \item Rational Approximation of Material Properties (RAMP): \mcode{interpolation = 'ramp';}
\end{enumerate}

\textbf{SIMP.} The most widely used material interpolation scheme is SIMP, defined as
\begin{equation} \label{eq_Einterpolation}
	E(\rho_e) = \rho_e^p E_0 ,
\end{equation}
where $E_0$ is the Young's modulus of the base material, and $p$ is the penalization parameter. A common choice is $p=3$, which penalizes intermediate densities and promotes near solid-void designs. To compute \eqref{eq_simpComplianceSens}, $\bK'$ is
\begin{equation} \label{eq_dKdrho_SIMP}
\frac{\partial \bk_e}{\partial \rho_e}
= p \rho_e^{\,p-1} \bk_0 ,
\end{equation}
where $\bk_0$ is the elemental stiffness matrix of the base material.

\textbf{RAMP.} Another commonly used interpolation scheme is RAMP, defined as
\begin{equation} \label{eq_EinterpolationRAMP}
	E(\rho_e) = \frac{\rho_e}{1 + q(1-\rho_e)} E_0 ,
\end{equation}
where $q$ controls the nonlinearity of the interpolation. Larger values of $q$ make intermediate densities less favorable. To compute \eqref{eq_simpComplianceSens}, $\bK'$ using this scheme is
\begin{equation} \label{eq_dKdrho_RAMP}
\frac{\partial \bk_e}{\partial \rho_e}
= \frac{1+q}{\left(1+q(1-\rho_e)\right)^2}\bk_0 .
\end{equation}
RAMP is often used when smoother interpolation behavior or improved numerical performance is desired, particularly in more challenging constrained optimization problems.

Typically, to improve convergence in SIMP, the penalty parameter $p$ is gradually increased throughout the optimization process, for example from 2.0 to 3.0 in increments of 0.05. This strategy is commonly referred to as a \textit{continuation method}. For example, in setting up a problem in STORX, one may write:
\begin{lstlisting}
interpolation = 'simp';
penaltyStruct = struct('min',2,'max',3,'inc',0.05);
\end{lstlisting}
which will be passed to the FEA solver upon construction as
\begin{lstlisting}
solver = feaClass(...,interpolation,penaltyStruct)
\end{lstlisting}

After the sensitivities are computed, the design variables are updated at each iteration using an optimization algorithm that seeks to satisfy the KKT conditions. In STORX, three update schemes are available:
\begin{enumerate}
    \item Optimality Criteria (OC): \\
    \mcode{update = 'OC';}
    \item Method of Moving Asymptotes (MMA): \\
    \mcode{update = 'MMA';}
    \item Globally convergent MMA: \\
    \mcode{update = 'GCMMA';}
\end{enumerate}

which will be passed to the optimizer upon construction as
\begin{lstlisting}
topopt = topoptClass(...,update, ...);
\end{lstlisting}

\begin{figure*}[t]
  \centering 
  \includegraphics[width=\linewidth]{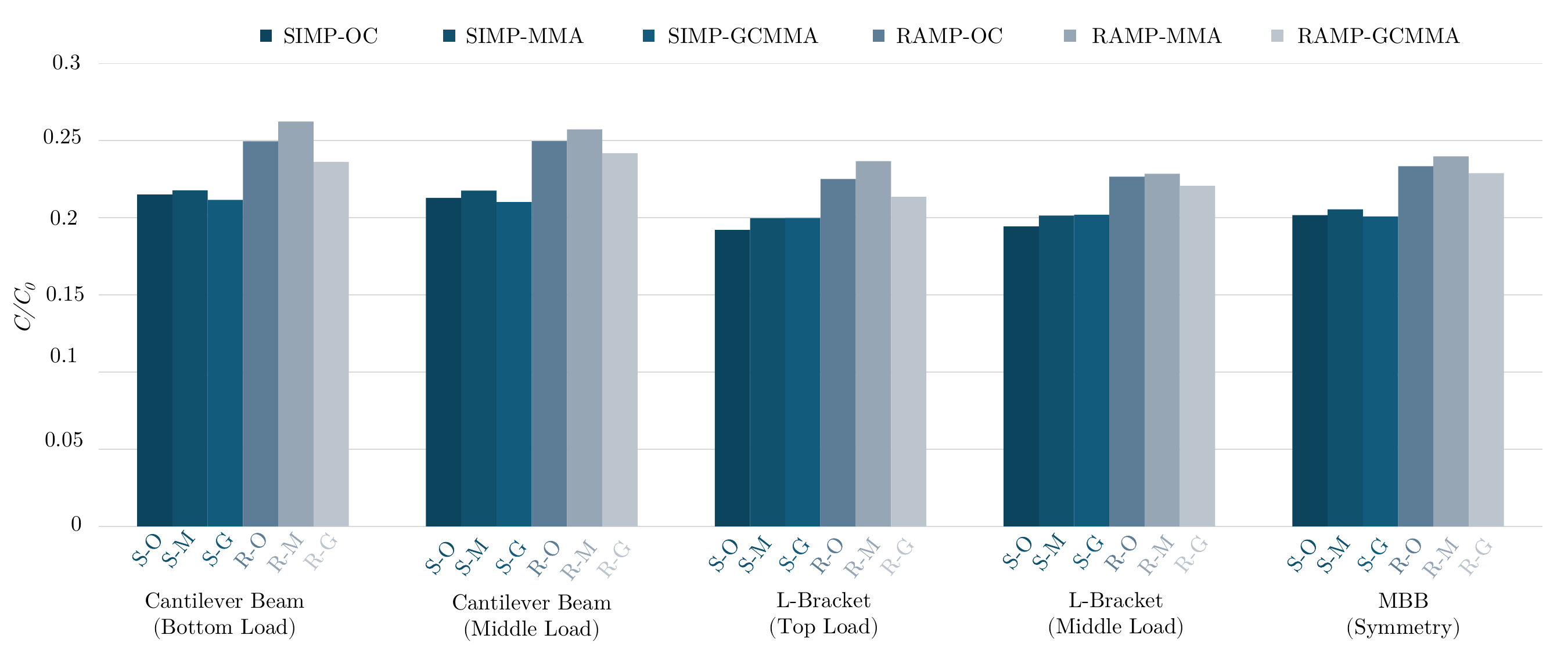}
  \caption{Comparison of density-based topology optimization results for compliance minimization under different interpolation schemes (SIMP and RAMP) and optimization update methods (OC, MMA, and GCMMA).}
  \label{fig:densityTO_summary}
\end{figure*}

\subsection{Comparative Study for Density-based TO}

\begin{table}[t]
\centering
\caption{Final performance metrics for the optimized designs obtained with Density-based optimization using SIMP and OC update for the illustrative examples: compliance \(C\), maximum displacement \(\delta_{\max}\), and maximum von Mises stress \(\sigma_{\mathrm{vm},\max}\).}
\label{tab:SIMP_example_metrics}

\small
\renewcommand{\arraystretch}{1.15}
\begin{tabular}{
>{\centering\arraybackslash}p{2.5cm}
>{\centering\arraybackslash}p{1cm}
>{\centering\arraybackslash}p{1.5cm}
>{\centering\arraybackslash}p{1cm}}
\toprule
\textbf{Example} &
{$C$} &
{$\delta_{\max}$} &
{$\sigma_{\text{vm},\max}$} \\
&
{(N.m)} &
{(m)} &
{(MPa)} \\
\midrule

{Cantilever Beam (bottom load)}            & {6.98} & {7.46e-05} & {6.94} \\ \addlinespace
{Cantilever Beam (middle load)} & {10.6} & {8.50e-05} & {8.83} \\
\addlinespace
{L-bracket (top load)}     & {22.9} & {2.46e-04} & {21.0} \\
\addlinespace
{L-bracket (mid load)}     & {21.7} & {2.42e-04} & {23.0} \\
\addlinespace
{MBB (symmetry)}     & {15.0} & {1.64e-04} & {6.98} \\
\bottomrule
\end{tabular}
\end{table}

 \begin{figure}[t]
	\centering
	\begin{subfigure}[b]{0.45\linewidth}
		\centering
		\includegraphics[width=0.9\linewidth]{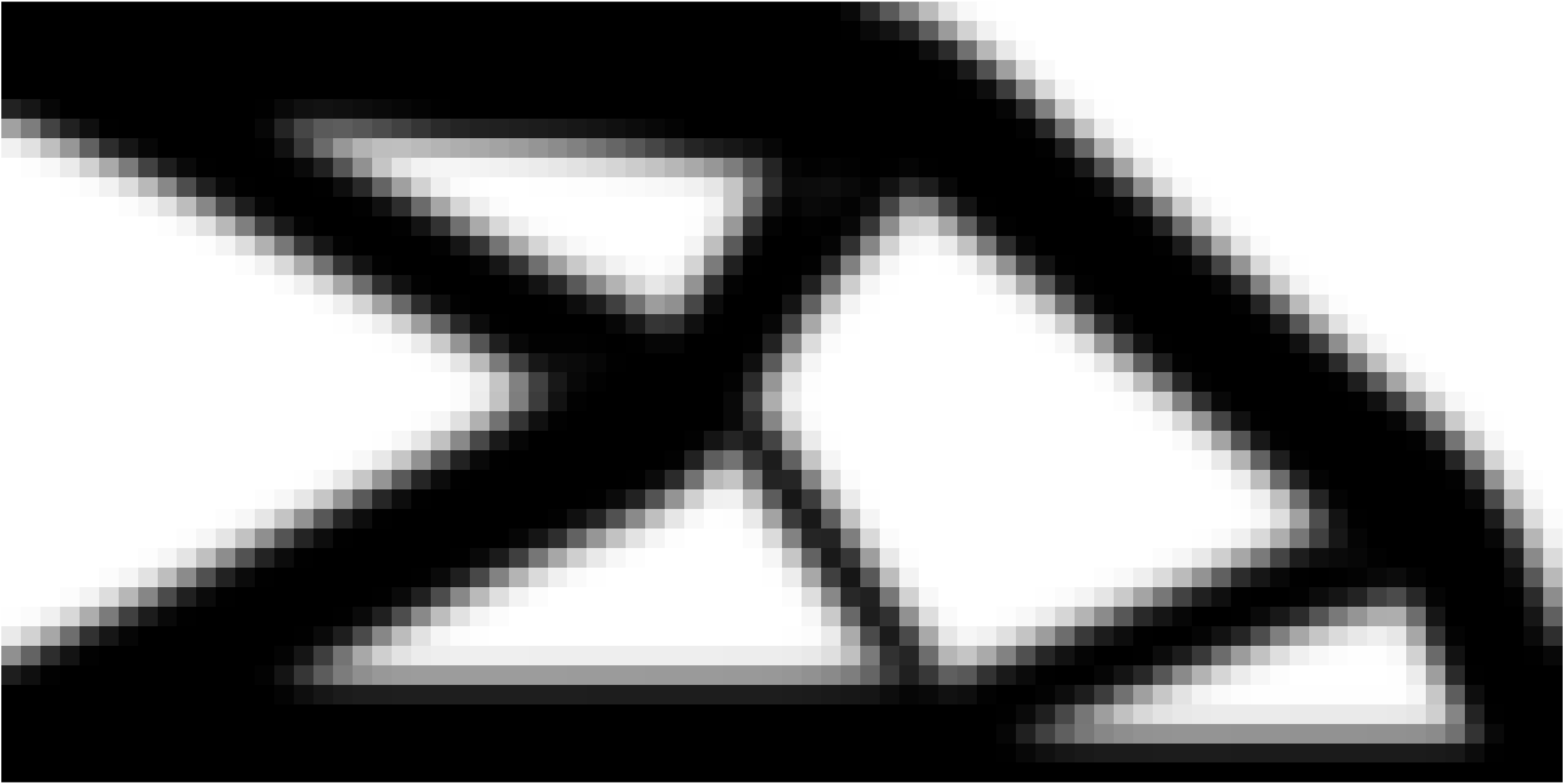}
		\caption{}
	\end{subfigure}
	\begin{subfigure}[b]{0.45\linewidth}
		\centering
		\includegraphics[width=0.9\linewidth]{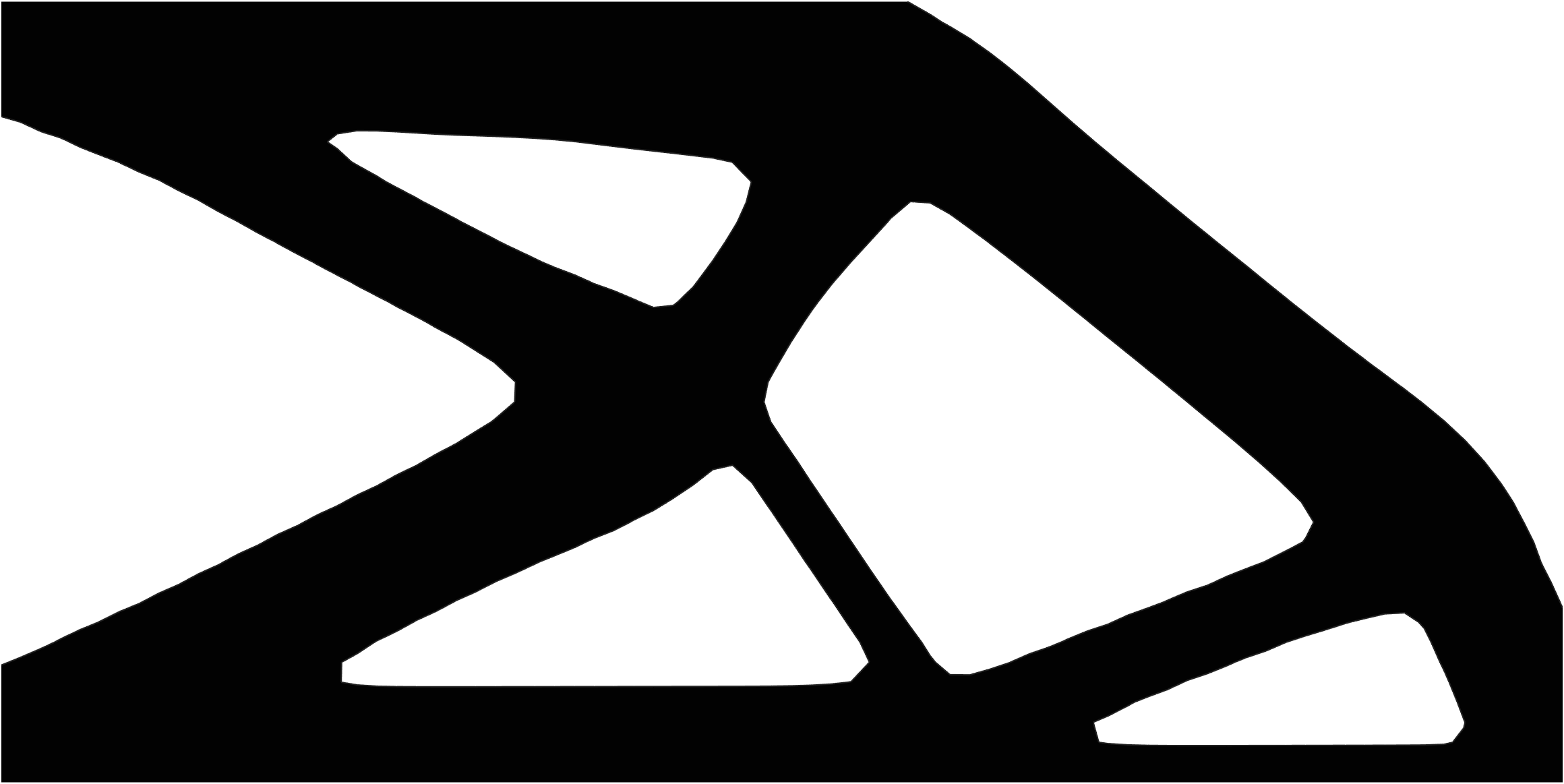}
		\caption{}
	\end{subfigure}

    \begin{subfigure}[b]{0.45\linewidth}
		\centering
		\includegraphics[width=0.9\linewidth]{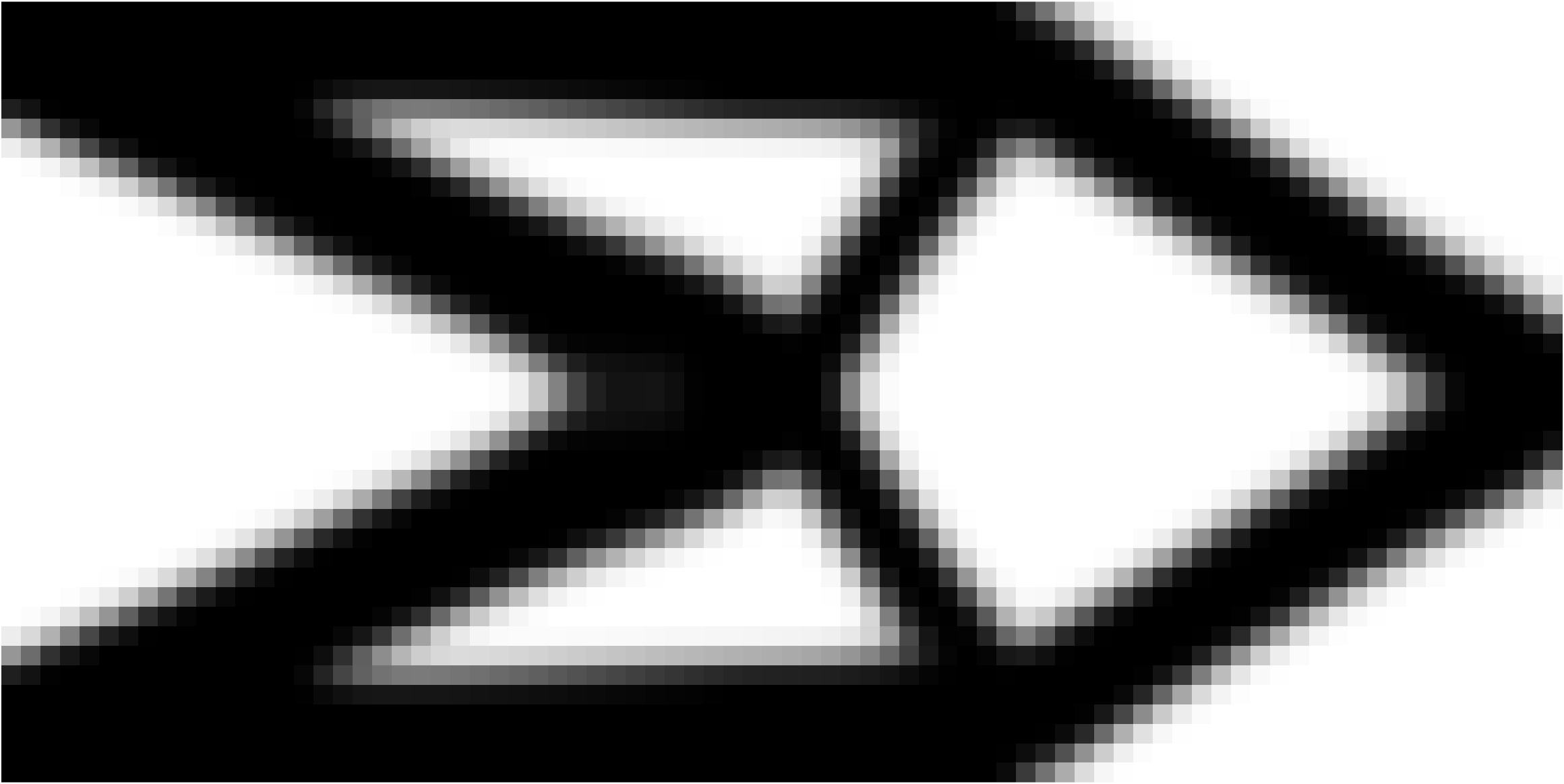}
		\caption{}
	\end{subfigure}
	\begin{subfigure}[b]{0.45\linewidth}
		\centering
		\includegraphics[width=0.9\linewidth]{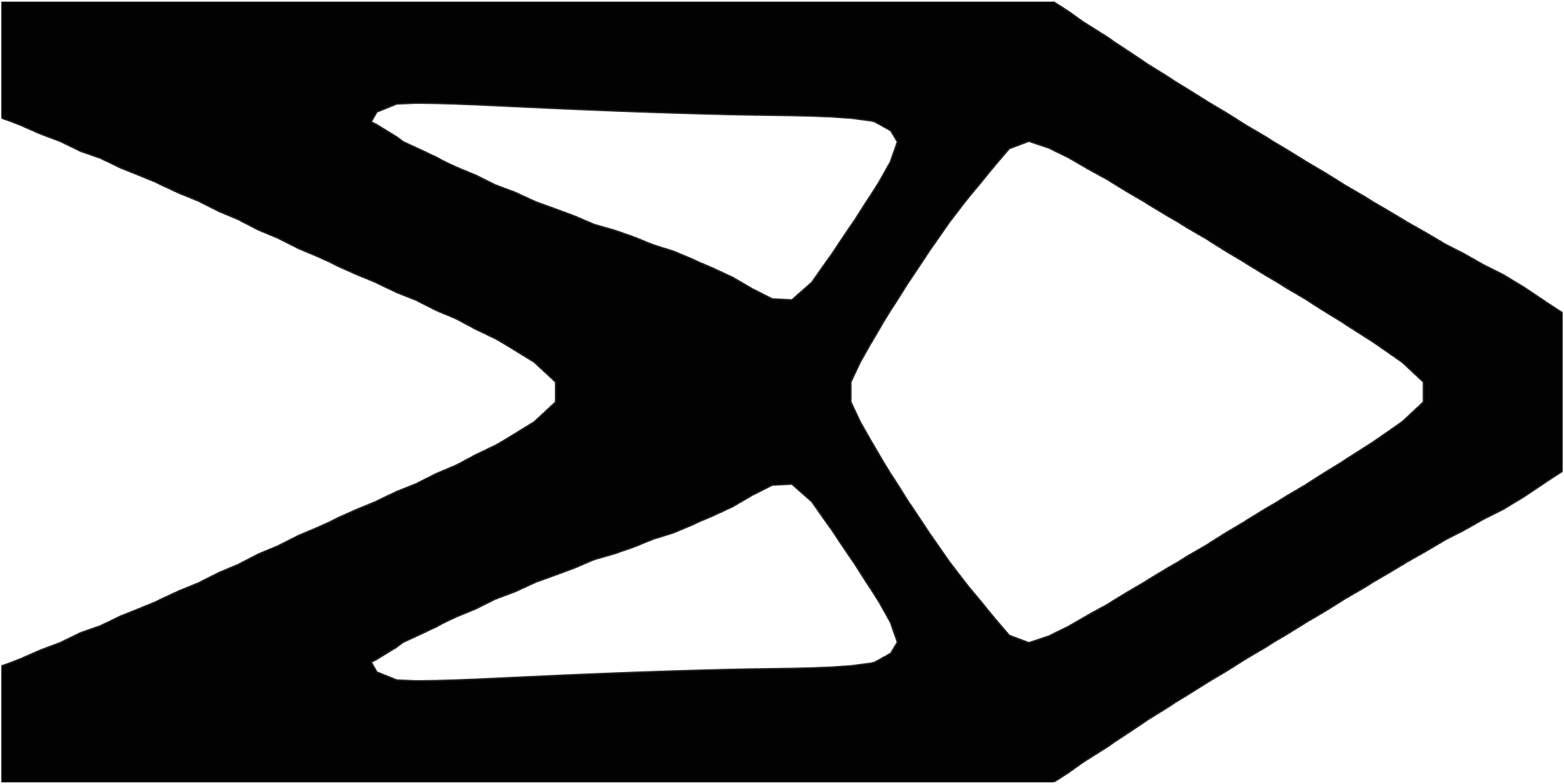}
		\caption{}
	\end{subfigure}

    \begin{subfigure}[b]{0.45\linewidth}
		\centering
		\includegraphics[width=0.9\linewidth]{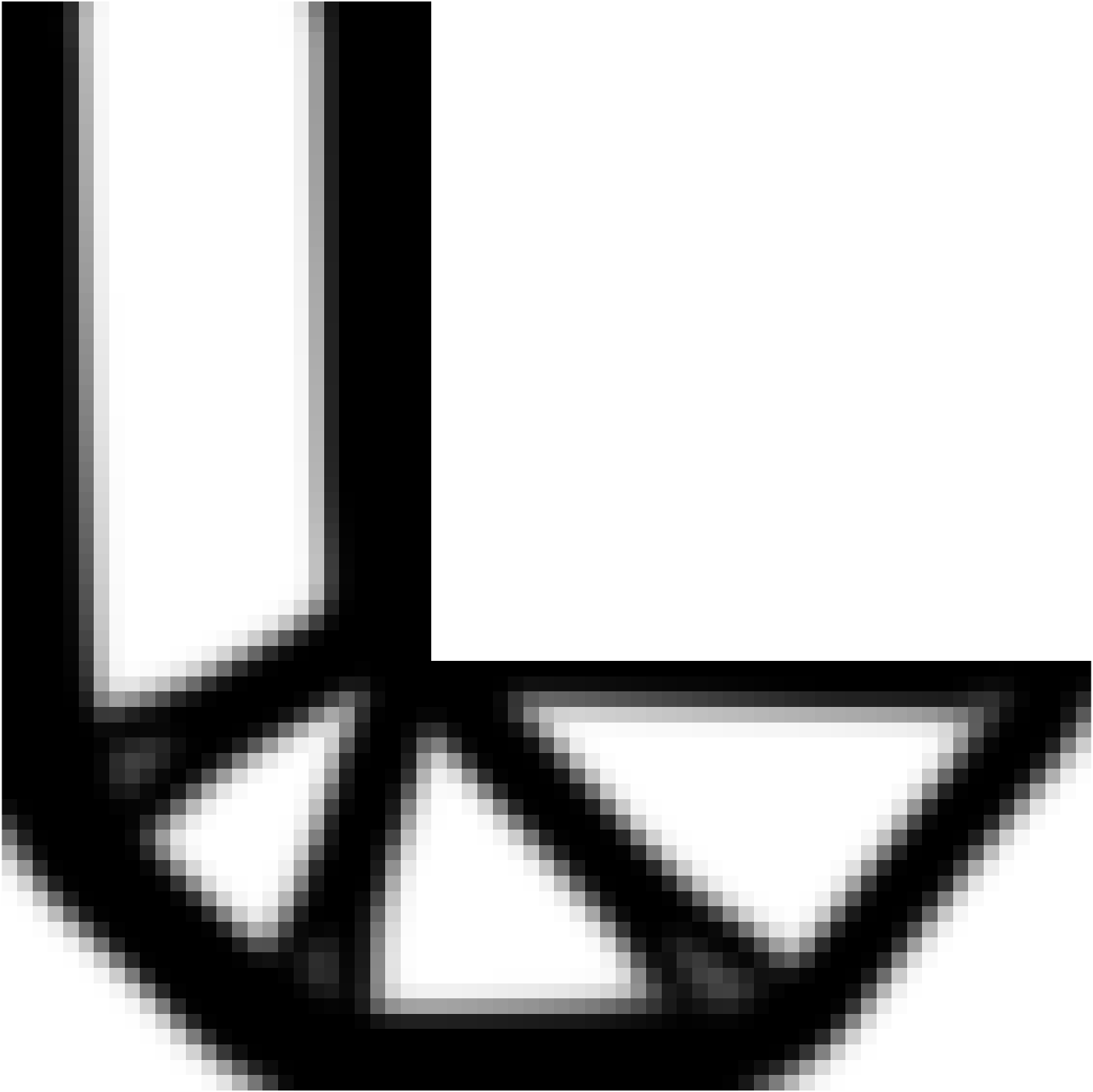}
		\caption{}
	\end{subfigure}
	\begin{subfigure}[b]{0.45\linewidth}
		\centering
		\includegraphics[width=0.9\linewidth]{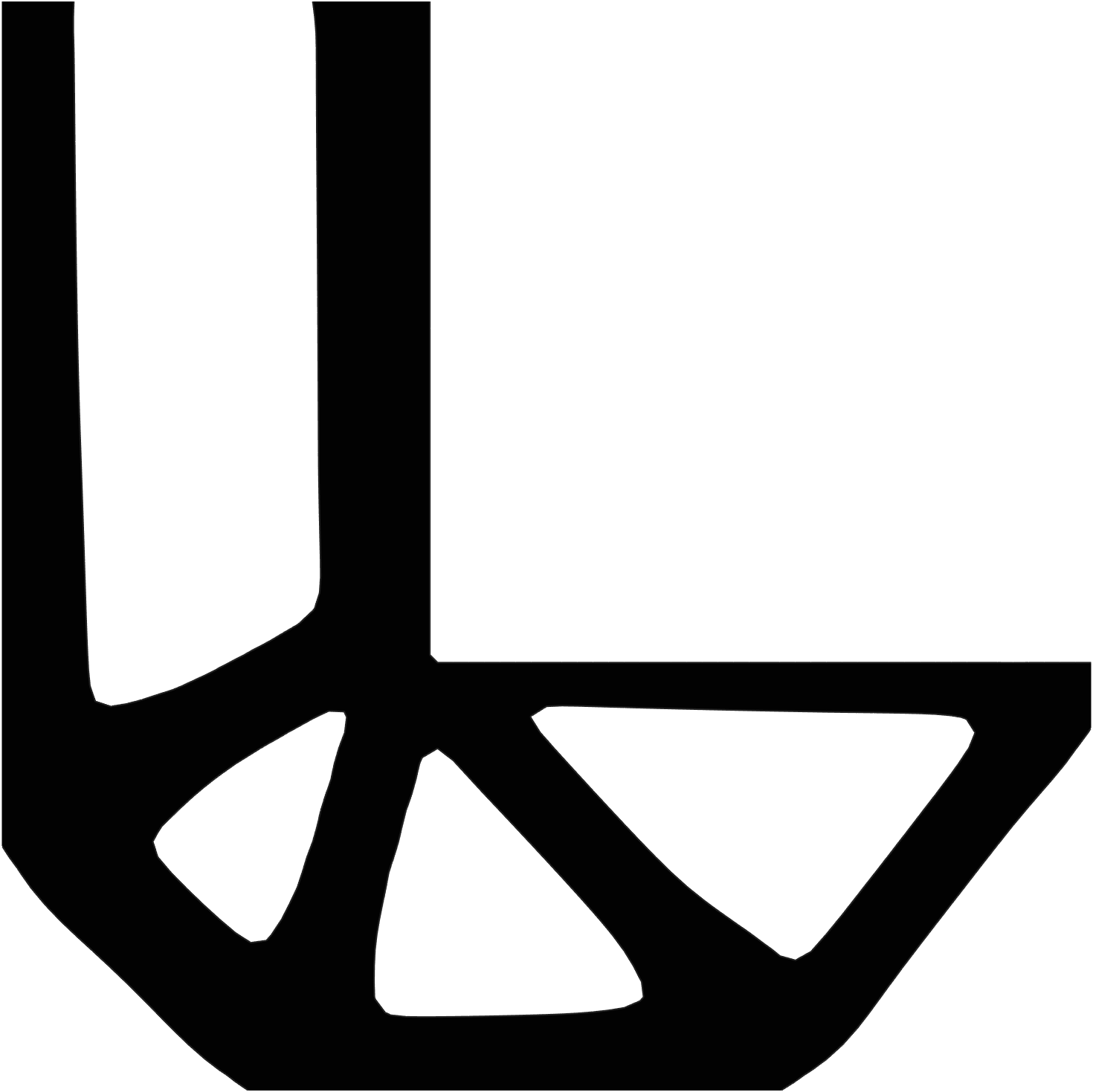}
		\caption{}
	\end{subfigure}

    \begin{subfigure}[b]{0.45\linewidth}
		\centering
		\includegraphics[width=0.9\linewidth]{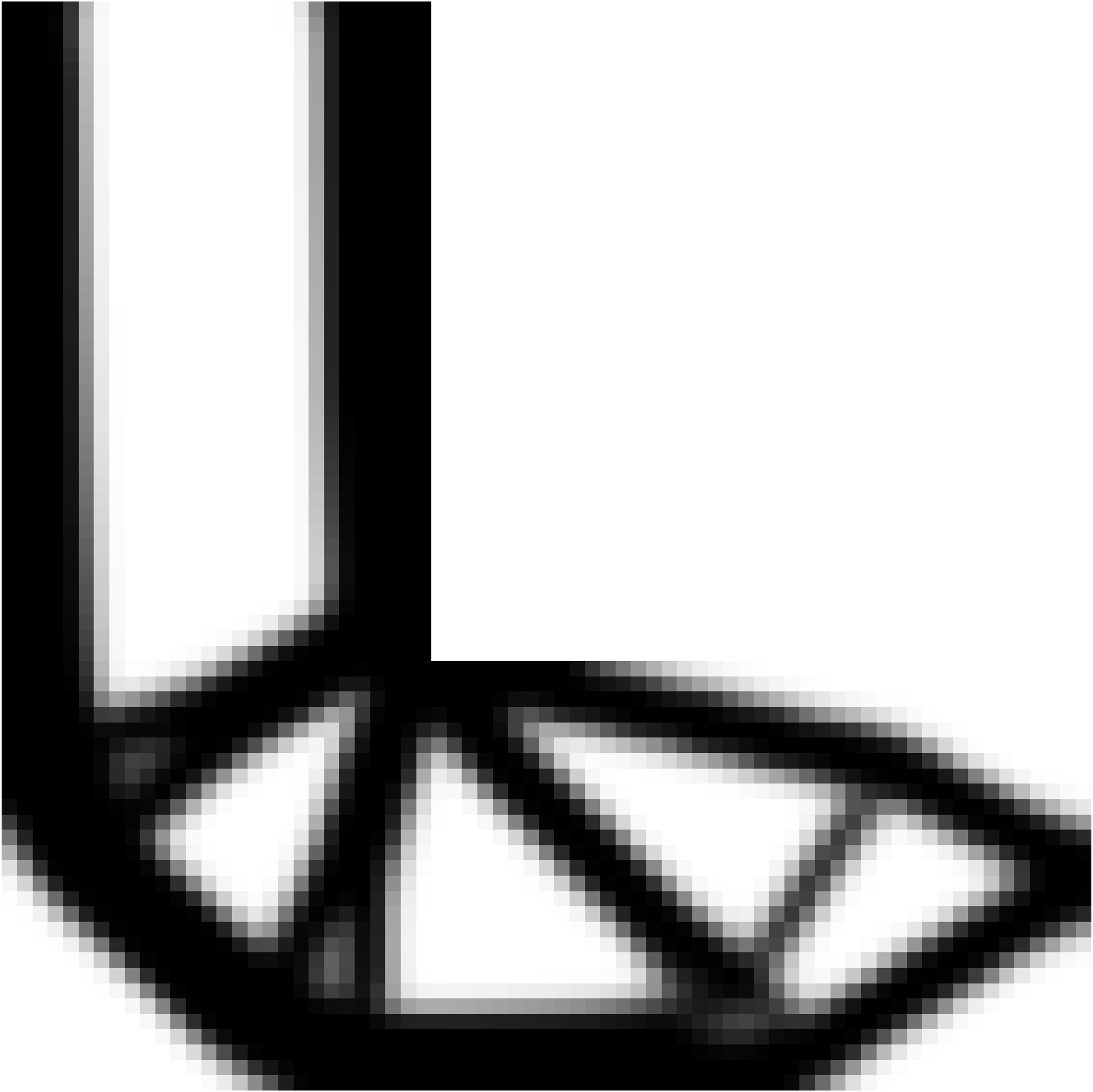}
		\caption{}
	\end{subfigure}
	\begin{subfigure}[b]{0.45\linewidth}
		\centering
		\includegraphics[width=0.9\linewidth]{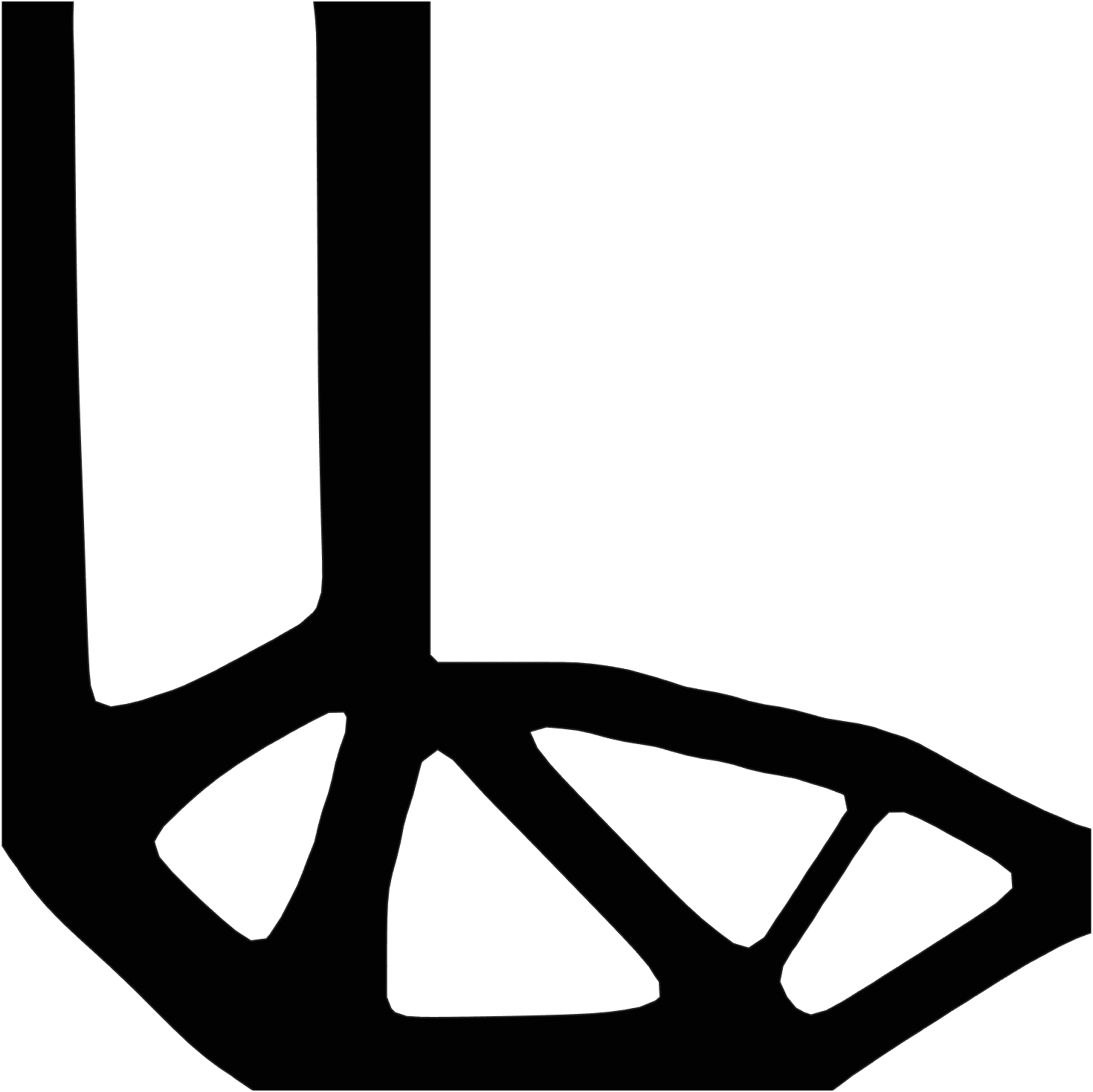}
		\caption{}
	\end{subfigure}

    \begin{subfigure}[b]{0.45\linewidth}
		\centering
		\includegraphics[width=0.9\linewidth]{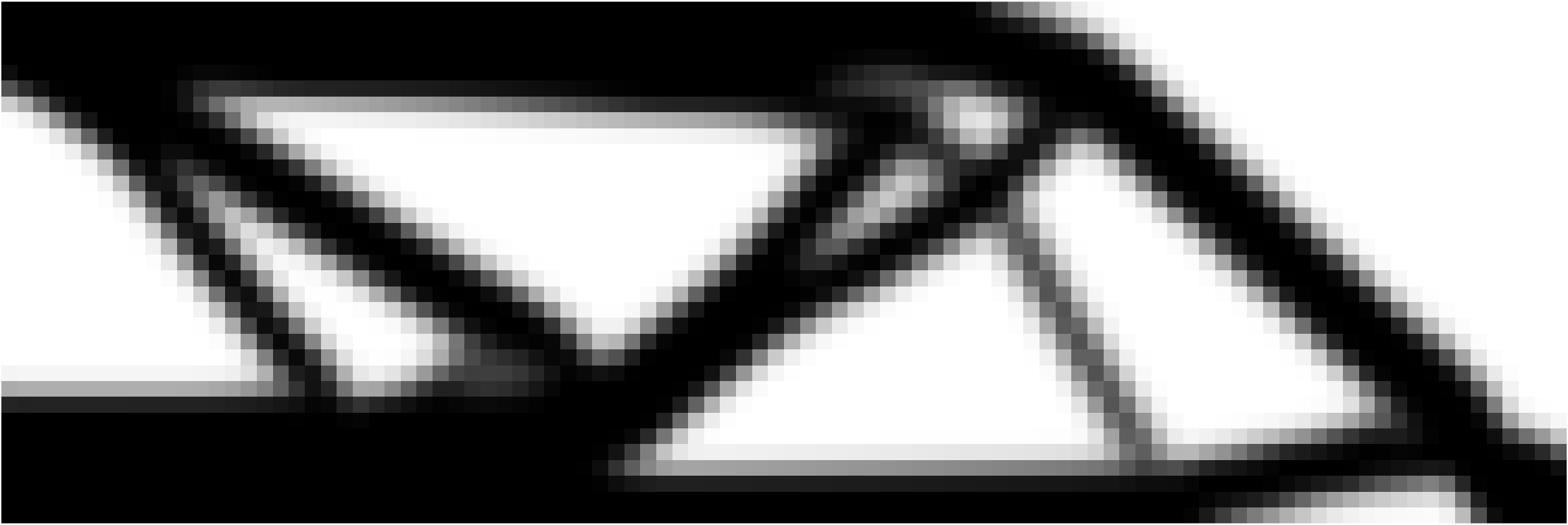}
		\caption{}
	\end{subfigure}
	\begin{subfigure}[b]{0.45\linewidth}
		\centering
		\includegraphics[width=0.9\linewidth]{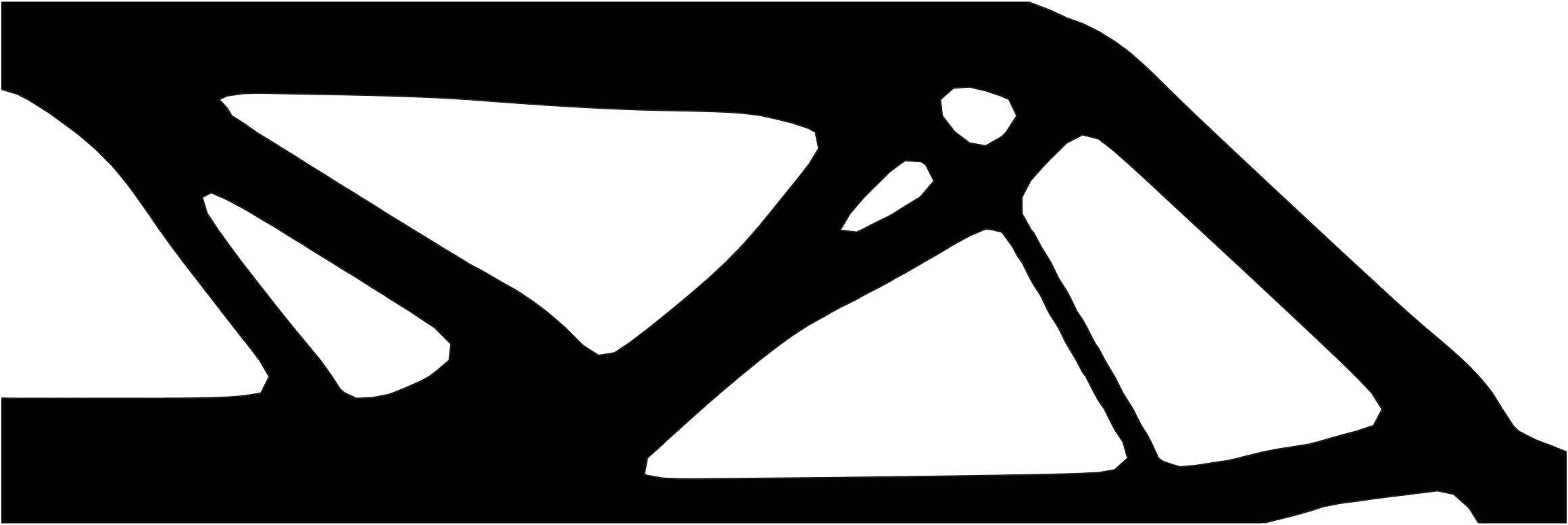}
		\caption{}
	\end{subfigure}
	\caption{Density-based TO via the OC update algorithm for the illustrative examples.}
	\label{fig_example_density}
\end{figure}

Figure~\ref{fig:densityTO_summary} compares density-based topology optimization results for compliance minimization under two interpolation schemes (SIMP and RAMP) and three update strategies (OC, MMA, and GCMMA), highlighting the trade-offs between computational cost and final objective value across common algorithmic choices. Building on this summary, we evaluate the density-based TO module in \textsc{STORX} on a set of standard two-dimensional {illustrative examples shown} in Fig.~\ref{fig_example_density}. In these examples the design variable is an element-wise pseudo-density field, which is mapped to material stiffness through the selected interpolation and updated iteratively subject to a volume constraint.

Table~\ref{tab:SIMP_example_metrics} reports compliance \(C\), maximum displacement \(\delta_{\max}\), and maximum von Mises stress \(\sigma_{\mathrm{vm},\max}\) for the designs obtained using SIMP with the OC update, which we use as a baseline configuration in the subsequent examples.

For each example, Fig.~\ref{fig_example_density} reports both the optimized pseudo-density field and the corresponding extracted solid/void geometry obtained by thresholding the density field, providing a consistent visual comparison of the resulting topologies.

For instance, to run the cantilever beam example, we first select the appropriate FEA and TO classes and set general parameters:
\begin{lstlisting}
%% Solvers
feaClass = @fea2d_elasticity;
topoptClass = @density2d_elasticity;
%% General Parameters
vectorize = true;
\end{lstlisting}

Next, we define the density-based TO parameters where we use SIMP interpolation with constant $p=3$ and OC update scheme up to 300 maximum iterations:
\begin{lstlisting}
%% Optimizer Parameters
interpolation = 'simp'; 
update = 'OC';
penaltyStruct = struct('min',3,'max',3,'inc',0.0);
maxNumIters = 300;
\end{lstlisting}

And construct the FEA solver:
\begin{lstlisting}
%% Problem Definition
brep = 'CantileverBeam.brep'; % geometry
numElements = 3200; % mesh
material.E = 100e9; material.nu = 0.3; % material
numScenarios = 1;
%% Construct FEA Solver
solver = feaClass(brep,numElements,material,vectorize,numScenarios, ...
    interpolation,penaltyStruct); % call superclass
solver = solver.fixEdge(5);
solver = solver.applyYForceOnEdge(2,-1e5);
solver = solver.preProcess(); % FEA pre-processing
\end{lstlisting}

The compliance objective, volume constraint at 0.5 target fraction, and minimum feature size filter with 1.5 voxel size radius are imposed via:
\begin{lstlisting}
%% Objective and Constraints
objective = densityComplianceElasticity(solver);
volumeFraction = 0.5;
constraints  = {volume(solver, volumeFraction)};
% manufacturing constraints
rmin = 1.5;
mfgConstraints = {minimumFeatureSize_dist(solver, rmin)}; 
\end{lstlisting}

Finally, we construct the optimizer and run the optimization:
\begin{lstlisting}
%% Construct Optimizer
topopt = topoptClass(solver,objective,constraints, ...
mfgConstraints,update,maxNumIters);
%% Optimize
topopt = topopt.optimize();
\end{lstlisting}

\begin{figure}[t]
  \centering 
  \includegraphics[width=\linewidth]{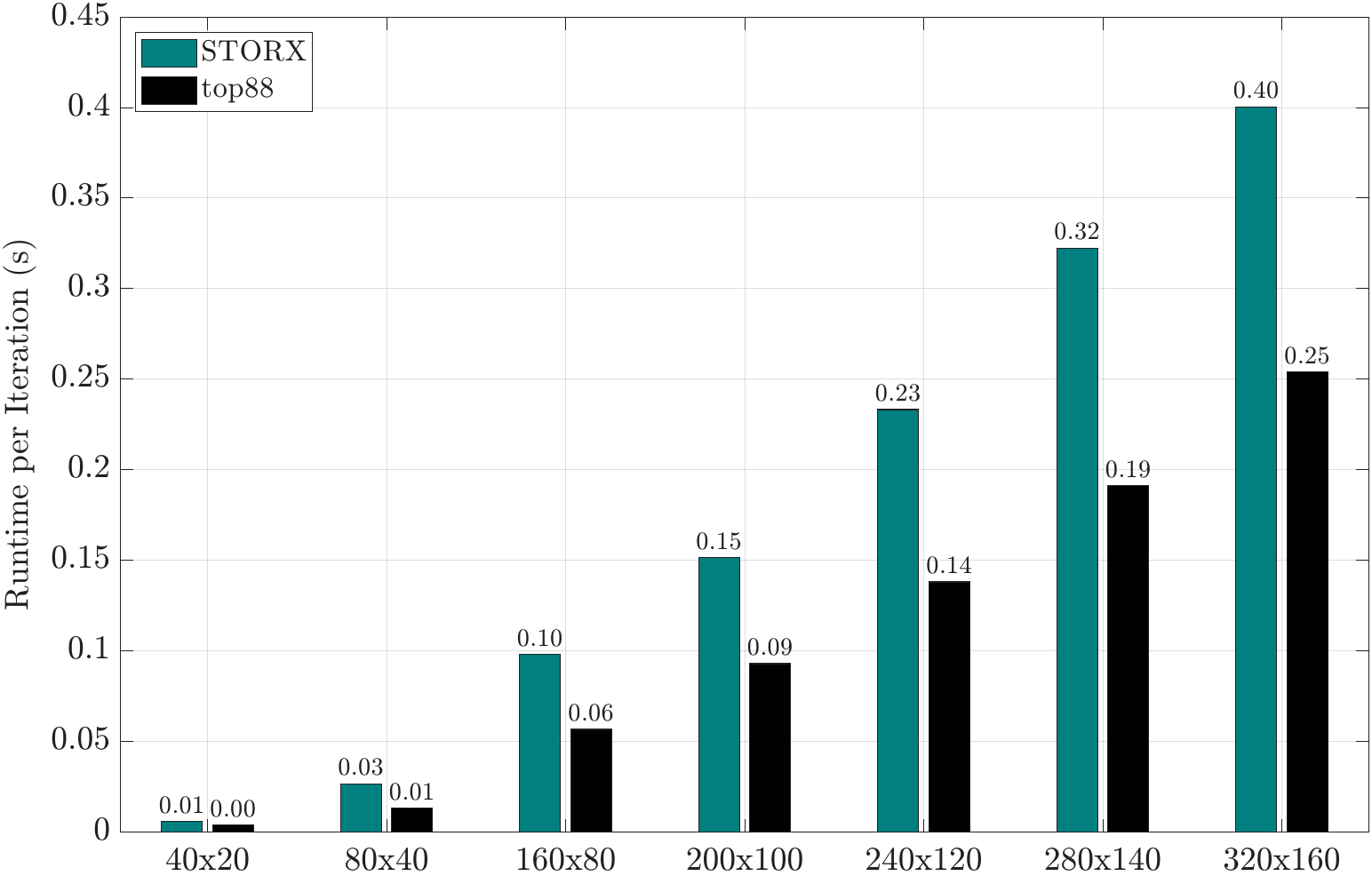}
  \caption{Runtime/iteration comparison of STORX and top88 code for the cantilever beam example at different resolutions.}
  \label{fig:benchmark_top88_storx}
\end{figure}
Figure~\ref{fig:benchmark_top88_storx} compares the runtime per iteration of the general-purpose STORX framework against the highly optimized \mcode{top88} \cite{Andreassen2011} SIMP code for the cantilever beam example across increasing mesh resolutions and without considering pre-processing and plotting times. As expected, the specialized \mcode{top88} implementation achieves lower computational cost due to its compact, problem-specific structure and minimal abstraction overhead. In contrast, STORX incurs additional runtime from its object-oriented architecture, modular solver interfaces, and extensible design. At medium resolutions, this overhead results in approximately 65-68\% longer runtimes per iteration which drops to about 60\% at the highest resolution of $320 \times 160$. Nevertheless, the performance remains competitive: for a problem with approximately 50{,}000 elements, each iteration (excluding visualization) requires about 0.4 seconds. All experiments were performed on an Intel(R) Core(TM) Ultra 9 285K (3.70~GHz) system with 128~GB RAM. While there is a measurable performance trade-off, STORX enables unified support for arbitrary geometries, multiple physics modules, and interchangeable optimization algorithms within a consistent computational framework.
\subsection{Level-Set Topology Optimization}\label{sec:LSTO}
One of the early methods of TO using level sets was based on shape optimization using the HJE. In this section, we will discuss level-set-based TO.

\subsubsection{Standard Hamilton-Jacobi}\label{sec:stHJE}

As was previously mentioned in Section \ref{sec:LSSO}, the standard HJE does not allow the nucleation of holes but can accommodate the merging of holes.  Thus, to find an optimized design within a design domain $D$ ($\Omega \subset D$) with internal voids, we can create more holes in the initial design and then merge them, if necessary. One simple way to create an LSF with periodic holes is as follows:
\begin{equation} \label{eq_holeInit}
	\psi(x,y) = \textbf{1}_D(x,y) \cos(  \dfrac{n_x\pi x}{l_x}) \cos(\dfrac{n_y\pi y}{l_y})
\end{equation}
where  $\textbf{1}_D(x,y): D\rightarrow \{0,1\}$ is the indicator function, which is 1 if $(x,y) \in D$ and 0 otherwise. $n_x$ and $n_y$ are number of holes along $x$ and $y$, while $l_x$ and $l_y$ are the dimensions of the bounding box along $x$ and $y$. In STORX, this function is implemented as follows:
\begin{lstlisting}
function obj = initializeHoles(obj,hx,hy,r)
    % Generate initial holes using the function:
    % Z = cos(X)*cos(Y)
    ...
    [X,Y] = meshgrid(1:obj.m_solver.m_nx,1:obj.m_solver.m_ny);
    Z = cos(2*X*hx*pi/obj.m_solver.m_nx).*cos(2*Y*hy*pi/obj.m_solver.m_ny)+(1-r);
    Z = Z.*obj.m_solver.m_existingElems;
    obj.m_x = zeros(obj.m_solver.m_ny,obj.m_solver.m_nx);
    obj.m_x(Z > 0) = 1;
    % Filter the density to ensure retain elements are not removed
    obj = obj.filterDensity();
    ...
end      
\end{lstlisting}

Compared to the cantilever beam example considered in Section \ref{sec:LSSO}, the main difference in setting up the problem in the domain initialization, where we introduce $4 \times 2$ holes with 0.5 radius:
\begin{lstlisting}
    nHolesX = 4; nHolesY = 2; r0 = 0.5;
\end{lstlisting}
To ensure that the critical edges (e.g., subject to load) remain solid after the initial holes are introduced, we impose a retain manufacturing constraint, here on the edge $\#2$:
\begin{lstlisting}
% manufacturing constraints
mfgConstraints = {minimumFeatureSize_conv(solver)
    retain_levelset(solver,2) };
\end{lstlisting}

 \begin{figure}[t]
	\centering
	\begin{subfigure}[b]{0.55\linewidth}
		\centering
		\includegraphics[width=\linewidth]{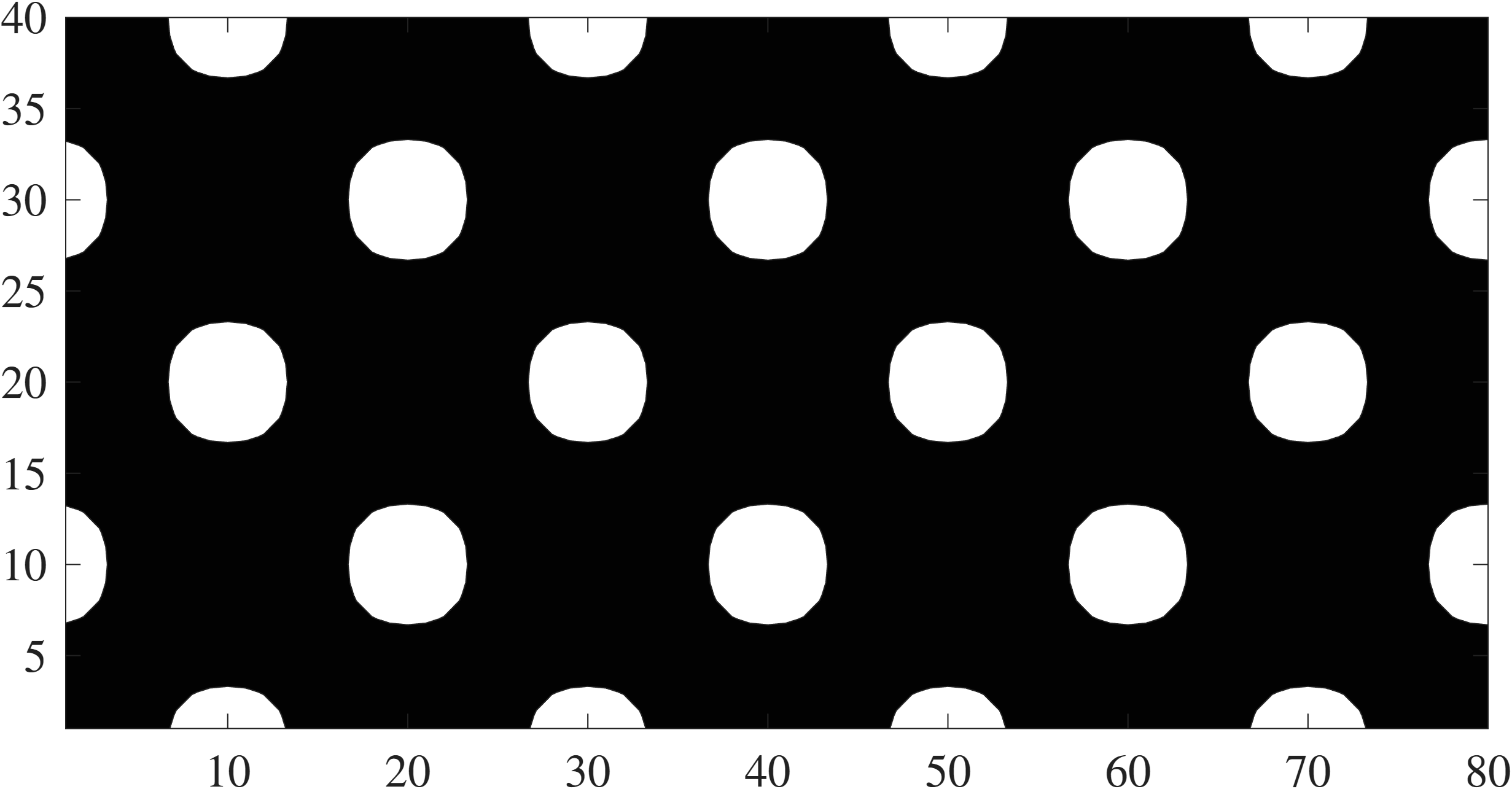}
		\caption{Cantilever beam}
		\label{fig_cantilever_initialHoles}
	\end{subfigure}
	\begin{subfigure}[b]{0.43\linewidth}
		\centering
		\includegraphics[width=\linewidth]{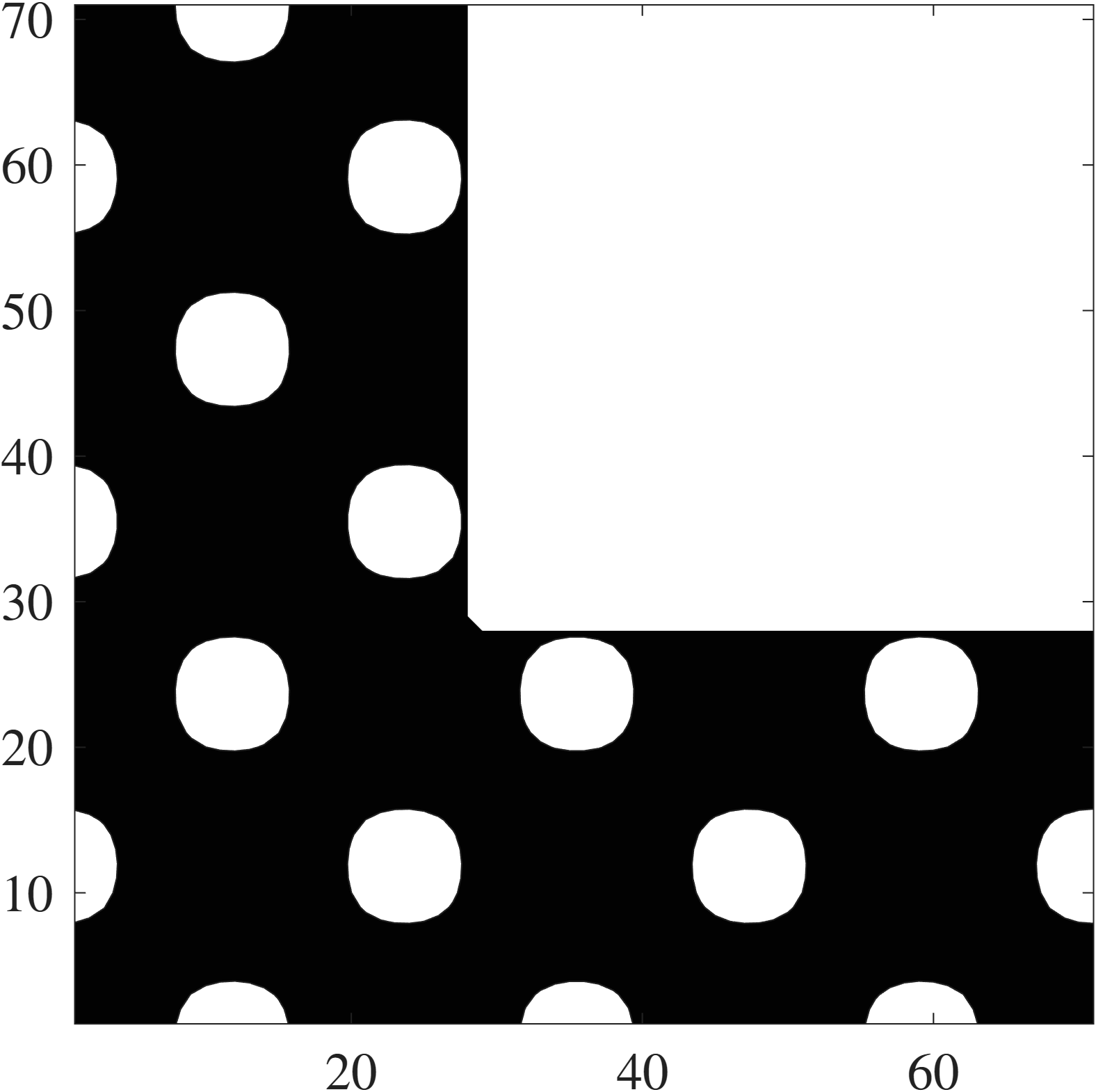}
		\caption{L-bracket}
	\end{subfigure}
	\begin{subfigure}[b]{0.7\linewidth}
		\centering
		\includegraphics[width=0.85\linewidth]{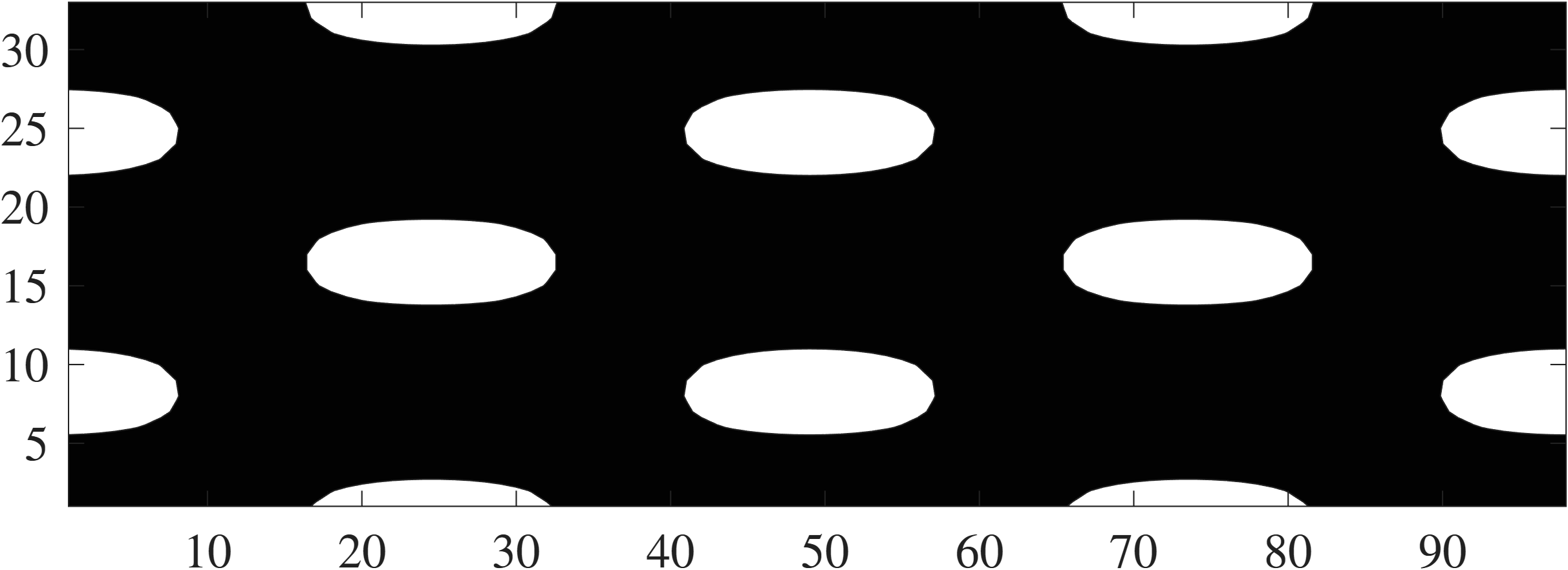}
		\caption{MBB (symmetry)}
	\end{subfigure}
	\caption{Initialization for TO via standard HJE.}
	\label{fig_initialHoles_stHJE}
\end{figure} 

\begin{table}[!h]
\centering
\caption{Final performance metrics for the optimized designs obtained with \textit{level-set TO using standard HJE}: compliance \(C\), maximum displacement \(\delta_{\max}\), and maximum von Mises stress \(\sigma_{\mathrm{vm},\max}\).}
\label{tab:stHJE_example_metrics}

\small
\renewcommand{\arraystretch}{1.15}
\begin{tabular}{
>{\centering\arraybackslash}p{2.5cm}
>{\centering\arraybackslash}p{1cm}
>{\centering\arraybackslash}p{1.5cm}
>{\centering\arraybackslash}p{1cm}}
\toprule
\textbf{Example} &
{$C$} &
{$\delta_{\max}$} &
{$\sigma_{\text{vm},\max}$} \\
&
{(N.m)} &
{(m)} &
{(MPa)} \\
\midrule

{Cantilever Beam (bottom load)}            & {6.38} & {6.85e-05} & {2.34} \\ \addlinespace
{Cantilever Beam (middle load)} & {9.68} & {7.84e-05} & {2.83} \\
\addlinespace
{L-bracket (top load)}     & {21.2} & {2.27e-04} & {25.2} \\
\addlinespace
{L-bracket (mid load)}     & {19.7} & {2.21e-04} & {36.1} \\
\addlinespace
{MBB (symmetry)}     & {14.0} & {1.53e-04} & {3.98} \\
\bottomrule
\end{tabular}
\end{table}

 \begin{figure}[!h]
	\centering
	\begin{subfigure}[b]{0.45\linewidth}
		\centering
		\includegraphics[width=0.9\linewidth]{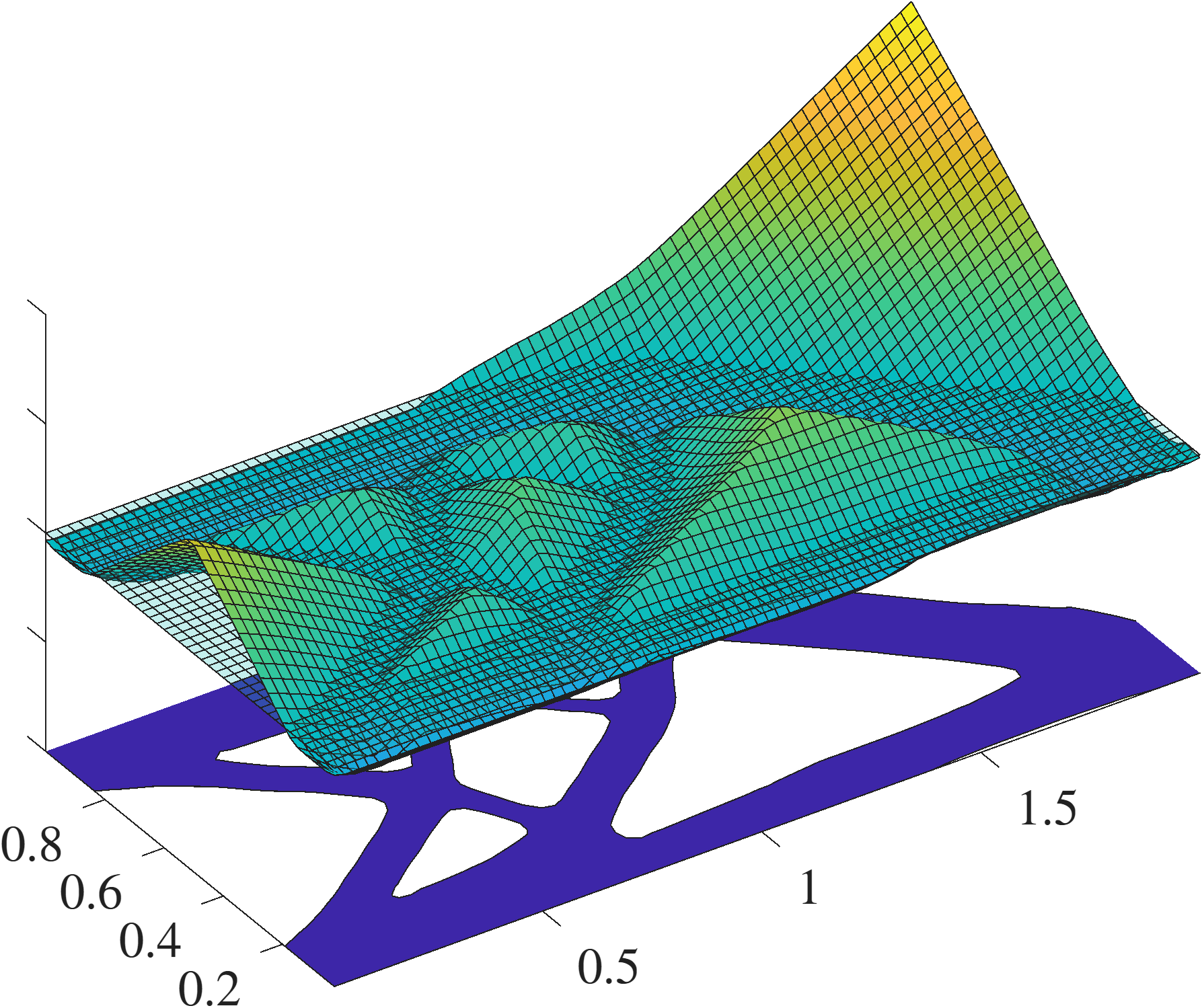}
		\caption{}
	\end{subfigure}
	\begin{subfigure}[b]{0.45\linewidth}
		\centering
		\includegraphics[width=0.9\linewidth]{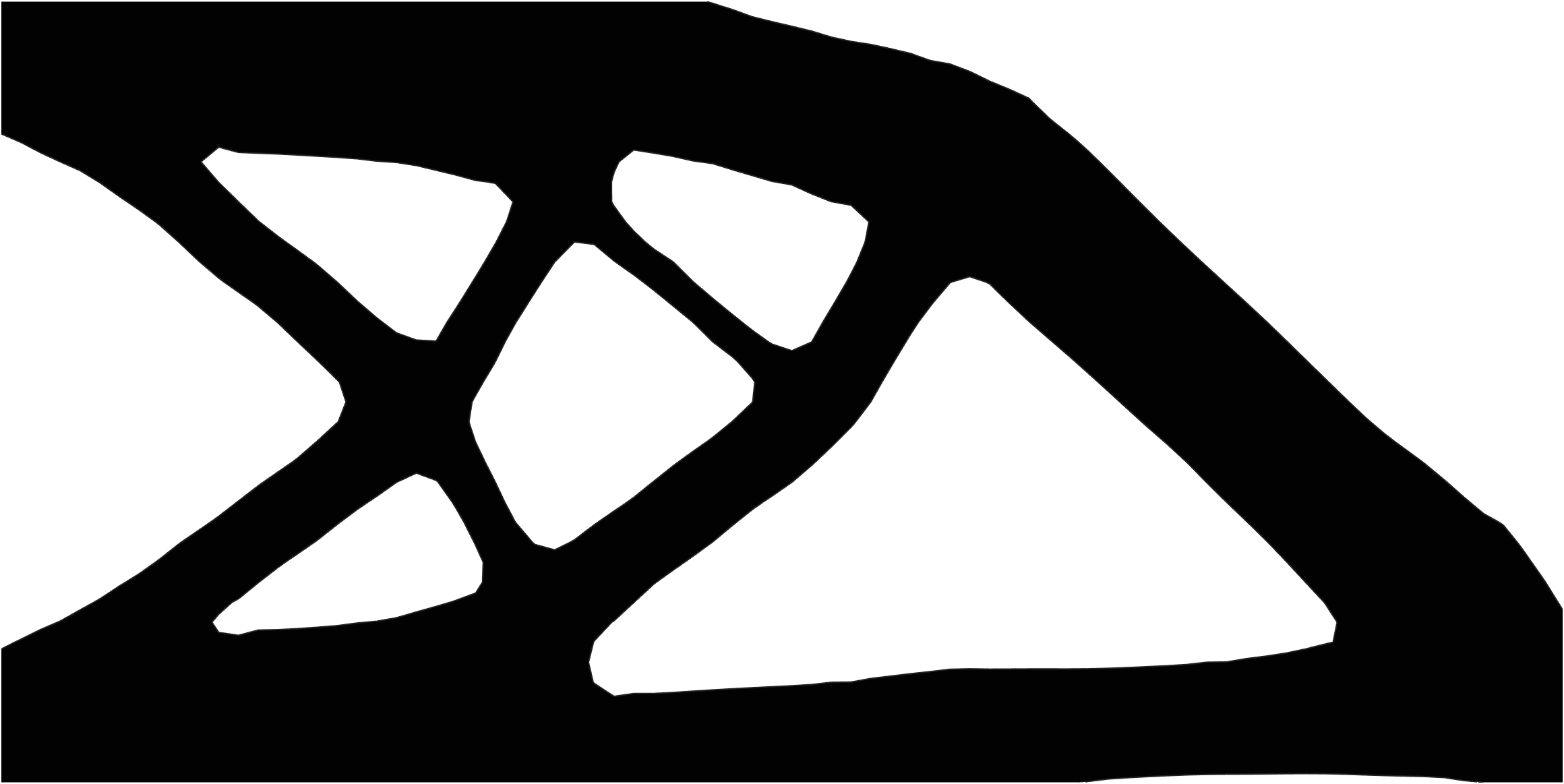}
		\caption{}
	\end{subfigure}

    \begin{subfigure}[b]{0.45\linewidth}
		\centering
		\includegraphics[width=0.9\linewidth]{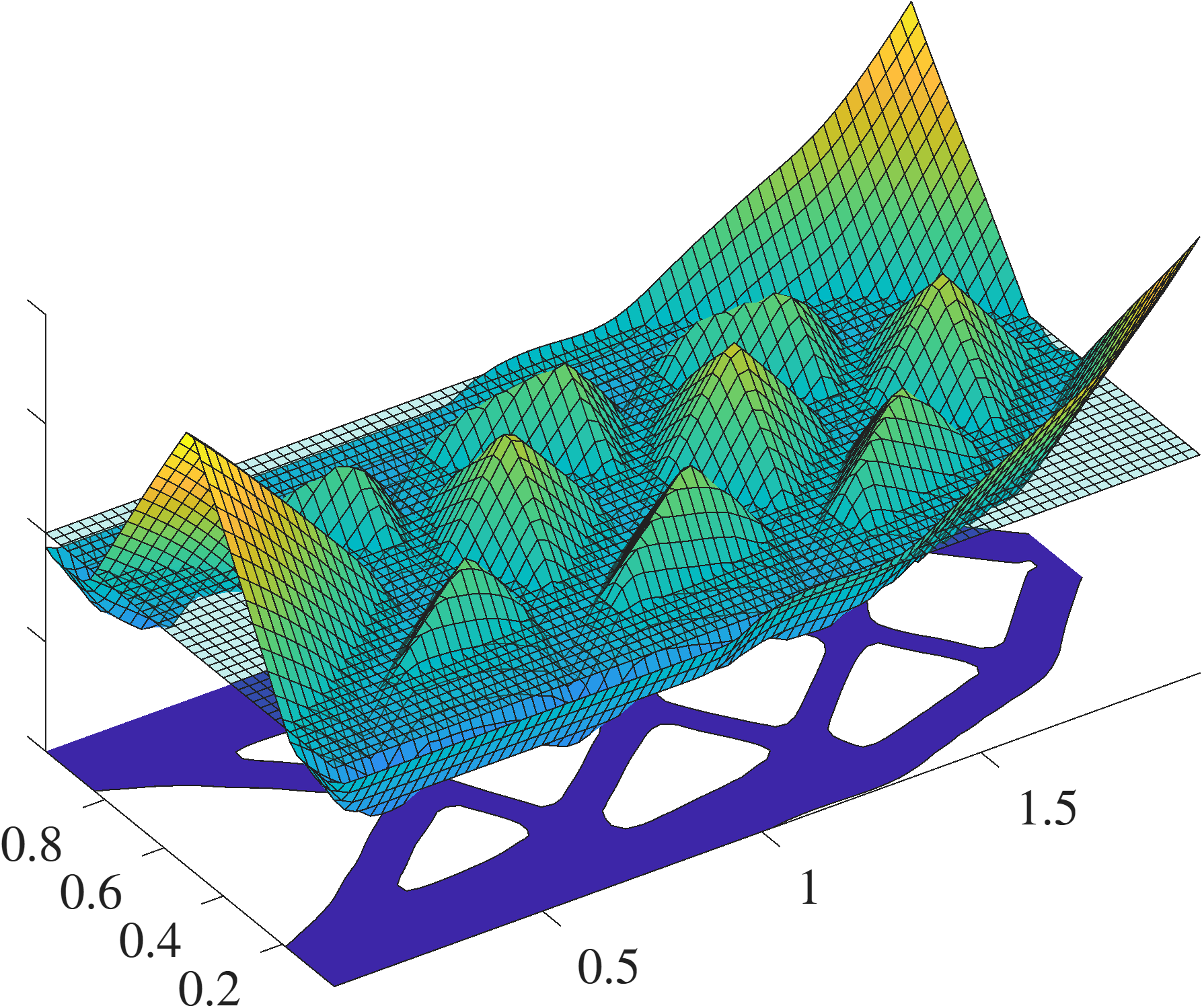}
		\caption{}
	\end{subfigure}
	\begin{subfigure}[b]{0.45\linewidth}
		\centering
		\includegraphics[width=0.9\linewidth]{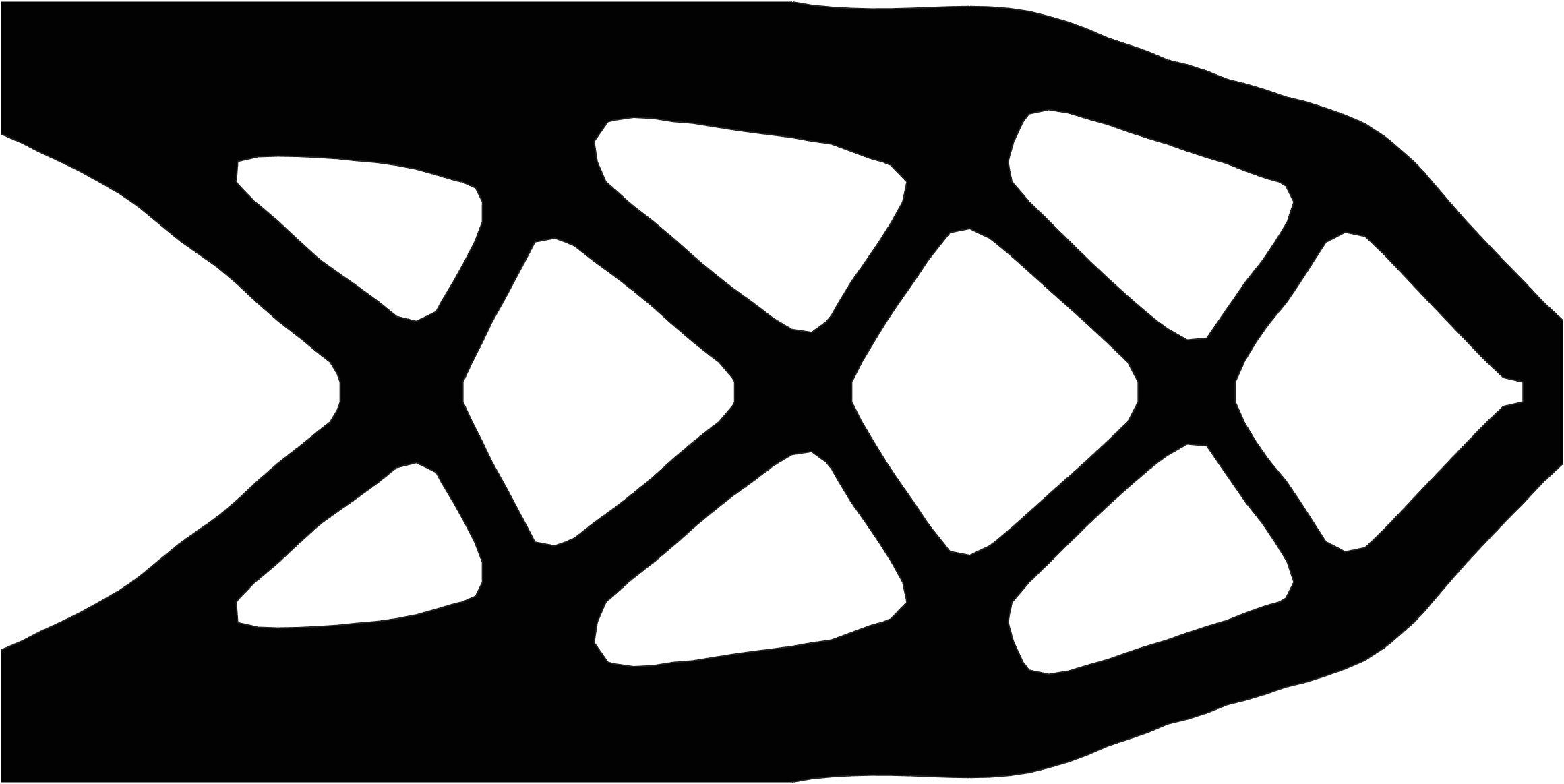}
		\caption{}
	\end{subfigure}

    \begin{subfigure}[b]{0.45\linewidth}
		\centering
		\includegraphics[width=0.9\linewidth]{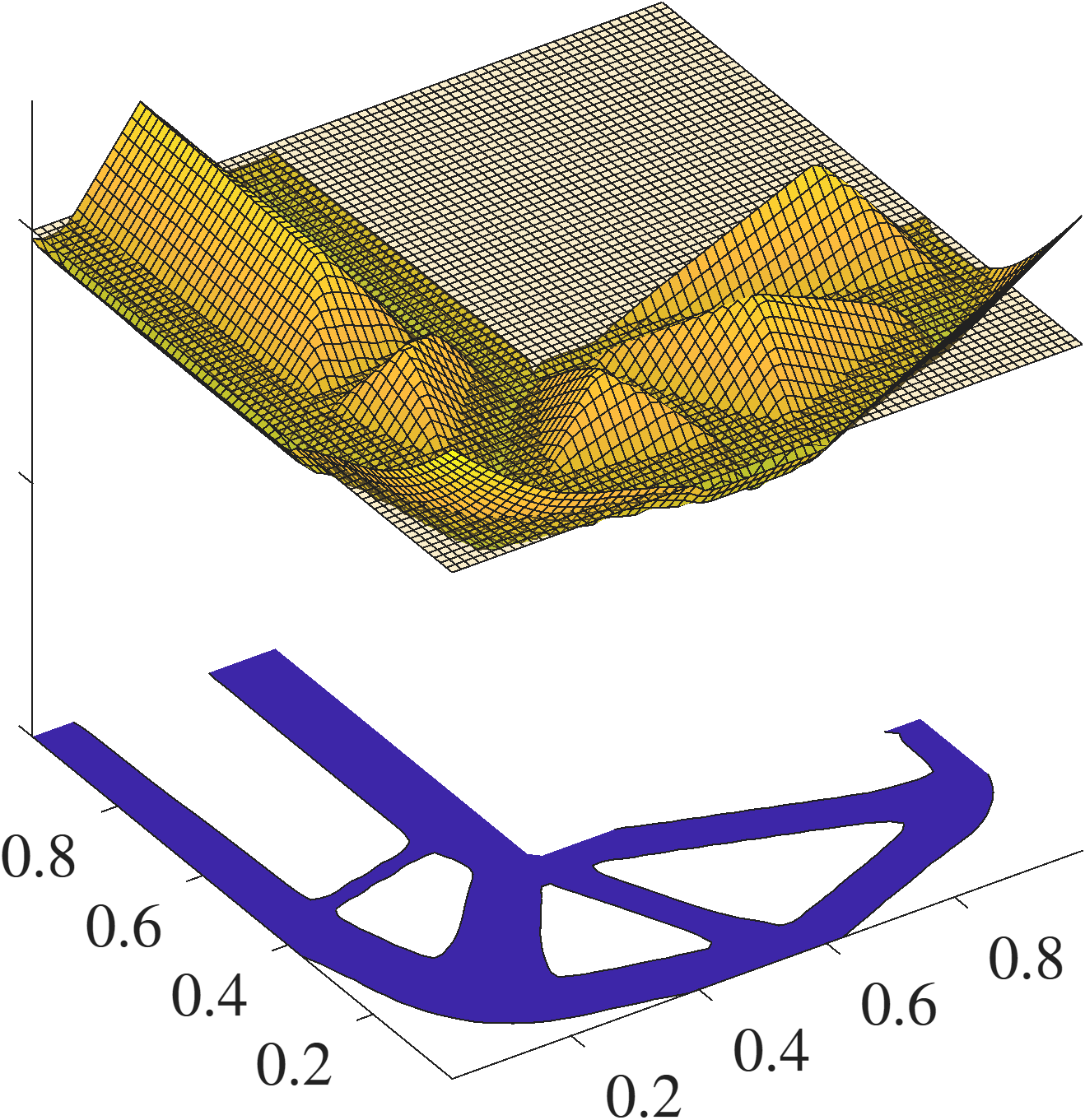}
		\caption{}
	\end{subfigure}
	\begin{subfigure}[b]{0.45\linewidth}
		\centering
		\includegraphics[width=0.9\linewidth]{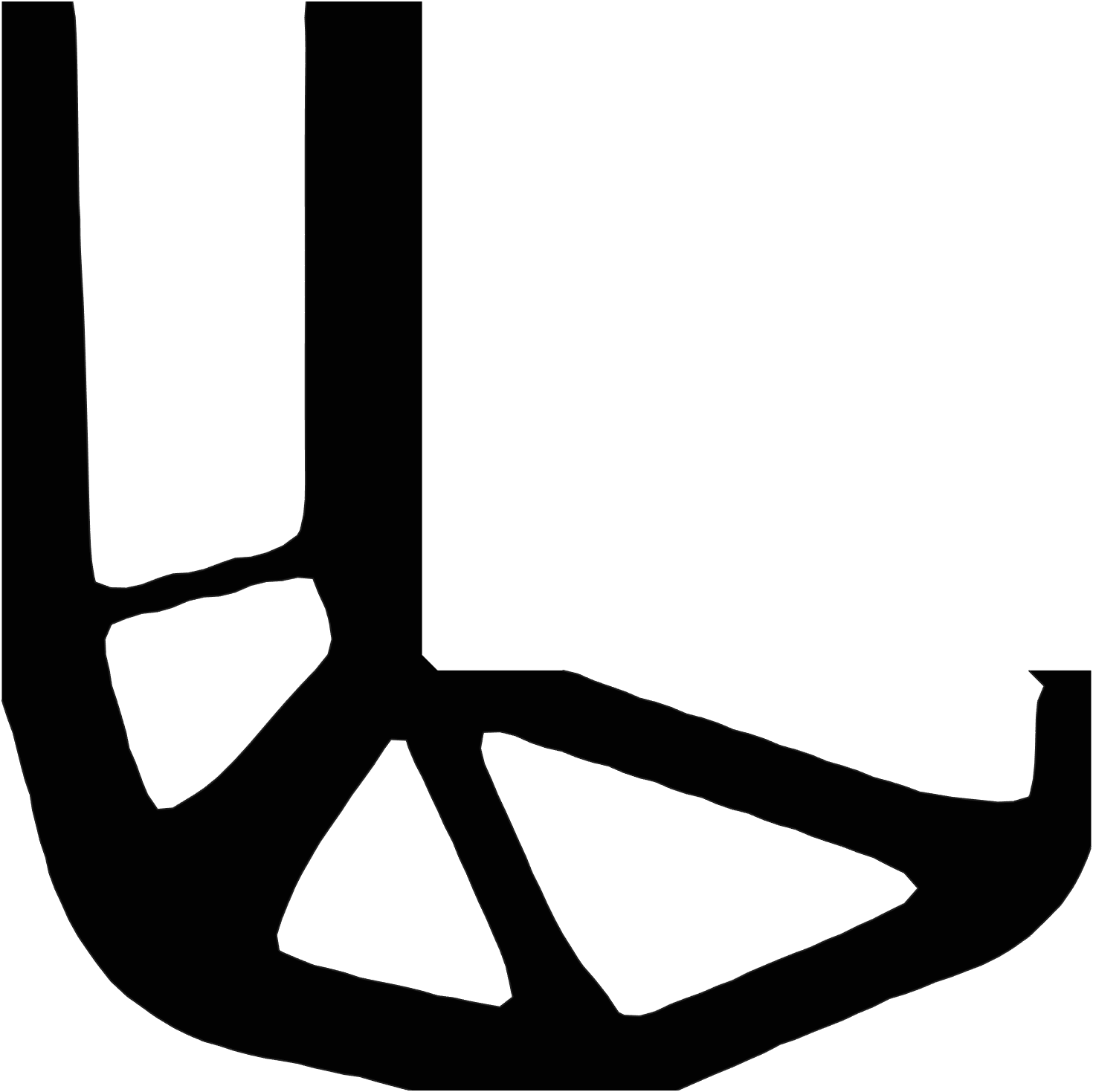}
		\caption{}
	\end{subfigure}

    \begin{subfigure}[b]{0.45\linewidth}
		\centering
		\includegraphics[width=0.9\linewidth]{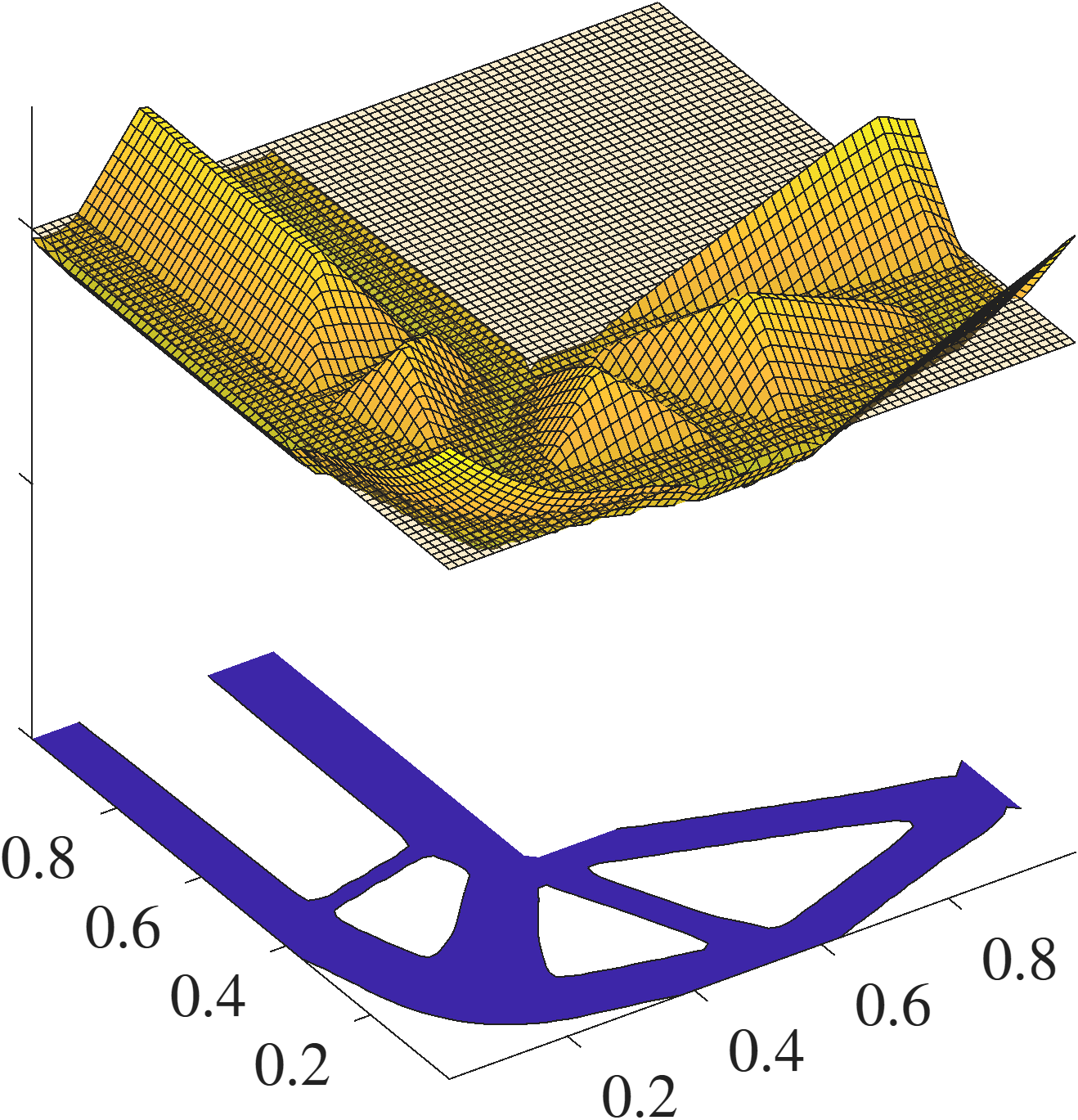}
		\caption{}
	\end{subfigure}
	\begin{subfigure}[b]{0.45\linewidth}
		\centering
		\includegraphics[width=0.9\linewidth]{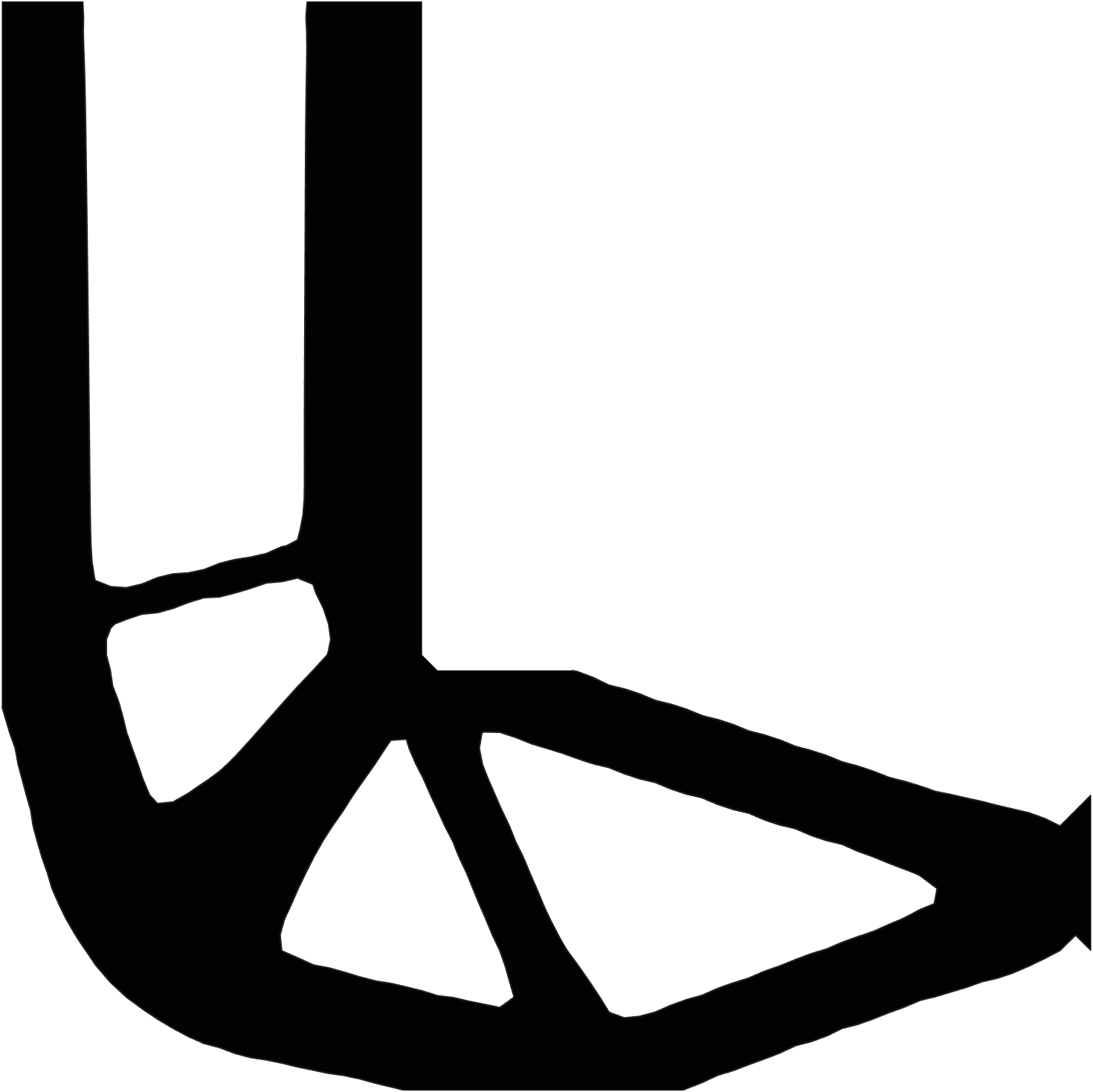}
		\caption{}
	\end{subfigure}

    \begin{subfigure}[b]{0.45\linewidth}
		\centering
		\includegraphics[width=0.9\linewidth]{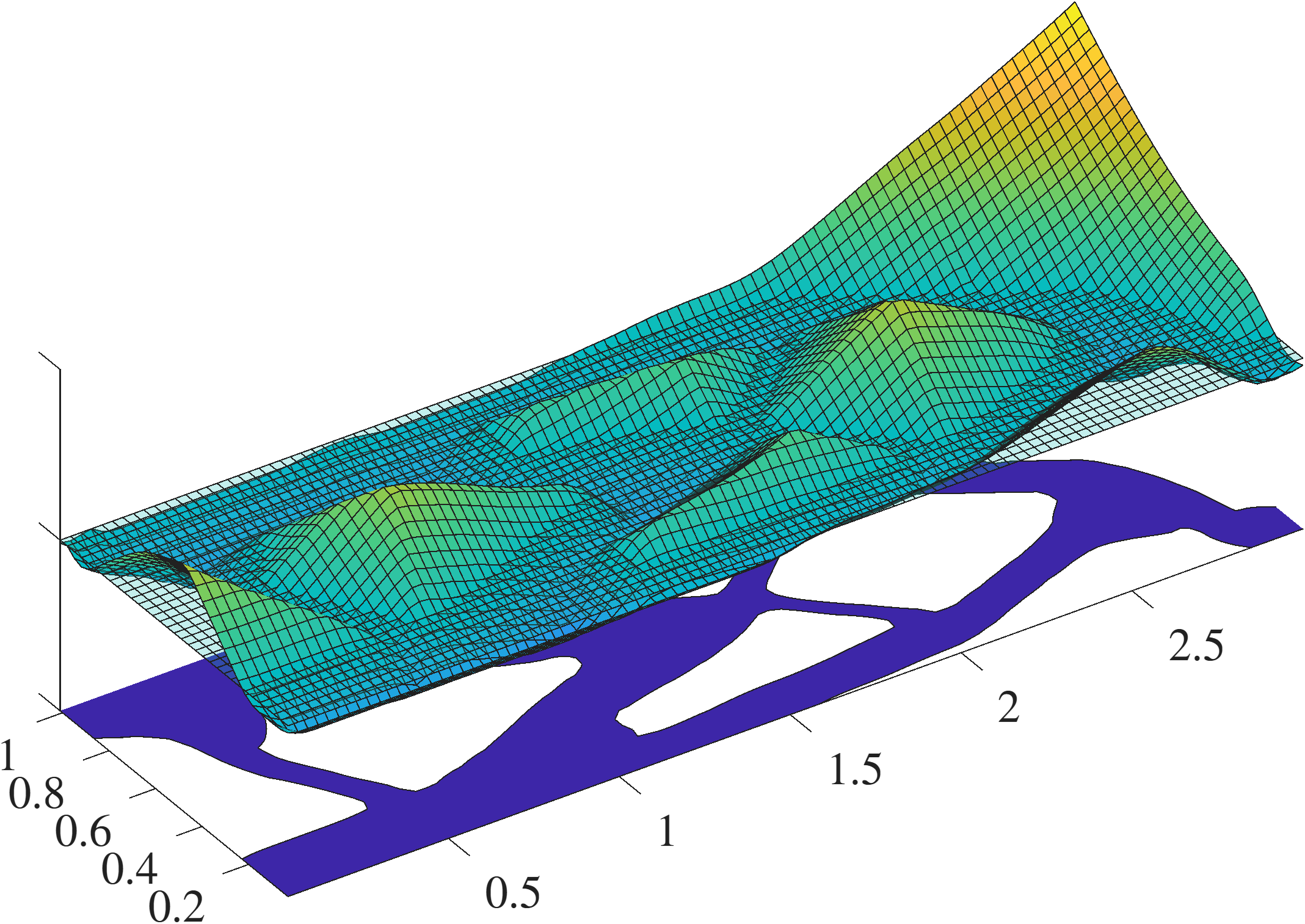}
		\caption{}
	\end{subfigure}
	\begin{subfigure}[b]{0.45\linewidth}
		\centering
		\includegraphics[width=0.9\linewidth]{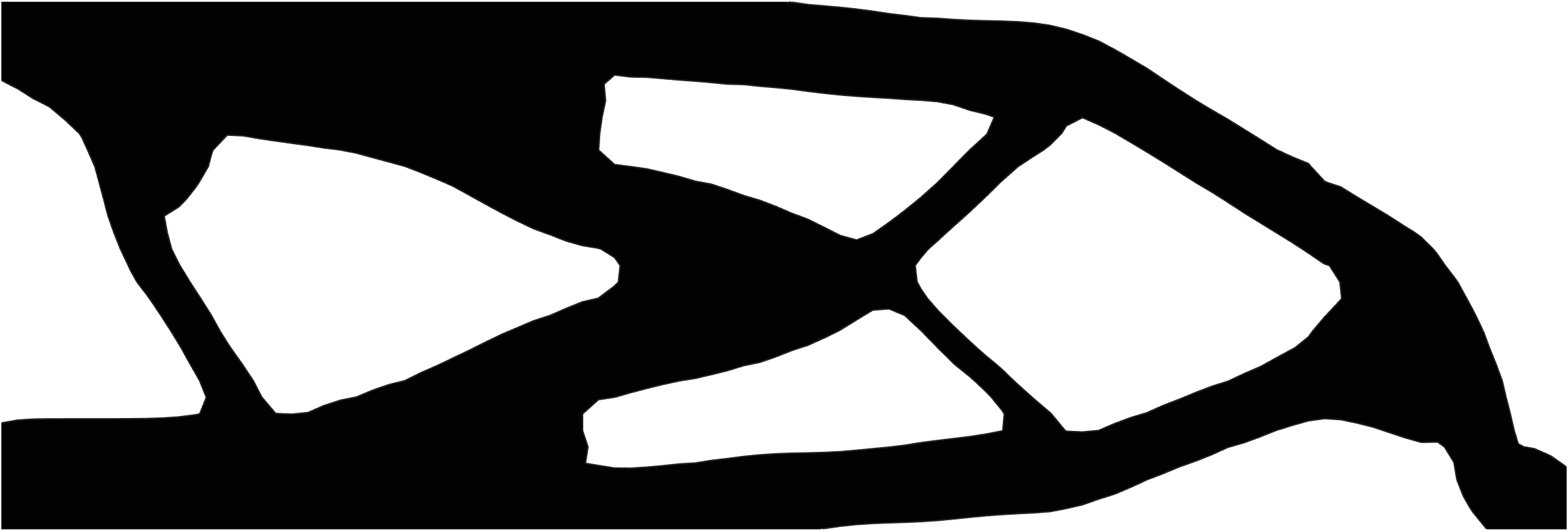}
		\caption{}
	\end{subfigure}

	\caption{Level-set TO via standard HJE for the illustrative examples.}
	\label{fig_example_stLSTO}
\end{figure}

Figure~\ref{fig_initialHoles_stHJE} shows the initial designs used for TO via the standard HJE, and Fig.~\ref{fig_example_stLSTO} presents the corresponding optimized results obtained by evolving the level-set function according to the standard HJE starting from these initializations and Table \ref{tab:stHJE_example_metrics} reports the final performance metrics.

\subsubsection{Modified Hamilton-Jacobi Equation}\label{sec:modHJE}

The LSSO discussed in the Section \ref{sec:LSSO} was incapable of introducing new holes into the domain. This limitation was alleviated by the development of new concepts such as \textit{topological sensitivity}.

\subsubsection{Topological Sensitivity}
\begin{figure}[t]
	\centering
	\includegraphics[width=\linewidth]{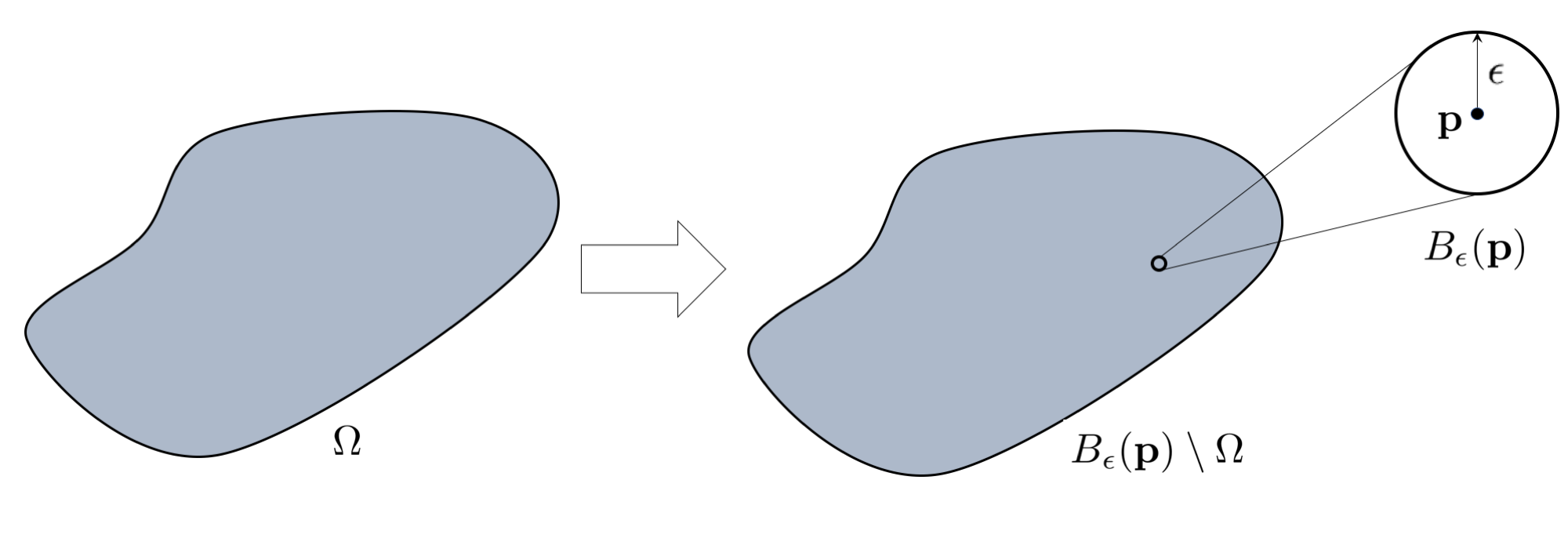}
	\caption{Topological sensitivity concept.}
	\label {fig:TSConcept}
\end{figure}

The topological sensitivity is an extension of the classic shape sensitivity found through the systemic approach explained above, called the topological shape sensitivity method. 
To illustrate the concept of topological sensitivity, consider inserting a small \textit{hypothetical} hole of radius $\epsilon$ at point $\bp \in \Omega$ denoted by $B_\epsilon (\bp)$ as shown in Fig. \ref{fig:TSConcept}.

Topological sensitivity $\TS$ in 2D is defined as the ratio of the first-order change in the objective $\varphi$ to the area of the \textit{hypothetical} infinitesimal hole in the design. Mathematically,
\begin{equation} \label{eq_TSF2D}
	{\rm {\mathcal T}}(\bp){\rm \; }\equiv \mathop{\lim }\limits_{\epsilon\to 0^+} \frac{\varphi (\Omega_\epsilon)-\varphi (\Omega) }{\pi \epsilon^2}
\end{equation}

It can be shown that the topological sensitivity field (TSF) can be used to represent and modify the domain using the positive real-valued parameter $\tau$, where for a sufficiently small $\tau$ and hole's inward normal $\textbf{n}$ we have:
\begin{equation} \label{eq_OmegaTau}
	\Omega_{\tau} = \left\{ \bp_\tau \in \R^2 ~\big\lvert ~ \bp_\tau^{} = \bp + \tau \textbf{n}, \bp \in  \Omega_\epsilon   \right\}
\end{equation}

Observe that $\bp_\tau \big\lvert_{\tau=0}^{} = \bp$ and $ \Omega_\tau \big\lvert_{\tau=0} ^{} = \Omega_\epsilon$. 
 
A closed-form expression for the topological derivative can be derived \cite{feijoo2005topological}.

For compliance, 
\begin{equation} \label{eq_CompTSF2D_plainStress}
	{\rm {\mathcal T}}(\bp){\rm \; }=\frac{4}{1+\nu } \sigma:\varepsilon-\frac{1-3\nu }{1-\nu ^{2} } tr(\sigma )tr(\varepsilon)
\end{equation}

where $\nu $ is the Poisson's ratio, $\sigma (\textbf{u})$ is the stress field associated with the displacement field $\textbf{u}$, and $\varepsilon (\blambda )$ is the strain field associated with an adjoint field $\blambda $ that depends on the quantity of interest. Observe that the stresses and strains are evaluated at point $\bp$  within the original domain before the hole is inserted.  

To generalize the classical level-set method, we first modify the standard HJE by introducing a forcing term $g = -\text{sign} (\psi) \TS$, where $w$ is a positive weight factor and $\psi$ denotes the LSF \cite{burger2004incorporating,challis2010discrete}.

 \begin{figure}[!h]
	\centering
	\begin{subfigure}[b]{0.45\linewidth}
		\centering
		\includegraphics[width=0.9\linewidth]{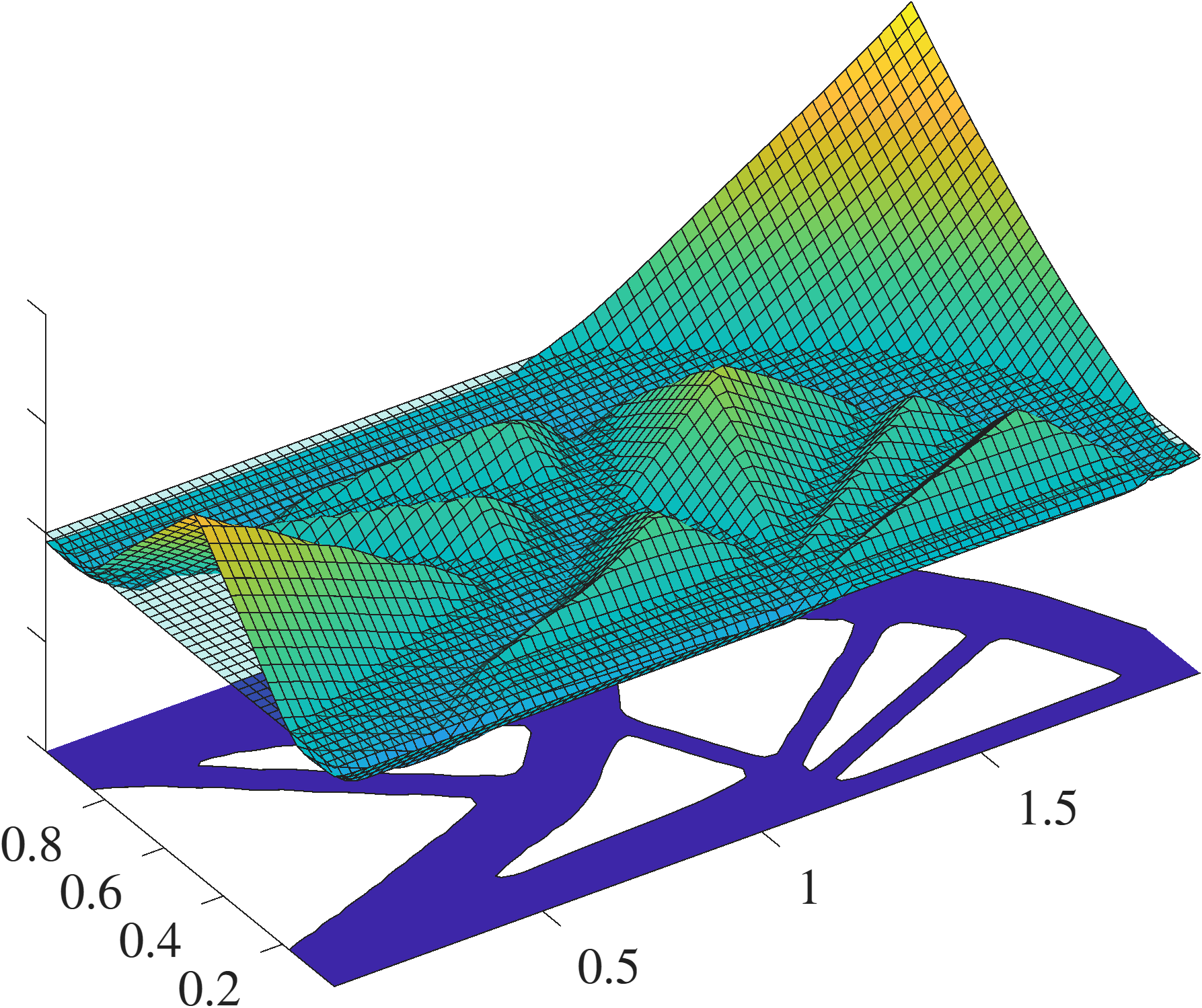}
		\caption{}
	\end{subfigure}
	\begin{subfigure}[b]{0.45\linewidth}
		\centering
		\includegraphics[width=0.9\linewidth]{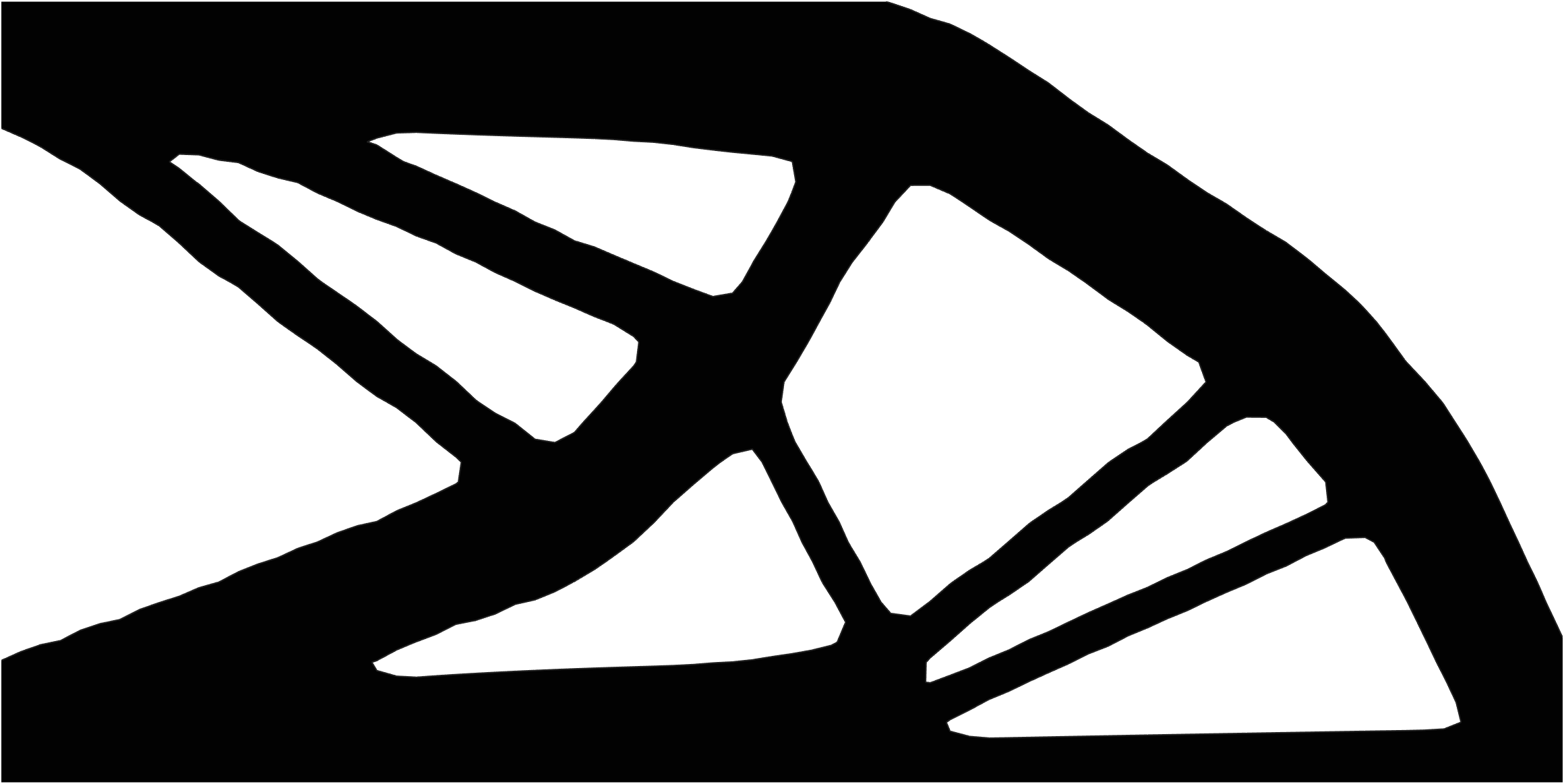}
		\caption{}
	\end{subfigure}

    \begin{subfigure}[b]{0.45\linewidth}
		\centering
		\includegraphics[width=0.9\linewidth]{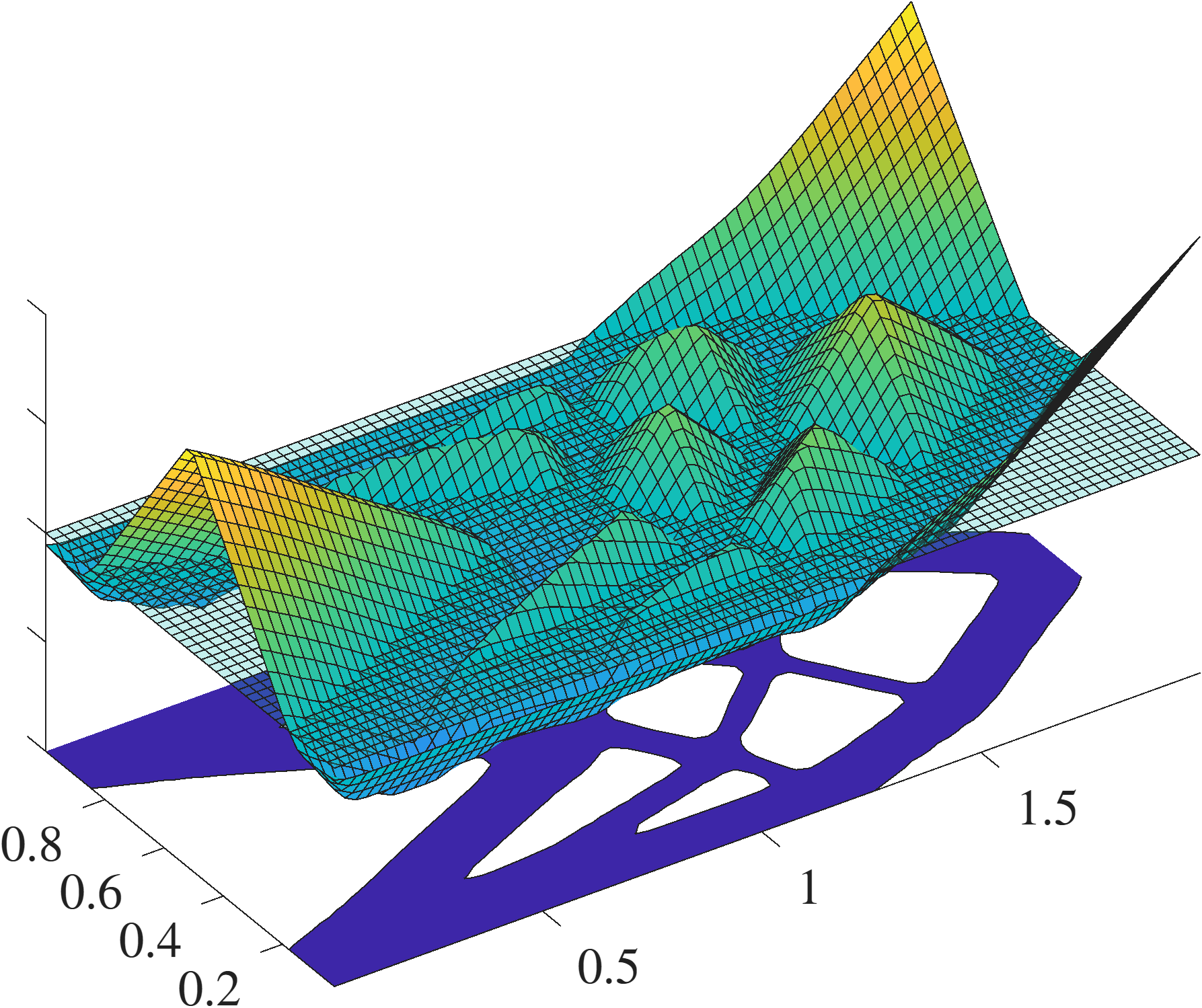}
		\caption{}
	\end{subfigure}
	\begin{subfigure}[b]{0.45\linewidth}
		\centering
		\includegraphics[width=0.9\linewidth]{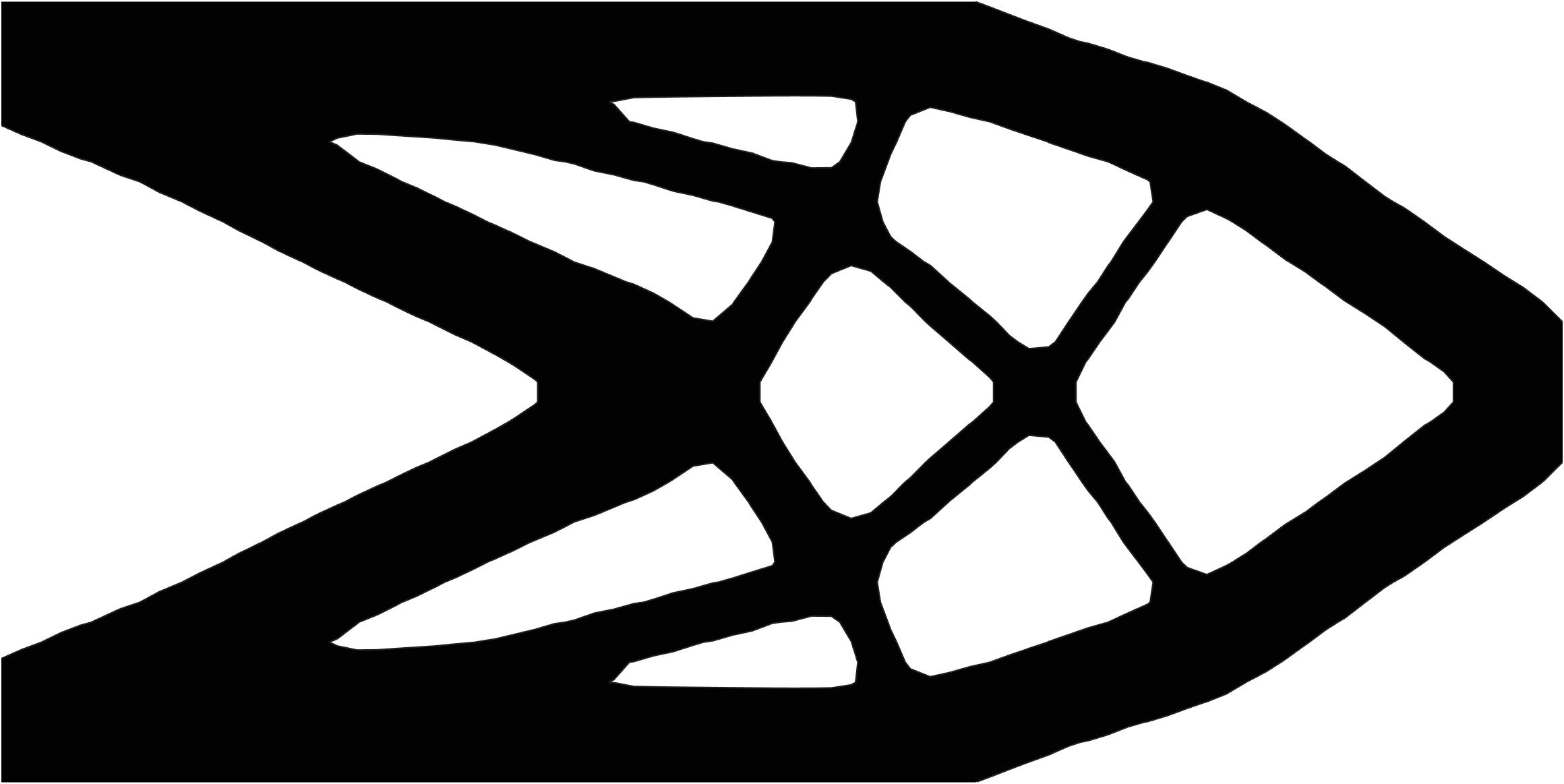}
		\caption{}
	\end{subfigure}

    \begin{subfigure}[b]{0.45\linewidth}
		\centering
		\includegraphics[width=0.9\linewidth]{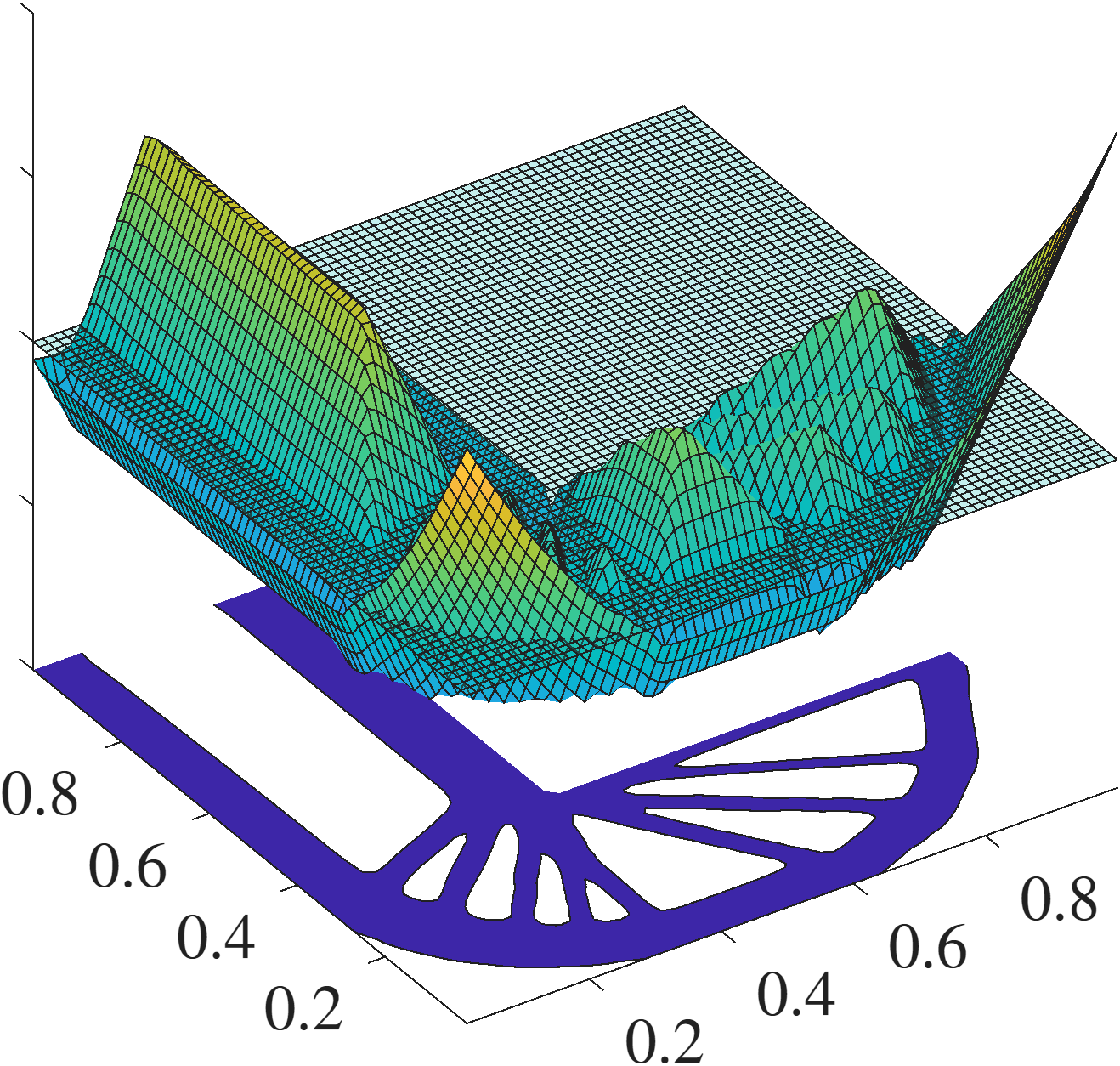}
		\caption{}
	\end{subfigure}
	\begin{subfigure}[b]{0.45\linewidth}
		\centering
		\includegraphics[width=0.9\linewidth]{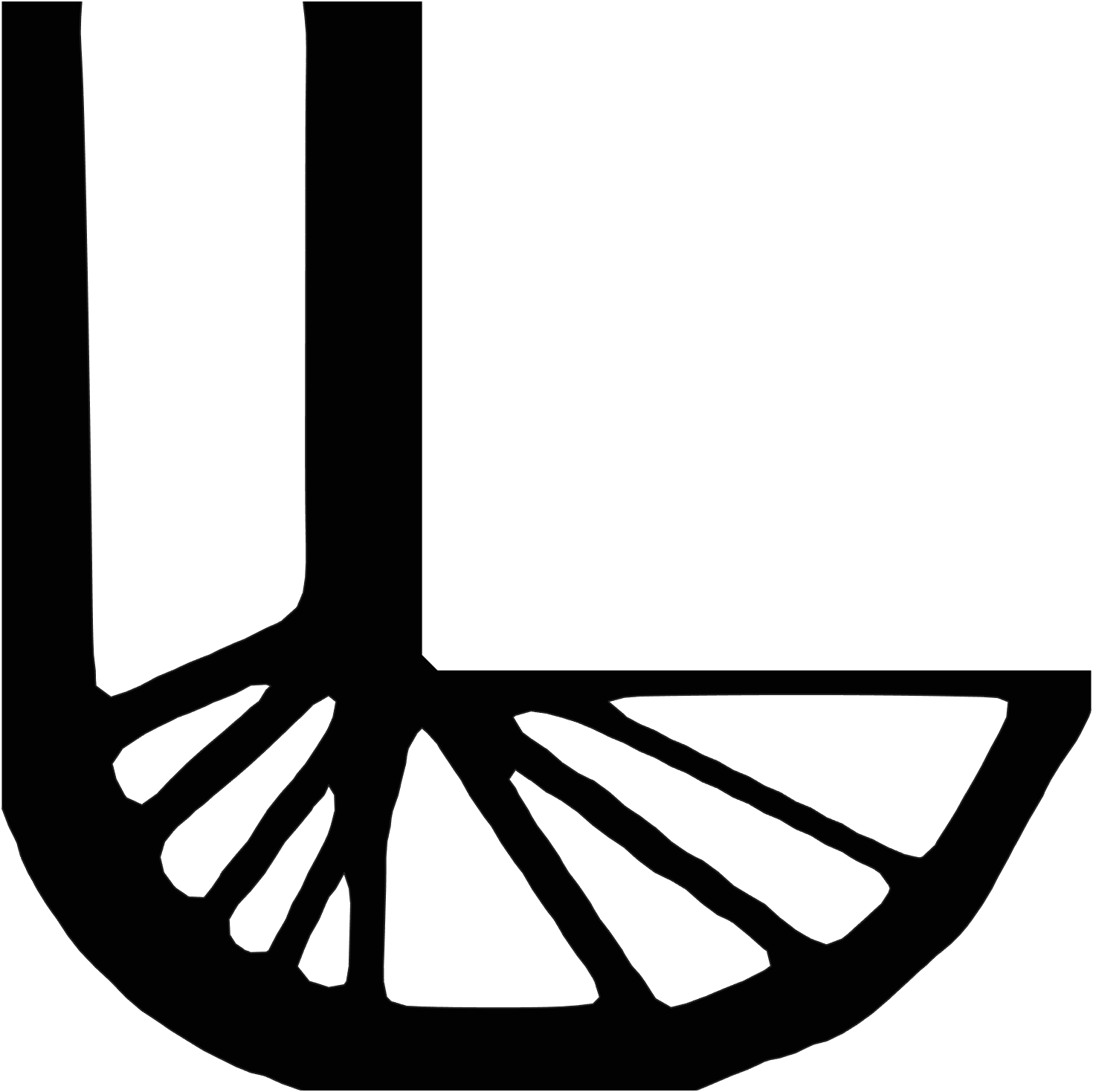}
		\caption{}
	\end{subfigure}

    \begin{subfigure}[b]{0.45\linewidth}
		\centering
		\includegraphics[width=0.9\linewidth]{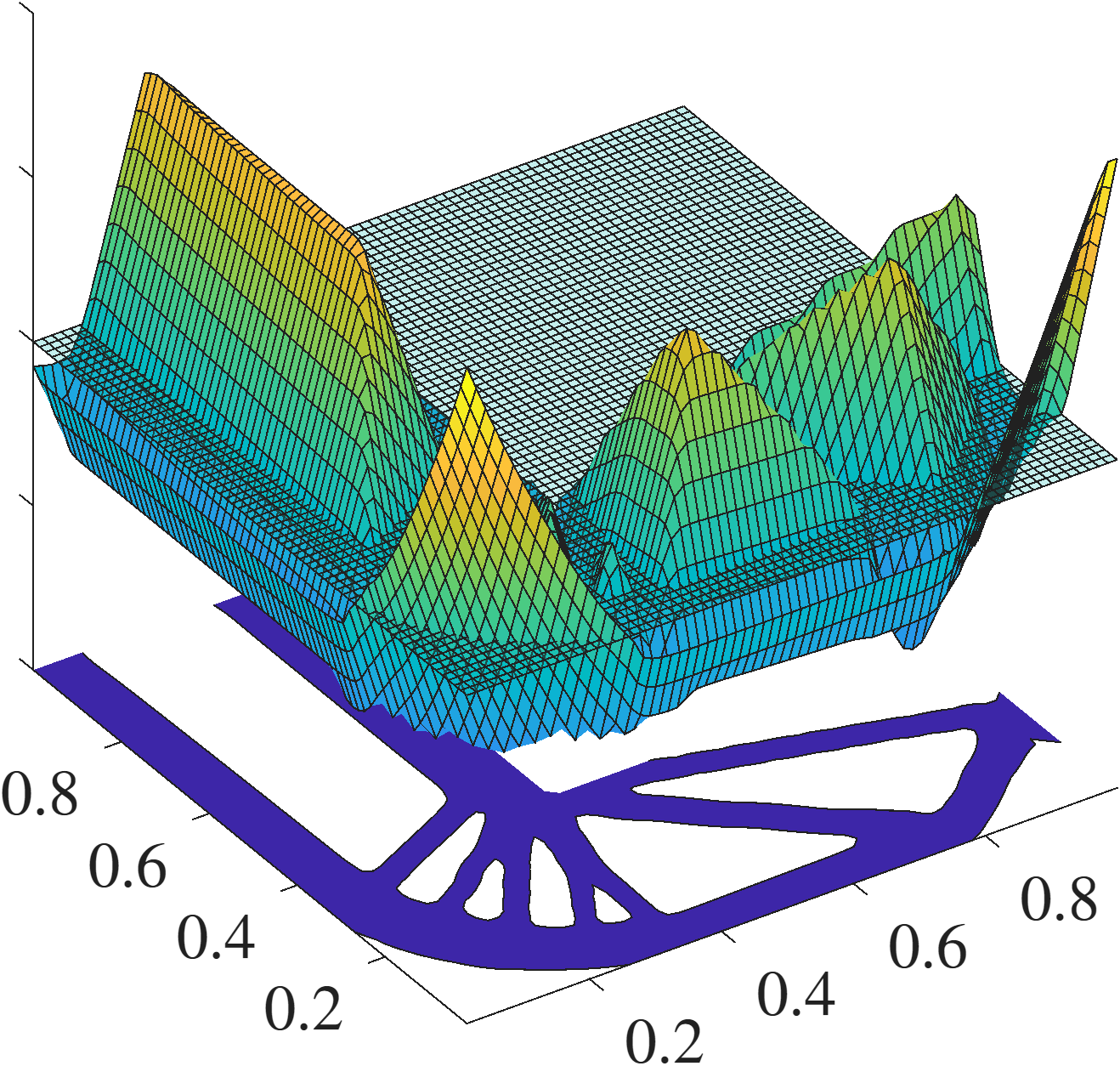}
		\caption{}
	\end{subfigure}
	\begin{subfigure}[b]{0.45\linewidth}
		\centering
		\includegraphics[width=0.9\linewidth]{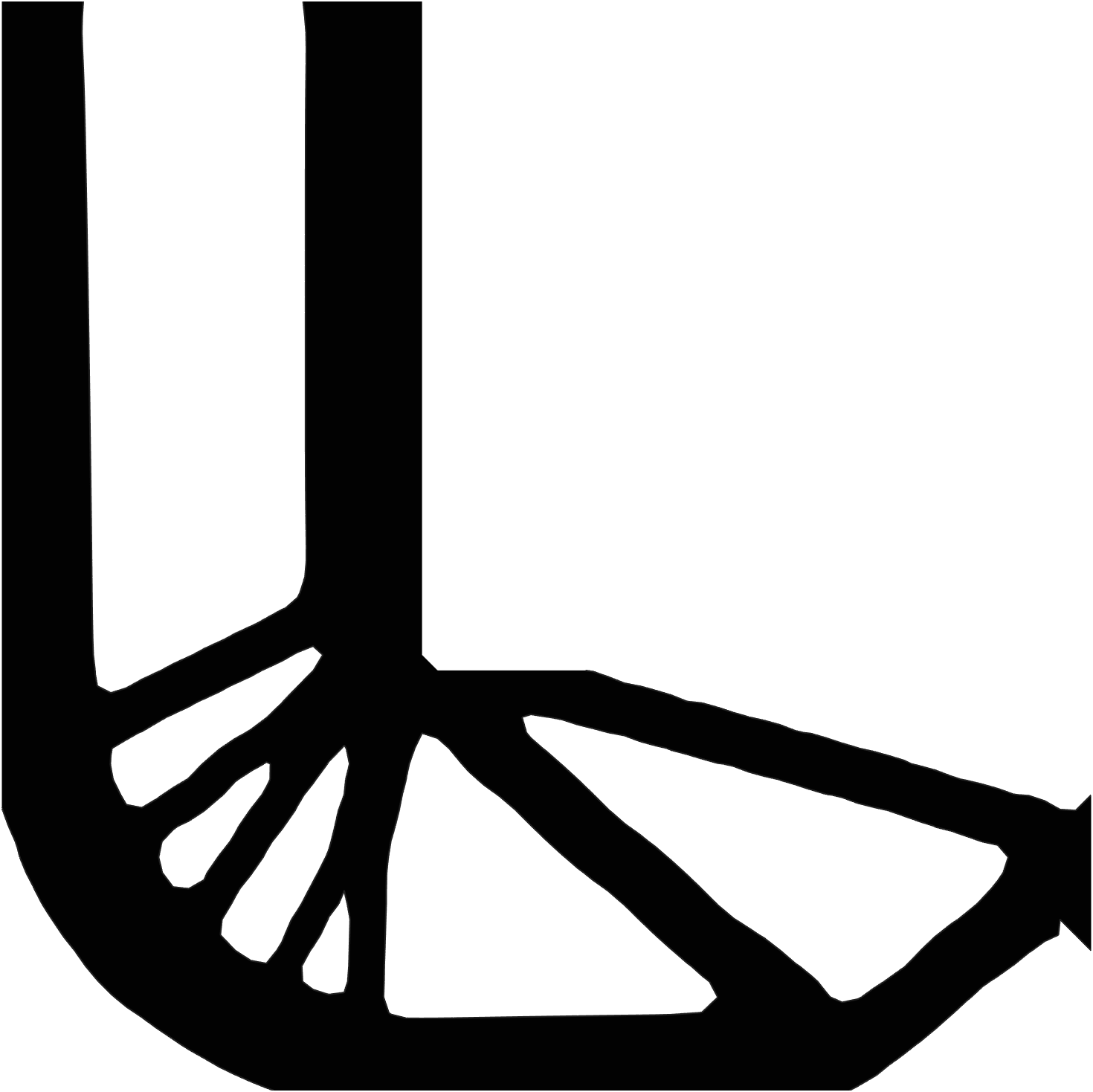}
		\caption{}
	\end{subfigure}

    \begin{subfigure}[b]{0.45\linewidth}
		\centering
		\includegraphics[width=0.9\linewidth]{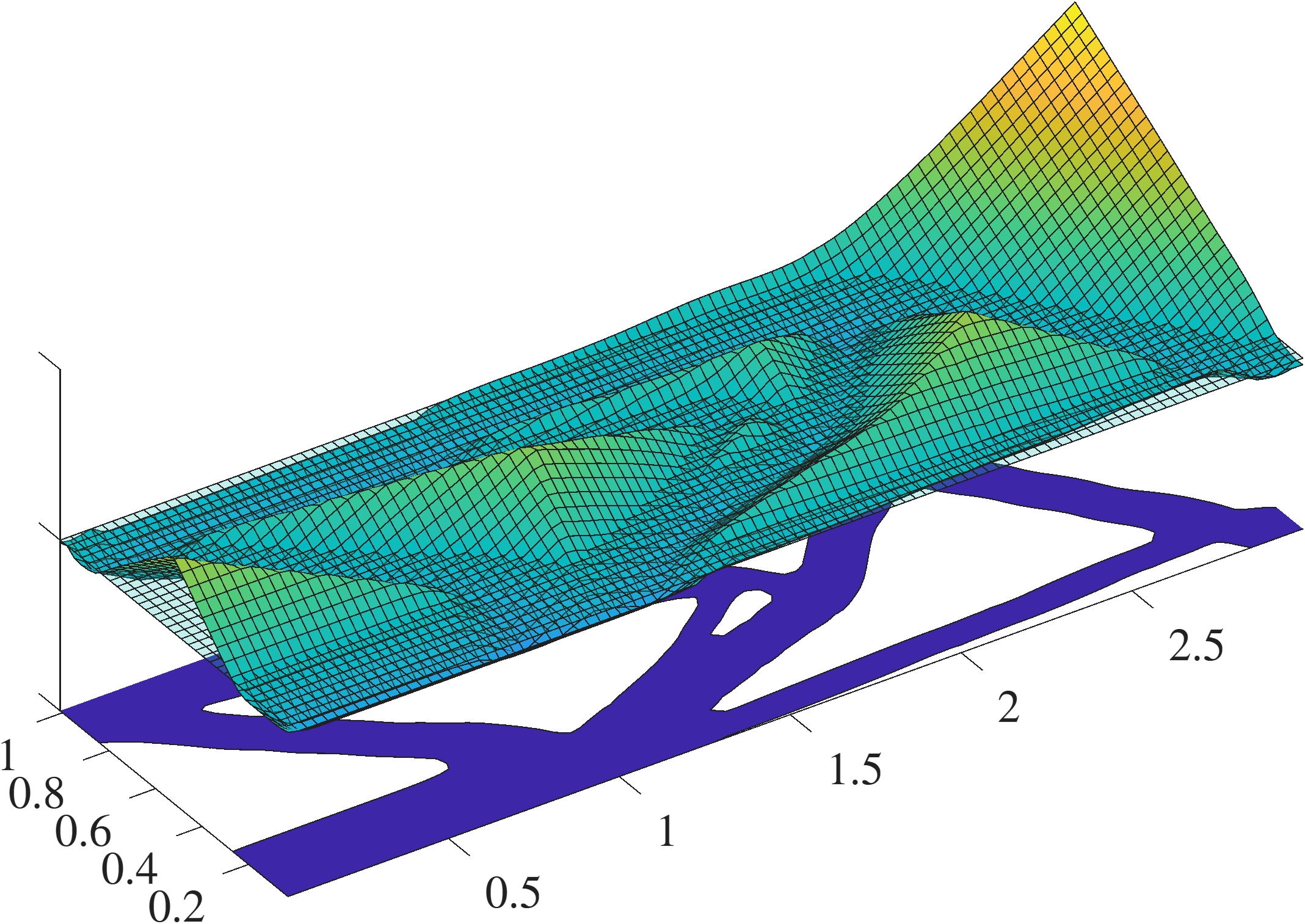}
		\caption{}
	\end{subfigure}
	\begin{subfigure}[b]{0.45\linewidth}
		\centering
		\includegraphics[width=0.9\linewidth]{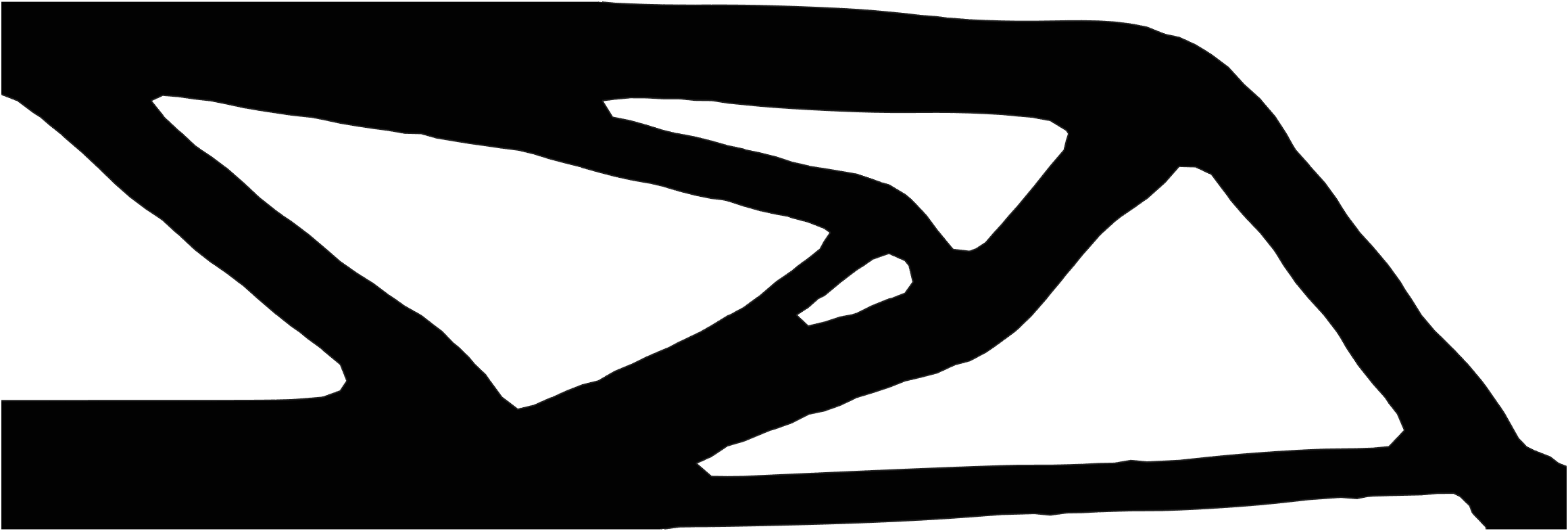}
		\caption{}
	\end{subfigure}

	\caption{Level-set TO via modified HJE for the illustrative examples.}
	\label{fig_example_LSTOmod}
\end{figure}

Figure~\ref{fig_example_LSTOmod} presents the optimized designs obtained by evolving the level-set function with the modified HJE, and Table~\ref{tab:modHJE_example_metrics} reports the corresponding final performance metrics.

\begin{table}[!h]
\centering
\caption{Final performance metrics for optimized designs obtained with \textit{level-set TO using modified HJE}: compliance \(C\), maximum displacement \(\delta_{\max}\), and maximum von Mises stress \(\sigma_{\mathrm{vm},\max}\).}
\label{tab:modHJE_example_metrics}

\small
\renewcommand{\arraystretch}{1.15}
\begin{tabular}{
>{\centering\arraybackslash}p{2.5cm}
>{\centering\arraybackslash}p{1cm}
>{\centering\arraybackslash}p{1.5cm}
>{\centering\arraybackslash}p{1cm}}
\toprule
\textbf{Example} &
{$C$} &
{$\delta_{\max}$} &
{$\sigma_{\text{vm},\max}$} \\
&
{(\si{\newton\meter})} &
{(\si{\meter})} &
{(\si{\mega\pascal})} \\
\midrule

{Cantilever Beam (bottom load)}            & {6.31} & {6.70e-05} & {2.59} \\ \addlinespace
{Cantilever Beam (middle load)} & {9.65} & {7.76e-05} & {3.06} \\
\addlinespace
{L-bracket (top load)}     & {20.1} & {2.20e-04} & {9.49} \\
\addlinespace
{L-bracket (mid load)}     & {19.3} & {2.19e-04} & {29.5} \\
\addlinespace
{MBB (symmetry)}     & {13.7} & {1.50e-04} & {4.74} \\
\bottomrule
\end{tabular}
\end{table}

\begin{figure*}[!h]
  \centering 
  \includegraphics[width=\linewidth]{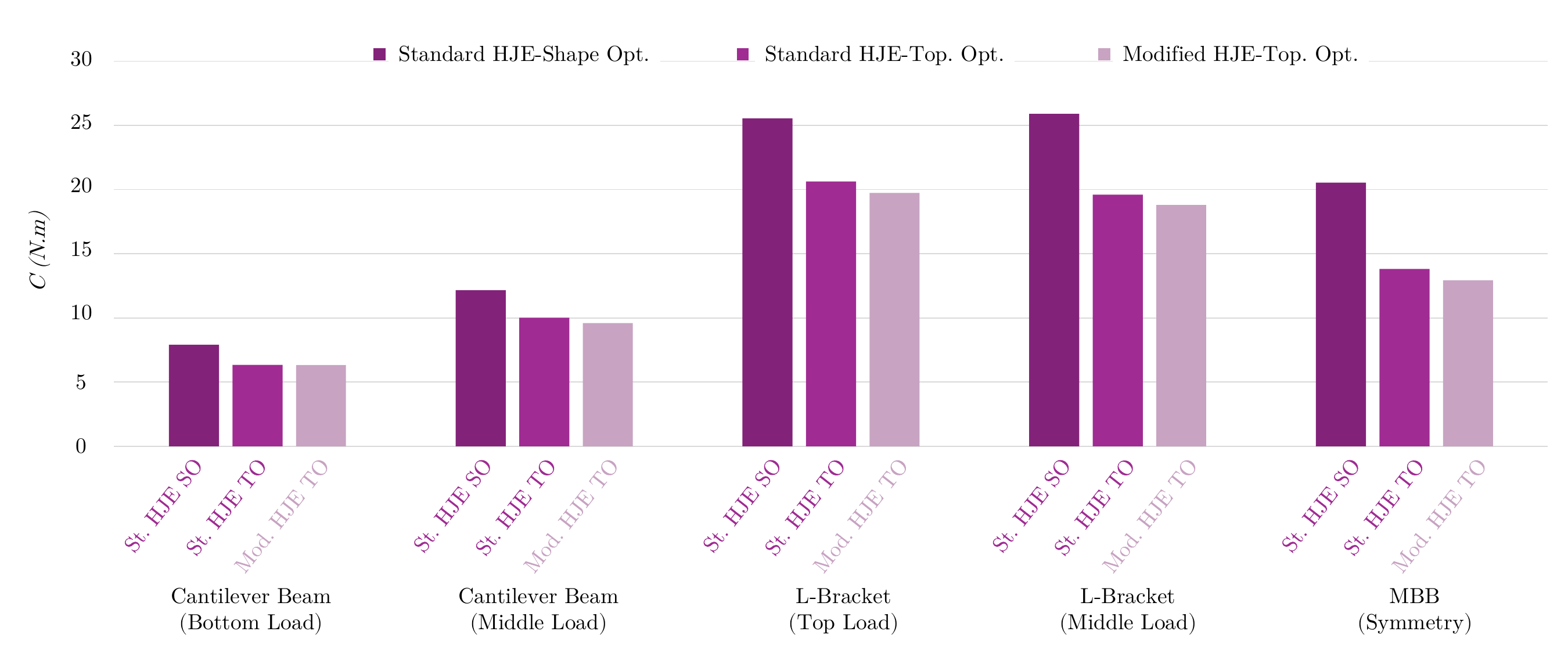}
  \caption{Comparison of level-set shape and topology optimization based standard and modified HJE.}
  \label{fig:LS_summary}
\end{figure*}

\subsection{Topological Sensitivity Methods}\label{sec:TopSensMethod}
Another approach to TO is the Evolutionary Structural Optimization (ESO) method. ESO removes inefficient elements iteratively to improve the design. The original method used low von Mises stress as the rejection criterion, while later versions used sensitivities to support more general objectives \cite{xie_simple_1993,xie1997basic}. Because ESO is irreversible, it can suffer from premature deletion and sensitivity to algorithmic parameters. BESO alleviates this issue by allowing elements to be both removed and reintroduced, generally improving robustness and convergence behavior \cite{xia2018bi}. Still, both ESO and BESO remain heuristic update schemes, so their intermediate designs are not necessarily locally optimal for a given volume fraction, and global optimality is not guaranteed.
In contrast, Pareto-tracing approaches explicitly solve a sequence of constrained optimization problems, yielding intermediate designs that are typically closer to locally optimal for their corresponding volume fractions.

In this section, we first define the element-wise sensitivity field directly on the discretized domain and relate it to the topological sensitivity discussed in Section~\ref{sec_DSF}. We then describe the evolutionary update rules for ESO, BESO, and Pareto-tracing methods in Sections~\ref{sec:eso}-\ref{sec:PareTO}, briefly summarize their implementation within \textsc{STORX}, and present representative illustrative examples.

\subsubsection{Discrete Sensitivity Field} \label{sec_DSF}

\begin{figure}[t]
	 \centering
    \begin{subfigure}[b]{.475\linewidth}
    	\centering
    	\includegraphics[width=\linewidth]{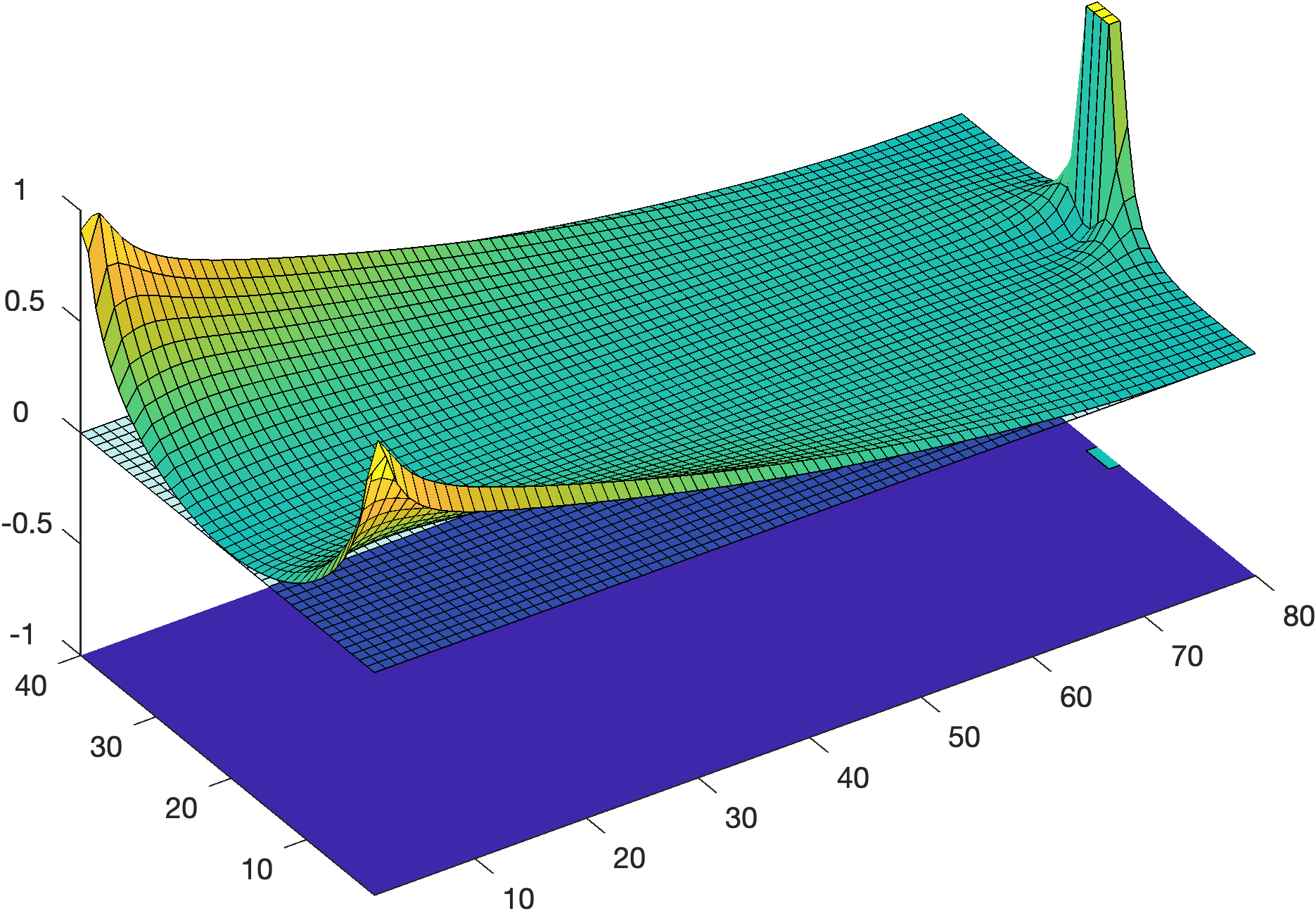}
    	\caption{TSF}
    	\label{fig_fig_TSBeamdomain_TSF}
    	\end{subfigure}
    \begin{subfigure}[b]{.475\linewidth}
    	\centering
    	\includegraphics[width=\linewidth]{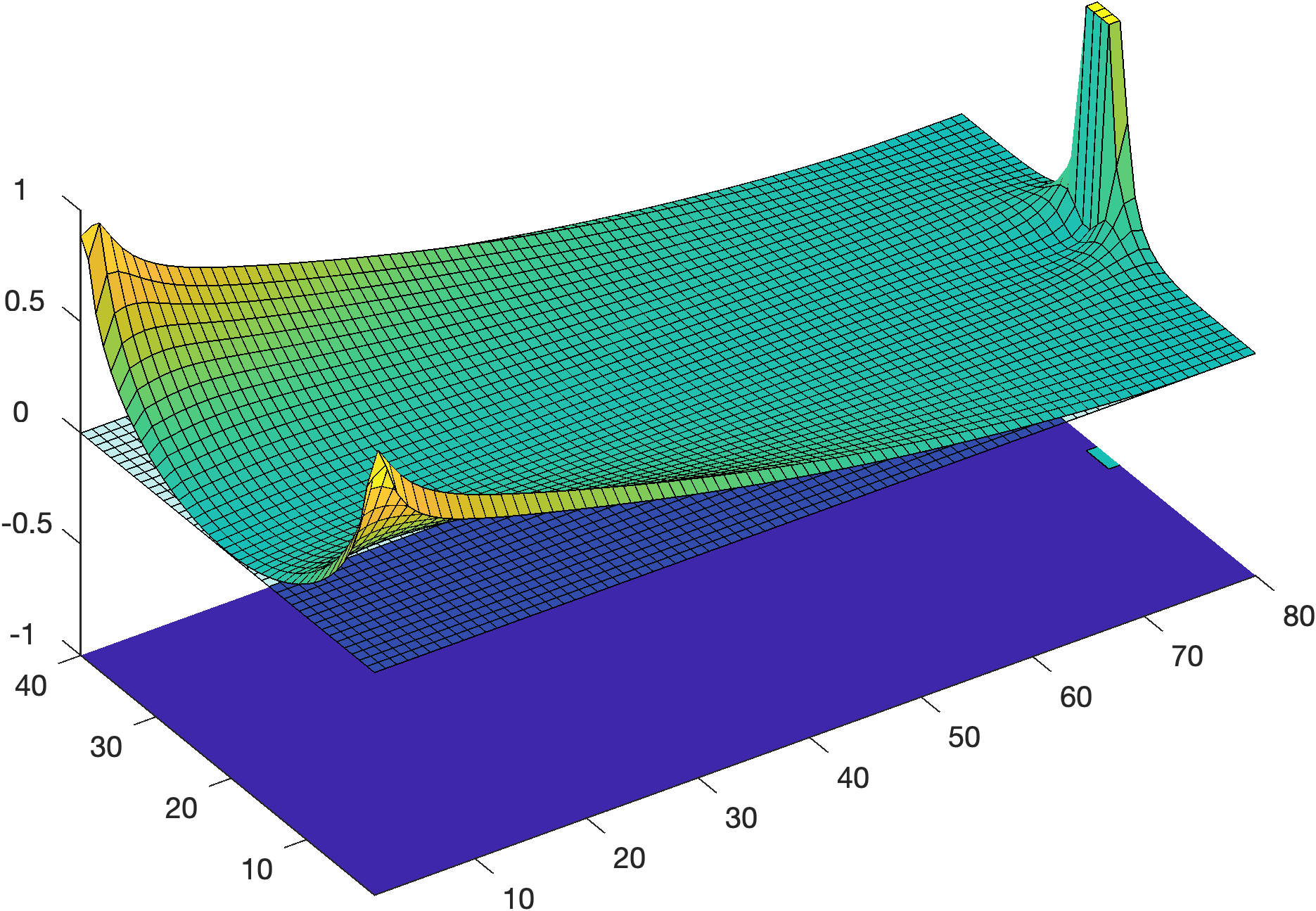}
    	\caption{DSF}
    	\label{fig_fig_TSBeamdomain_DSF}
    \end{subfigure}
	\caption{Initial design (a) TSF and (b) DSF.}
	\label{fig_TSBeamdomain}
\end{figure}

For many quantities of interest other than compliance, deriving a closed-form expression for the TSF might become challenging. We will therefore also consider an alternate \textit{discrete sensitivity field} (DSF), defined as the change in any quantity of interest when a single finite element \textit{e} is deleted from the mesh. Figure \ref{fig_TSBeamdomain} shows a comparison between TSF and DSF for the cantilever beam example.

Consider a cantilever beam example of Fig. \ref{fig_BC_BeamMid}. To compute the TSF, we (hypothetically) insert an infinitesimal hole in the design, which gives the TSF of \ref{fig_fig_TSBeamdomain_TSF}. 
On the other hand, to compute the discrete sensitivity field, we (hypothetically) remove a (sufficiently small) finite element from the discretized domain, which gives the DSF of Fig. \ref{fig_fig_TSBeamdomain_DSF}. Observe that for a reasonable discretization, the two fields exhibit the same behavior and are comparable.

\subsubsection{Evolutionary Structural Optimization} \label{sec:eso}

\begin{figure}[t]
	\centering
	\begin{subfigure}[b]{.475\linewidth}
		\centering
		\includegraphics[width=\linewidth]{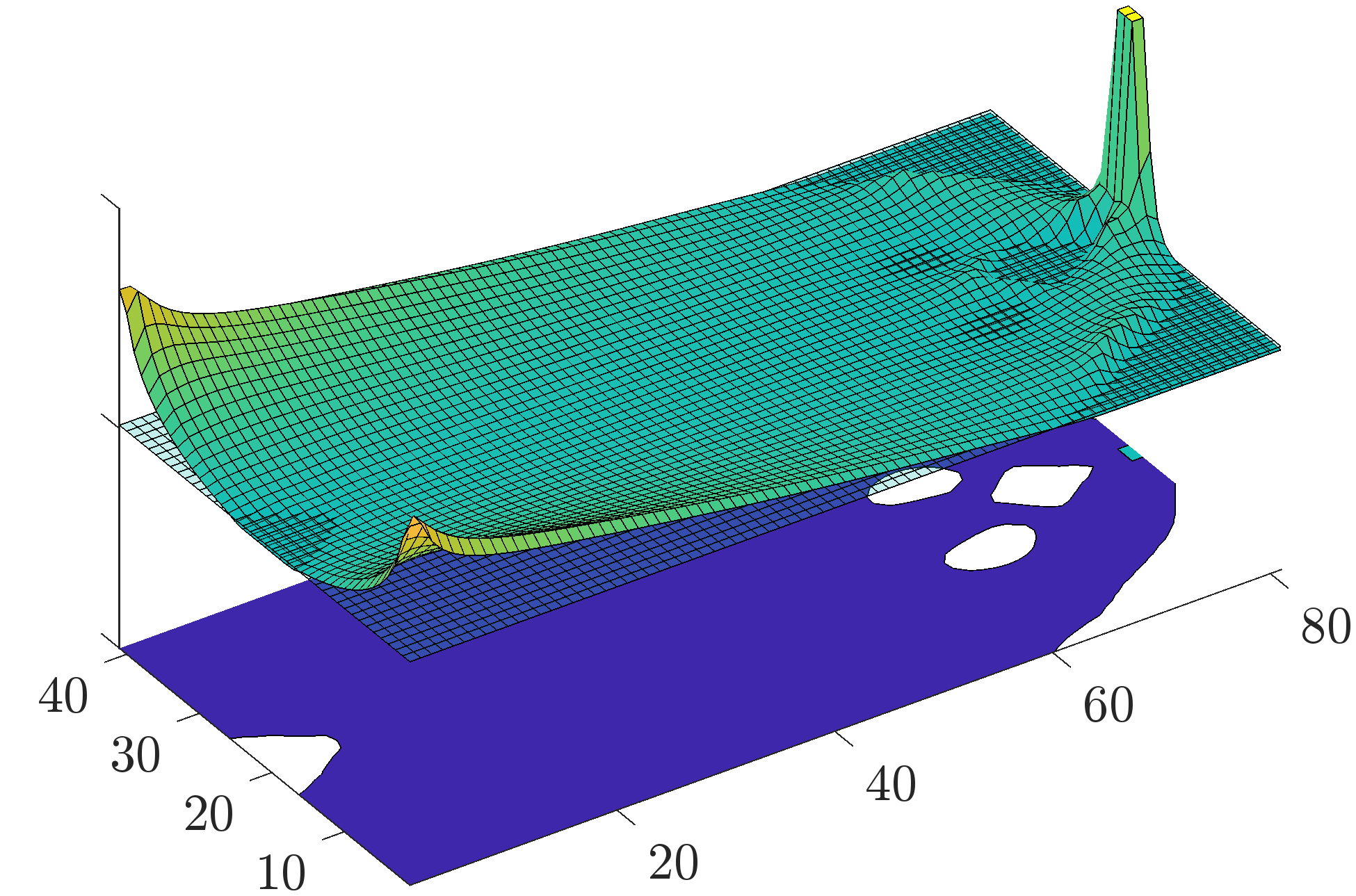}
		\caption{TSF}
	\end{subfigure}
	\begin{subfigure}[b]{.475\linewidth}
	\centering
	\includegraphics[width=\linewidth]{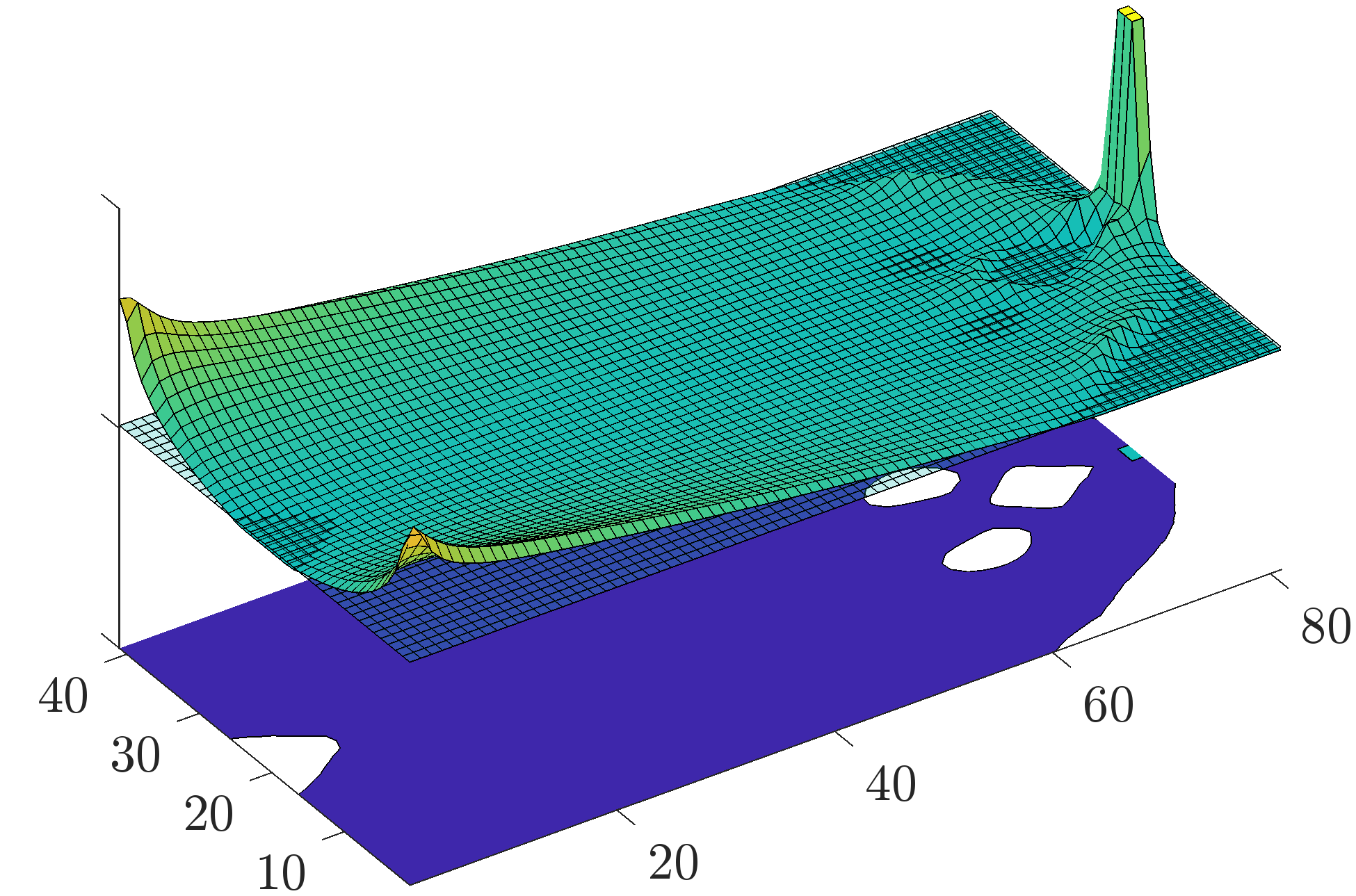}
	\caption{DSF}
\end{subfigure}
	\caption{Updated design at 0.9 volume fraction by taking the level-set of  (a) TSF and (b) DSF.}
	\label{fig_TS_tau}
\end{figure}
\begin{figure*}[!h]
	\centering
	\includegraphics[width=\textwidth]{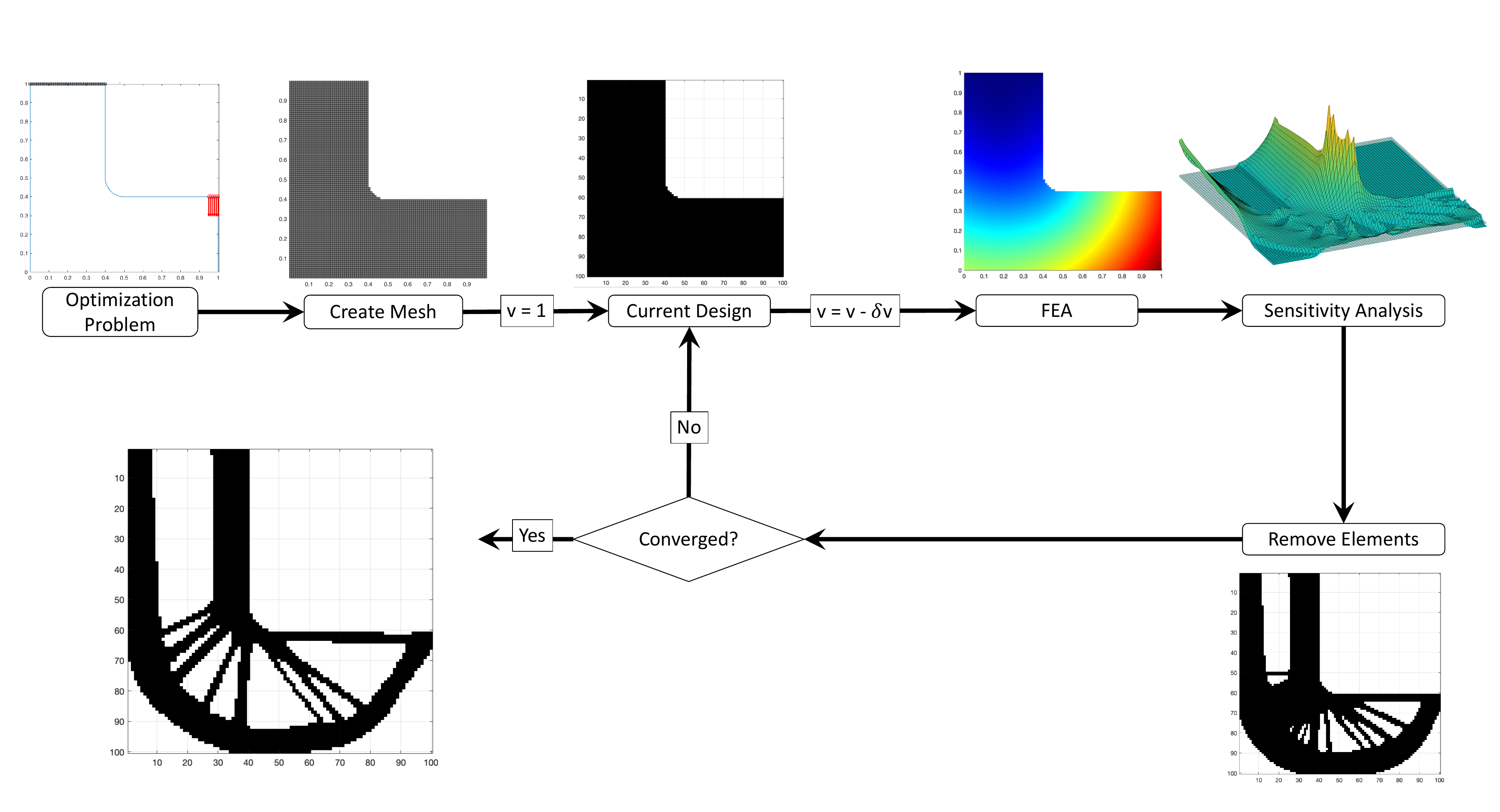}
	\caption{ESO workflow.}
	\label{fig_esoWorkflow}
\end{figure*}

In this section, we will describe the hard-kill ESO method, where once an element is removed, it stays deleted in subsequent iterations. At every step, we perform FEA, compute the sensitivity field, reject a few elements with \textit{lowest} sensitivity value, and repeat until we converge to the target volume fraction.
 
Consider now a domain $\Omega _{\tau } $ as the set of all points where the sensitivity field exceeds the value $\tau $, defined per:
\begin{equation} \label{eq_levelset}
	\Omega _{\tau } \equiv \{ \bp ~\big\lvert~ \DS_\varphi (\bp)>\tau \}
\end{equation}

In STORX, this operation is implemented as:
\begin{lstlisting}
function obj = update(obj,volFrac)
    % Find the level-set value such that the contour has given vol fraction
    obj.m_tau = obj.findContourValueWithVolumeFraction(volFrac);
    index = find(obj.m_dfdx < obj.m_tau); % eliminate all elements less than this value
    obj.m_x = obj.m_solver.m_existingElems; % start with the full domain
    obj.m_x(ind2sub(size(obj.m_dfdx),index)) = 0; % remove elements
    obj.m_solver = obj.m_solver.setDesign(obj.m_x);
end
\end{lstlisting}

Recalling the above example in Fig. \ref{fig_TSBeamdomain},  Fig. \ref{fig_TS_tau} illustrates both TSF and DSF together with a cutting plane at the threshold value $\tau $ and the resulting shapes that correspond to topologies of reduced volume fraction such that the material contributing least to the stiffness of the structure is removed. The cutting-plane value $\tau $ can be chosen such that, say, 10\% of the volume is removed.

Figure \ref{fig_esoWorkflow} illustrates the ESO workflow to gradually remove inefficient elements.

For the cantilever beam example, we set the objectives and constraints as:
\begin{lstlisting}
%% Objective and Constraints
objective = topologicalSensitivityComplianceElasticity(solver);
volumeFraction = 0.5;
constraints  = {volume(solver, volumeFraction)};
% manufacturing constraints
mfgConstraints = {minimumFeatureSize_gaussian(solver)};
\end{lstlisting}

And construct the optimizer with $2.5\%$ volume decrement at each iteration:
\begin{lstlisting}
topoptClass = @eso2d_elasticity;
%% Construct Optimizer
volDecrement = 0.025;
topopt = topoptClass(solver, ...
    objective,constraints,mfgConstraints,volDecrement);
\end{lstlisting}

\subsubsection{Bi-directional Evolutionary Structural Optimization} \label{sec:beso}
BESO extends the hard-kill ESO concept by allowing \emph{both} removal and re-introduction of elements. The class in inherited from ESO as \mcode{(Abstract) beso2d  < eso2d}. In contrast to ESO, where deleted elements are permanently removed, BESO maintains a binary design field $x_e\in\{0,1\}$ on a fixed background mesh and updates it by \textit{rejecting} inefficient material and \textit{adding back} material in regions where it is predicted to be most beneficial. This bi-directional mechanism reduces the risk of irreversible early mistakes and typically yields more robust evolution histories.

Implementation-wise, aside from the constructor, the only modification is in the \mcode{obj = update(obj, volFrac)} routine where simple element removal is replaced by an OC-type loop.

One may interpret the update as applying two cutting planes to the filtered sensitivity field: a lower threshold $\tau^{-}$ that identifies material to be removed and an upper threshold $\tau^{+}$ that identifies void regions to be refilled, i.e.,
\begin{align}
\Omega^{\text{keep}}_{\tau^-} &\equiv \{ \bp\in\Omega \mid \alpha(\bp)>\tau^- \},
\\
\Omega^{\text{add}}_{\tau^+} &\equiv \{ \bp\in\Omega \mid \alpha(\bp)>\tau^+ \},
\end{align}
with the understanding that $\Omega^{\text{add}}_{\tau^+}$ is applied only on the current void region. In practice, $\tau^{-}$ and $\tau^{+}$ are chosen (via sorting/bisection) so that the number of removed and added elements matches the desired volume update while preserving a binary design.
Implementation-wise, aside from the constructor the only modification is in the \mcode{obj = update(obj, volFrac)} routine where simple element removal is replaced by an OC-type loop.
Finally, to mitigate oscillations due to discrete add/remove operations, BESO implementations commonly employ stabilization strategies such as filtering and/or sensitivity averaging across iterations. Overall, BESO provides a simple yet effective mechanism to evolve topologies while retaining the ability to correct earlier decisions through bi-directional updates.

For the cantilever beam example, we only need to change the optimizer as:
\begin{lstlisting}
topoptClass = @beso2d_elasticity;
\end{lstlisting}

\subsubsection{Pareto-tracing Topology Optimization}\label{sec:PareTO}
\begin{figure} [t]
	\begin{subfigure}[t]{0.48\linewidth}
		\centering
		\includegraphics[width=0.9\linewidth]{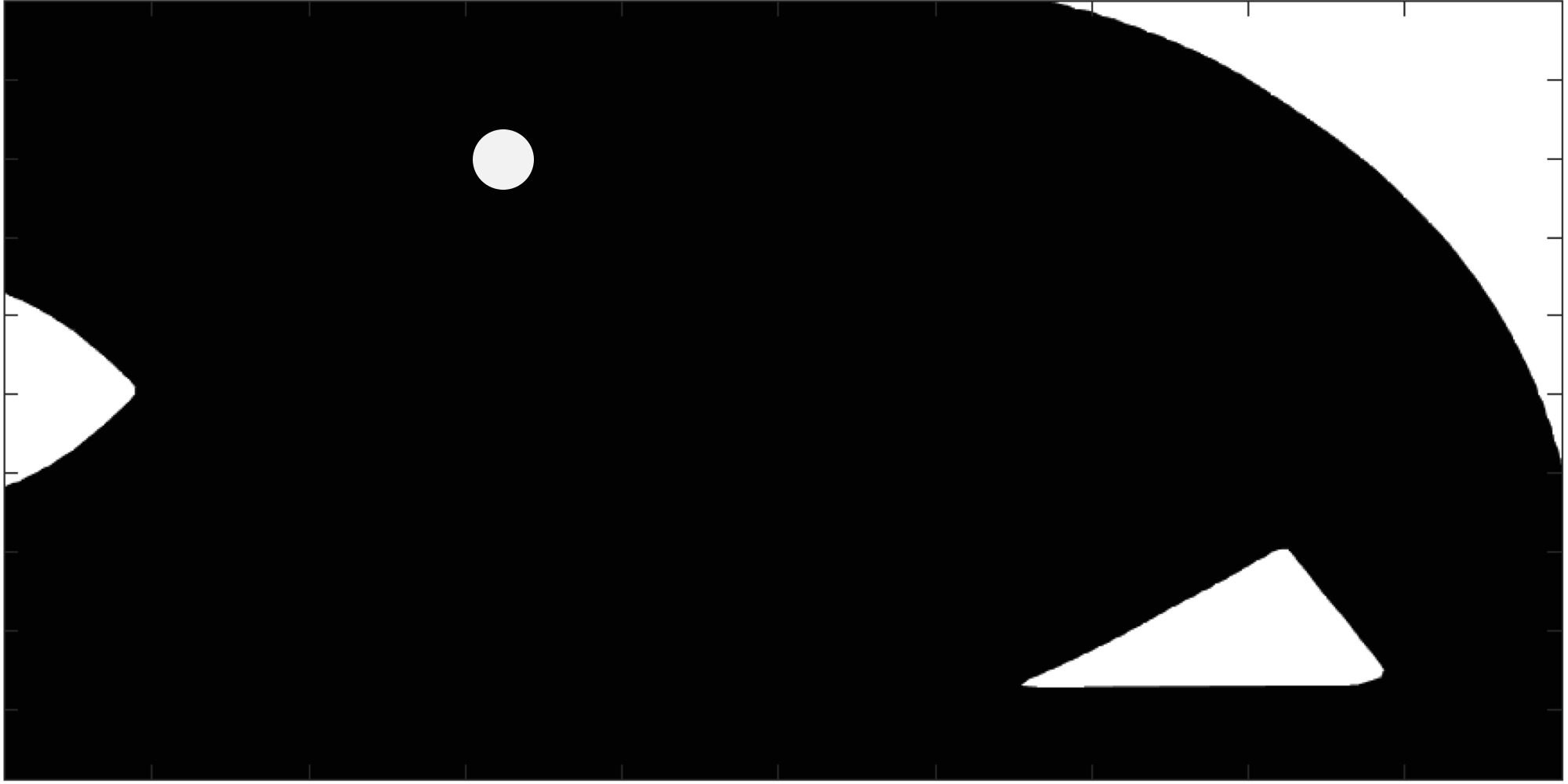}%
		\caption{}
	\end{subfigure}
	\begin{subfigure}[t]{0.48\linewidth}
		\centering
		\includegraphics[width=0.9\linewidth]{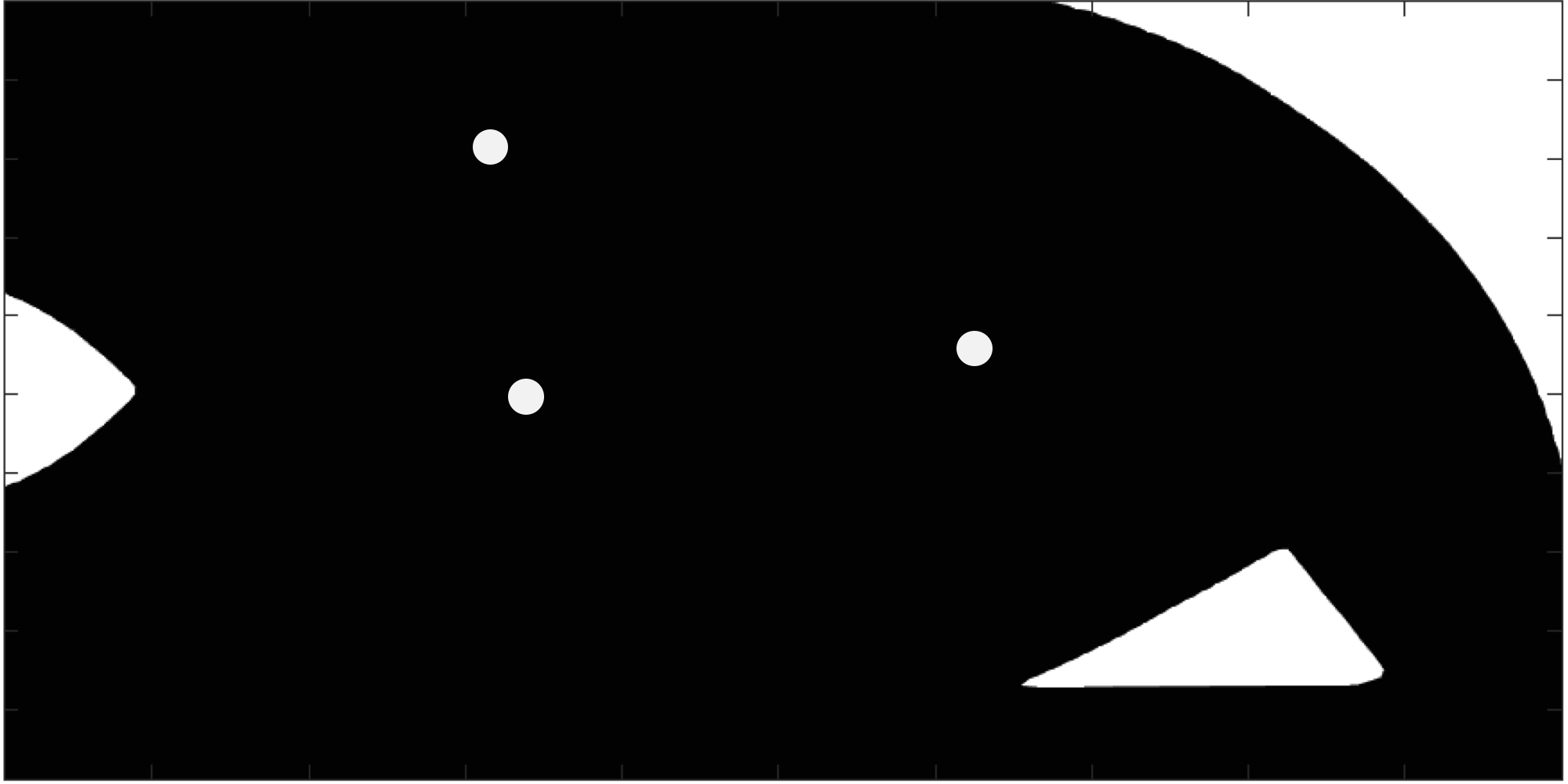}%
		\caption{}
	\end{subfigure}
	
	\begin{subfigure}[t]{0.48\linewidth}
		\centering
		\includegraphics[width=0.9\linewidth]{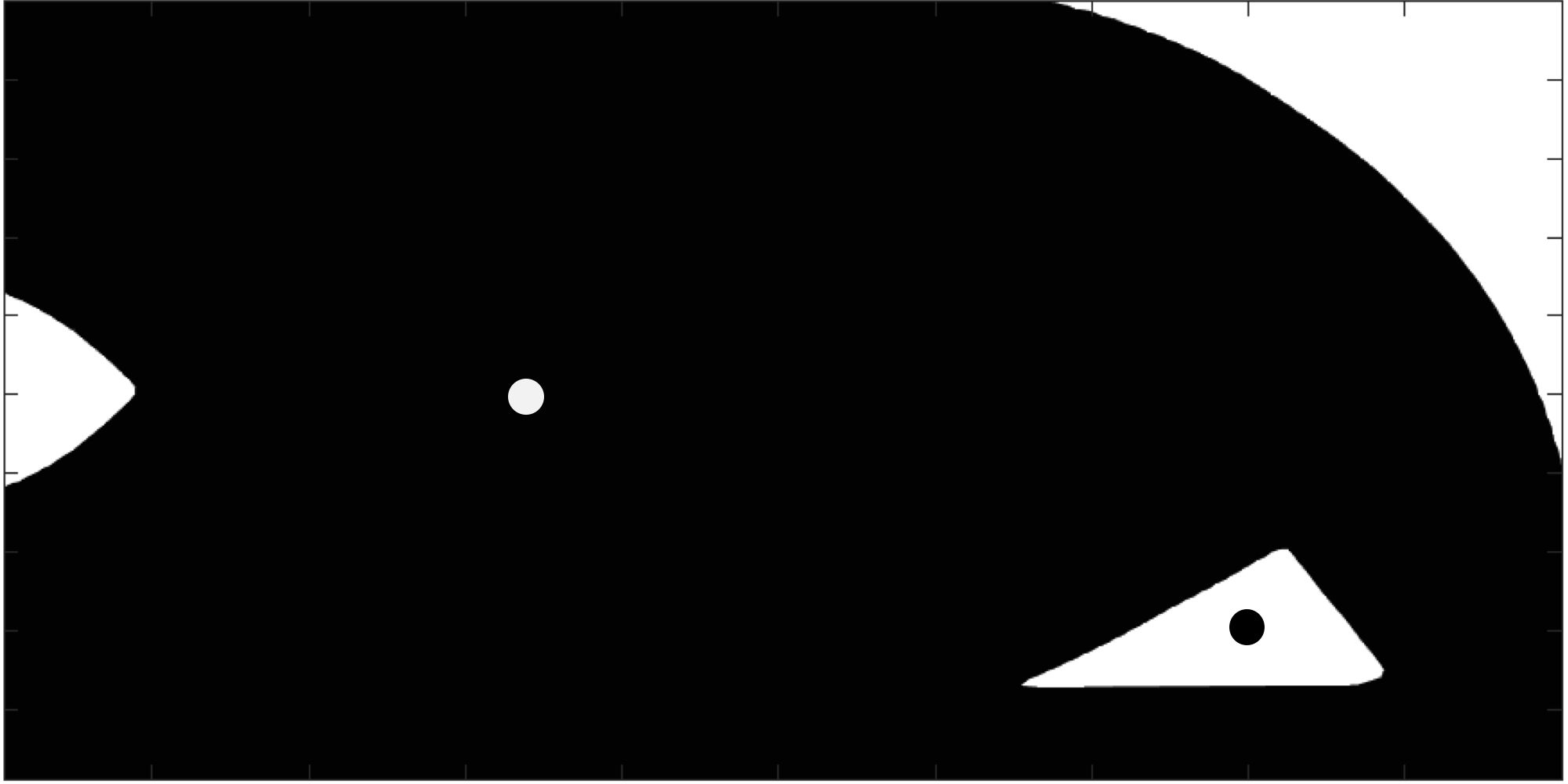}%
		\caption{}
	\end{subfigure}
	\begin{subfigure}[t]{0.48\linewidth}
		\centering
		\includegraphics[width=0.9\linewidth]{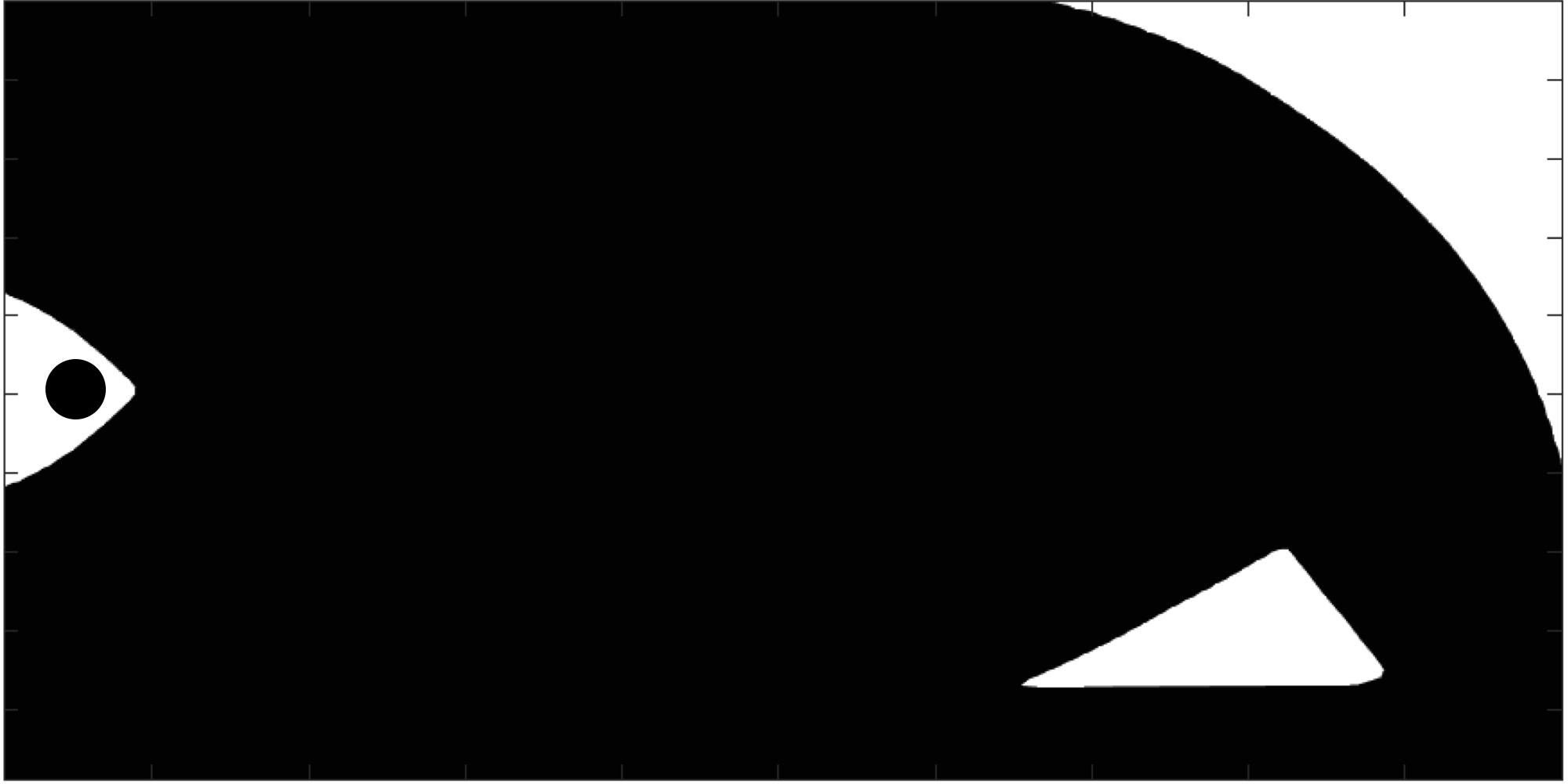}%
		\caption{}
	\end{subfigure}
	\caption{Examples of nearby topologies. } \label{fig_nearbyTopologies}
\end{figure}

\begin{figure*}[t]
\centering\includegraphics[width=\linewidth]{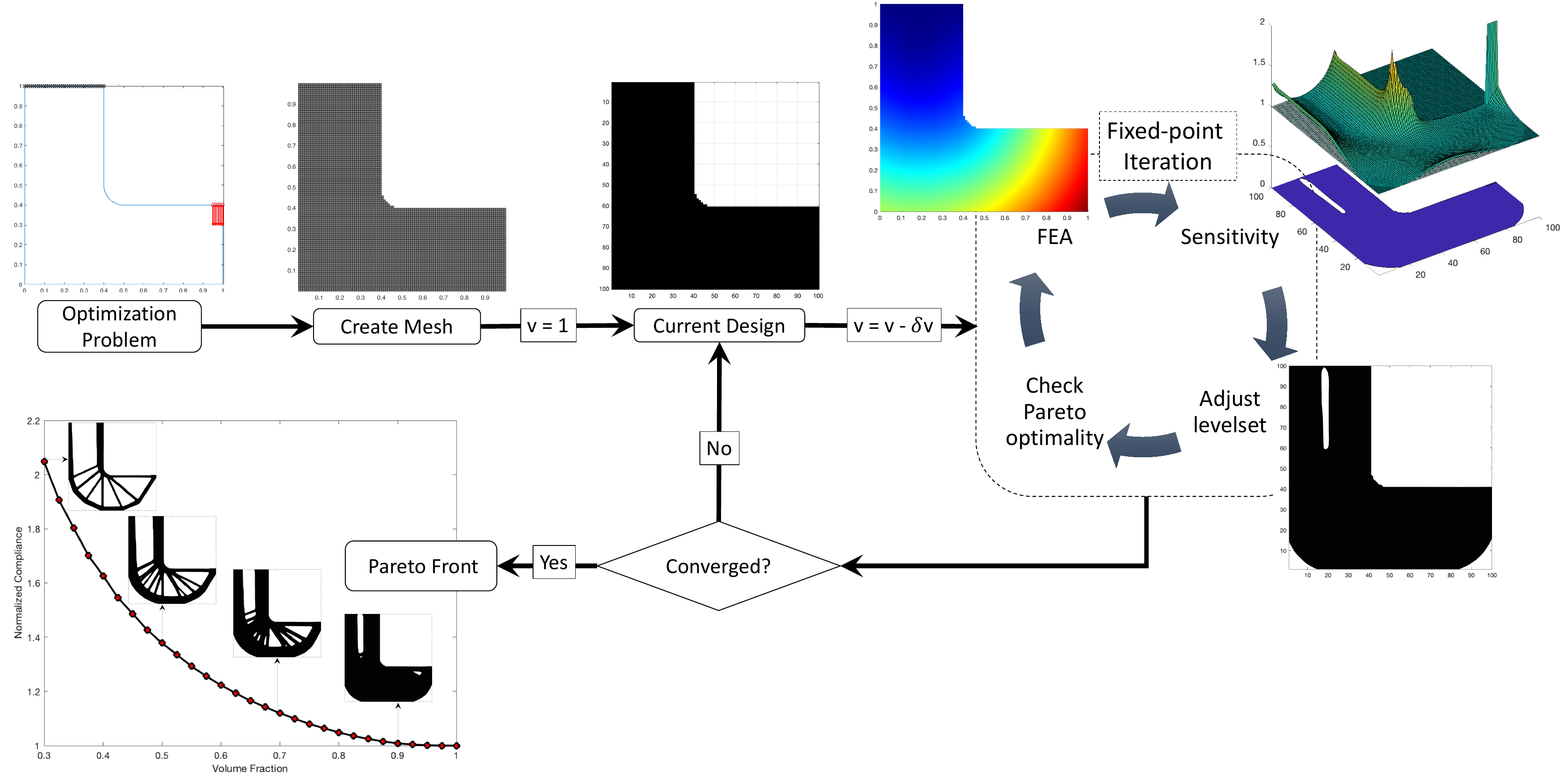}
	\caption{PareTO method workflow.}
	\label{fig_paretoWorkflow}
\end{figure*}

In PareTO, we aim to find designs that offer the best trade-off, at least locally, between multiple objectives. Such design is called a `Pareto-optimal design' and a set of Pareto-optimal designs is referred to as a `Pareto front' \cite{suresh2010199}.  In other words, we want to directly trace the Pareto front of multiple objective functions.

For simplicity, we will focus on two objectives 1) volume fraction and 2) a performance objective (e.g., compliance). Mathematically, we seek to solve the following bi-objective problem:

\begin{equation} \label{eq_paretoTOProblem}
	\begin{aligned}
		\minimize\limits_{\Omega (\bx) \subseteq D} \quad & \left\{V, \varphi(\bd)   \right\}  \\  
		\textrm{s.t.} \quad & V(\bx) \le V^*\\	
		&{R}_{el}(\mathbf{d}) \coloneqq \mathbf{K}_{el}\mathbf{d}-\mathbf{f}_{el}=\mathbf{0}
	\end{aligned}
\end{equation}

where $V$ is the design volume fraction and $\varphi$ is the performance objective. 
As was previously mentioned, finding global Pareto optimal solutions is an open problem, thus we focus on local Pareto optimality, meaning we want to find the designs that have the best performance among their nearby topologies. 

To this end, we first need to define the notion of nearby topologies.

Topologies $\Omega_1$ and $\Omega_2$  are $\delta$-apart if their symmetric volume difference is less than $\delta$, i.e.,:
\begin{equation}
      \Delta V(\Omega_1,\Omega_2) = V(\Omega_1 \setminus \Omega_2 ) +V(\Omega_2 \setminus \Omega_1 ) \le \delta
\end{equation}
      
$\Omega_1$ and $\Omega_2$  are nearby if and only if $\delta$ is sufficiently small.

In the above definition, $\delta$ is essentially the volume of disk-type inclusions. Figure \ref{fig_nearbyTopologies} illustrates different nearby topologies: 

A topology $\Omega$ is locally Pareto-optimal if it is Pareto-optimal with respect to all topologies that are within a distance $\delta$ apart from it, where $\delta$ is sufficiently small. 

Figure \ref{fig_paretoWorkflow} illustrates the workflow of the PareTO method, in which a fixed-point iteration is performed at each outer iteration to ensure that the intermediate designs remain locally Pareto optimal.

PareTO class inherits the primary logic and outer-loop from ESO, \mcode{(Abstract) pareto2d < eso2d}. Similar to BESO, the main modification occurs in the \mcode{update} function, where:

\begin{lstlisting}
function obj = update(obj,volFrac)
    iter = 0;
    isParetoOptimal = 0;
    while (iter < 20) % to avoid cycles typically a few iterations is sufficient
        if ((iter > 0)&&(isParetoOptimal)) % done with current vol
            break
        end
        % Find the level-set value such that the contour has given vol fraction
        obj.m_tau = obj.findContourValueWithVolumeFraction(volFrac);
        index = find(obj.m_dfdx < obj.m_tau); % eliminate all elements less than this value
        obj.m_x = obj.m_solver.m_existingElems; % start with the full domain
        obj.m_x(ind2sub(size(obj.m_dfdx),index)) = 0; % remove elements
        obj.m_solver = obj.m_solver.setDesign(obj.m_x);
        % solve, new sensitivity, ...
        
        isParetoOptimal = obj.analyzeTopology();
        iter= iter+1;
    end
end
\end{lstlisting}
where we check if the design is Pareto-optimal w.r.t. a specified aggressiveness, a scalar value between 0 and 1, where higher value keeps the designs closer the Pareto front:
\begin{lstlisting}
function isParetoOptimal = analyzeTopology(obj)
    T_InMin = min(obj.m_dfdx(obj.m_x==1)); % Min of topological field inside the domain
    T_OutMax = max(obj.m_dfdx(obj.m_x==0 & obj.m_solver.m_existingElems == 1)); % Max of topological field outside the domain
    if (T_InMin > obj.m_paretoAggressiveness*T_OutMax)
        isParetoOptimal = 1;else, isParetoOptimal = 0; end
end
\end{lstlisting}

For the cantilever beam example, we only need to make two changes:
\begin{lstlisting}
topoptClass = @pareto2d_elasticity;
...
%% Construct Optimizer
paretoAggressiveness = 0.65;
topopt = topoptClass(..., paretoAggressiveness);
\end{lstlisting}

 \begin{figure*}[!h]
	\centering
	\begin{subfigure}[b]{0.49\linewidth}
		\centering
		\includegraphics[width=0.99\linewidth]{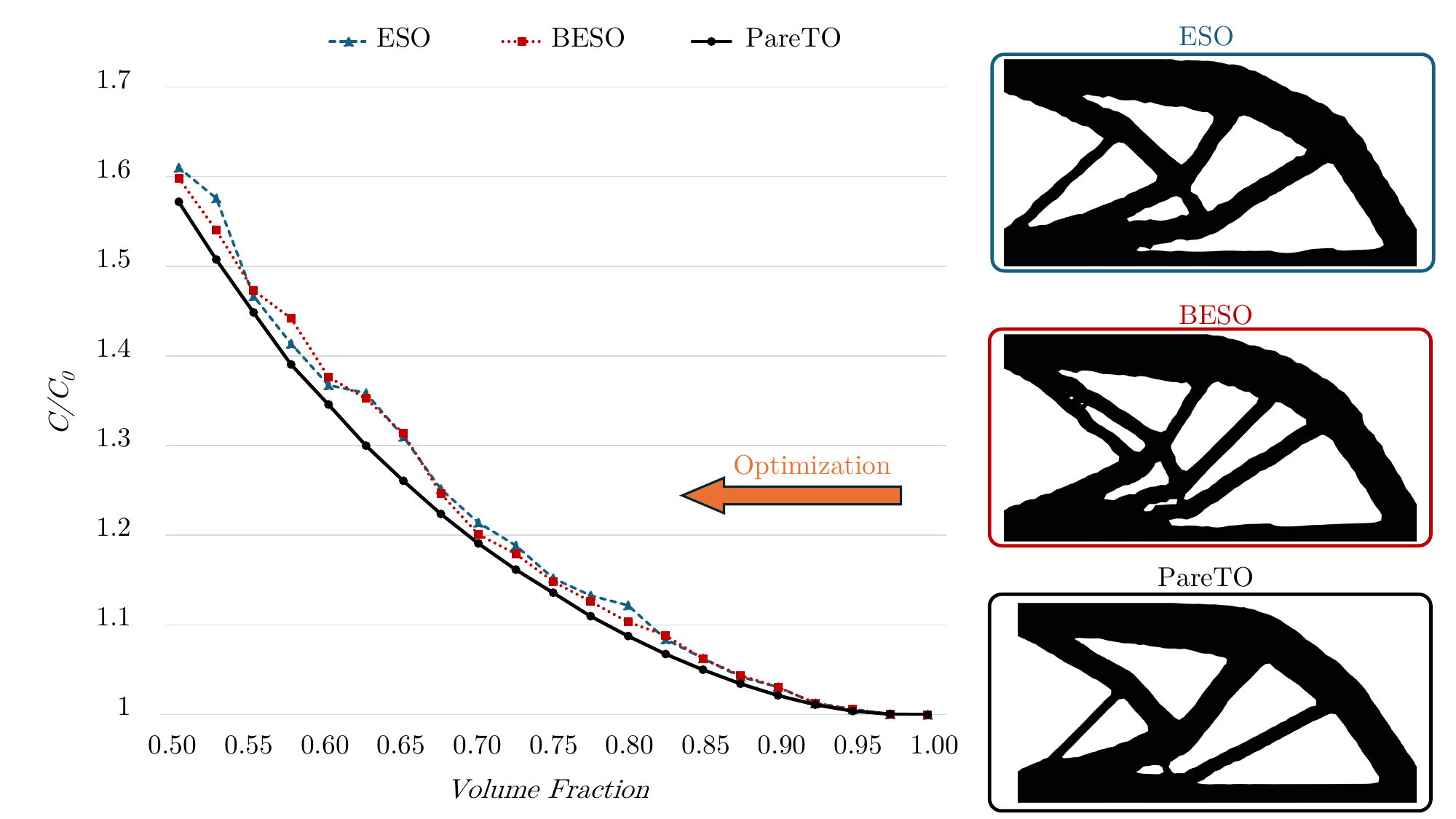}
		\caption{}
	\end{subfigure}
	\begin{subfigure}[b]{0.49\linewidth}
		\centering
		\includegraphics[width=0.99\linewidth]{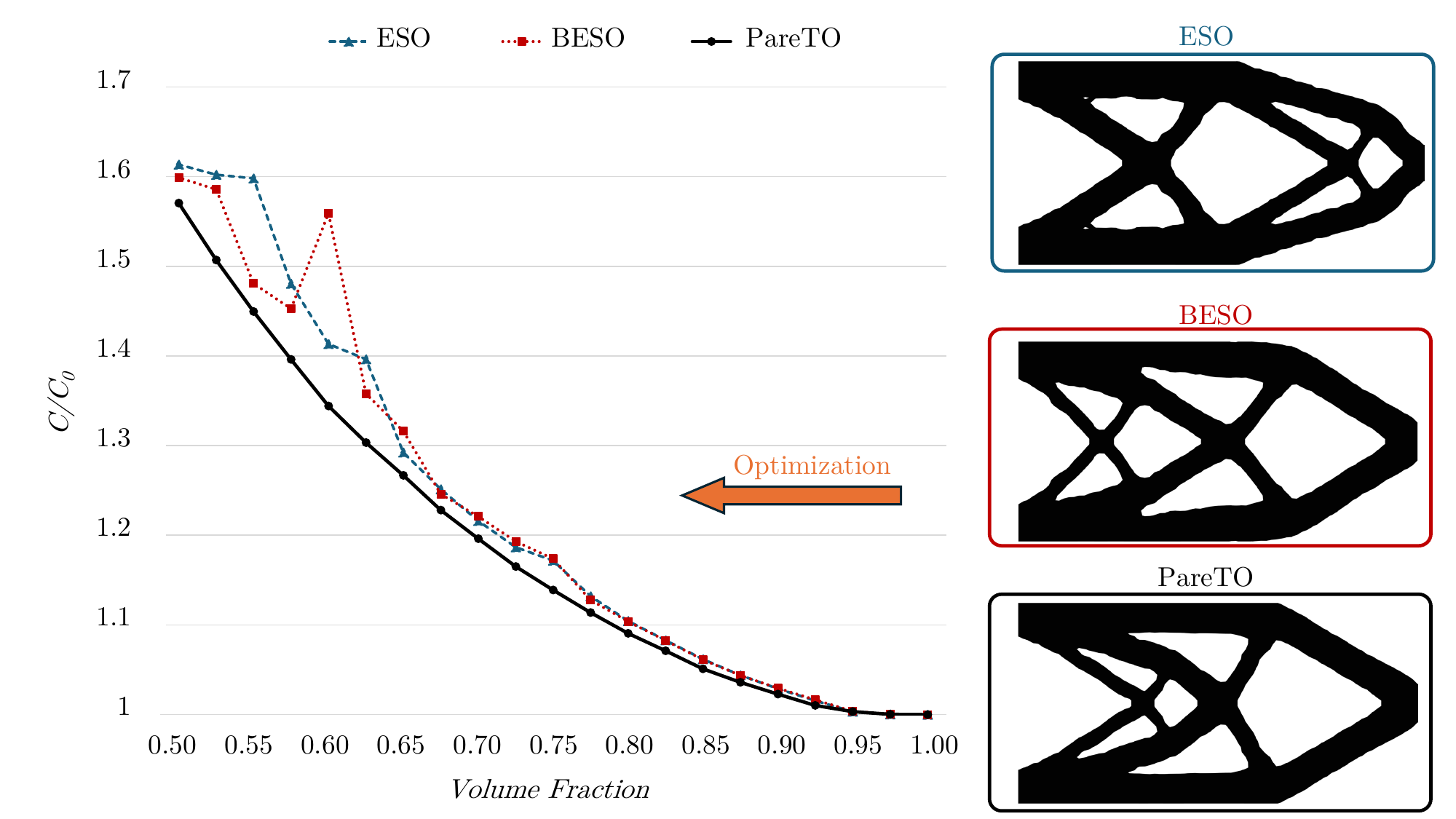}
		\caption{}
	\end{subfigure}
    
	\begin{subfigure}[b]{0.49\linewidth}
		\centering
		\includegraphics[width=0.99\linewidth]{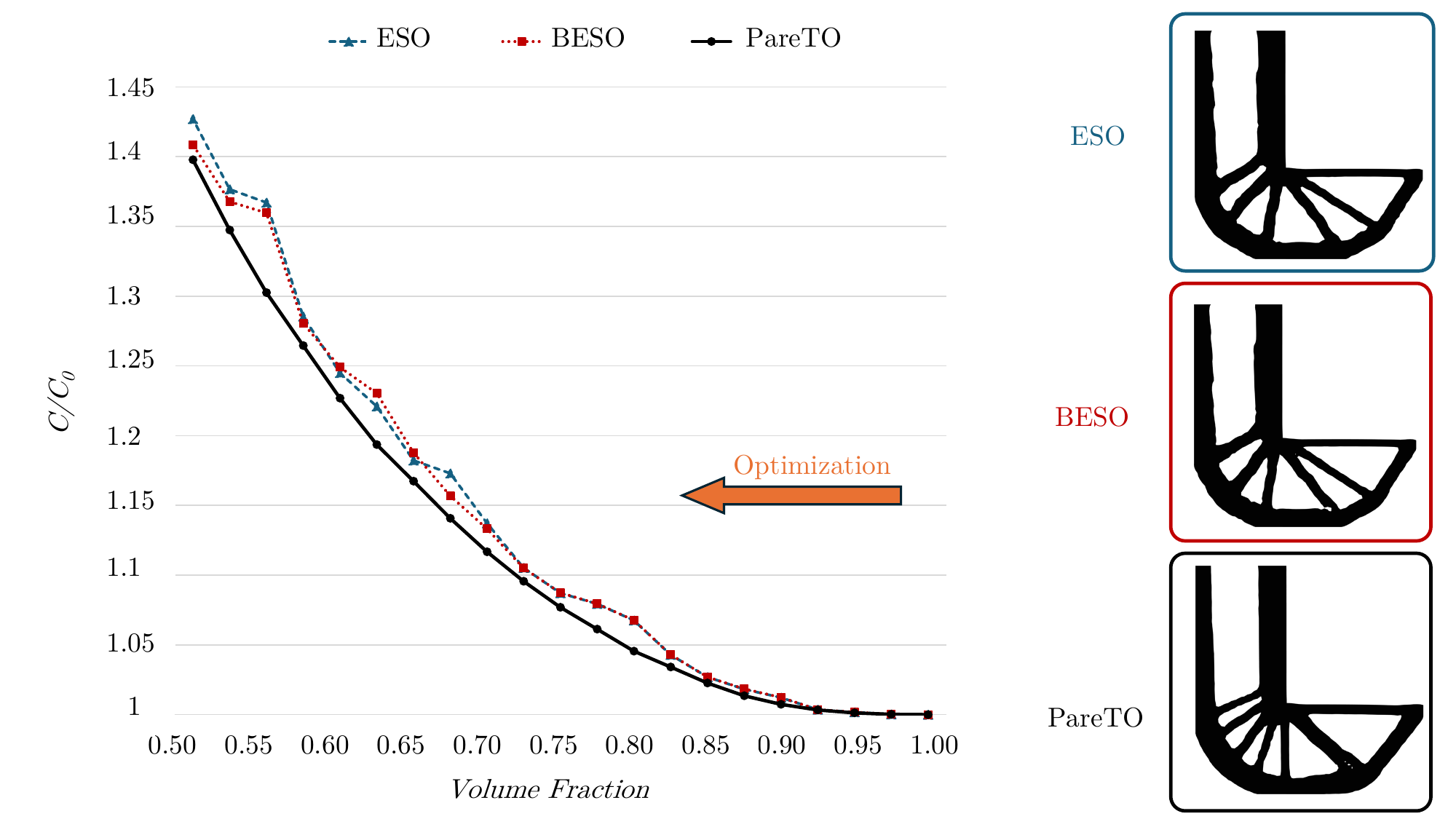}
		\caption{}
	\end{subfigure}
    \begin{subfigure}[b]{0.49\linewidth}
		\centering
		\includegraphics[width=0.99\linewidth]{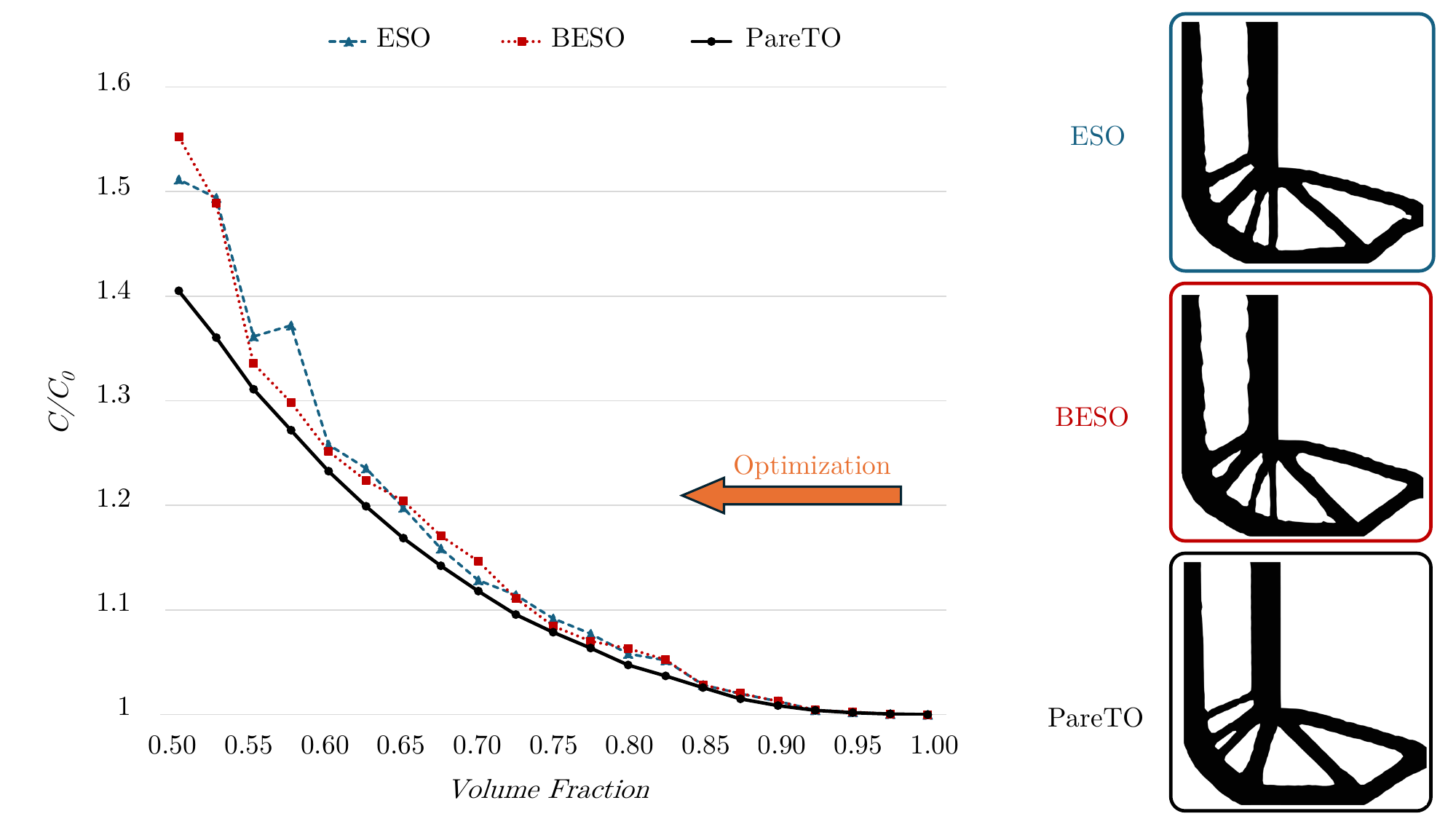}
		\caption{}
	\end{subfigure}
    
	\begin{subfigure}[b]{\linewidth}
		\centering
		\includegraphics[width=0.45\linewidth]{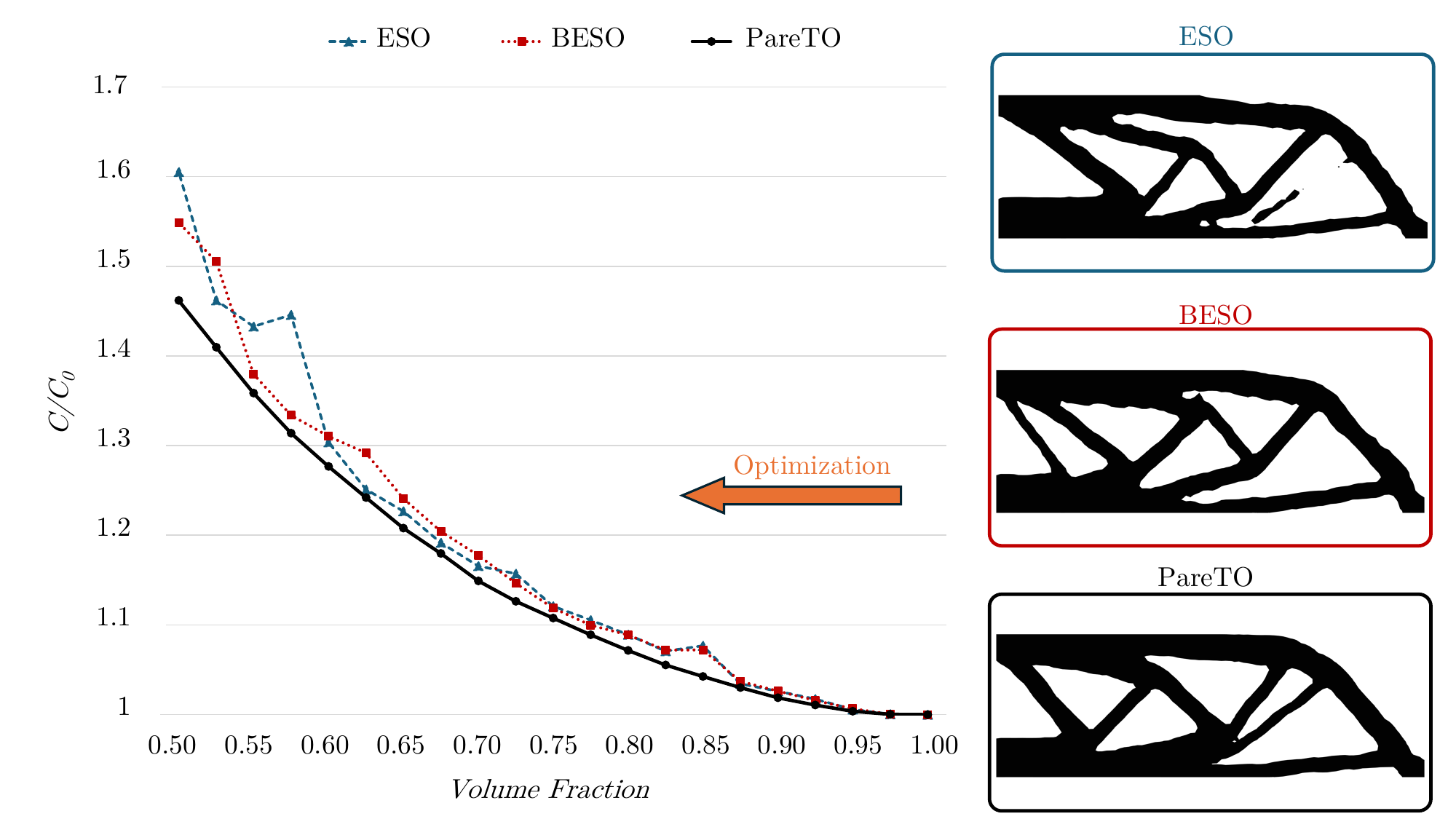}
		\caption{}
	\end{subfigure}

	\caption{Comparison of optimization progress curves for topological sensitivity methods (i.e., ESO, BESO, and PareTO) with compliance minimization using DSF and TSF.}
	\label{fig_topSens_comparison}
\end{figure*} 

\begin{figure*}[!h]
  \centering 
  \includegraphics[width=\linewidth]{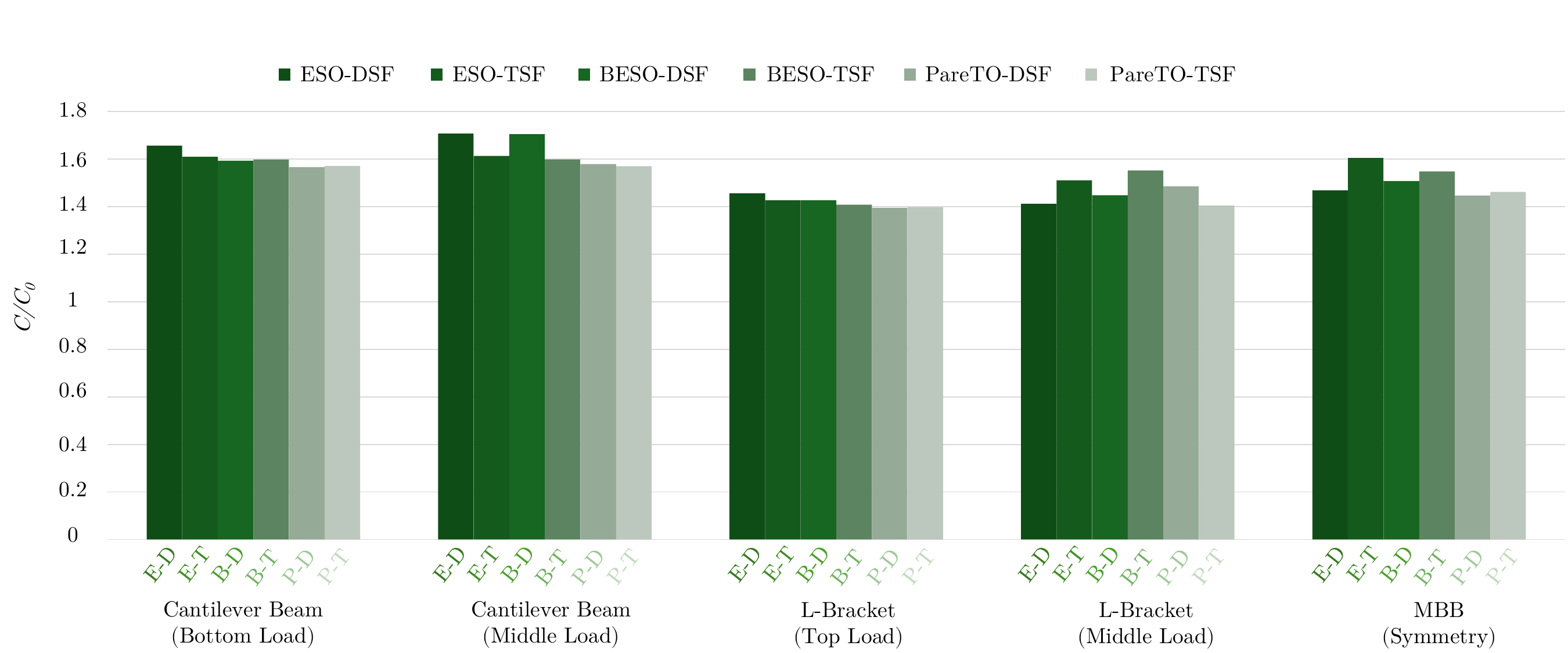}
  \caption{Comparison of topological sensitivity methods (i.e., ESO, BESO, and PareTO) for compliance minimization using DSF and TSF.}
  \label{fig:topSens_summary}
\end{figure*}

\subsubsection{Comparative Study for Topological Sensitivity Methods}\label{sec:compStudyTS}
\begin{table*}[t]
\centering
\caption{Final optimized performance for topological-sensitivity methods.}
\label{tab:topSens_example_metrics}
\scriptsize
\setlength{\tabcolsep}{4pt}
\renewcommand{\arraystretch}{0.92}
\begin{tabular*}{0.88\textwidth}{@{\extracolsep{\fill}} l l c c c @{}}
\toprule
{Example} & {Method}
& {$C$} & {$\delta_{\max}$} & {$\sigma_{\mathrm{vm},\max}$} \\
& & {(N.m)} & {(m)} & {(MPa)} \\
\midrule
\multirow{3}{*}{Cantilever, bottom load}
& ESO    & 6.48 & 6.89e-05 & 4.52 \\
& BESO   & 6.43 & 6.86e-05 & 5.79 \\
& PareTO & 6.32 & 6.74e-05 & 3.24 \\
\\
\multirow{3}{*}{Cantilever, middle load}
& ESO    & 9.76 & 7.83e-05 & 6.69 \\
& BESO   & 9.76 & 7.83e-05 & 6.69 \\
& PareTO & 9.59 & 7.69e-05 & 3.28 \\
\\
\multirow{3}{*}{L-bracket, top load}
& ESO    & 20.4 & 2.23e-04 & 7.99 \\
& BESO   & 20.4 & 2.23e-04 & 7.99 \\
& PareTO & 20.3 & 2.21e-04 & 8.01 \\
\\
\multirow{3}{*}{L-bracket, mid load}
& ESO    & 21.2 & 3.20e-04 & 490 \\
& BESO   & 21.2 & 3.20e-04 & 490 \\
& PareTO & 19.2 & 2.16e-04 & 19.6 \\
\\
\multirow{3}{*}{MBB, symmetry}
& ESO    & 14.2 & 1.55e-04 & 19.4 \\
& BESO   & 14.2 & 1.55e-04 & 19.4 \\
& PareTO & 13.4 & 1.46e-04 & 5.39 \\
\bottomrule
\end{tabular*}
\end{table*}

In this section, we compare the three topological-sensitivity-driven strategies, namely ESO, BESO, and PareTO, on standard compliance-minimization examples, and contrasts between DSF and TSF.

Figure \ref{fig_topSens_comparison} reports the optimization progress curves for the cantilever beam (including a “mid-load” variant), the L-bracket (including a “mid” variant), and the MBB beam, allowing a direct comparison of convergence behavior and achieved compliance across methods and sensitivity choices. 

Figure \ref{fig:topSens_summary} shows the comparative study by summarizing the final designs/performance trends across all cases, highlighting how ESO vs. BESO vs. PareTO differ in their evolution characteristics and how DSF vs. TSF influences both the trajectory and the resulting topologies. 

As expected for the examples considered, the Pareto-based approach attains the best progress curves and the lowest final compliance values, as it explicitly constructs Pareto-optimal solutions across constraint levels. Its intermediate designs are also maintained to be approximately locally optimal for their corresponding volume constraints, which helps explain its systematically improved performance relative to ESO and BESO in these experiments. Table~\ref{tab:topSens_example_metrics} summarizes the structural metrics across all cases. In general, PareTO achieves equal or lower compliance than ESO and BESO, indicating improved global stiffness for the same material budget, and this is accompanied by slightly reduced maximum displacements, reflecting more efficient load paths. Performance differences between ESO and BESO are typically small, whereas Pareto provides modest but consistent gains in stiffness and, in most cases, improved stress control.

\subsection{Design and Manufacturing Constraints} \label{sec:mfgConstraint}

Design and manufacturing constraints are essential in SO/TO because mathematically optimal layouts often contain fine-scale features, disconnected fragments, or grayscale material distributions that are impossible, or prohibitively expensive, to fabricate and verify. Practical design must therefore bridge the gap between idealized continuum formulations and real manufacturing processes, where tool size, deposition width, structural continuity, assembly interfaces, and inspection requirements impose hard geometric and physical limits. The operators in the {\small\texttt{07-mfgConstraints/}} module address this gap by acting directly on design variables and/or sensitivity fields to regularize the design space toward physically realizable solutions. Representative constraints include enforcing a minimum feature size to avoid mesh-dependent microstructures, retaining prescribed regions to protect functional interfaces or load paths, applying physical density filtering or projection to eliminate non-physical intermediate densities, and imposing symmetry when required by loading, assembly, or manufacturing considerations.

A key design principle of the framework is that such constraints are implemented as modular operators that integrate seamlessly into the optimization workflow. 
All manufacturing constraints inherit from the abstract base class {\small\texttt{mfgConstraints}}, which defines a unified interface to the solver. To introduce a new constraint, only two methods are required: (1) {\small\texttt{filterDesign}}, which transforms the design variable field prior to analysis (e.g., filtering, projection, masking, or symmetry enforcement), and (2) {\small\texttt{filterSensitivity}}, which applies the consistent transformation to the sensitivity field during gradient evaluation. This abstraction allows manufacturability operations to function as plug-in operators on the design and gradient fields without modifying the underlying physics solver or optimization algorithm. The solver instance is provided through the constructor, enabling access to mesh information, domain masks, and state variables while preserving separation between physical modeling and constraint enforcement. This object-oriented structure enables rapid extension to new manufacturing constraints while maintaining consistency across different physics modules and optimization formulations.

\subsubsection{Minimum Feature Size Filter}
A well-known issue in density-based TO is that if we directly use the sensitivity field,
the optimizer converges to a non-manufacturable material distribution with \textit{checkerboard pattern} (Fig. \ref{fig_L0.3_checkerboard}).
 
The checkerboard patterns are essentially an artifact of FEA overestimating the stiffness of these structures. In other words,  FEA, using bilinear quadrilateral elements, predicts artificially high stiffness for this type of material layout.

To prevent mesh-dependent checkerboards and unrealistically thin members, a minimum length scale is enforced using a density filter. The filter operates only on elements that belong to the physical design domain, denoted by the indicator field
\begin{equation}
\chi_e =
\begin{cases}
1, & e \in \Omega_\text{design}, \\
0, & \text{otherwise},
\end{cases}
\end{equation}
which corresponds to \texttt{m\_existingElems} property in the \texttt{gridMesher} class implementation.

The filtered density at element $e$ is defined as
\begin{equation}
\tilde{\rho}_e
=
\frac{\displaystyle \sum_{i=1}^{N} H_{e i}\,\chi_i\,\rho_i}
     {\displaystyle \sum_{i=1}^{N} H_{e i}\,\chi_i},
\label{eq:density_filter_masked}
\end{equation}
where $H_{e i}$ is the distance-based weight
\begin{equation}
H_{e i} =
\begin{cases}
r_{\min} - \mathrm{dist}(e,i), & \mathrm{dist}(e,i) \le r_{\min}, \\
0, & \text{otherwise}.
\end{cases}
\end{equation}
This masking ensures that filtering is restricted to the active design domain and avoids artificial averaging with void or non-design regions. After filtering, the density field is again multiplied by $\chi_e$ so that $\tilde{\rho}_e=0$ outside $\Omega_\text{design}$.

Let $\varphi$ be an objective or constraint depending on $\tilde{\rho}_e$. Using the chain rule,
\begin{equation}
\frac{\partial \varphi}{\partial \rho_i}
=
\sum_{e=1}^{N}
\frac{\partial \varphi}{\partial \tilde{\rho}_e}
\frac{\partial \tilde{\rho}_e}{\partial \rho_i}.
\end{equation}
From Eq.~\eqref{eq:density_filter_masked},
\begin{equation}
\frac{\partial \tilde{\rho}_e}{\partial \rho_i}
=
\frac{H_{e i}\,\chi_i}
     {\displaystyle \sum_{j=1}^{N} H_{e j}\,\chi_j}.
\label{eq:filter_derivative}
\end{equation}

In practice, sensitivities are filtered in a form consistent with density filtering,
\begin{equation}
\widehat{\left(\frac{\partial \varphi}{\partial \rho_i}\right)}
=
\frac{1}{\max(\rho_i,\varepsilon)}
\sum_{e=1}^{N}
H_{e i}\,
\rho_i\,
\frac{\partial \varphi}{\partial \tilde{\rho}_e},
\qquad \varepsilon \ll 1,
\label{eq:sens_filter}
\end{equation}
where a small lower bound $\varepsilon$ (here $10^{-3}$) is introduced for numerical regularization when $\rho_i \to 0$. This avoids ill-conditioning associated with divisions by vanishing densities while having negligible influence on the optimization trajectory, since $\varepsilon$ is several orders of magnitude smaller than physically meaningful density values.

The optimized L-bracket at 0.3 volume fraction after applying the minimum feature size filter is shown in Fig. \ref{fig_L0.3MFS}:

\begin{figure} [t]
\centering
    \begin{subfigure}[t]{0.48\linewidth}
    \centering
    \includegraphics[width=\linewidth]{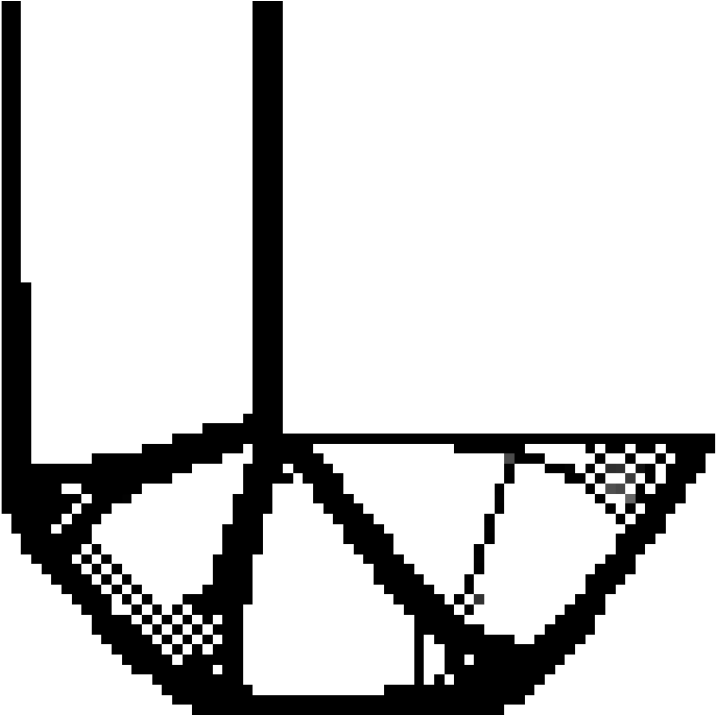}%
    \caption{No filter} 
    \label{fig_L0.3_checkerboard}
    \end{subfigure}
    \begin{subfigure}[t]{0.48\linewidth}
    \centering
    \includegraphics[width=\linewidth]{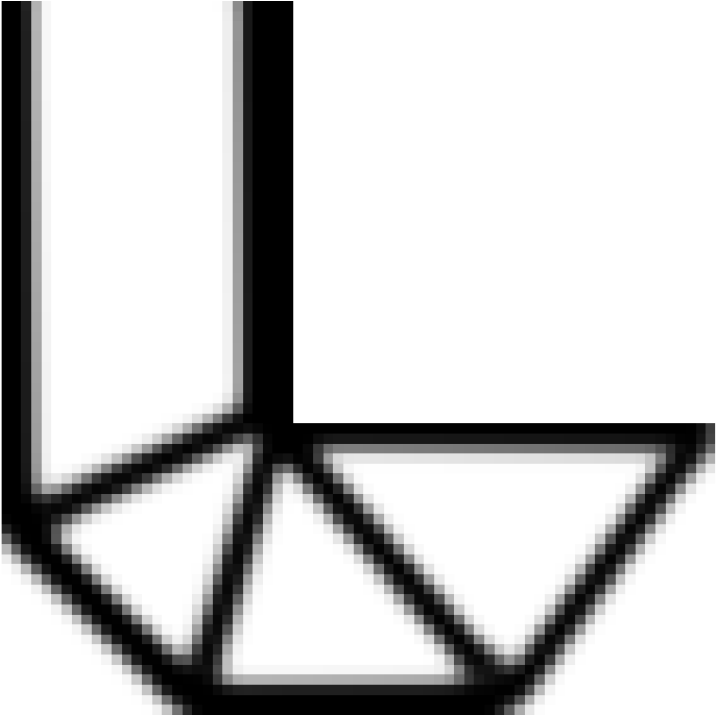}%
    \caption{Filter with $r_{\text{min}}=1.5$ }
    \label{fig_L0.3MFS}
    \end{subfigure}
\caption{Optimized L-bracket at 0.3 volume fraction with and without checkerboard pattern. } \label{fig_L0.3_checkerboardFilter}
\end{figure}

The density filter used here is one of several established techniques for enforcing minimum feature size in SIMP-based topology optimization. Several alternative strategies have been proposed in the literature to enforce a minimum length scale within SIMP-based topology optimization. Sensitivity filtering \cite{sigmund200199} operates by smoothing the gradient field rather than the design variables themselves, indirectly suppressing high-frequency oscillations. Heaviside projection methods \cite{guest2004achieving,sigmund2007morphology} combine density filtering with a nonlinear projection to generate binary solid/void designs while retaining explicit control over the minimum feature size. PDE-based filters \cite{lazarov2011filters} introduce a length scale by solving a Helmholtz-type equation, yielding a mesh-independent and physically interpretable filtering radius. More recently, morphological and robust formulation approaches based on erosion/dilation operators have been employed to simultaneously control both minimum and maximum feature sizes and to improve manufacturability under geometric uncertainties.

All of these approaches introduce a controllable geometric length scale; the present implementation adopts the convolution-based density filter due to its simplicity, computational efficiency, and compatibility with multiple physics modules within the framework.

Similar methods can be used to impose minimum feature size constraint by filtering the design and sensitivity fields for level-set \cite{challis2010discrete} and topological sensitivity methods \cite{suresh2010199}.

\subsubsection{Heaviside Projection for Physical Density}

To suppress intermediate (gray) densities and obtain physically meaningful 0-1 designs, a smooth Heaviside projection is incorporated directly into the optimization loop to map design variables $\rho_e$ to physical densities $\hat{\rho}_e$:
\begin{equation}\label{eq_HeavisidePrj}
\hat{\rho}_e =
\dfrac{\tanh (\beta \eta) + \tanh\!\big(\beta (\rho_e - \eta)\big)}
      {\tanh (\beta \eta) + \tanh\!\big(\beta (1 - \eta)\big)} ,
\end{equation}
where $\beta$ controls the sharpness of the projection and $\eta$ is the threshold parameter, typically chosen as $\eta=0.5$. As $\beta$ increases, the mapping approaches a step function resulting in crisp black and white designs while maintaining differentiability. Figure \ref{fig_H} illustrates the Heaviside projection with different values of $\beta$ and $\eta$:

\begin{figure} [t]
	\centering
	\includegraphics[width=0.9\linewidth]{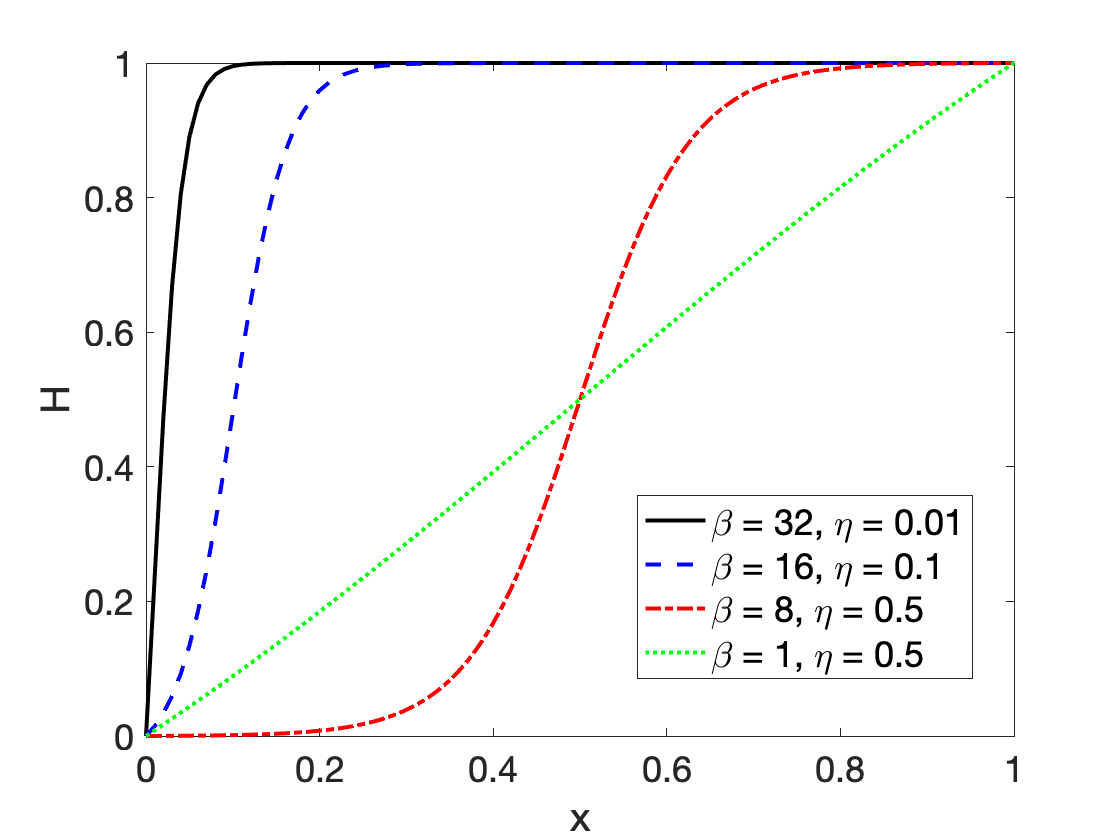}%
	\caption{Heaviside projection with different $\beta$ and $\eta$ values.} 
	\label{fig_H}
\end{figure}

The derivative of the projection is
\begin{equation}
\frac{\partial \hat{\rho}_e}{\partial \rho_e}
=
\frac{\beta(1 - \tanh(\beta(\rho_e - \eta))^2)}{\tanh(\beta\eta) + \tanh\!\big(\beta(1-\eta)\big)}.
\label{eq:heaviside_sens}
\end{equation}
For an objective or constraint $\varphi(\hat{\rho})$, the chain rule gives
\begin{equation}
\frac{\partial \varphi}{\partial \rho_e}
=
\frac{\partial \varphi}{\partial \hat{\rho}_e}
\frac{\partial \hat{\rho}_e}{\partial \rho_e}.
\end{equation}

When both minimum feature size filtering and Heaviside projection are applied, the design mapping becomes
\[
\rho \;\xrightarrow{\text{filter}}\; \tilde{\rho}
\;\xrightarrow{\text{projection}}\; \hat{\rho}.
\]

The \mcode{filterDensity} method in \mcode{density} class imposes these constraints on the pseudo-density automatically: 
\begin{lstlisting}
function obj = filterDensity(obj)
    obj.m_stageDesign = cell(obj.m_numMfgConstraints+1,1);
    obj.m_stageDesign{1} = obj.m_x .* obj.m_solver.m_existingElems;  % x^(0)

    for k = 1:obj.m_numMfgConstraints
        obj.m_stageDesign{k+1} = obj.m_mfgConstraints{k}.filterDesign(obj.m_stageDesign{k});
    end
    obj.m_xPhys = obj.m_stageDesign{end}; % x^(M)
end
\end{lstlisting}
Observe that \mcode{obj.m_mfgConstraints} can contain \textit{any} manufacturing constraint implementation (e.g., symmetry, retain) as long it inherits from the \mcode{mfgConstraint} abstract class.

The corresponding gradient propagation follows the reverse path (from physical densities back to design variables):
\[
\frac{\partial  \varphi}{\partial \rho}
\;\xleftarrow{\;\frac{\partial \tilde{\rho}}{\partial \rho}\;}
\frac{\partial  \varphi}{\partial \tilde{\rho}}
\;\xleftarrow{\;\frac{\partial \hat{\rho}}{\partial \tilde{\rho}}\;}
\frac{\partial  \varphi}{\partial \hat{\rho}}.
\]

The complete sensitivity follows from the chain rule:
\begin{equation}
\frac{\partial  \varphi}{\partial \rho_i}
=
\sum_{e=1}^{N}
\frac{\partial  \varphi}{\partial \hat{\rho}_e}
\frac{\partial \hat{\rho}_e}{\partial \tilde{\rho}_e}
\frac{\partial \tilde{\rho}_e}{\partial \rho_i},
\label{eq:full_chain_rule}
\end{equation}

Thus, the minimum feature size filter smooths the spatial distribution of design variables, while the Heaviside projection sharpens material transitions; their sensitivities combine multiplicatively through~\eqref{eq:full_chain_rule}.

The \mcode{filterSensitivity} method in \mcode{density} class imposes these constraints automatically. For example, the following code snipet demonstrates this for the objective function(s): 
\begin{lstlisting}
...
% Objectives
for fId = 1:num_fIds
    dfdx = obj.m_dfdx((fId-1)*numElems+1:fId*numElems,1);
    dfdx = reshape(dfdx, obj.m_solver.m_ny, obj.m_solver.m_nx);  % df/dx^(M)
    sens = dfdx;
    for k = obj.m_numMfgConstraints:-1:1
        design_in = obj.m_stageDesign{k};% x^(k-1)
        sens = obj.m_mfgConstraints{k}.filterSensitivity(design_in, sens);
    end
    sens = sens .* obj.m_solver.m_existingElems;
    obj.m_dfdx((fId-1)*numElems+1:fId*numElems,1) = sens(:);
end
...
\end{lstlisting}
Observe that the sensitivity constraints are imposed in the backward direction, consistent with \eqref{eq:full_chain_rule}.

\begin{figure} [t]
	\begin{subfigure}[t]{0.3\linewidth}
		\centering
		\includegraphics[width=0.99\linewidth]{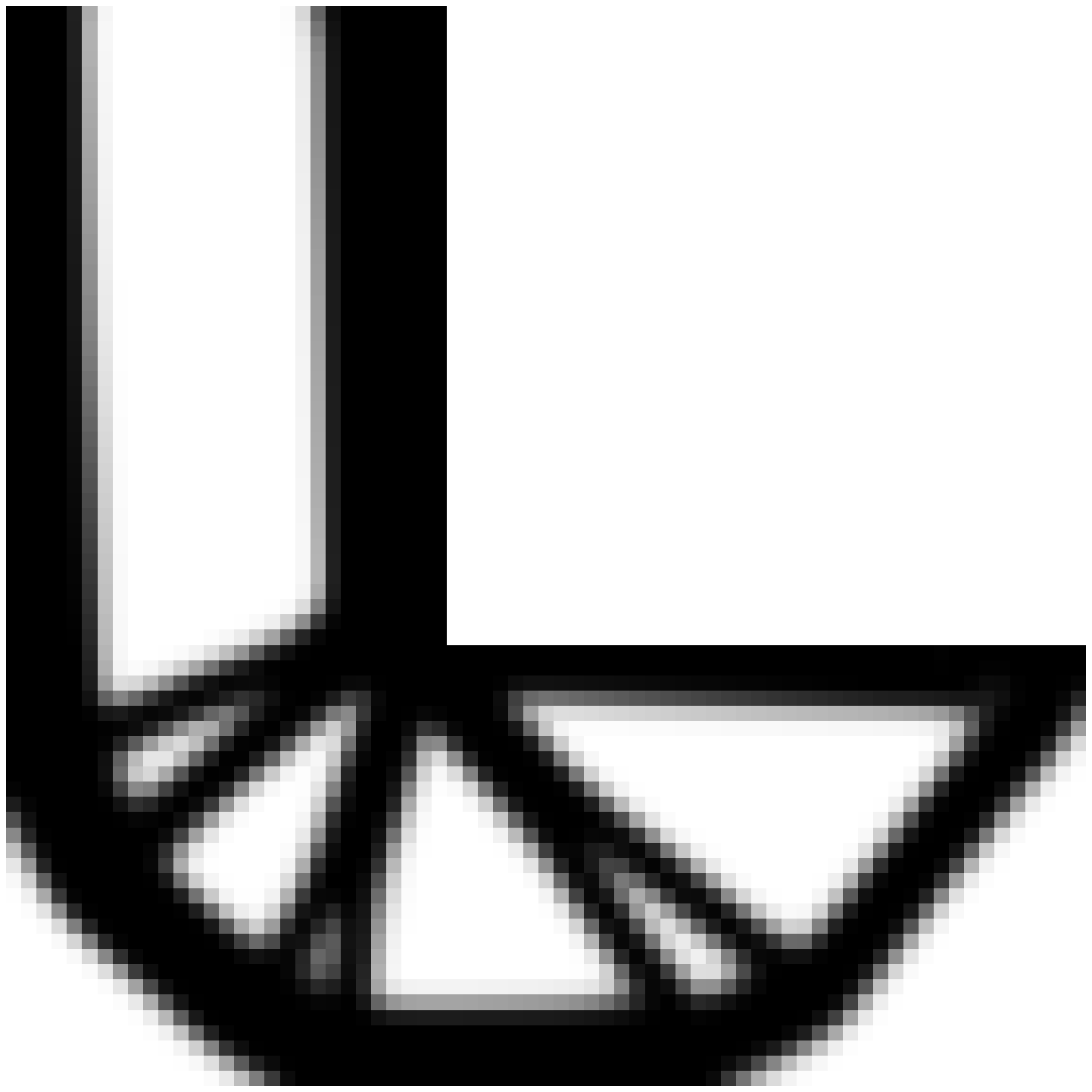}%
		\caption{$\beta=1$}
	\end{subfigure}
	\begin{subfigure}[t]{0.3\linewidth}
		\centering
		\includegraphics[width=0.99\linewidth]{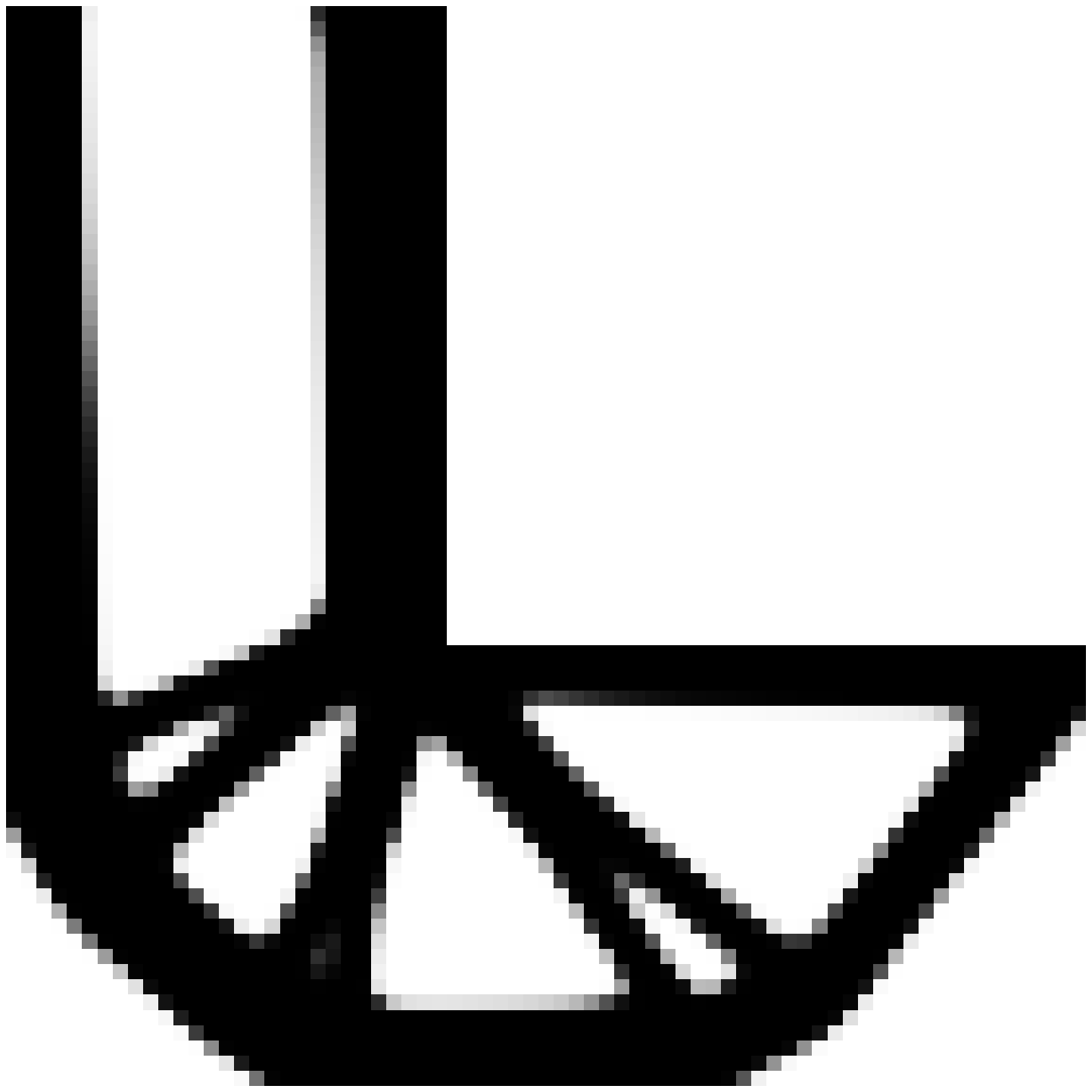}%
		\caption{$\beta=8$}
	\end{subfigure}
	\begin{subfigure}[t]{0.3\linewidth}
		\centering
		\includegraphics[width=0.99\linewidth]{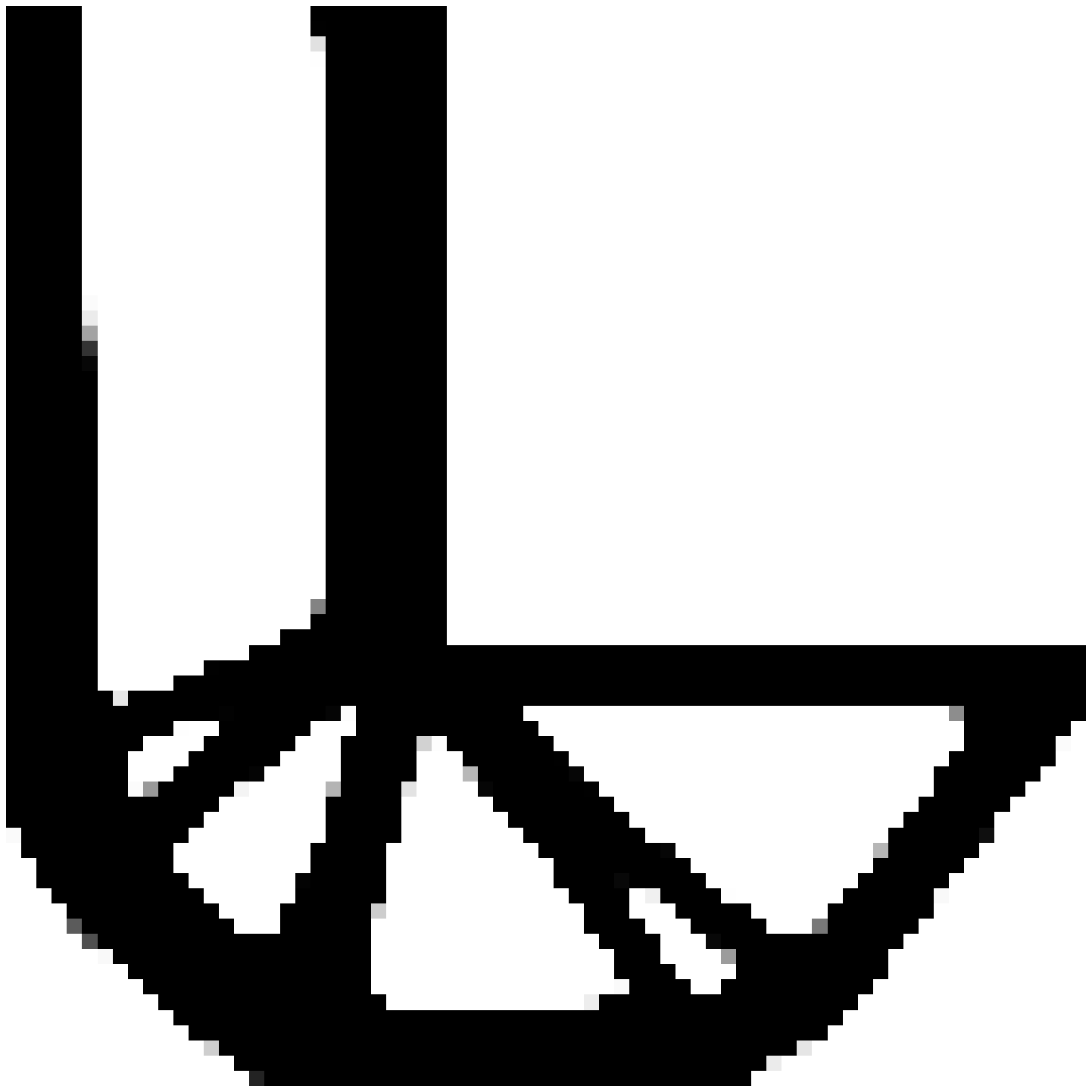}%
		\caption{$\beta=64$}
	\end{subfigure}
	\caption{Impact of the Heaviside projection at different $\beta$ values. } \label{fig_L_Heaviside}
\end{figure}

Figure \ref{fig_L_Heaviside} shows the impact of projection with different $\beta$ values ($\eta=0.5$) on the physical density. Increasing $\beta$ penalizes the intermediate density values and results in a closer to 0-1 design.
   
\subsubsection{Retained Regions} \label{sec:retainedRegions}

\begin{figure}[t]
    \centering
    \includegraphics[width=0.7\linewidth]{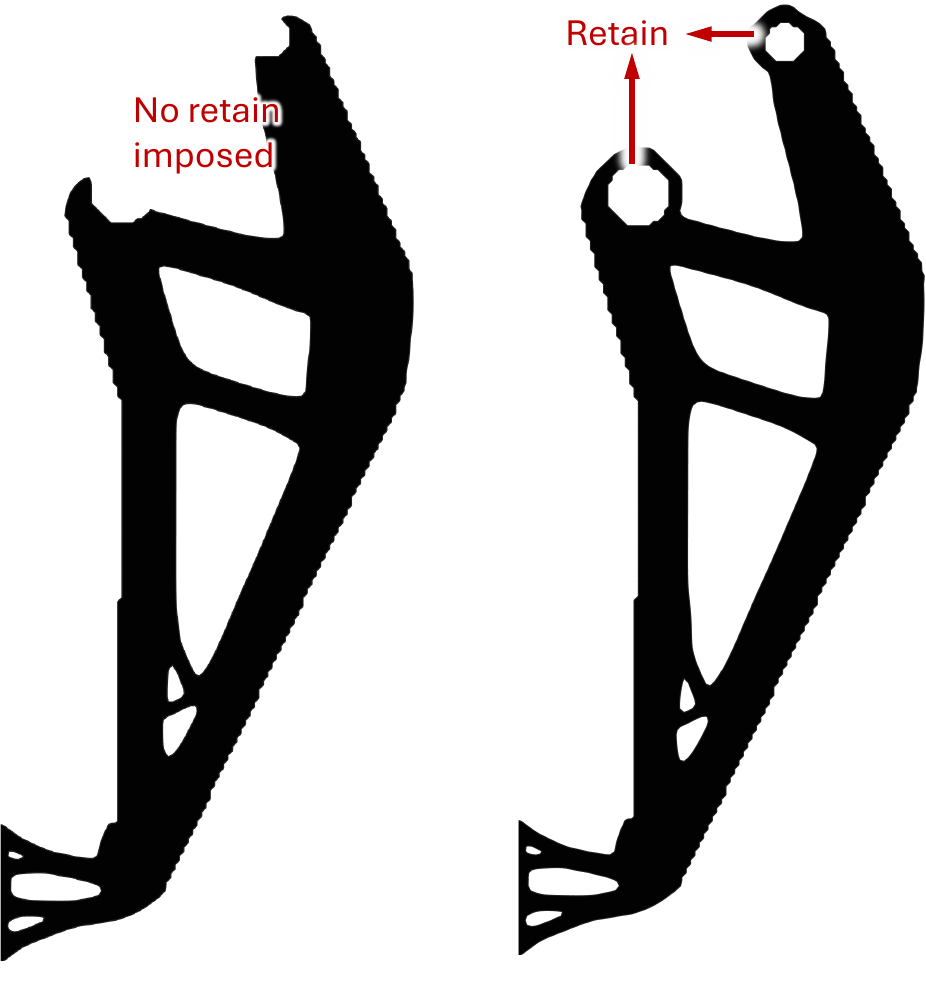}%
    \caption{Gripper example optimized using SIMP interpolation with the OC method. Left: boundary condi tions. Middle: optimized design at volume fraction $0.65$ without retained regions. Right: optimized design at volume fraction $0.65$ with a retain constraintimposed on the circular edge regions.}
    \label{fig_gripper_retain}
\end{figure}

\begin{figure*}[t]
    \centering
    \includegraphics[width=0.7\linewidth]{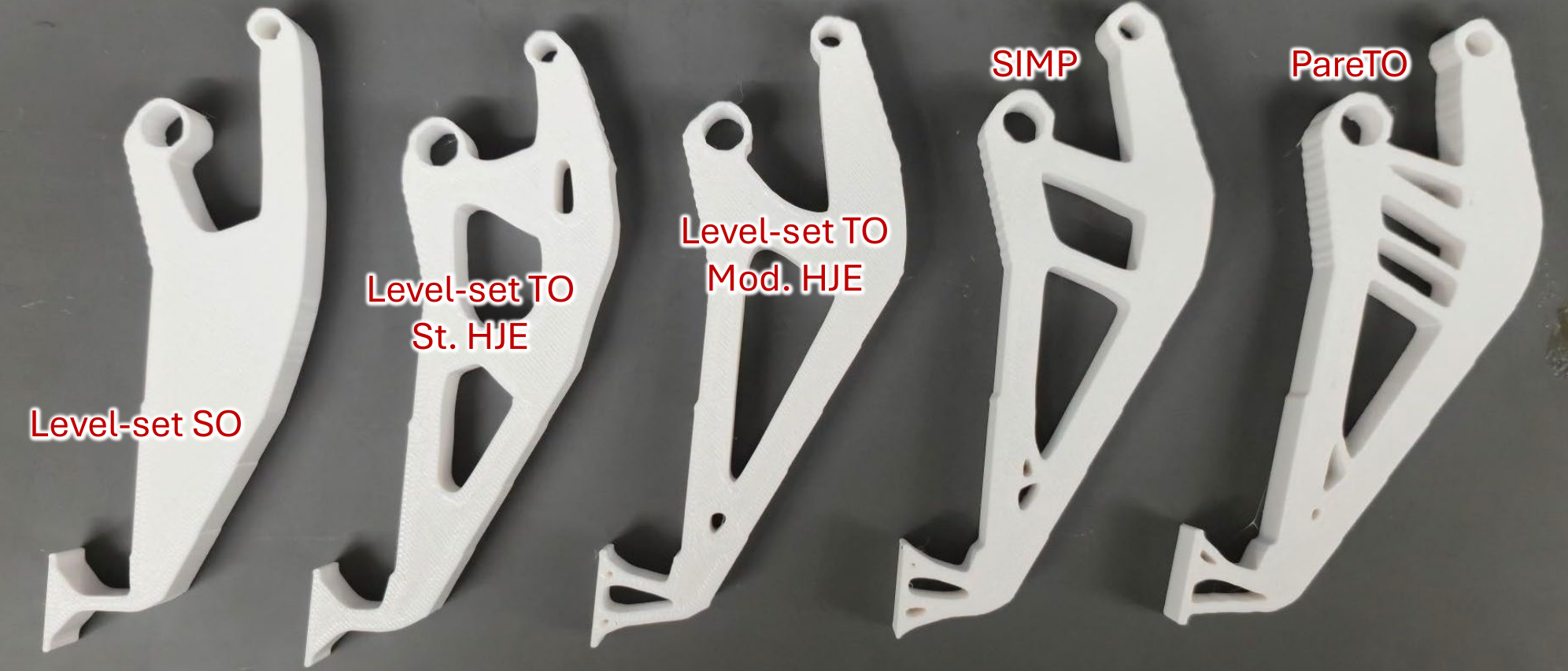}%
    \caption{3D-printed optimized grippers at volume fraction $0.65$ using different optimization approaches, with circular edge regions retained.}
    \label{fig_gripper_printed}
\end{figure*}

In many practical problems, portions of the domain must remain solid to preserve interfaces, supports, or functional boundaries. This is enforced through a retain mask $\chi_e^{\text{ret}}\in\{0,1\}$, where $\chi_e^{\text{ret}}=1$ denotes elements that must remain material. The design update operators are modified so that these elements are excluded from material removal.

For density-based methods, retained elements are forced to full material,
\begin{equation}
\rho_e^{\text{filtered}} = 1 \quad \text{if } \chi_e^{\text{ret}}=1,
\end{equation}
and their sensitivities are set to the minimum value in the field,
\begin{equation}
\left.\frac{\partial \varphi}{\partial \rho_e}\right|_{\chi_e^{\text{ret}}=1}
= \min_{i} \left(\frac{\partial \varphi}{\partial \rho_i}\right),
\end{equation}
which prevents the optimizer from driving these elements toward void.

For level-set methods, the geometry is fixed by nullifying the shape sensitivity in retained regions,
\begin{equation}
\left.\frac{\partial \varphi}{\partial \phi_e}\right|_{\chi_e^{\text{ret}}=1} = 0,
\end{equation}
so the interface does not move locally.

For topological sensitivity methods (e.g., ESO/BESO), retained elements are biased toward material retention by assigning the maximum sensitivity value,
\begin{equation}
\left.\frac{\partial \varphi}{\partial \rho_e}\right|_{\chi_e^{\text{ret}}=1}
= \max_{i} \left(\frac{\partial \varphi}{\partial \rho_i}\right),
\end{equation}
thereby preventing element removal during evolutionary updates.

This masking strategy provides a simple and robust way to enforce non-design or protected regions while remaining consistent with the update logic of each optimization formulation.

Figure~\ref{fig_gripper_retain} illustrates the optimized gripper example and the effect of enforcing retained regions. The boundary conditions are shown on the left; the optimized design at a volume fraction of $0.65$ is shown without a retain constraint (middle) and with a retain constraint applied to the circular edge regions (right). The problem is discretized with $10{,}000$ elements and the material is PETG (polymer) with $E=2~\mathrm{GPa}$ and $\nu=0.35$ under an applied force of $10~\mathrm{N}$. The normalized compliance values are $C_\text{NoRetain}/C_0=0.32$ (without retain) and $C_\text{Retain}/C_0=0.33$ (with retain), indicating that (for this example) retaining the circular edges has a modest impact on performance while enforcing a manufacturable and functionally required interface.

All designs were obtained at a volume fraction of $0.65$ while retaining the circular edge regions.
The designs were fabricated using an Elegoo Centauri Carbon printer with white PETG. From left to right, the optimized grippers correspond to the following approaches and final compliance values:
\begin{enumerate}
    \item level-set SO: $C_\text{LSSO} = 1.04\times10^{-5}\,\mathrm{N.m}$,
    \item level-set TO with standard HJE: $C_\text{stHJE} = 8.49\times10^{-6}\,\mathrm{N.m}$,
    \item level-set TO: modified HJE and $C_\text{modHJE} = 8.33\times10^{-6}\,\mathrm{N.m}$,
    \item SIMP-based TO: $C_\text{SIMP} = 8.66\times10^{-6}\,\mathrm{N.m}$, and 
    \item Pareto-based TO: $C_\text{PareTO} = 8.29\times10^{-6}\,\mathrm{N.m}$.
\end{enumerate}

These values are reported for the specific examples, parameter settings, and stopping criteria used here; they are not intended to establish a general ranking or comparative superiority among the methods.

\section{Additional Problem Formulations} \label{sec:advTopics}

This section describes additional modeling features supported by STORX that broaden the range of design problems that can be addressed. These include FEA and TO under multiple load cases (Section~\ref{sec:multiLoad}), incorporation of body forces such as self-weight (Section~\ref{sec:selfWeight}), and TO under thermal conduction (Section~\ref{sec:thermal}). These capabilities enable the treatment of more realistic loading conditions and multi-physics design scenarios within the same unified framework.

\subsection{Multiple Load Scenarios} \label{sec:multiLoad}

\begin{figure} [t]
		\centering
                \begin{subfigure}[t]{\linewidth}
			\centering
			\includegraphics[width=0.7\linewidth]{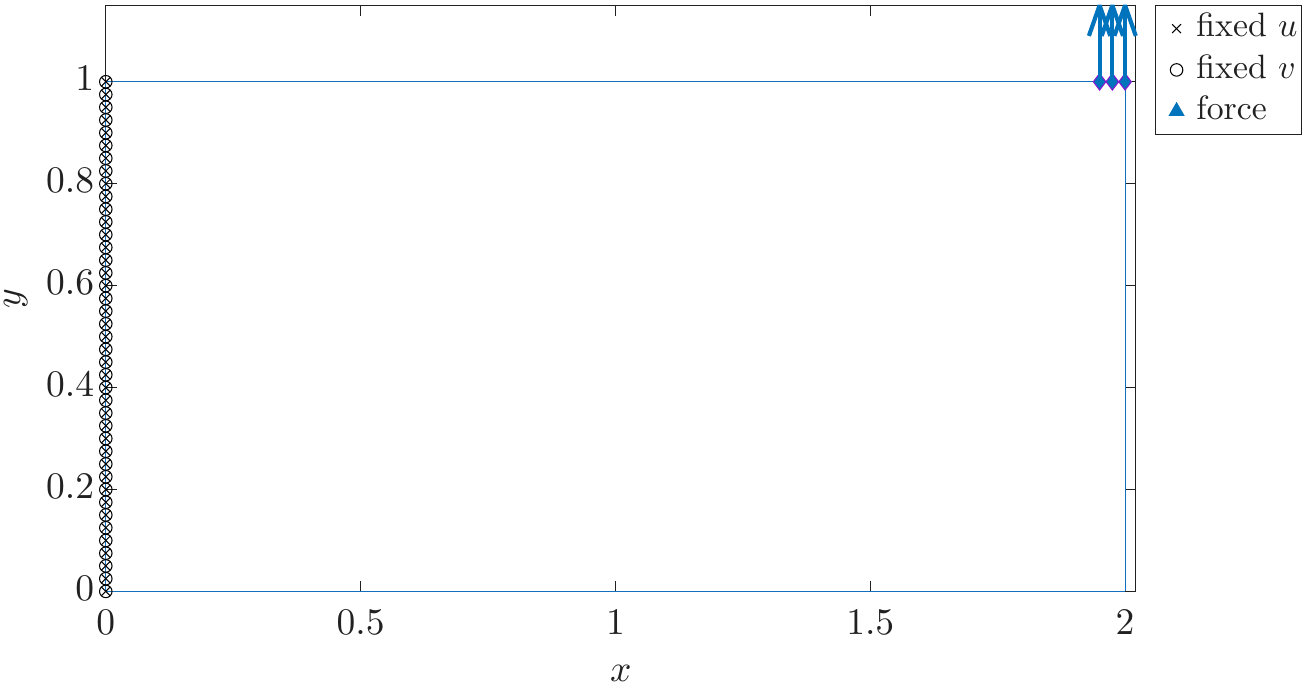}%
			\caption{Scenario 1: upward force $F_1=200\;kN$}
		\end{subfigure}
        
		\begin{subfigure}[t]{\linewidth}
			\centering
	\includegraphics[width=0.7\linewidth]{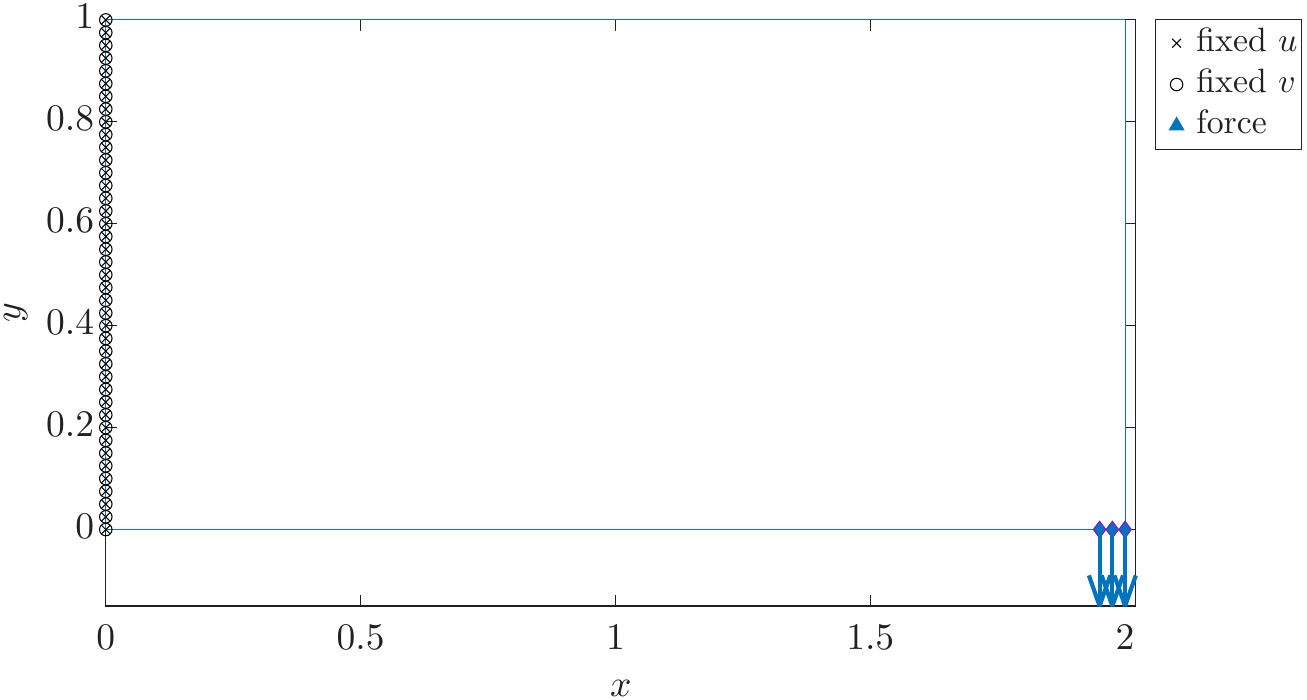}%
			\caption{Scenario 2: downward force $F_2=-100\;kN$}
		\end{subfigure}
		\caption{Cantilever beam under two load scenarios. } \label{fig_beam_multiLoad_BC}
\end{figure}

Many practical structures are subjected to multiple service load cases rather than a single load. Designs optimized for only one loading condition often perform poorly under other scenarios, and a worst-case formulation may result in invalid solutions that completely ignore other loading conditions . Multi-load TO therefore seeks material layouts that provide robust performance across all relevant operating conditions. A common choice is to minimize the \emph{average compliance} over $N_L$ load cases,
\begin{equation}
C_{\text{avg}} = \frac{1}{N_L}\sum_{\ell=1}^{N_L} C^{(\ell)}
= \frac{1}{N_L}\sum_{\ell=1}^{N_L} \mathbf{F}^{(\ell)T}\mathbf{d}^{(\ell)},
\label{eq:avg_compliance}
\end{equation}
where $\mathbf{F}^{(\ell)}$ and $\mathbf{d}^{(\ell)}$ denote the load vector and displacement solution for load case $\ell$, respectively. This formulation promotes designs that distribute stiffness efficiently across load scenarios rather than overfitting a single case \cite{martin2004topology}.

\begin{figure} [t]
		\centering
		\begin{subfigure}[t]{0.45\linewidth}
			\centering
	           \includegraphics[width=\linewidth]{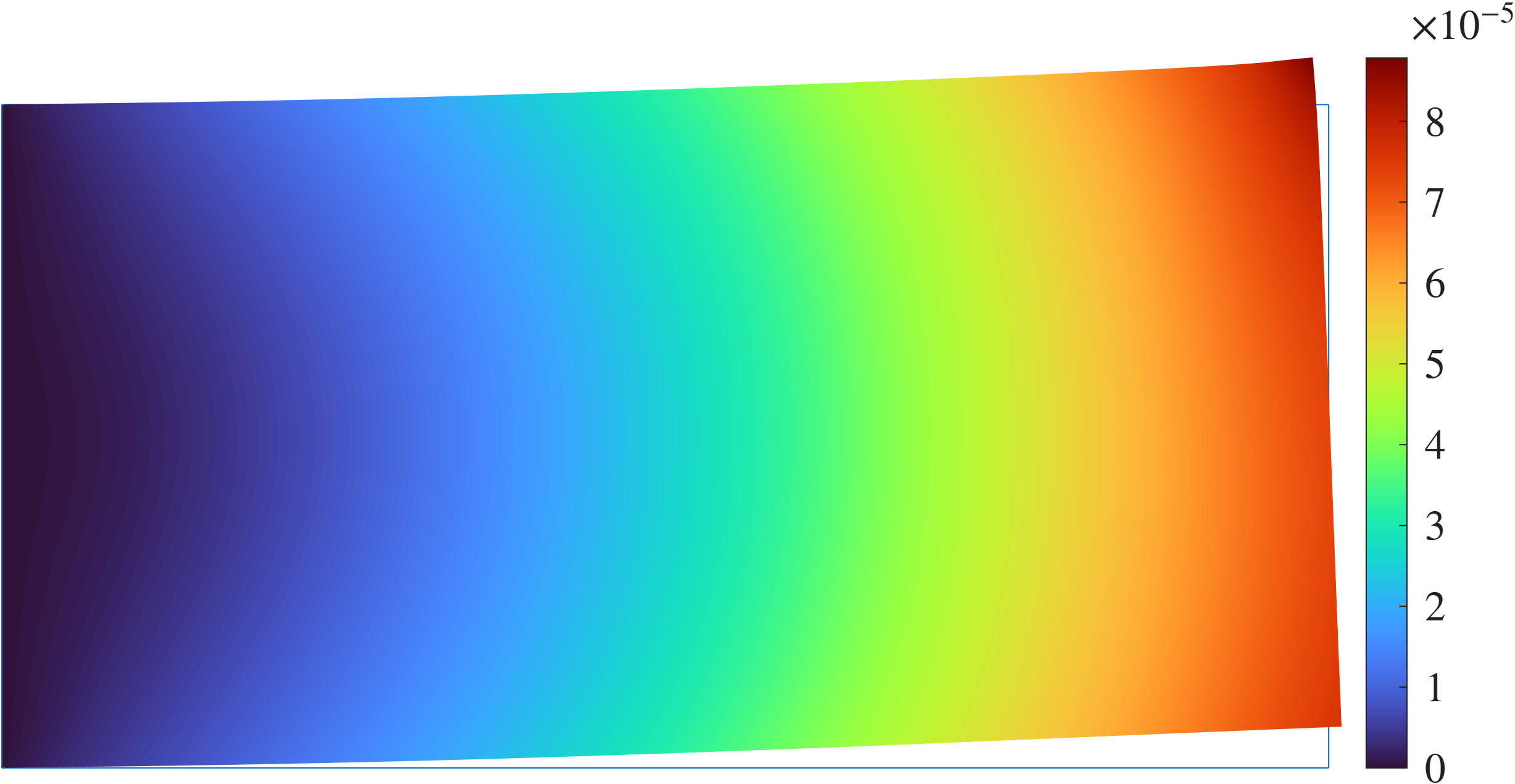}%
			\caption{Deformation 1}
		\end{subfigure}
        \begin{subfigure}[t]{0.45\linewidth}
			\centering
			\includegraphics[width=\linewidth]{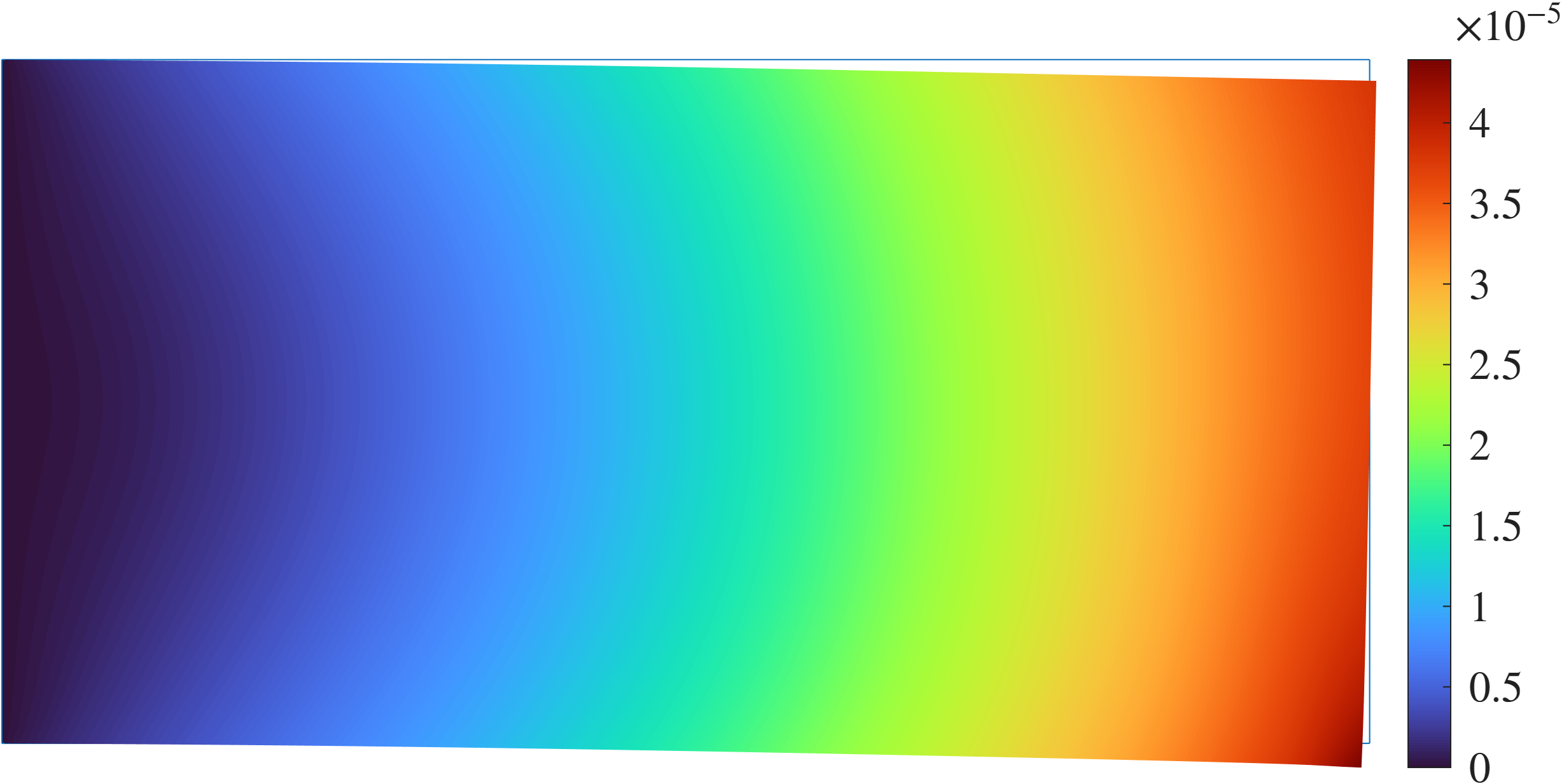}%
			\caption{Deformation 2}
		\end{subfigure}

        \begin{subfigure}[t]{0.45\linewidth}
			\centering
	           \includegraphics[width=\linewidth]{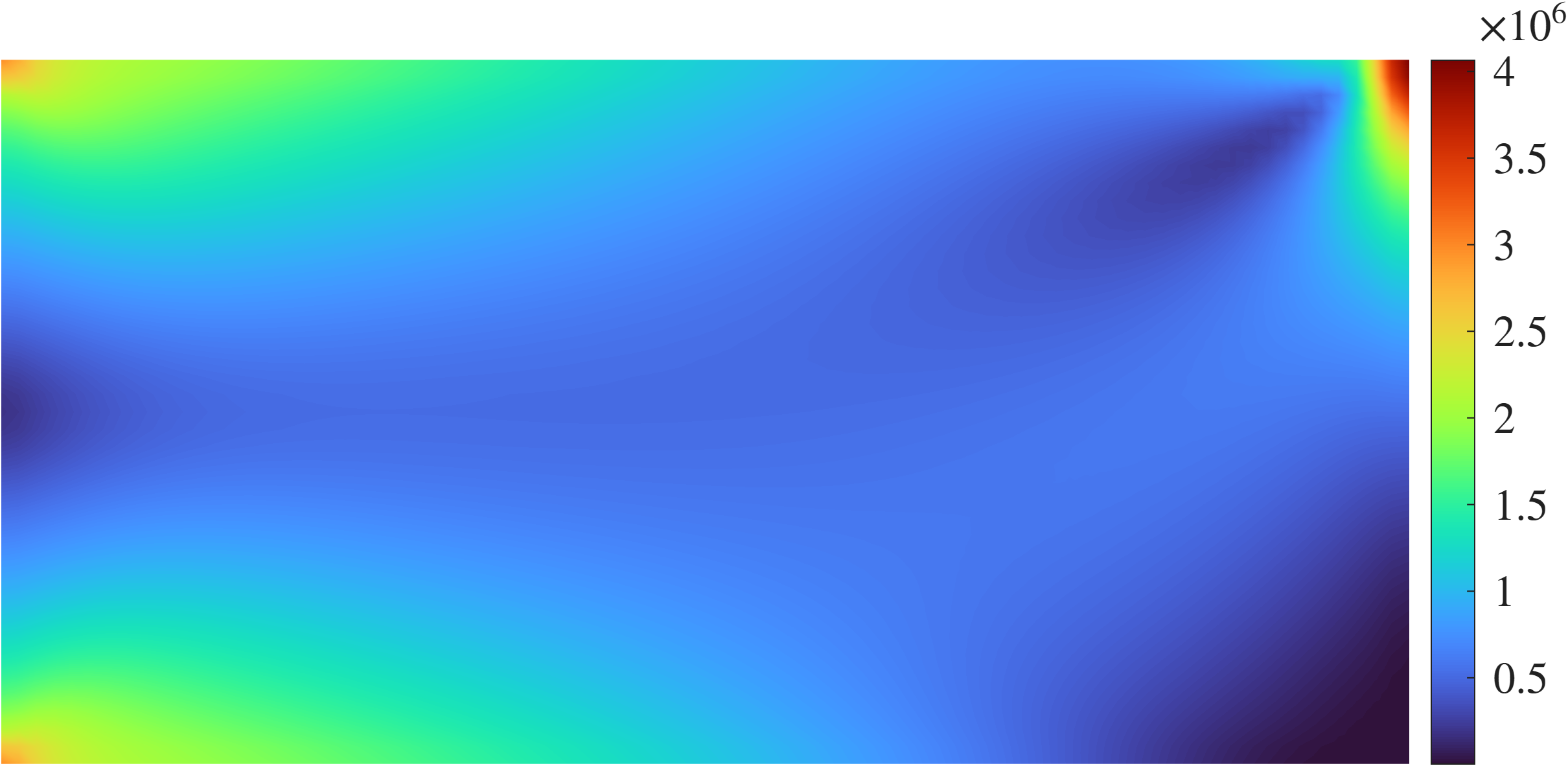}%
			\caption{Stress 1}
		\end{subfigure}
        \begin{subfigure}[t]{0.45\linewidth}
			\centering
			\includegraphics[width=\linewidth]{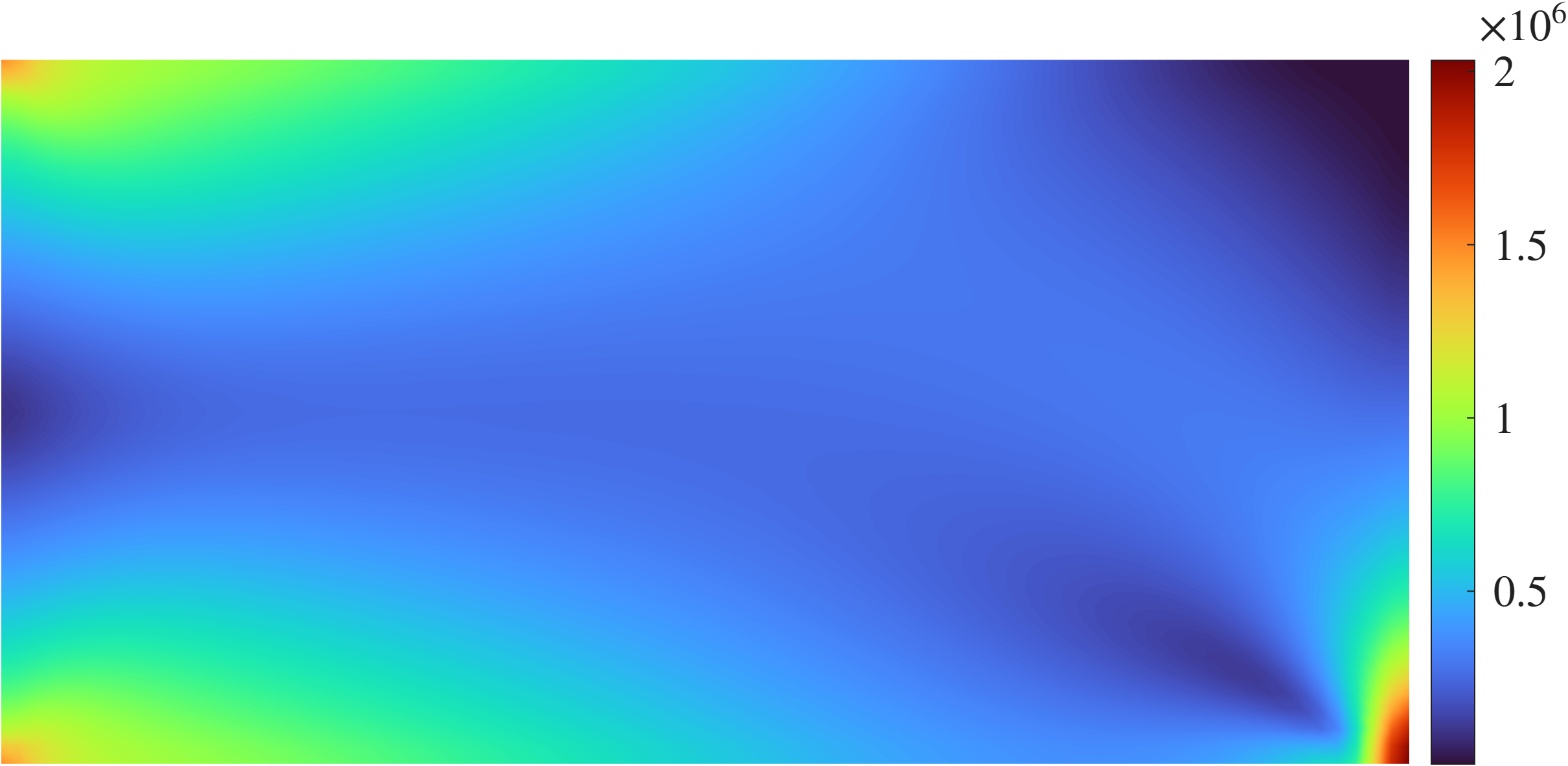}%
			\caption{Stress 2}
		\end{subfigure}
        
		\caption{Deformation and von Mises stress plots for the cantilever beam under two load scenarios. } \label{fig_beam_multiLoad_fields}
\end{figure}

We consider a cantilever beam subjected to two load scenarios. The material properties are $E = 100~\mathrm{GPa}$ and $\nu = 0.3$, and the domain is discretized into $3{,}200$ elements. The maximum number of optimization iterations is set to 500, and the volume fraction is $0.5$. The two loading conditions are $F_1 = 200~\mathrm{kN}$ and $F_2 = 100~\mathrm{kN}$, applied at different locations as shown in Fig.~\ref{fig_beam_multiLoad_BC}. Figure~\ref{fig_beam_multiLoad_fields} presents representative deformation and von Mises stress fields under the two load cases, illustrating the distinct structural response patterns that must be accommodated simultaneously.

\begin{figure} [t]
		\centering
		\begin{subfigure}[t]{0.45\linewidth}
			\centering
	\includegraphics[width=\linewidth]{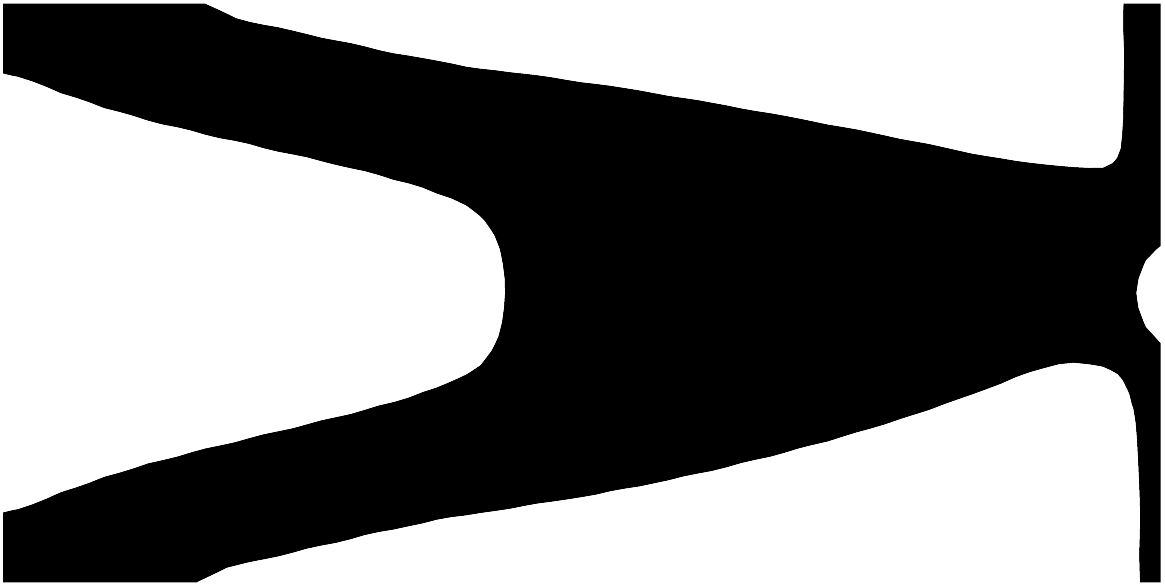}%
			\caption{LSSO}
		\end{subfigure}
        \begin{subfigure}[t]{0.45\linewidth}
			\centering
			\includegraphics[width=\linewidth]{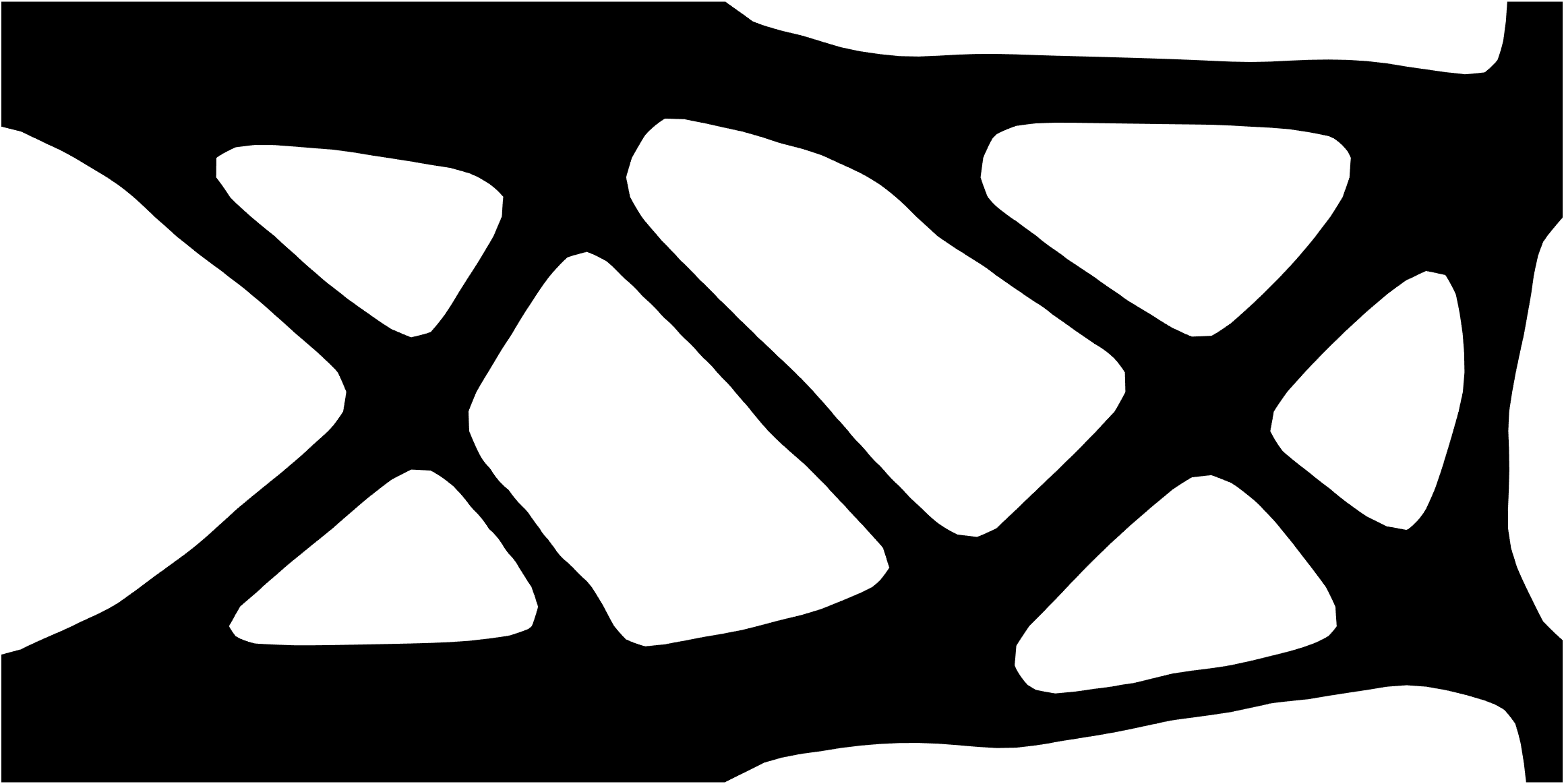}%
			\caption{Standard HJE}
		\end{subfigure}
        \begin{subfigure}[t]{0.45\linewidth}
			\centering
			\includegraphics[width=\linewidth]{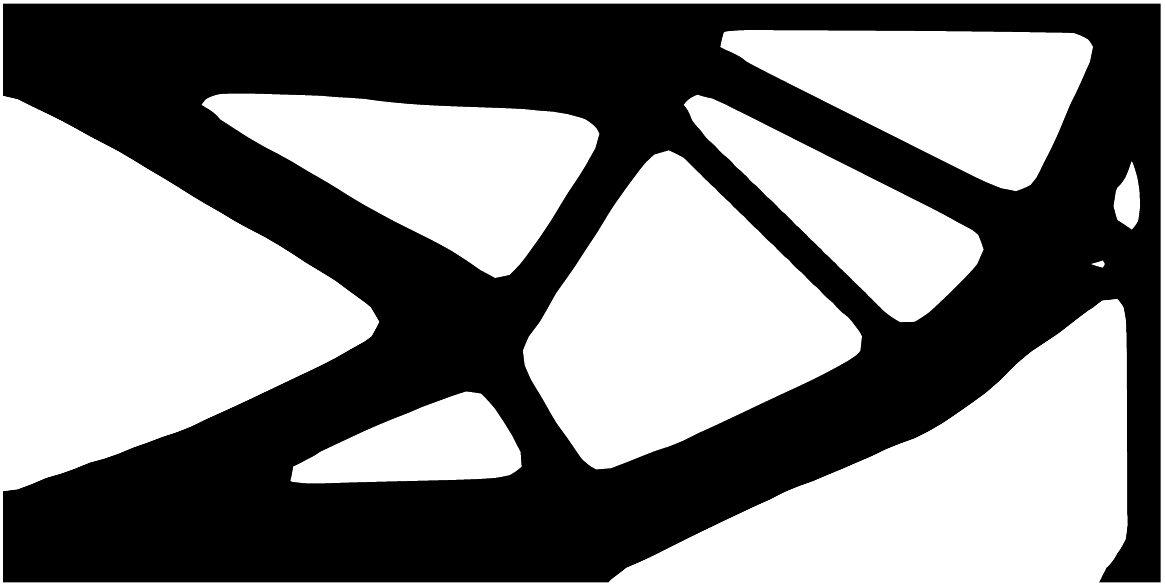}%
			\caption{Modified HJE}
		\end{subfigure}
        \begin{subfigure}[t]{0.45\linewidth}
			\centering
			\includegraphics[width=\linewidth]{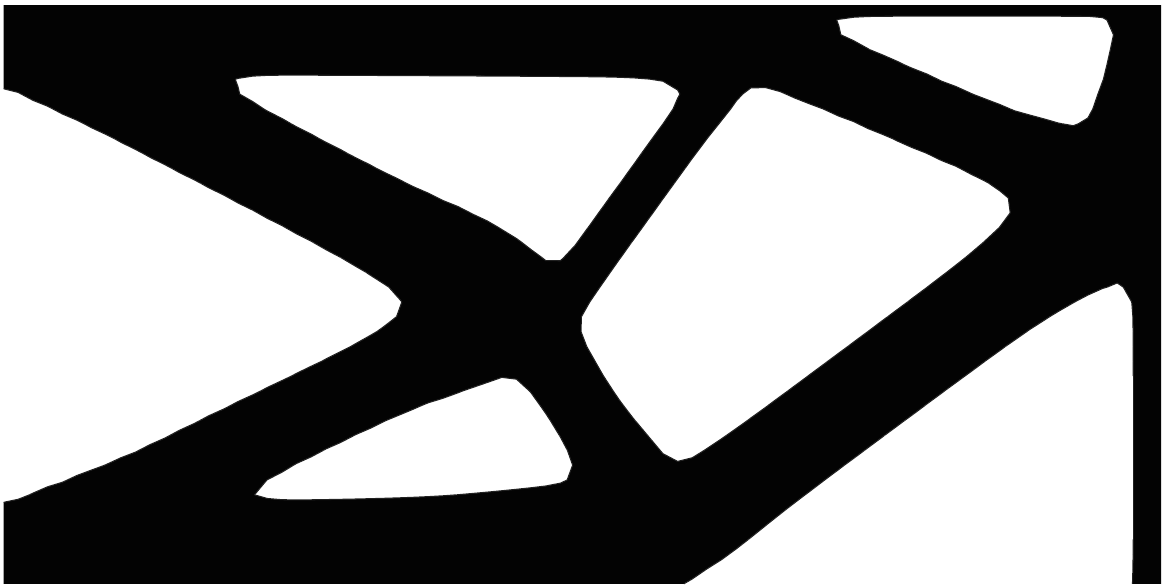}%
			\caption{SIMP}
		\end{subfigure}

        \begin{subfigure}[t]{0.45\linewidth}
			\centering
			\includegraphics[width=\linewidth]{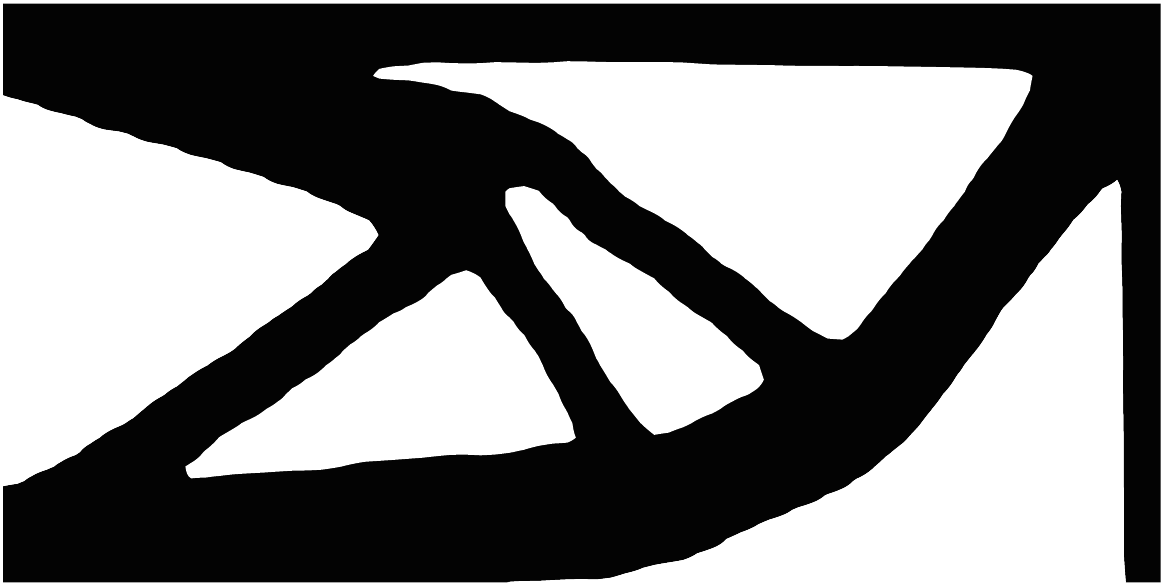}%
			\caption{PareTO}
		\end{subfigure}
        
		\caption{Optimized cantilever beam under two load scenarios with different SO/TO approaches. } \label{fig_beam_multiLoad_optimized}
\end{figure}

Figure~\ref{fig_beam_multiLoad_optimized} compares optimized designs obtained using different SO/TO approaches under the multi-load objective. The resulting average compliance values are: level-set SO ($C_\text{LSSO}=2.11\,\mathrm{N.m}$), level-set TO with standard HJE ($C_\text{stHJE}=1.84\,\mathrm{N.m}$), level-set TO with modified HJE ($C_\text{modHJE}=1.68\,\mathrm{N.m}$), SIMP-based TO ($C_\text{SIMP}=1.89\,\mathrm{N.m}$), and Pareto-based TO ($C_\text{PareTO}=1.7\,\mathrm{N.m}$). As in the single-load cases, topology optimization methods outperform pure shape optimization, and approaches that better explore the design space (modified HJE and Pareto) achieve lower average compliance. The resulting structures exhibit multiple load paths and reinforced regions corresponding to both loading scenarios, highlighting the importance of multi-load formulations in practical design problems.

\subsection{Self-Weight} \label{sec:selfWeight}

 \begin{figure}[t]
	\centering
	\begin{subfigure}[b]{0.54\linewidth}
		\centering
		\includegraphics[width=0.98\linewidth]{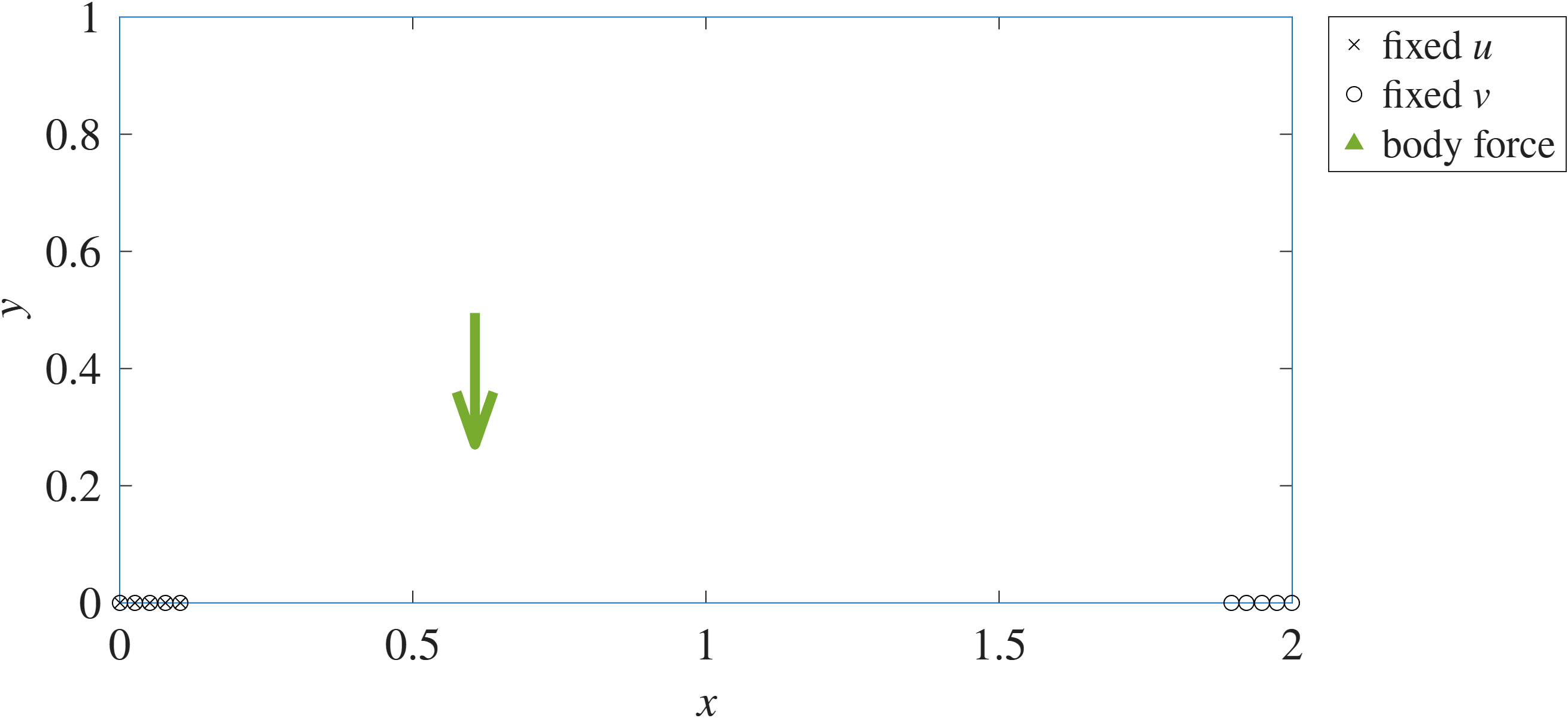}
		\caption{Boundary condition}\label{fig:selfW_bc}
	\end{subfigure}
        \begin{subfigure}[b]{0.44\linewidth}
		\centering
		\includegraphics[width=0.99\linewidth]{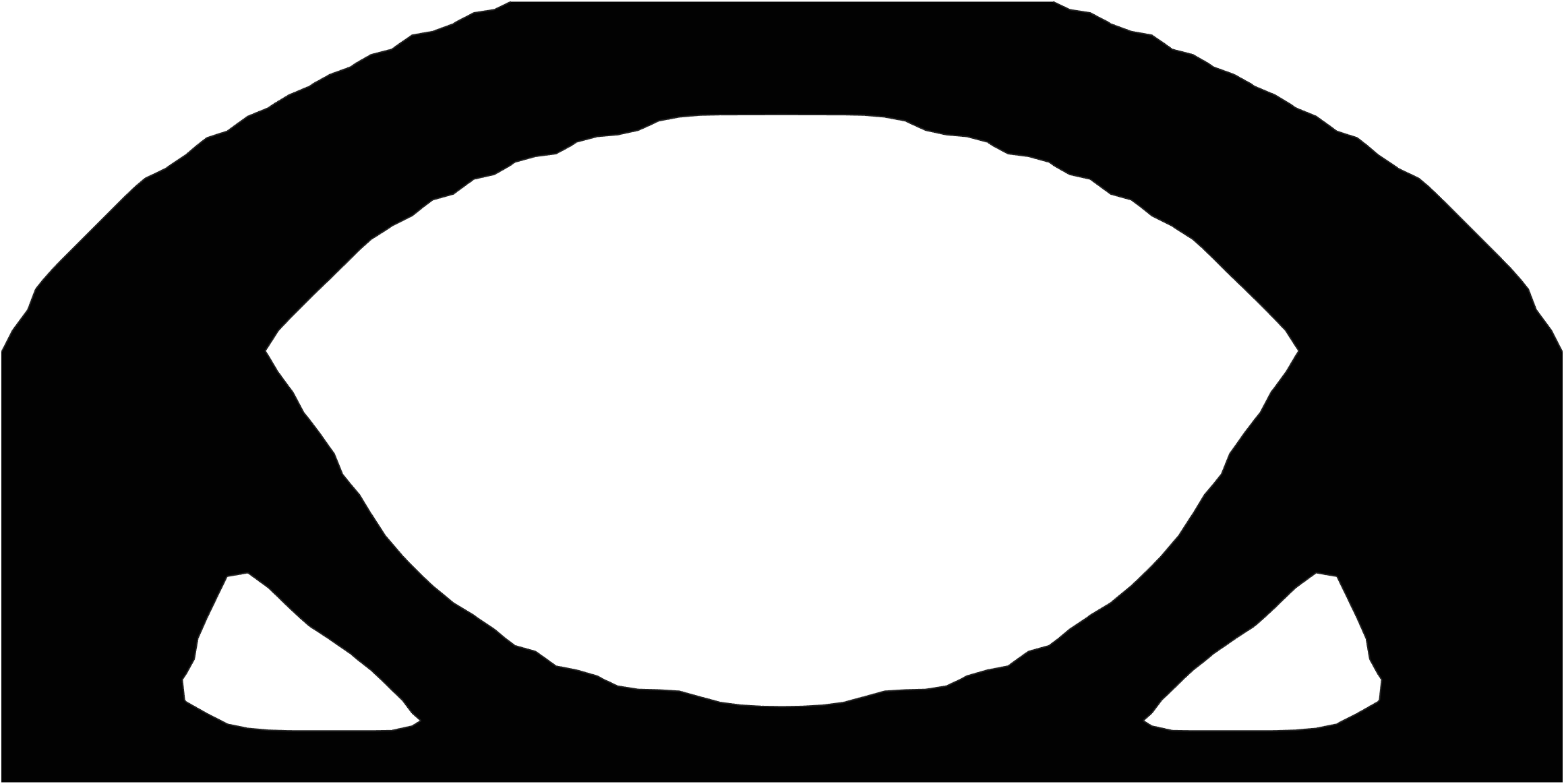}
		\caption{Optimized design}
	\end{subfigure}

    \begin{subfigure}[b]{0.54\linewidth}
		\centering
		\includegraphics[width=\linewidth]{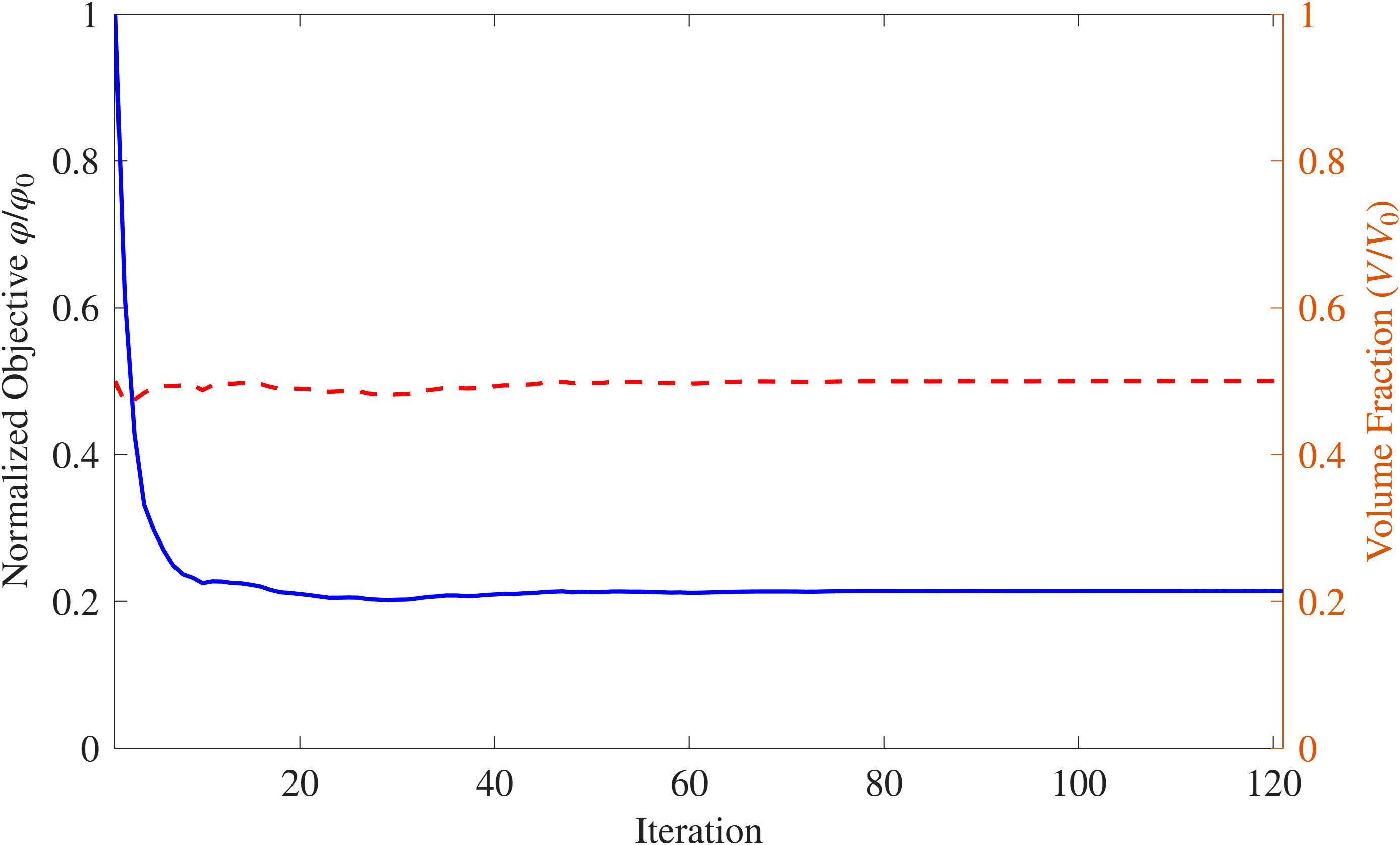}
		\caption{Convergence}
	\end{subfigure}
	\begin{subfigure}[b]{0.44\linewidth}
		\centering
		\includegraphics[width=\linewidth]{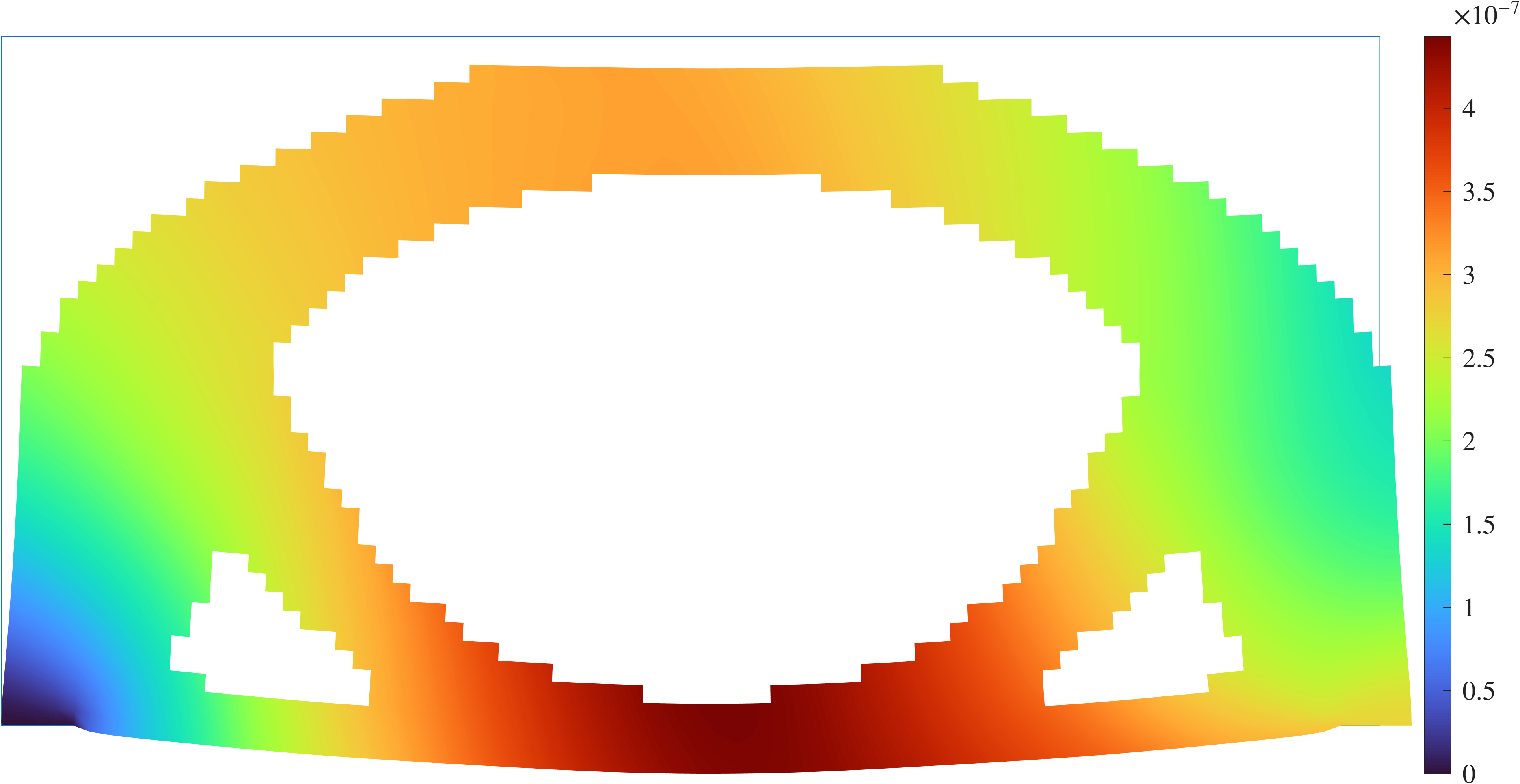}
		\caption{Deformation}
	\end{subfigure}

    \begin{subfigure}[b]{0.44\linewidth}
		\centering
		\includegraphics[width=0.98\linewidth]{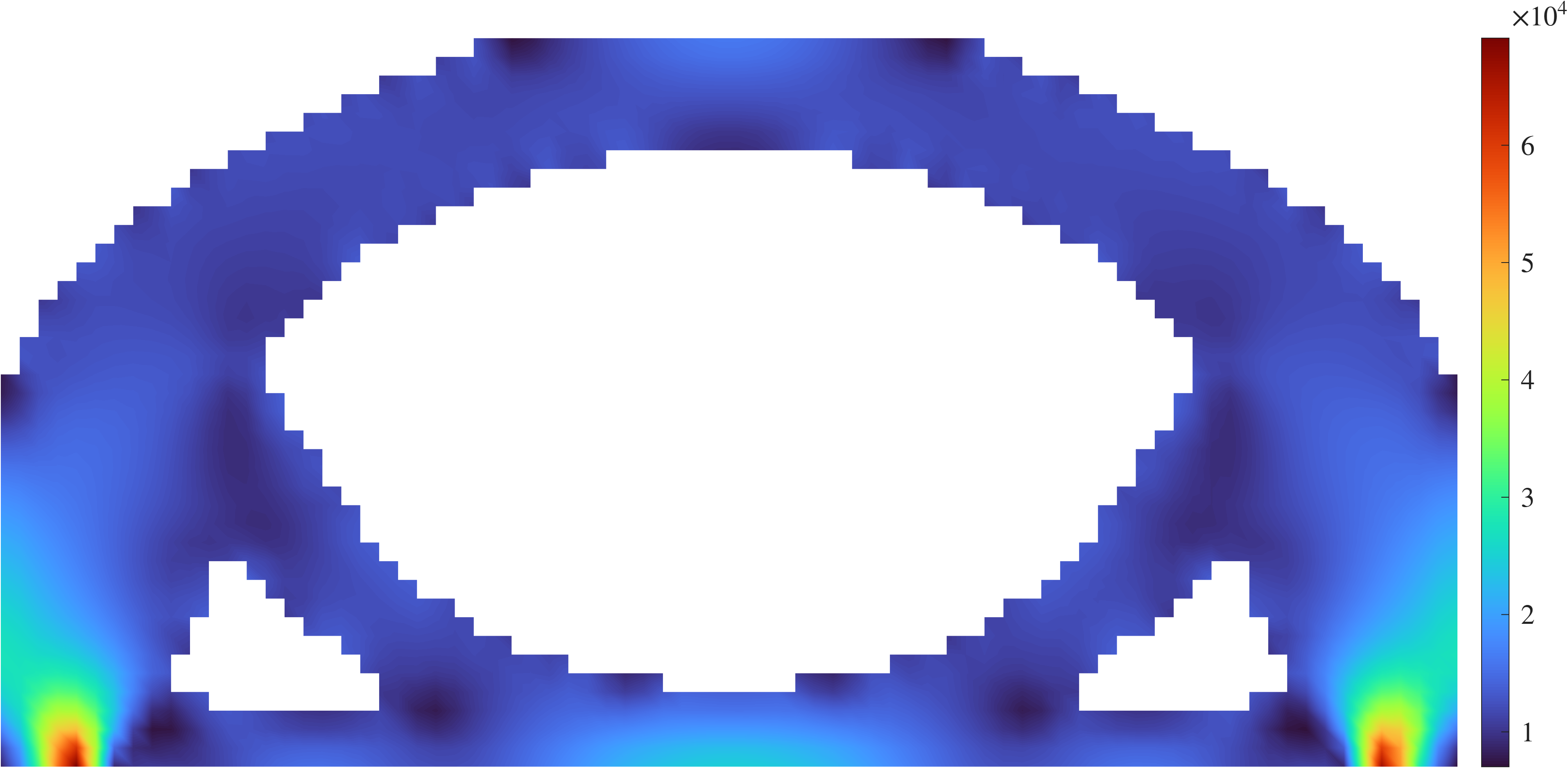}
		\caption{Von Mises stress}
	\end{subfigure}
	\begin{subfigure}[b]{0.44\linewidth}
		\centering
		\includegraphics[width=\linewidth]{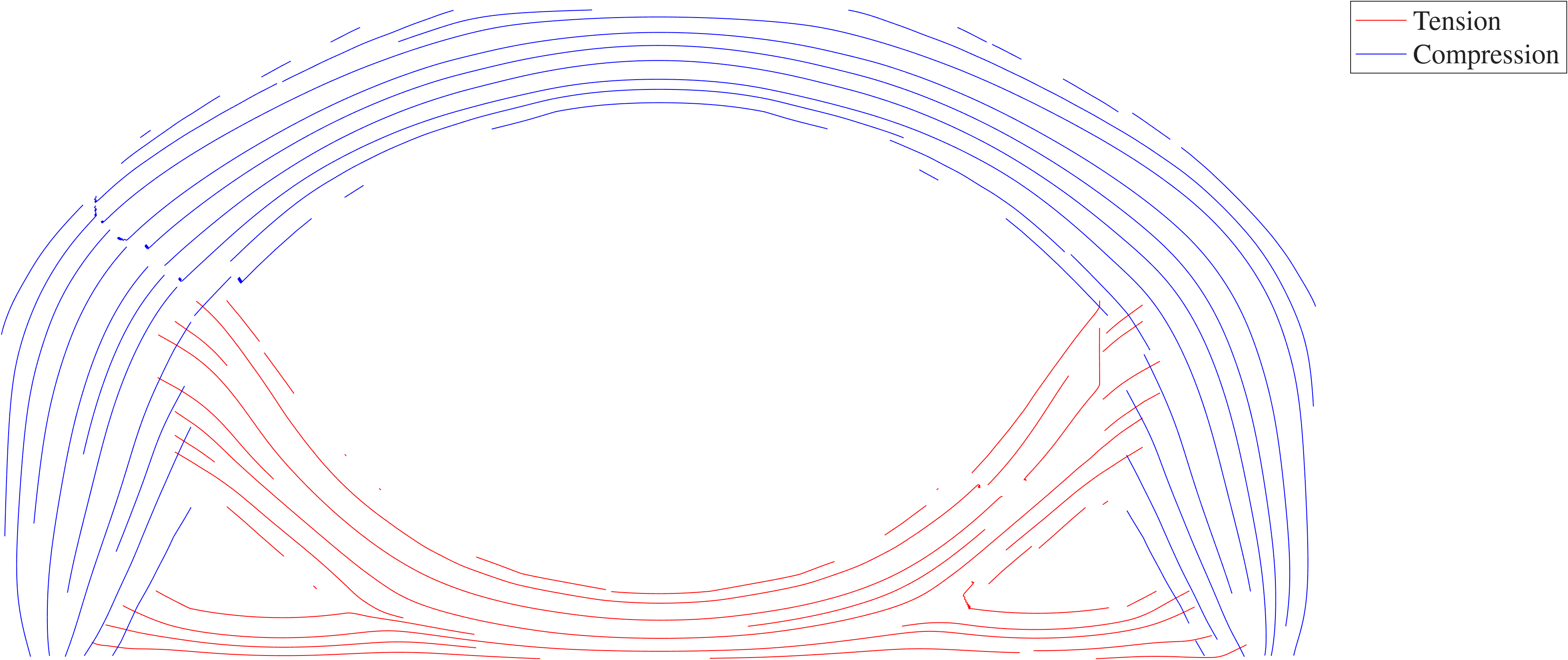}
		\caption{Principal stress}
	\end{subfigure}

	\caption{Optimized bridge under self-weight load.}
	\label{fig:selfWeightBridge}
\end{figure}

Let us  revisit the optimization problems of \eqref{eq_simpTOProblem}, where the external force comprises a constant external force and a body force (e.g., weight of the current design as shown in \eqref{eq_simpTOProblemBodyForce} 

\begin{subequations} \label{eq_simpTOProblemBodyForce}
	\begin{align}
		\minimize\limits_{\boldsymbol{\rho}}& \quad  \varphi (\bd;\boldsymbol{\rho}) \label{seq_simp_obj_weight}\\  
		 \textrm{s.t.} \quad  g &\coloneqq \sum\limits_{e}\rho_e v_e -  V^* \le 0\label{seq_simp_g_weight}\\	
		\tilde{R}_{el}(\mathbf{d;\boldsymbol{\rho}}) &\coloneqq \mathbf{K}_{el} (\boldsymbol{\rho})\mathbf{d}-\mathbf{f}_{el}-\mathbf{b}(\boldsymbol{\rho})=\mathbf{0} \label{seq_simp_fea_weight}\\
		& 0<\rho_{min} \le \rho_e \le 1; \quad \forall e \label{seq_simp_rho_weight}
	\end{align}
\end{subequations}

In other words, given the effective material density at element
\begin{equation}
     \rho_{e}^{\text{eff}} = \rho_e \rho^{\text{mat}}_e,
\end{equation}
where $\rho_e$ and $\rho^{\text{mat}}_e$ denote the pseudo-density and material density at element $e$. Thus, the body force $\mathbf{b}$ is
\begin{equation}
\mathbf{b}= \sum_e \rho_{e}^{\text{eff}} \mathbf{a}_e v_e,
\end{equation}
where $a_e$ and $v_e$ are the acceleration and volume of element $e$.
In the special case of self-weight for a grid mesh, $\mathbf{a}_e=\mathbf{g}$ and $v_e$ are constant and equal for all elements.
Thus, we have at each element
\begin{equation} \label{eq:selfWeightDer}
	\mathbf{b}' =  \rho^{\text{mat}}_e \mathbf{a}_e v_e.
\end{equation}

To apply this in STORX, once the \mcode{solver} is created, the acceleration \mcode{acceleration = [0,-10];} can be specified through:
\begin{lstlisting}
solver = solver.applyAcceleration(acceleration);
\end{lstlisting}

Figure \ref{fig:selfW_bc} illustrates the design domain boundary condition for a bridge example under self-weight load. 
In this example we use a $3000$-element mesh, RAMP penalty through continuation $p=3$ to 5 with 0.2 increments, and isotropic linear elasticity with $E=100\,\mathrm{GPa}$ and $\nu=0.3$, and $\rho^{\text{mat}}=1000 \,\mathrm{kg}/\mathrm{m}^{3}$. The acceleration $\mathbf{g} = [0, -10]\,\mathrm{m}/\mathrm{s}^{2}$.

\subsection{Thermal Conduction}\label{sec:thermal}
In many industrial applications, such as aerospace, automotive, and consumer electronics, thermal management plays a crucial role. We now consider TO under steady-state and isotropic thermal conduction.

 \begin{figure}[t]
	\centering
	\begin{subfigure}[b]{0.54\linewidth}
		\centering
		\includegraphics[width=0.98\linewidth]{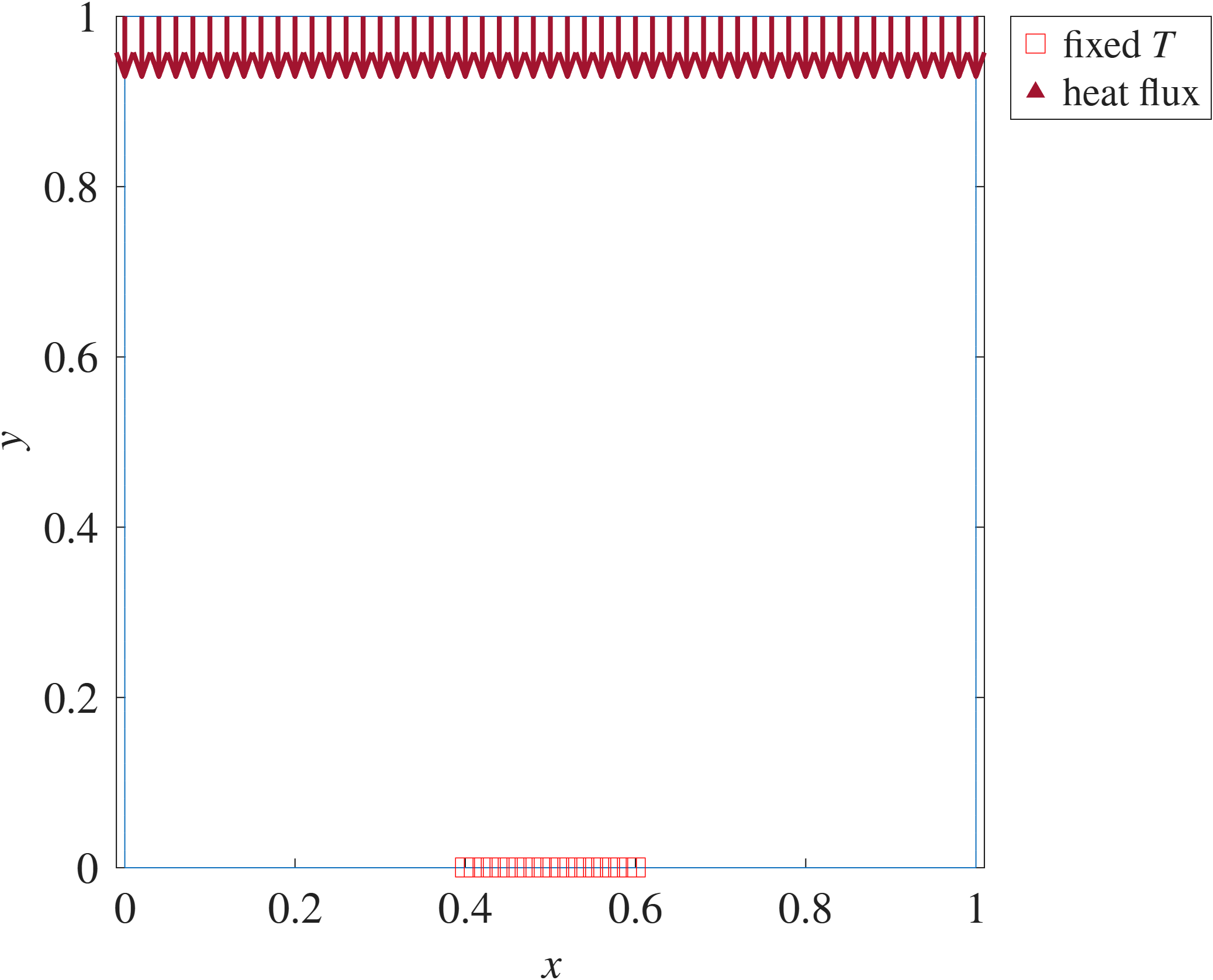}
		\caption{Boundary condition}\label{fig:thermalFlux_bc}
	\end{subfigure}
        \begin{subfigure}[b]{0.44\linewidth}
		\centering
		\includegraphics[width=0.9\linewidth]{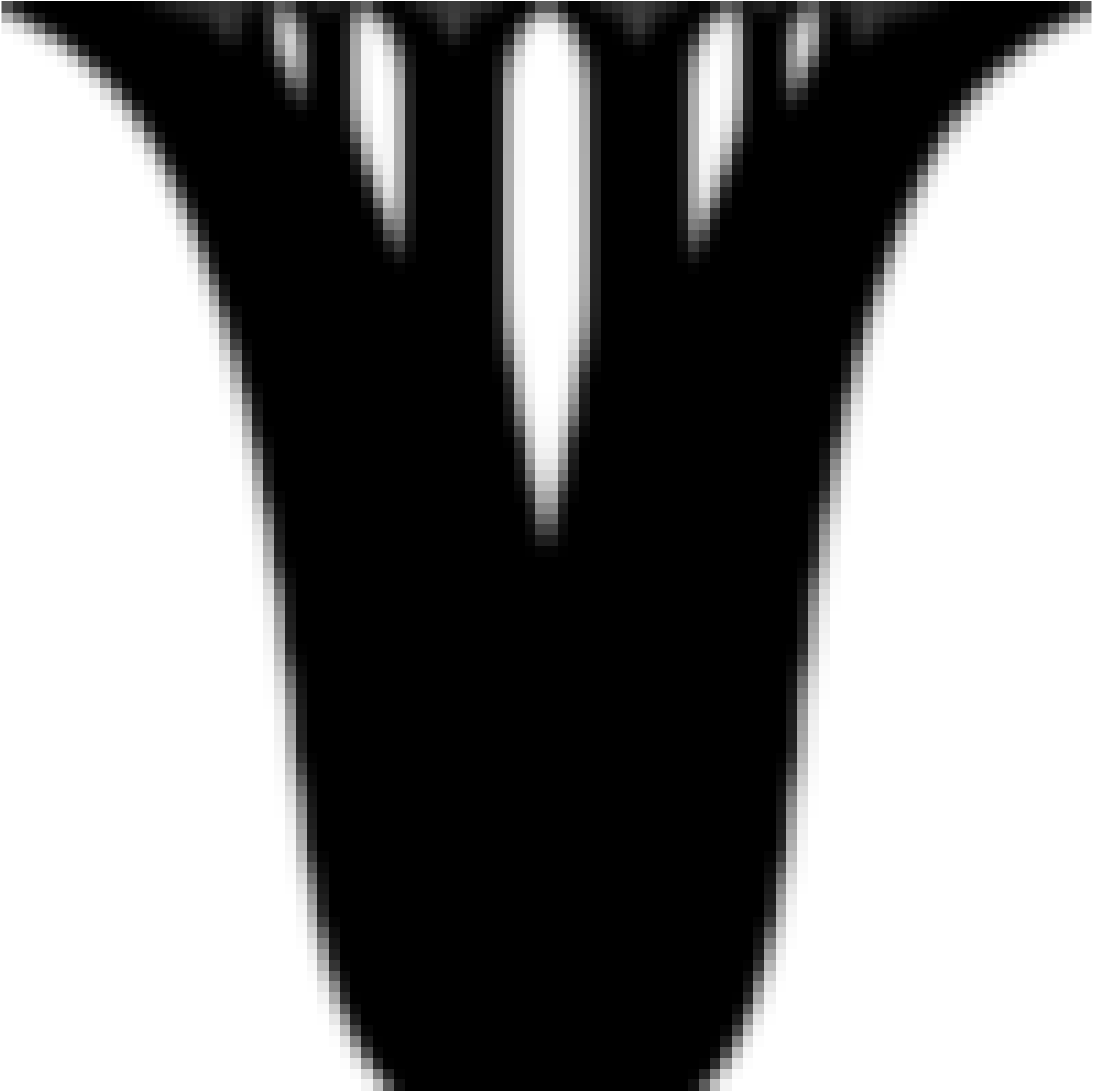}
		\caption{Optimized density}
	\end{subfigure}

    \begin{subfigure}[b]{0.45\linewidth}
		\centering
		\includegraphics[width=\linewidth]{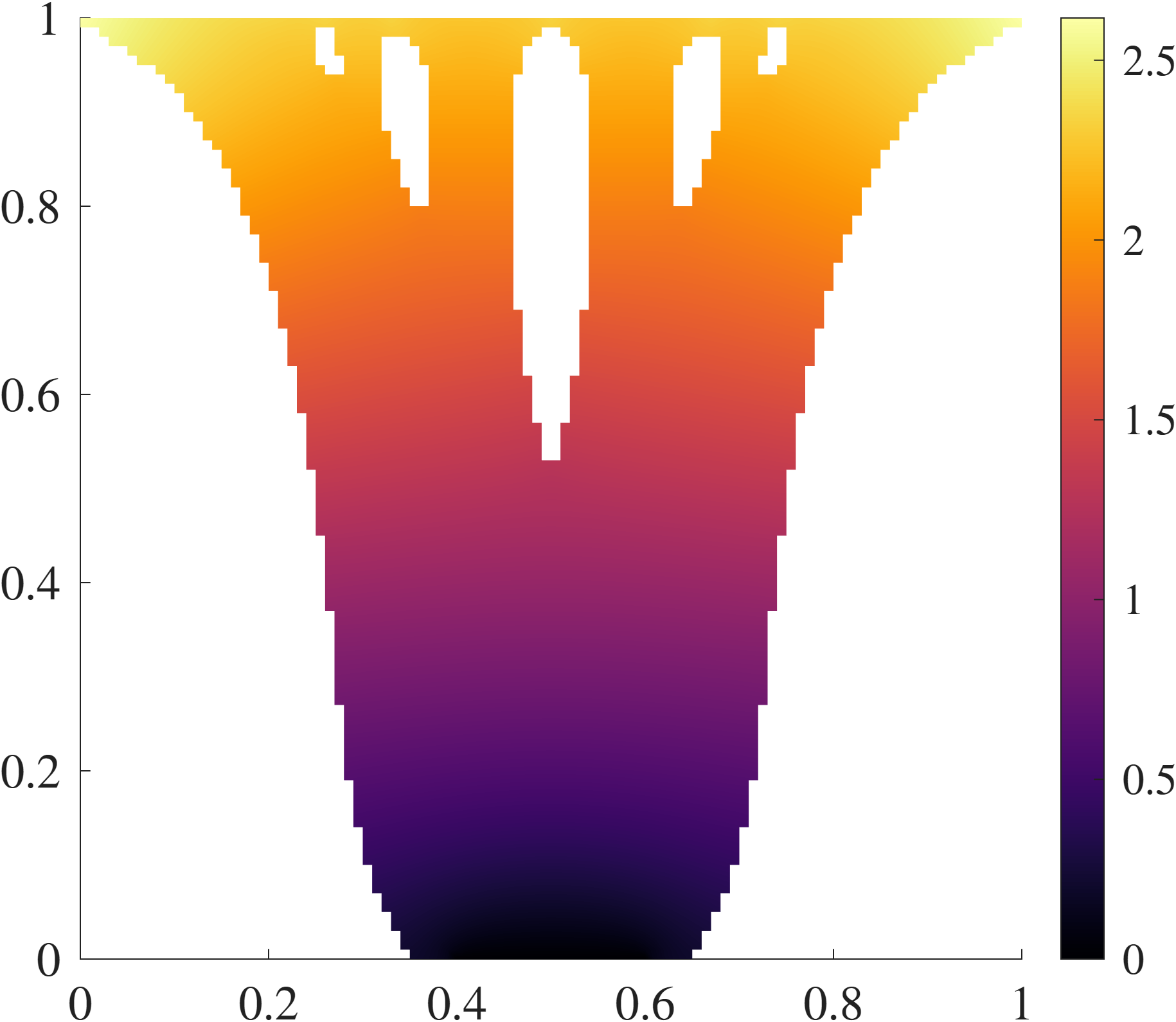}
		\caption{Temperature}
	\end{subfigure}
	\begin{subfigure}[b]{0.5\linewidth}
		\centering
		\includegraphics[width=0.8\linewidth]{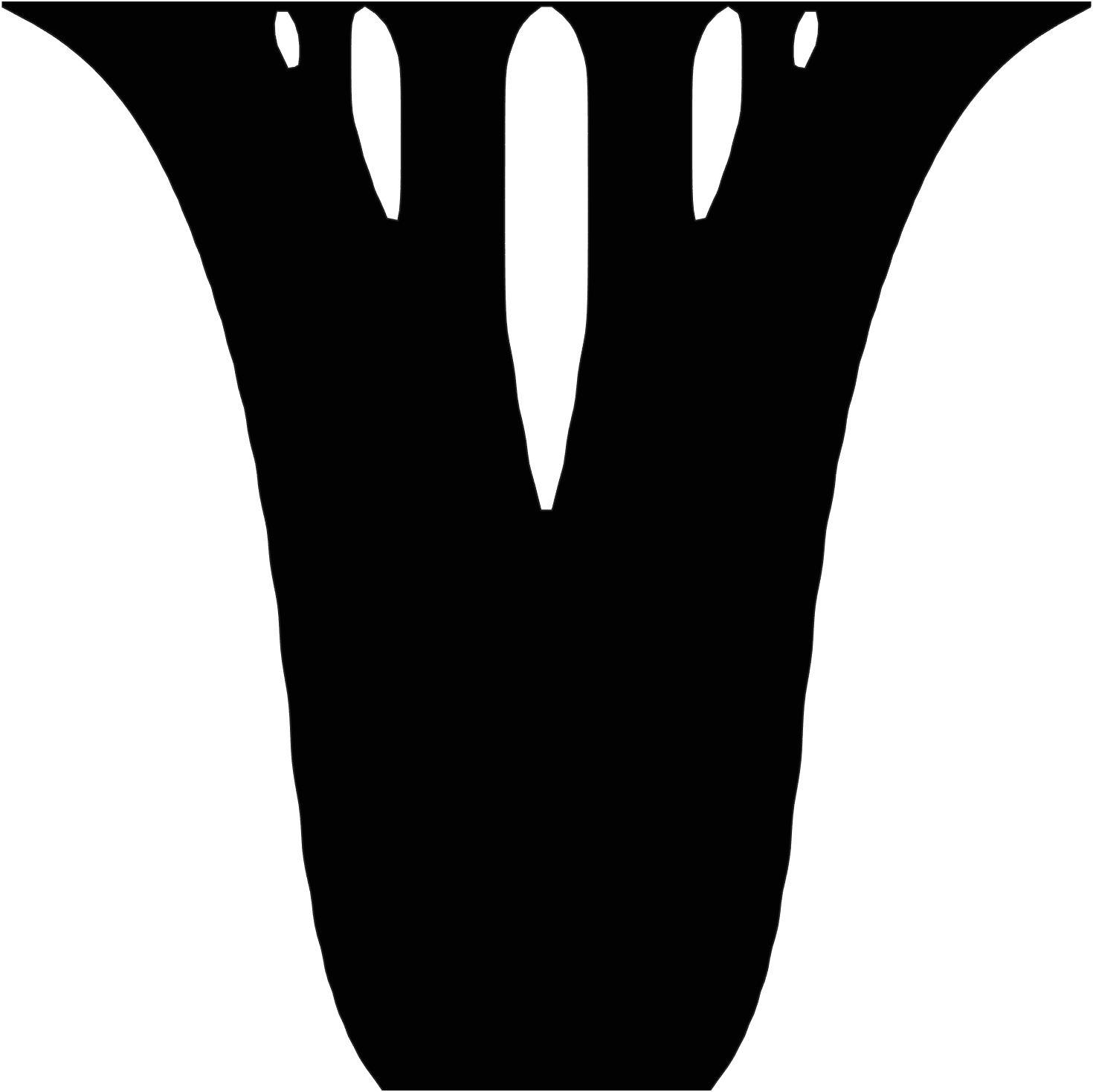}
		\caption{Optimized iso-surface}
	\end{subfigure}

	\caption{Optimized square under heat flux.}
	\label{fig:squreflux}
\end{figure}

Consider the design domain $\Omega$, where the temperature $T$ is fixed at the Dirichlet boundary $\Gamma_D$ and Neumann boundary $\Gamma_N$ is an adiabatic wall and restricts heat flux. 

The steady-state heat conduction problem can be expressed as:

\begin{subequations}
\label{eq:residual_form_heat}
\begin{empheq}[left={R_{th} \coloneqq \empheqlbrace}]{align}
  \nabla \cdot \big(\kappa \nabla T\big) + f &= 0,
   \bm{x} \in \Omega,
  \label{eq:residual_form_heat_pde} \\[0.3em]
  T - T_0 &= 0,
   \bm{x} \in \Gamma_D,
  \label{eq:residual_form_heat_dirichlet} \\[0.3em]
  \big(\kappa \nabla T\big)\cdot \bm{n} - q_n(\bm{x}) &= 0,
   \bm{x} \in \Gamma_N.
  \label{eq:residual_form_heat_neumann}
\end{empheq}
\end{subequations}

The strong-form residual $R_{th}$ of \eqref{eq:residual_form_heat_neumann} is projected onto a finite element basis and integrated using a Galerkin procedure, transforming the PDE into the algebraic system.

The constrained optimization problem for thermal conduction is
\begin{subequations} 
	\begin{align}
		\minimize\limits_{\boldsymbol{\rho}}& \quad  C_{th} (\mathbf{T};\boldsymbol{\rho} )\coloneqq  \mathbf{T}^\top \bK_{th}(\boldsymbol{\rho}) \mathbf{T}\label{seq_th_simp_obj}\\  
		 \textrm{s.t.} \quad & g \coloneqq \sum\limits_{e}\rho_e v_e -  V^* \le 0\label{seq_th_simp_g}\\	
		&\tilde{R}_{th}(\mathbf{T};\boldsymbol{\rho})=\bK_{th}(\boldsymbol{\rho})\mathbf{T}-\mathbf{f}_{th} = \mathbf{0} \label{seq_th_simp_fea}\\
		& 0<\rho_{min} \le \rho_e \le 1; \quad \forall e \label{seq_th_simp_rho}
	\end{align}
    \label{eq_thermalTOProblem}
\end{subequations}

We illustrate the thermal optimization of \eqref{eq_thermalTOProblem} using two canonical examples on a square domain discretized with $10{,}000$ finite elements and a volume fraction constraint of $0.5$.

\begin{figure}[t]
	\centering
	\begin{subfigure}[b]{0.54\linewidth}
		\centering
		\includegraphics[width=0.98\linewidth]{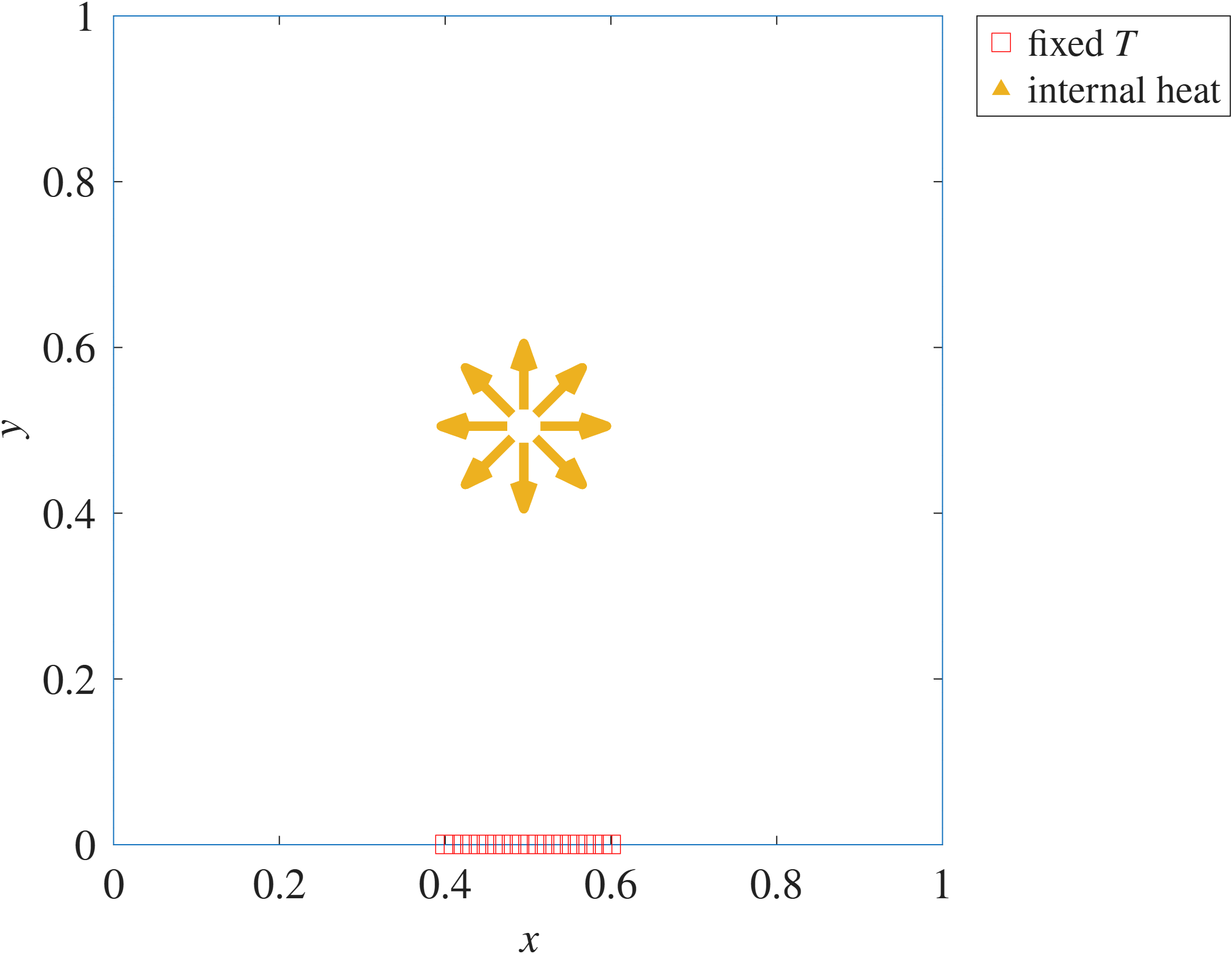}
		\caption{Boundary condition}\label{fig:thermal_bc}
	\end{subfigure}
        \begin{subfigure}[b]{0.44\linewidth}
		\centering
		\includegraphics[width=0.9\linewidth]{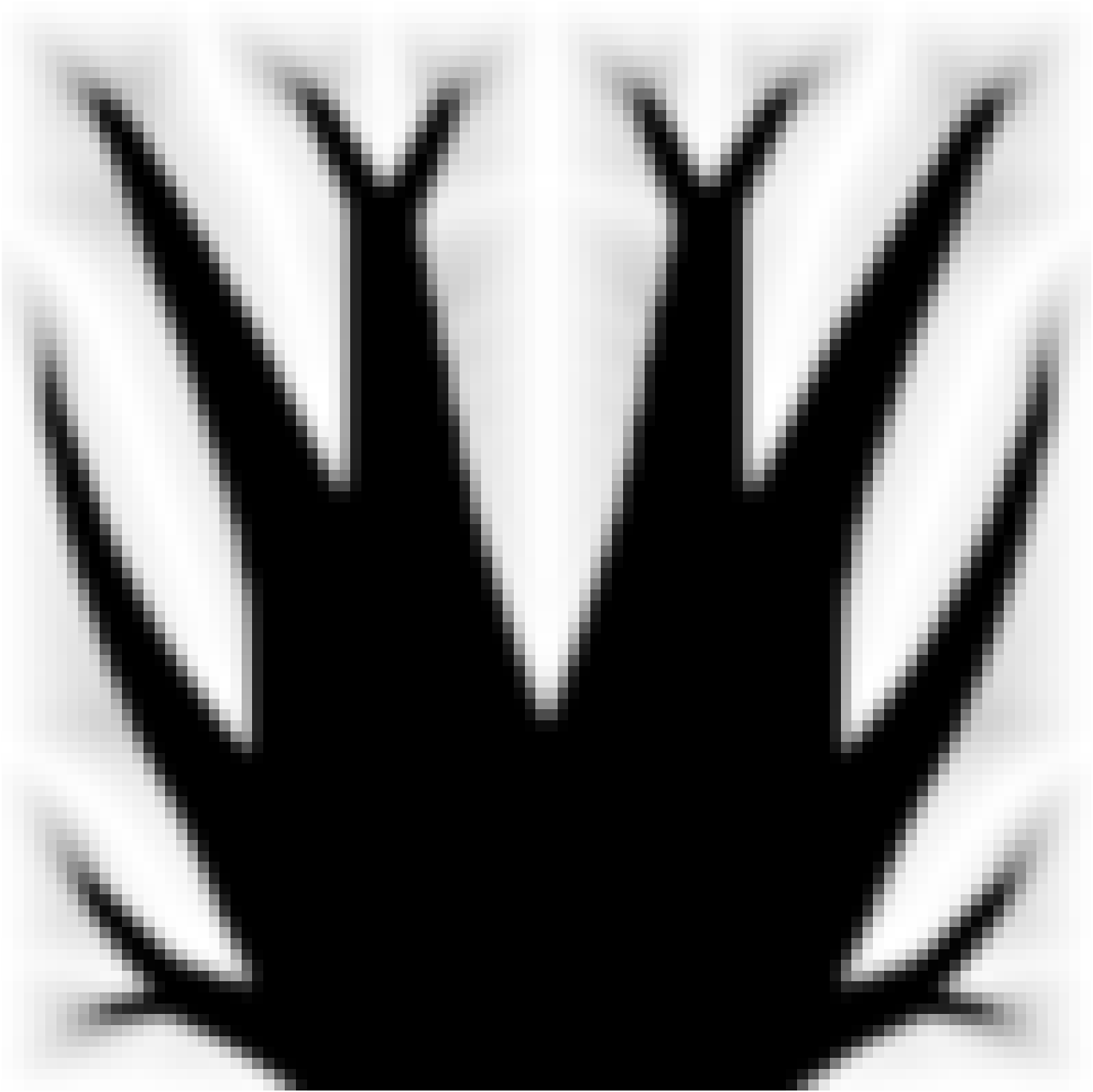}
		\caption{Optimized density}
	\end{subfigure}

    \begin{subfigure}[b]{0.45\linewidth}
		\centering
		\includegraphics[width=\linewidth]{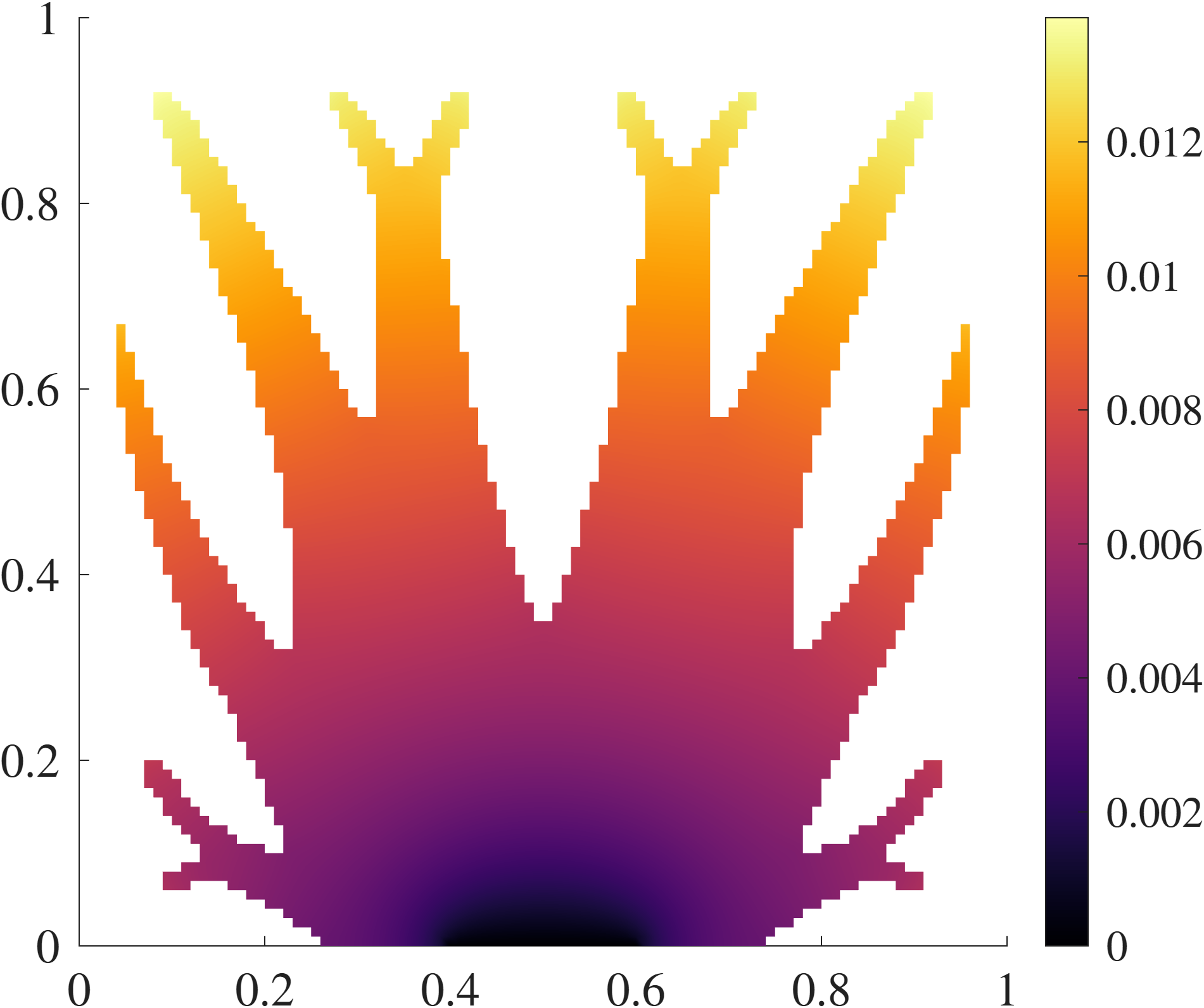}
		\caption{Temperature}
	\end{subfigure}
	\begin{subfigure}[b]{0.5\linewidth}
		\centering
		\includegraphics[width=0.75\linewidth]{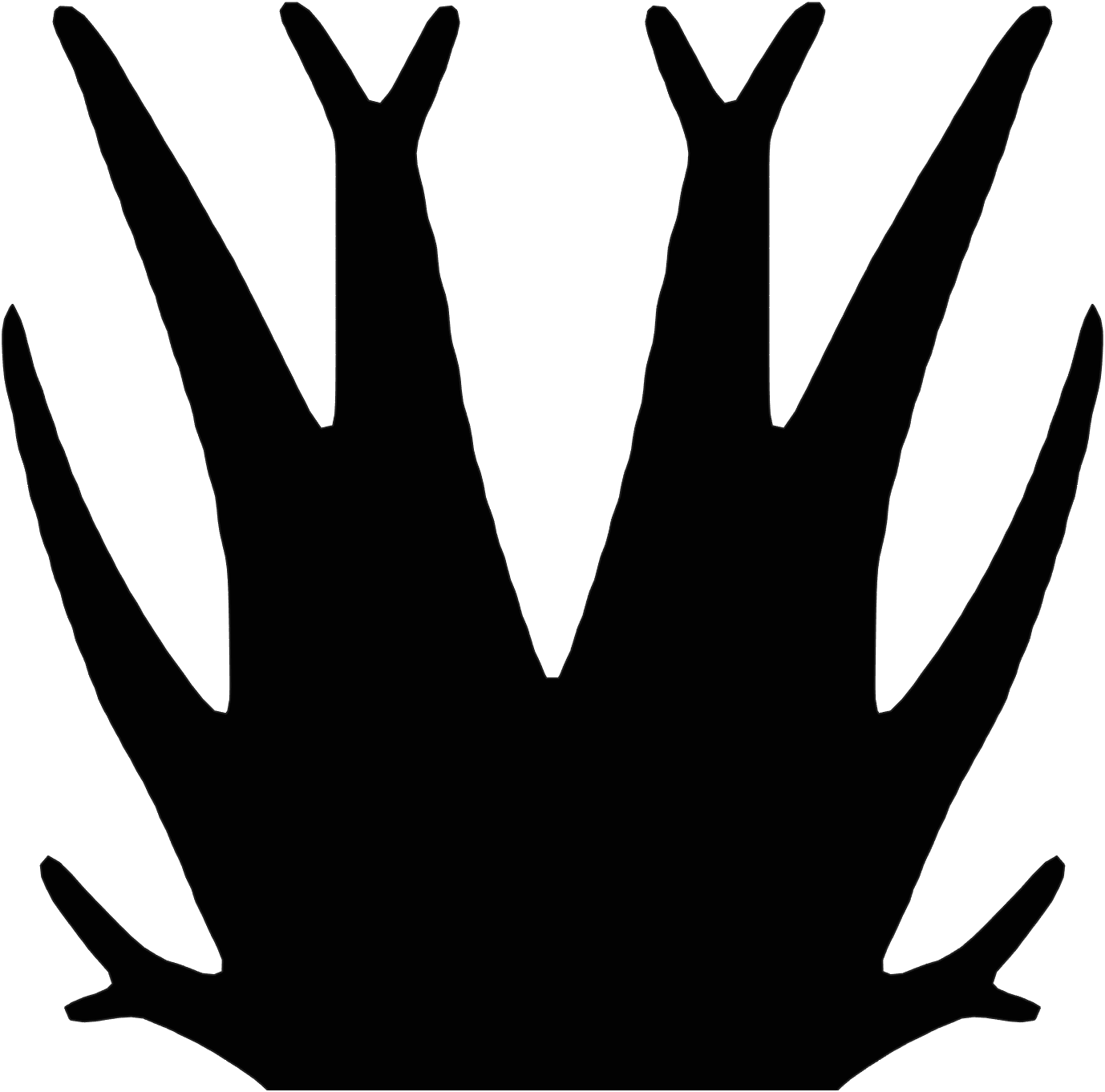}
		\caption{Optimized iso-surface}
	\end{subfigure}

	\caption{Optimized square under internal heat source.}
	\label{fig:squreInternalHeat}
\end{figure}
\subsubsection{Boundary heat flux loading}
A unit square domain with conductivity $\kappa=1$ is subjected to a prescribed Neumann boundary flux $q_n=1\,\mathrm{W}/\mathrm{m}^2$ on $\Gamma_N$, with homogeneous Dirichlet conditions $T=0\;K$ on $\Gamma_D$. No internal heat generation is present ($f=0\,\mathrm{W}/\mathrm{m}^3$). The objective is to distribute material to enhance heat conduction under boundary-driven loading. The design is parameterized using SIMP interpolation with a fixed penalization factor $p=3$, and the optimization is performed using the OC update scheme.

\subsubsection{Internal heat generation}
The same square domain and material properties are considered, but the loading arises from a uniform volumetric heat source $f=0.01\,\mathrm{W}/\mathrm{m}^3$ within $\Omega$, with boundary conditions as defined in ~\eqref{eq:residual_form_heat}. In this case, material distribution must manage internally generated heat rather than boundary flux. The RAMP interpolation scheme is used with a fixed penalization parameter $q=5$, and the OC method is again employed for design updates.

\section{Extending STORX} \label{sec:extendingSTORX}

STORX is extended by registering new objects at the optimizer interface, rather than modifying the optimization loop. New objectives and constraints are implemented as subclasses of \mcode{functional}: 1) \mcode{evaluate} returns the response value and 2) \mcode{gradient} returns its design sensitivity; passing an upper bound to the constructor uses the same class as a normalized inequality constraint, \(g=f/\bar{f}-1\le0\). 

Manufacturing constraints are subclasses of \mcode{mfgConstraints}; they map design variables to physical variables and propagate sensitivities through \mcode{filterDesign} and \mcode{filterSensitivity}.

A minimal STORX extension follows the two interfaces shown below: define a new functional for the response and sensitivity, and define any manufacturing rule as a \mcode{mfgConstraints} filter.
\begin{lstlisting}
objective      = myFunctional(solver);
constraints    = {  volume(solver,Vmax), ...
                    myFunctional(solver,gmax)};
mfgConstraints = {myMfgConstraint(solver)};
topopt = density2d_elasticity(  solver, objective, ...
                                constraints, ...
                                mfgConstraints,'MMA');
\end{lstlisting}

\begin{lstlisting}
classdef myFunctional < functional
    methods
        function [obj,f] = evaluate(obj),  obj.m_fx = ...; f = obj.m_fx; end
        function [obj,df] = gradient(obj), obj.m_dfdx = ...; df = obj.m_dfdx; end
    end
end
\end{lstlisting}

\begin{lstlisting}
classdef myMfgConstraint < mfgConstraints
    methods
        function x = filterDesign(obj,x),       x=...; end
        function s = filterSensitivity(obj,x,s),s=...; end
    end
end
\end{lstlisting}

The following examples use this interface to add stress minimization in density-based TO (Section~\ref{sec:stress}), local volume fraction constraints (Section~\ref{sec:localVolFrac}), and fluid TO (Section~\ref{sec:fluid}).
\subsection{Extending Optimization Objective: \textit{Stress Minimization}} \label{sec:stress}

We minimize a global measure of von Mises stress using a relaxed elementwise stress field and a smooth aggregation. Let $\rho_e\in[0,1]$ denote the pseudo-density in element $e$, and let $\sigma_{\mathrm{vm},e}(x,u)$ be the element von Mises stress computed from the FE stress tensor. To reduce singular behavior in low-density regions, we use the relaxed stress
\begin{equation} \label{eq:elemRelaxedStress}
\tilde{\sigma}_{\mathrm{vm},e} \;=\; \rho_e^{\,q_{\mathrm{vm}}}\,\sigma_{\mathrm{vm},e},
\end{equation}
where $q_{\mathrm{vm}}$ is the stress relaxation exponent (say 0.5).
We aggregate the elemental values of \eqref{eq:elemRelaxedStress} via a $p$-norm objective
\begin{equation}
\sigma_{PN}^{}(\bd;\boldsymbol{\rho}) \;=\; \Big(\sum_{e\in\Omega}\tilde{\sigma}_{\mathrm{vm},e}^{\,p_{\mathrm{vm}}}\Big)^{1/p_{\mathrm{vm}}}.
\end{equation}
The constrained optimization problem is
\begin{subequations} \label{eq_simpTOProblem_stress}
	\begin{align}
		\minimize\limits_{\boldsymbol{\rho}} \quad & \sigma_{PN} (\bd;\boldsymbol{\rho}) \label{seq_simp_obj_stress}\\  
		 \textrm{s.t.} \quad & g \coloneqq \sum\limits_{e}\rho_e v_e -  V^* \le 0\label{seq_simp_g_stress}\\	
		& {R}_{el}(\bd;\boldsymbol{\rho}) = \textbf{0} \label{seq_simp_fea_stress}\\
		& 0<\rho_{min} \le \rho_e \le 1; \quad \forall e \label{seq_simp_rho_stress}
	\end{align}
\end{subequations}

We enforce manufacturability by mapping the design variables through a minimum feature size filter and a physical density projection, so that from \eqref{eq:full_chain_rule} the chain rule for sensitivities becomes
\begin{equation}
\frac{d\sigma_{PN}}{d \rho_e}
\;=\;
\frac{\partial \sigma_{PN}}{\partial \hat{\rho}_e}\,
\frac{\partial \hat{\rho}_e}{\partial \bar{\rho}_e}\,
\frac{\partial \bar{\rho}_e}{\partial \rho_e}.
\end{equation}

 \begin{figure}[t]
	\centering
	\begin{subfigure}[b]{0.4\linewidth}
		\centering
		\includegraphics[width=0.98\linewidth]{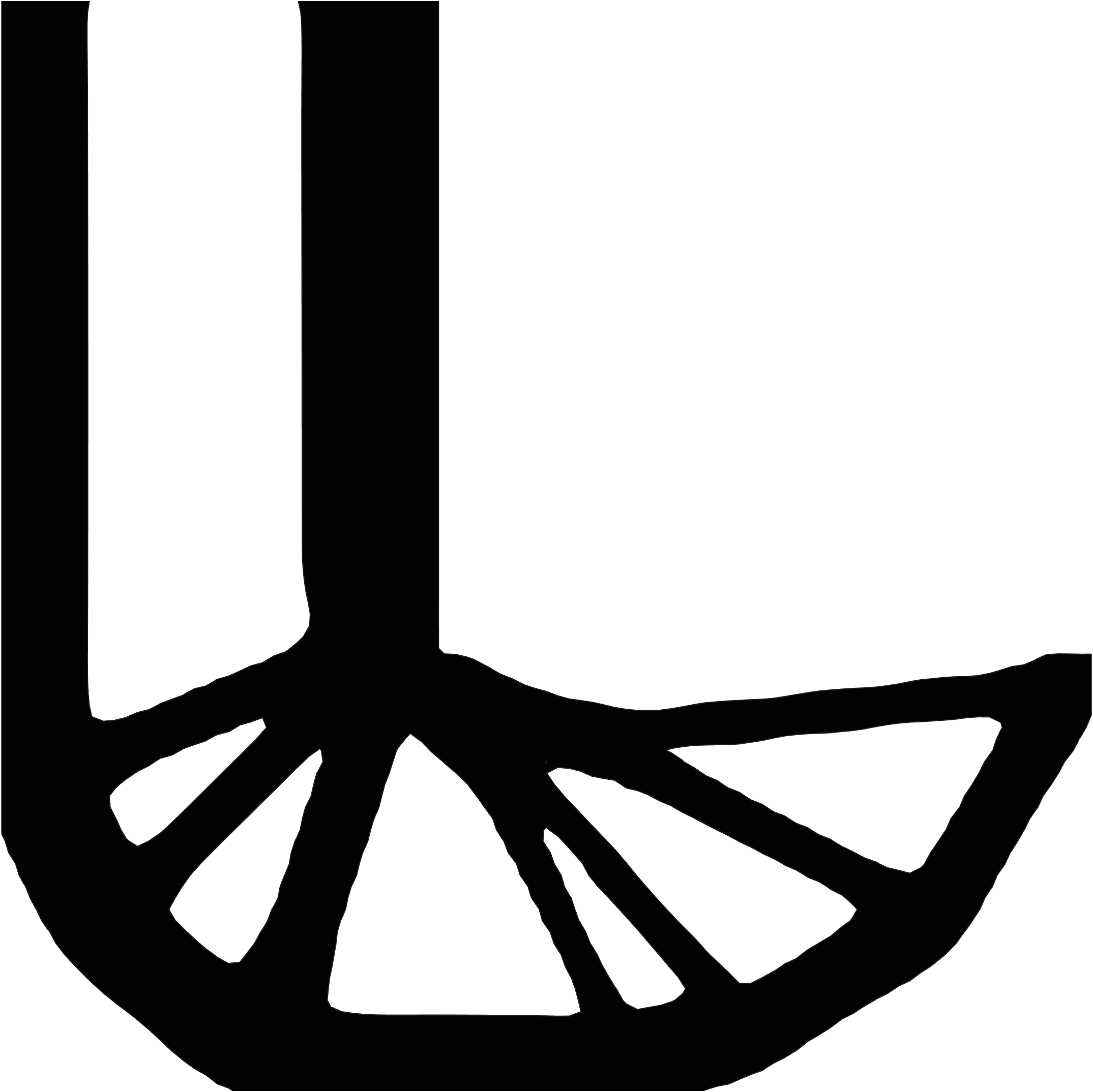}
		\caption{Optimized design}
	\end{subfigure}
        \begin{subfigure}[b]{0.58\linewidth}
		\centering
		\includegraphics[width=0.99\linewidth]{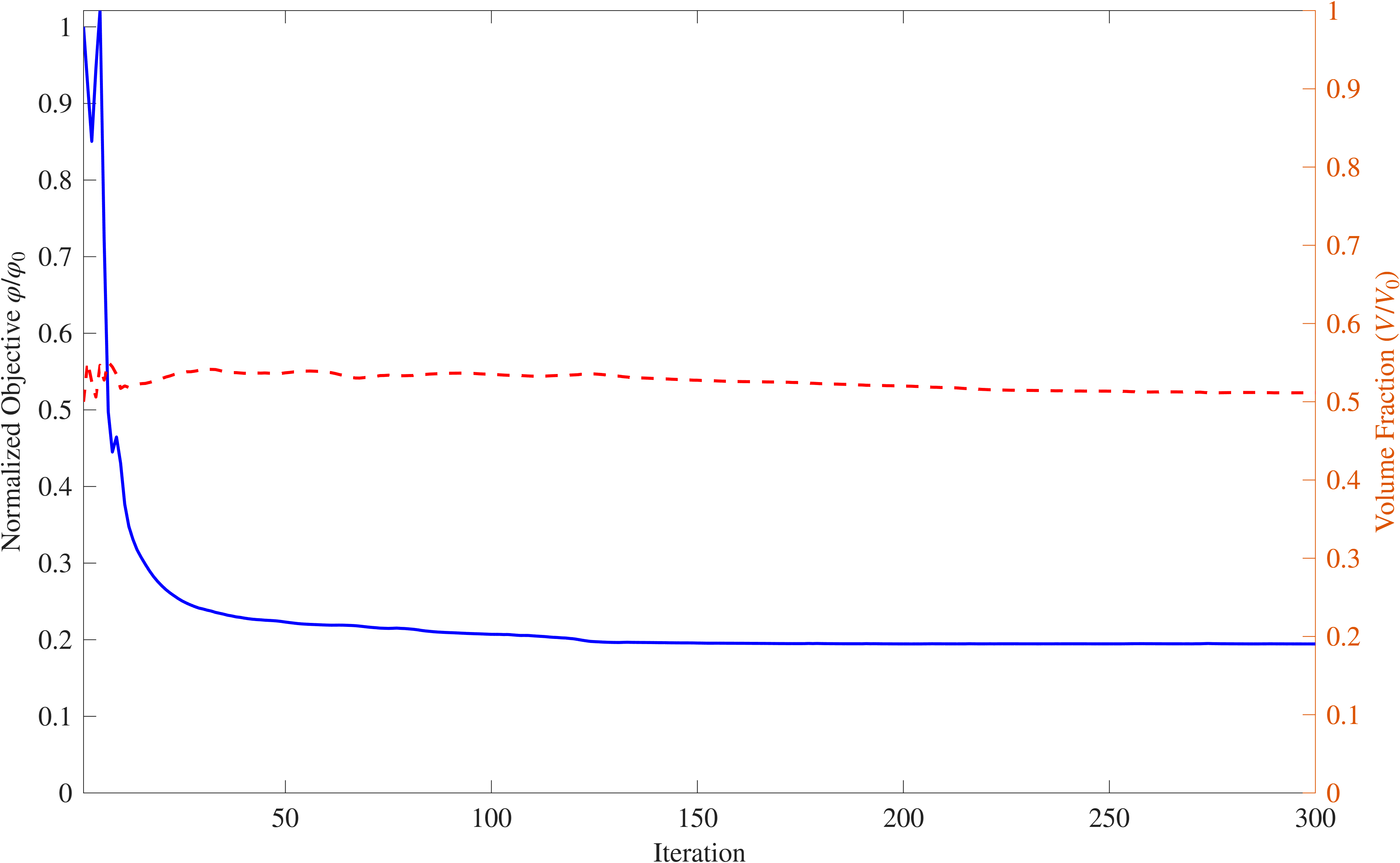}
		\caption{Convergence}
	\end{subfigure}

    \begin{subfigure}[b]{0.44\linewidth}
		\centering
		\includegraphics[width=\linewidth]{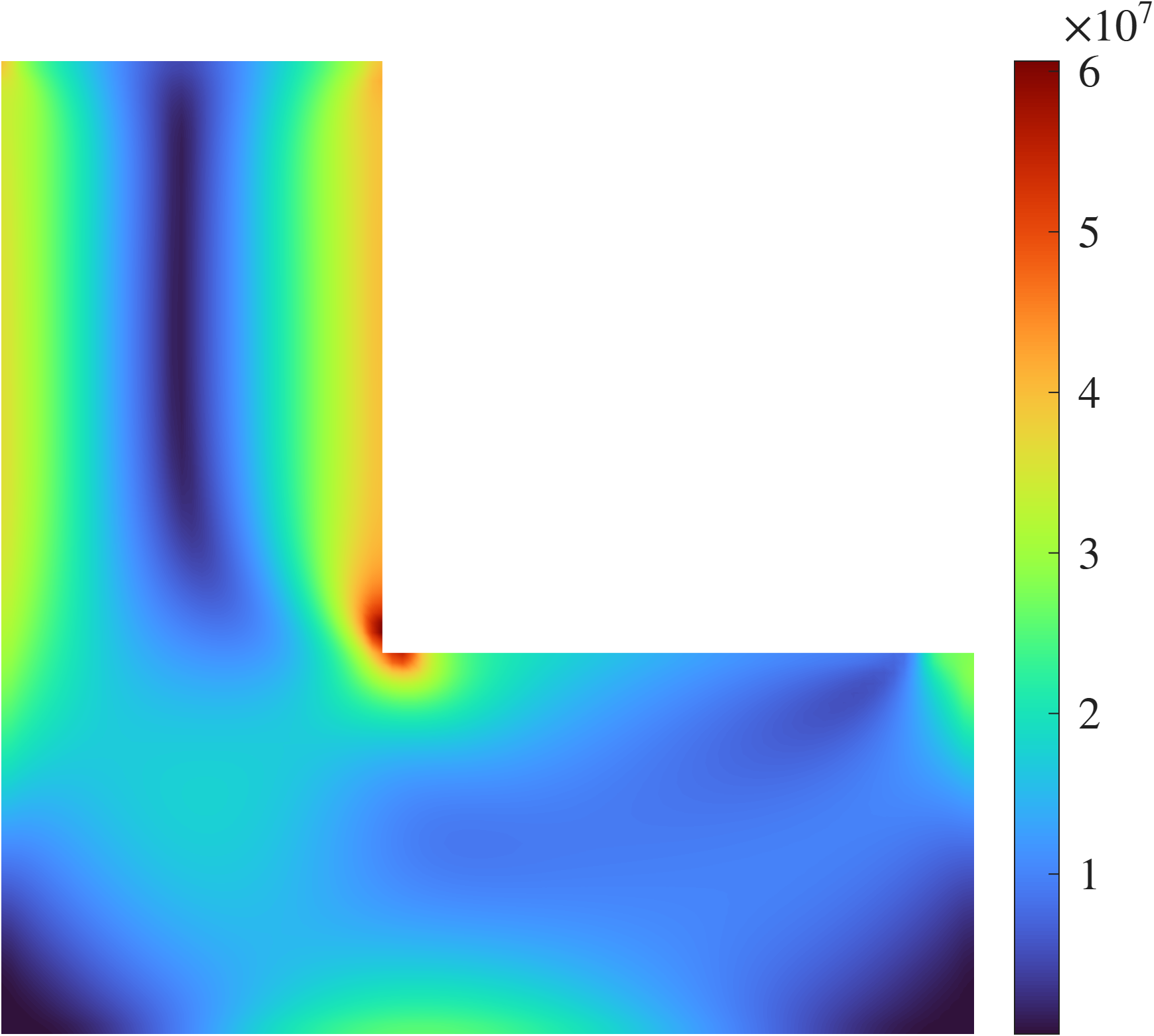}
		\caption{Initial von Mises}
	\end{subfigure}
	\begin{subfigure}[b]{0.44\linewidth}
		\centering
		\includegraphics[width=\linewidth]{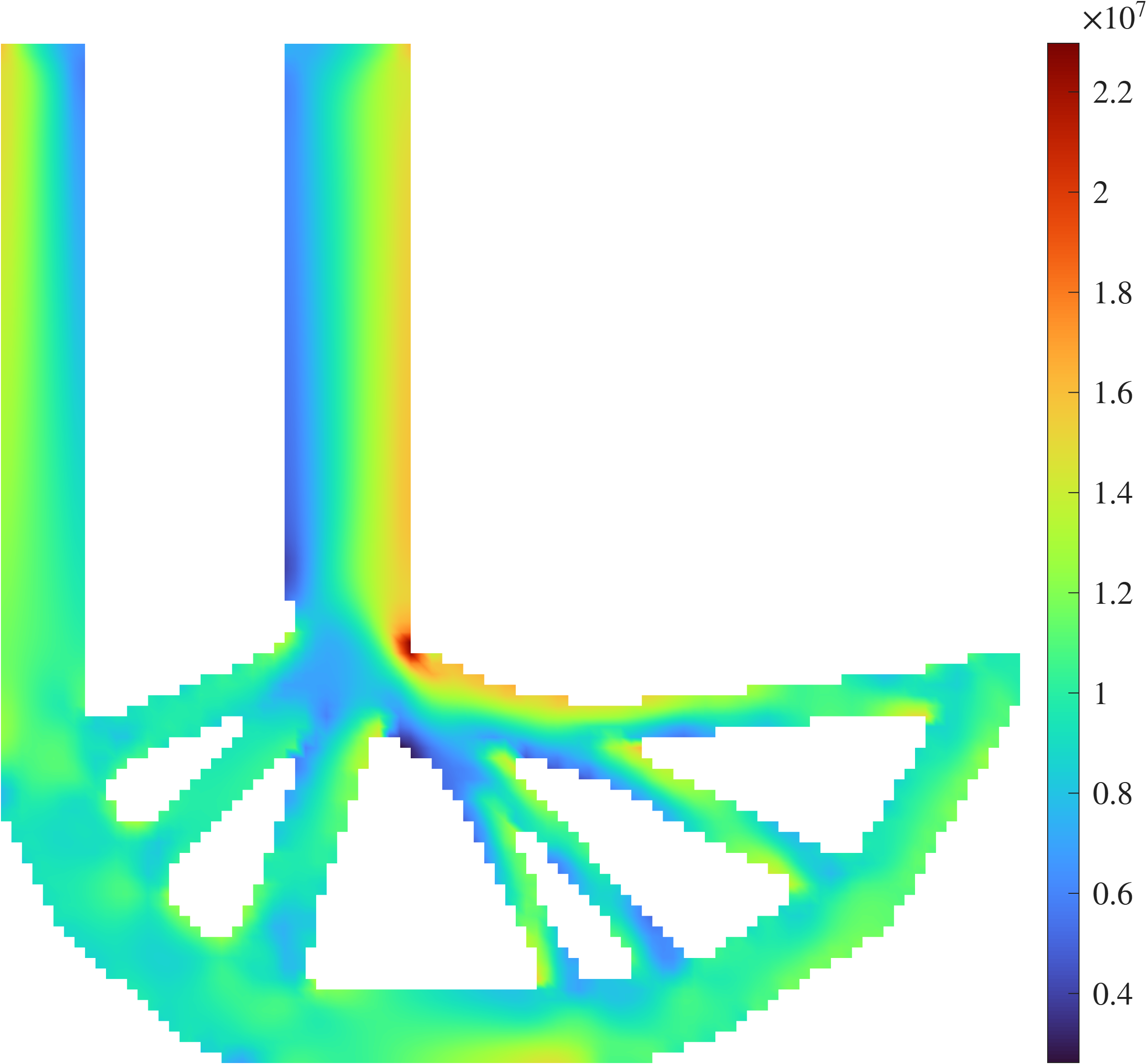}
		\caption{Final von Mises}
	\end{subfigure}

    \begin{subfigure}[b]{0.44\linewidth}
		\centering
		\includegraphics[width=\linewidth]{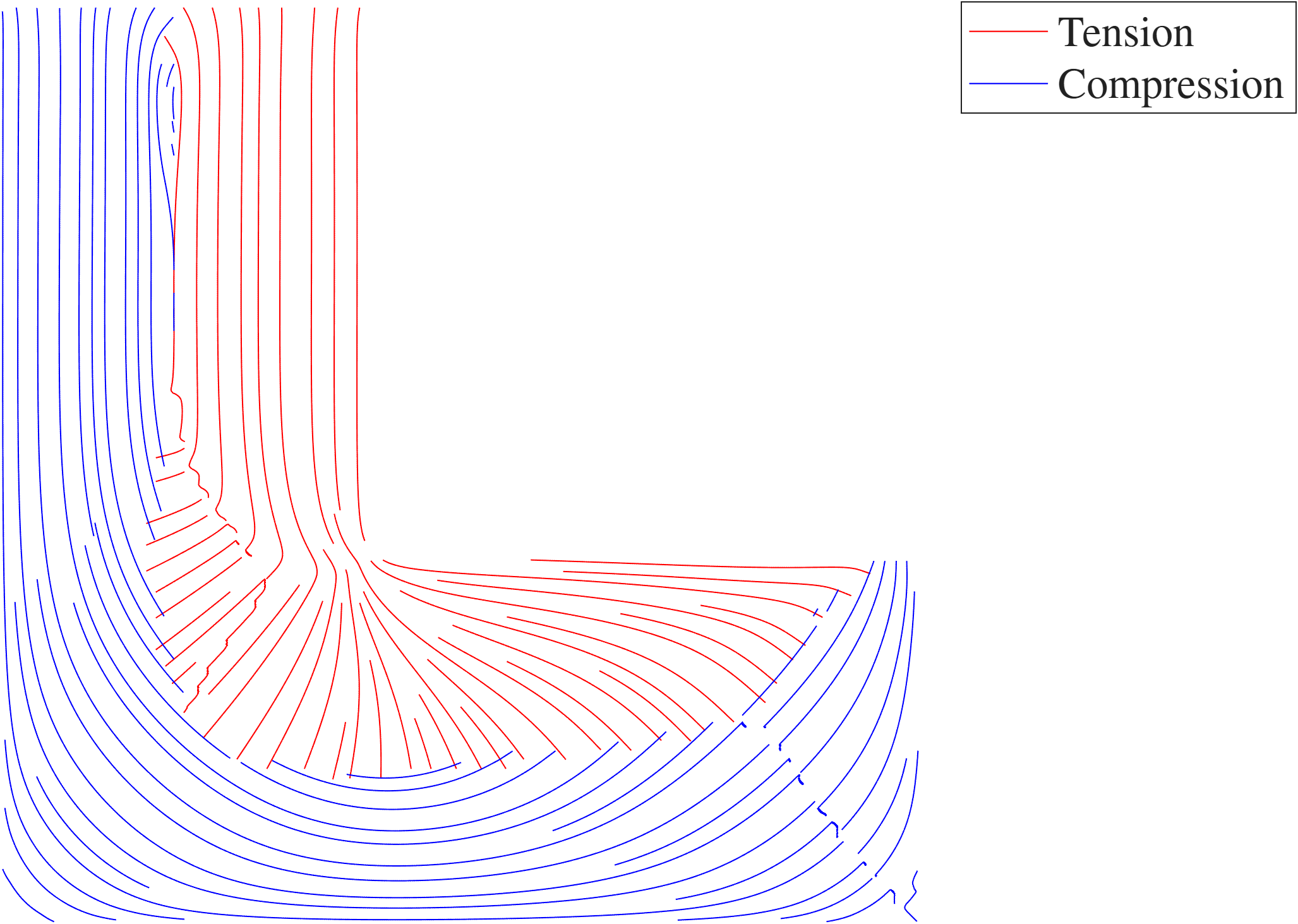}
		\caption{Initial principal stress}
	\end{subfigure}
	\begin{subfigure}[b]{0.44\linewidth}
		\centering
		\includegraphics[width=\linewidth]{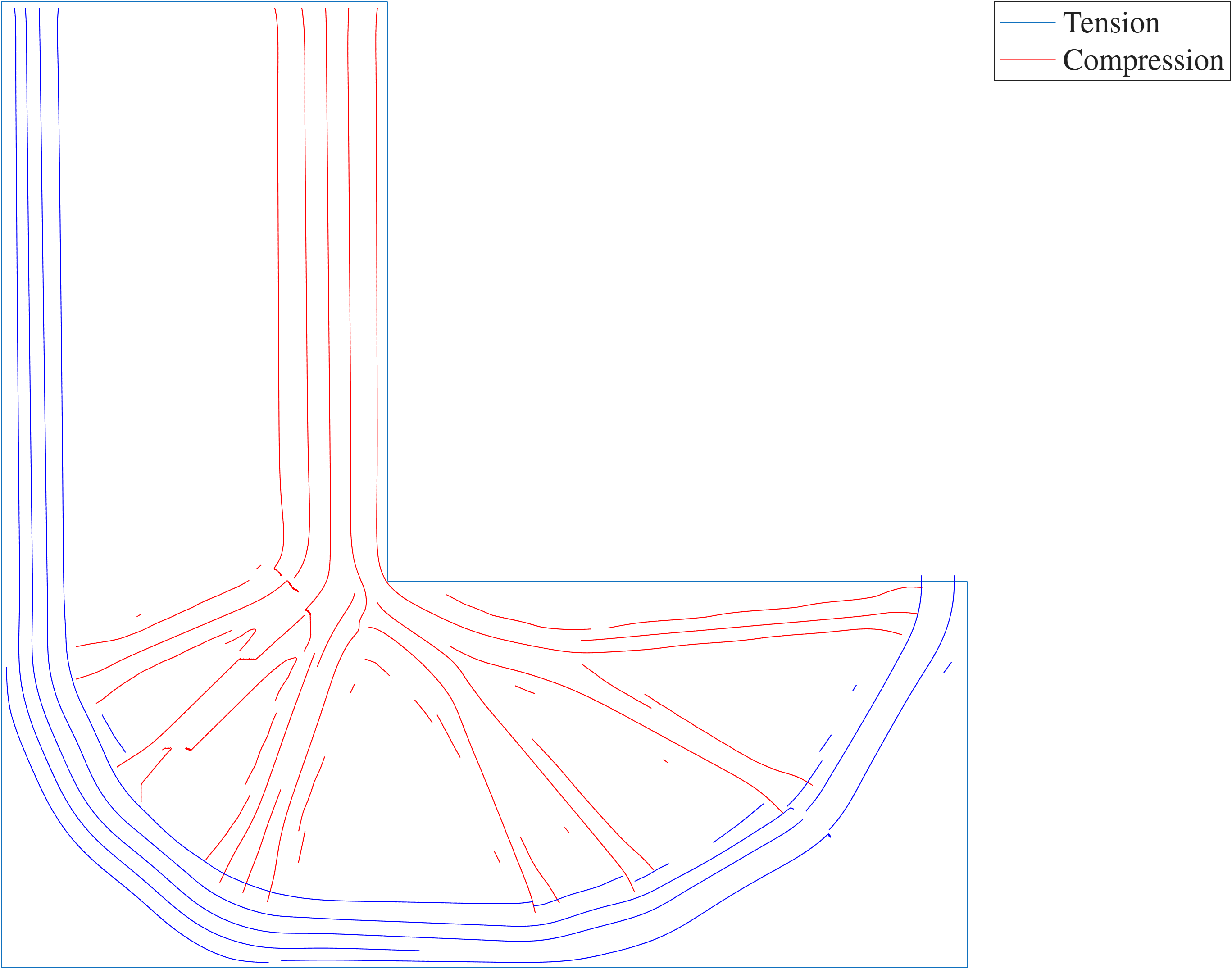}
		\caption{Final principal stress}
	\end{subfigure}

	\caption{Optimized L-bracket for the stress minimization problem.}
	\label{fig:stressMin}
\end{figure} 
Gradients are computed with a standard adjoint approach. Writing $\sigma_{PN}(\rho,u)$ and the equilibrium constraint ${R}_{el}(\bd;\boldsymbol{\rho}) = \textbf{0}$, the total derivative is obtained from
\begin{align}
\frac{d\sigma_{PN}}{d\boldsymbol{\rho}} &\;=\; \frac{\partial \sigma_{PN}}{\partial\boldsymbol{\rho}} \;+\; \lambda^{\mathsf T}\frac{\partial F}{\partial \boldsymbol{\rho}},
\\
\bK(\boldsymbol{\rho})^{\mathsf T}\lambda &\;=\; \frac{\partial \sigma_{PN}}{\partial \bd}.
\end{align}
Here, $\partial \sigma_{PN}/\partial \boldsymbol{\rho} = \left[\partial \sigma_{PN}/\partial \rho_e \right] $ follows from the chain rule through $\tilde{\sigma}_{\mathrm{vm},e}$ and the stress tensor, and $\partial R/\partial \boldsymbol{\rho} = \left[ \partial R/\partial \rho_e \right]$ is given by the stiffness interpolation derivative.

In this example we use a $6000$-element mesh, SIMP penalty $p=3$, and isotropic linear elasticity with $E=100\,\mathrm{GPa}$ and $\nu=0.3$, and $p_{vm}=6$. The resulting designs redistribute material to suppress peak stress while maintaining minimum-length-scale features imposed by $r_{\min}$.

The $p$-norm aggregation concentrates the optimization on high-stress regions from the initial values of 60.6 MPa to 23.0 MPa and yields a visibly more uniform von Mises field compared to compliance-driven layouts.

\subsection{Extending Constraints: \textit{Local Volume Fraction} } \label{sec:localVolFrac}

To demonstrate the extensibility of the STORX framework, we incorporate a \emph{local volume fraction} constraint following the formulation of Wu et al.~\cite{wu2017infill}. Unlike the standard global volume constraint, which allows material to cluster in a few regions, this approach limits the material fraction within a neighborhood around each element, leading to spatially distributed porosity and infill-like structures. \textit{The purpose of this section is to show that existing constraint formulations from the literature can be integrated into STORX with minimal implementation effort.}

The local volume constraint is implemented through the \texttt{localVolume} class located in \texttt{06-optFunctional/localVolume}, which conforms to the same interface as other constraint operators in the framework. As with all optimization functionals, only two methods are required: one to evaluate the constraint value and one to compute its sensitivities. No modifications to the physics solver or optimizer are necessary; the new constraint is simply passed to the list of active constraints.

The cantilever beam with mid-load example of Section \ref{sec:benchmarkExamples} is discretized into  $~80{,}000$ elements.
Instead of global volume constraint, we impose a local volume fraction of $0.5$, a neighborhood radius covering six adjacent elements, and a $p$-norm aggregation parameter $p=6$ with:

\begin{lstlisting}[language=Matlab, basicstyle=\small\ttfamily]
volumeFraction = 0.5;
localRadius    = 6;
localPNorm     = 16;
constraints = {localVolume(solver,... 
                localRadius,...
                localPNorm, ...
                volumeFraction)};
\end{lstlisting}
For manufacturing constraints, we impose minimum feature size with $r_{min}=1.5$, physical density, and symmetry along the $y$-axis. To do so, the \texttt{mfgConstraints} cell array is created by passing instantiations of each manufacturing constraint class.
\begin{lstlisting}[language=Matlab, basicstyle=\small\ttfamily]
% manufacturing constraints
rmin = 1.5;
mfgConstraints = {
  minimumFeatureSize_dist(solver, rmin)
  physicalDensity(solver) 
  symmetry_density(solver,1)% 0:x,1:y
};
\end{lstlisting}

 \begin{figure}[t]
	\centering
	\begin{subfigure}[b]{0.65\linewidth}
		\centering
		\includegraphics[width=\linewidth]{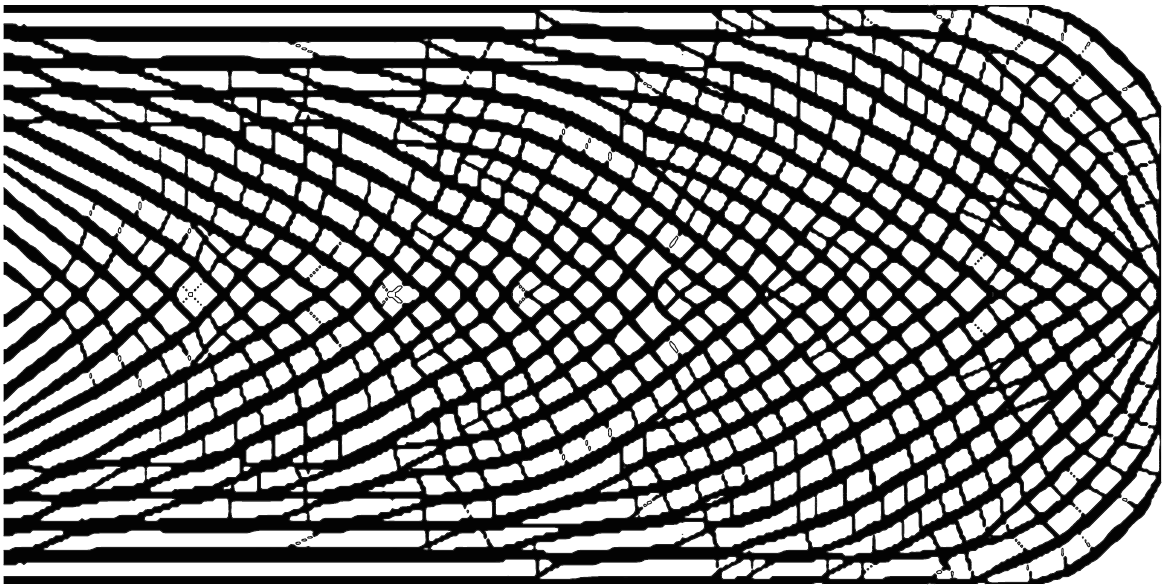}
        \caption{}\label{fig:beamLocalVol}
	\end{subfigure}
    \begin{subfigure}[b]{0.3\linewidth}
		\centering
		\includegraphics[width=\linewidth]{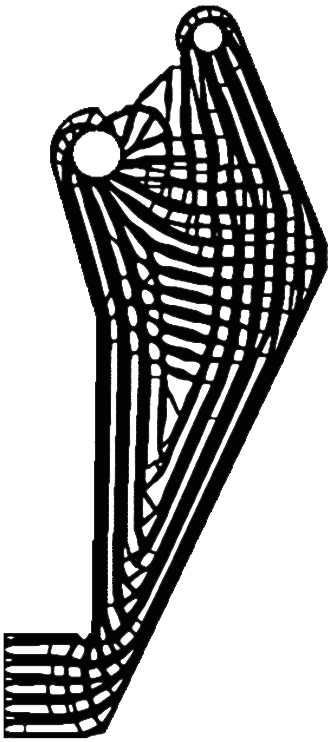}
		\caption{ }\label{fig:gripperLocalVol}
	\end{subfigure}

	\caption{Optimized (a) beam mid-load with 0.5 local volume fraction at 6 element radius neighborhood and (b) gripper with 0.65 local volume fraction at 10 element radius neighborhood.}
	\label{fig:localVolFracExamples}
\end{figure}

To ensure convergence, we use the MMA algorithm. The optimized cantilever beam is shown in Fig.~\ref{fig:localVolFracExamples}a.

Next, consider the gripper example of Section \ref{sec:retainedRegions}, where the design domain is discretized into  $~80{,}000$ active (inside the B-Rep) elements.
To improve manufacturability, we impose a local volume fraction of $0.65$, a neighborhood radius covering ten adjacent elements, and a $p$-norm aggregation parameter $p=6$. For manufacturing constraints, we impose minimum feature size with $r_{min}=2.5$, physical density, and retaining the circular cutouts. The optimized design is illustrated in Fig. \ref{fig:localVolFracExamples}b.

These examples illustrate how advanced manufacturability or morphology constraints from the literature can be readily adopted within STORX through its modular operator design.

\subsection{Extending through Wrapper Class: \textit{Fluid}} \label{sec:fluid}

In STORX, fluid TO is implemented as a \textit{wrapper class} (see Alexandersen~\cite{Alexandersen2022} for derivation and implementation details) that couples the Navier-Stokes solver with the density-based design update loop (filtering, interpolation/penalization, sensitivity evaluation, and constraint handling); specifically, we consider steady, laminar, incompressible flow using the Brinkman penalization (immersed solid) approach, where a continuous design field $\gamma(\mathbf{x})\in[0,1]$ represents the fluid-solid distribution (fluid: $\gamma=1$, solid: $\gamma=0$) and the solid region is modeled by adding a momentum-sink term to the Navier-Stokes equations with an interpolated penalty factor $\alpha(\gamma)$ that suppresses velocities in solid regions while keeping the PDE posed over the full computational domain.

\begin{figure}[!h]
	\centering
	\begin{subfigure}[b]{\linewidth}
		\centering
		\includegraphics[width=0.8\linewidth]{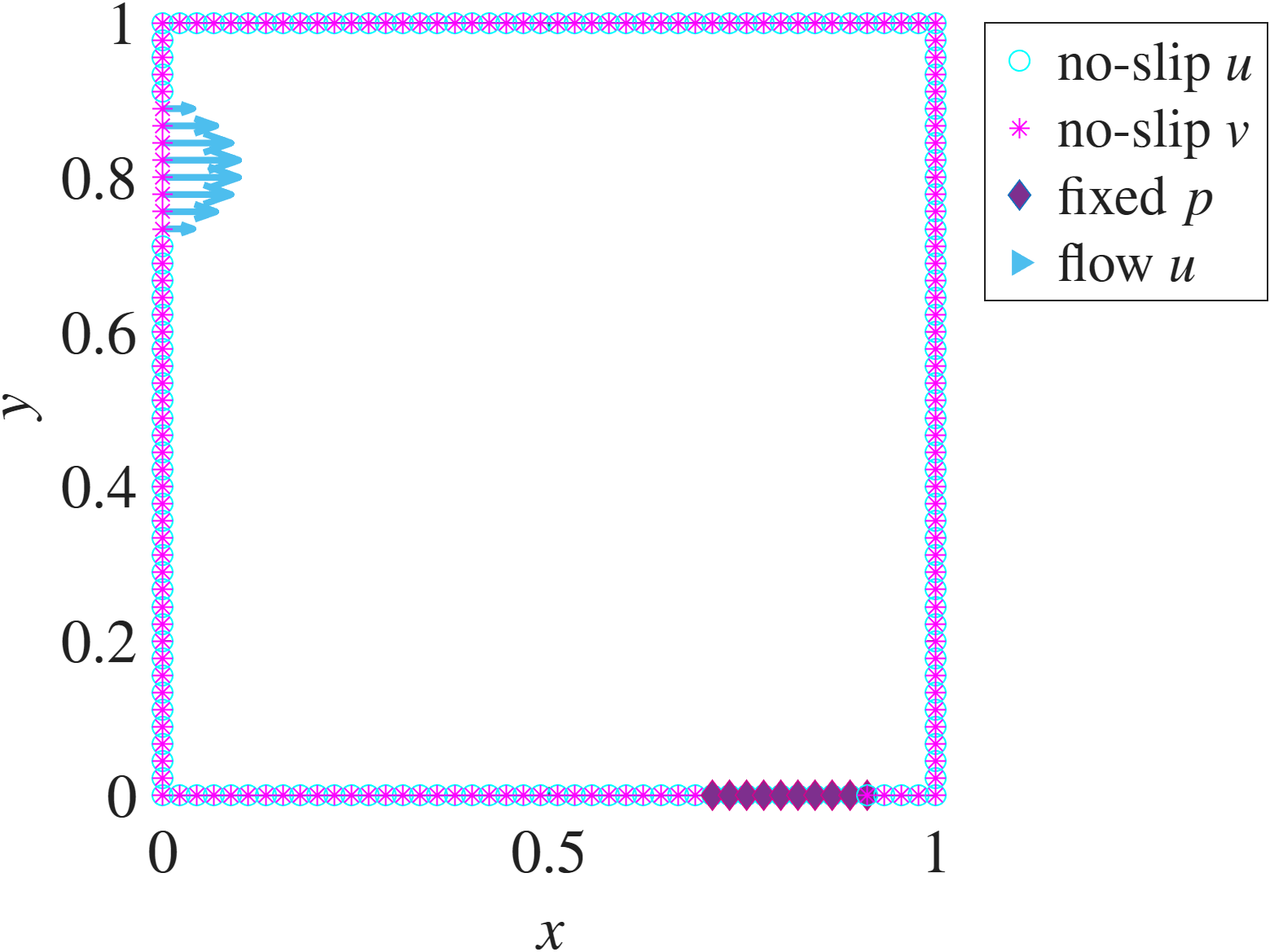}
\caption{Pipe bend B.C.}
	\end{subfigure}
        \begin{subfigure}[b]{0.7\linewidth}
		\centering
		\includegraphics[width=\linewidth]{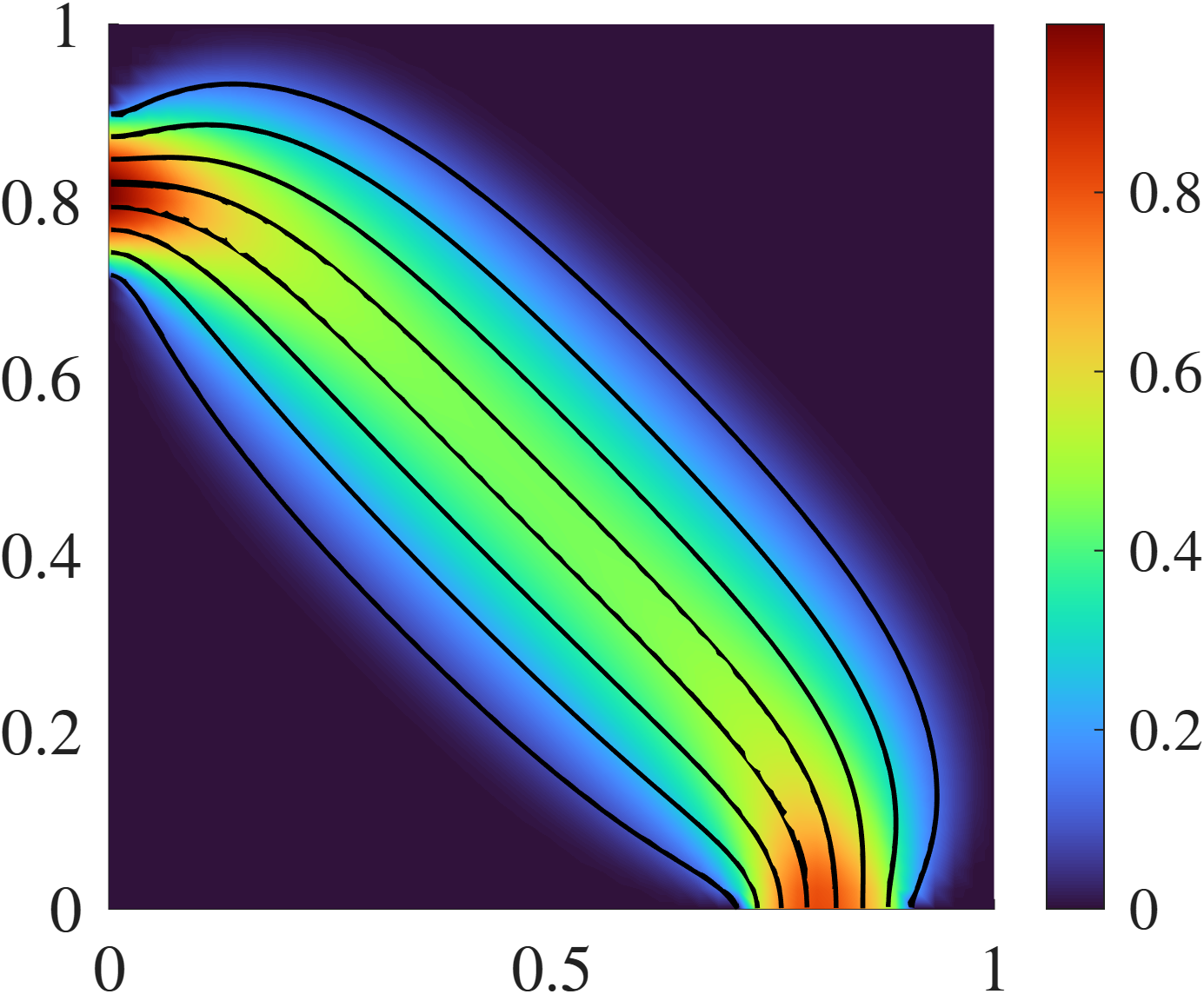}
		\caption{ Pipe bend optimized velocity}
	\end{subfigure}

	\begin{subfigure}[b]{\linewidth}
		\centering
		\includegraphics[width=0.9\linewidth]{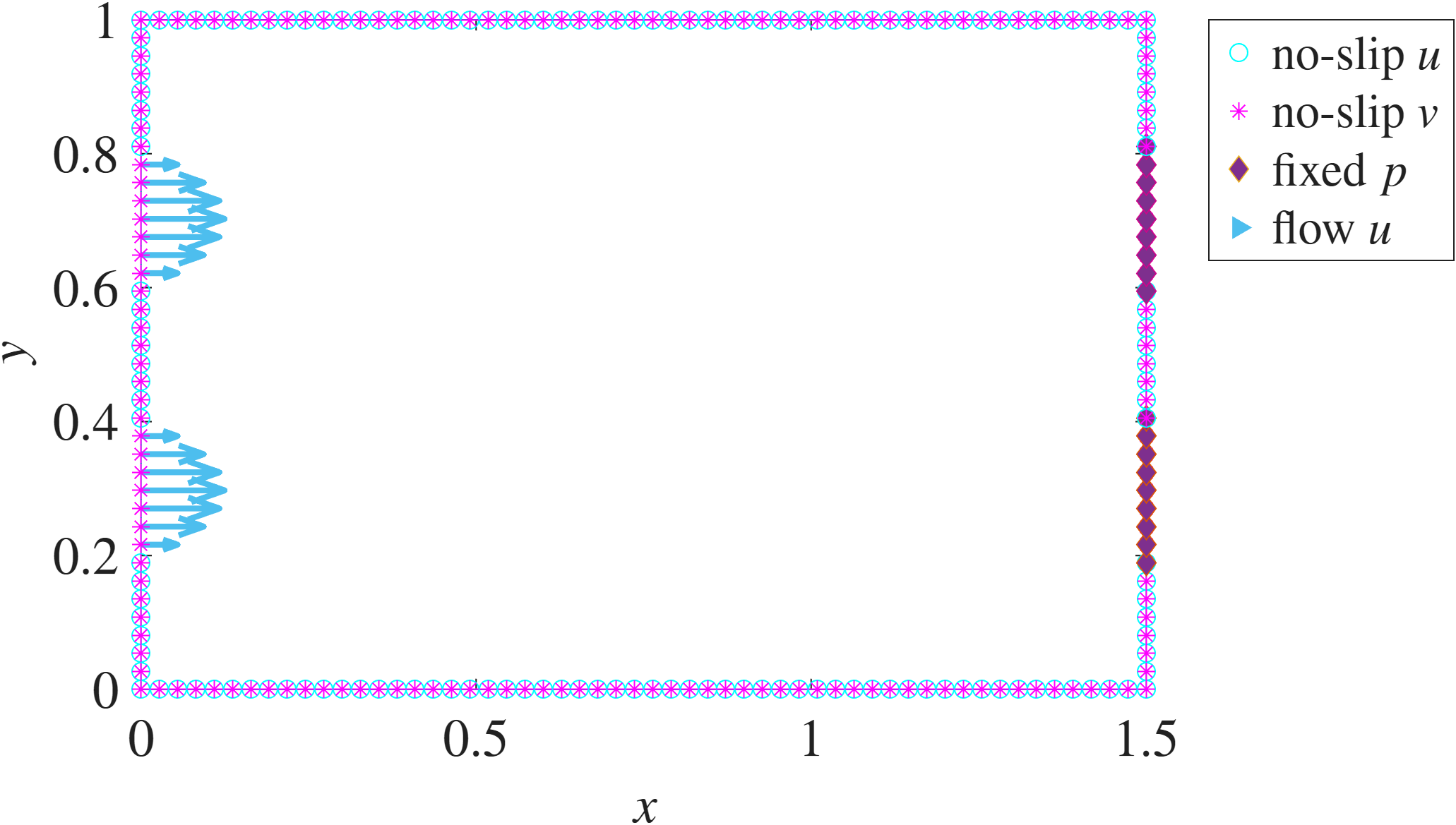}
\caption{Double-pipe B.C.}
	\end{subfigure}
        \begin{subfigure}[b]{0.8\linewidth}
		\centering
		\includegraphics[width=\linewidth]{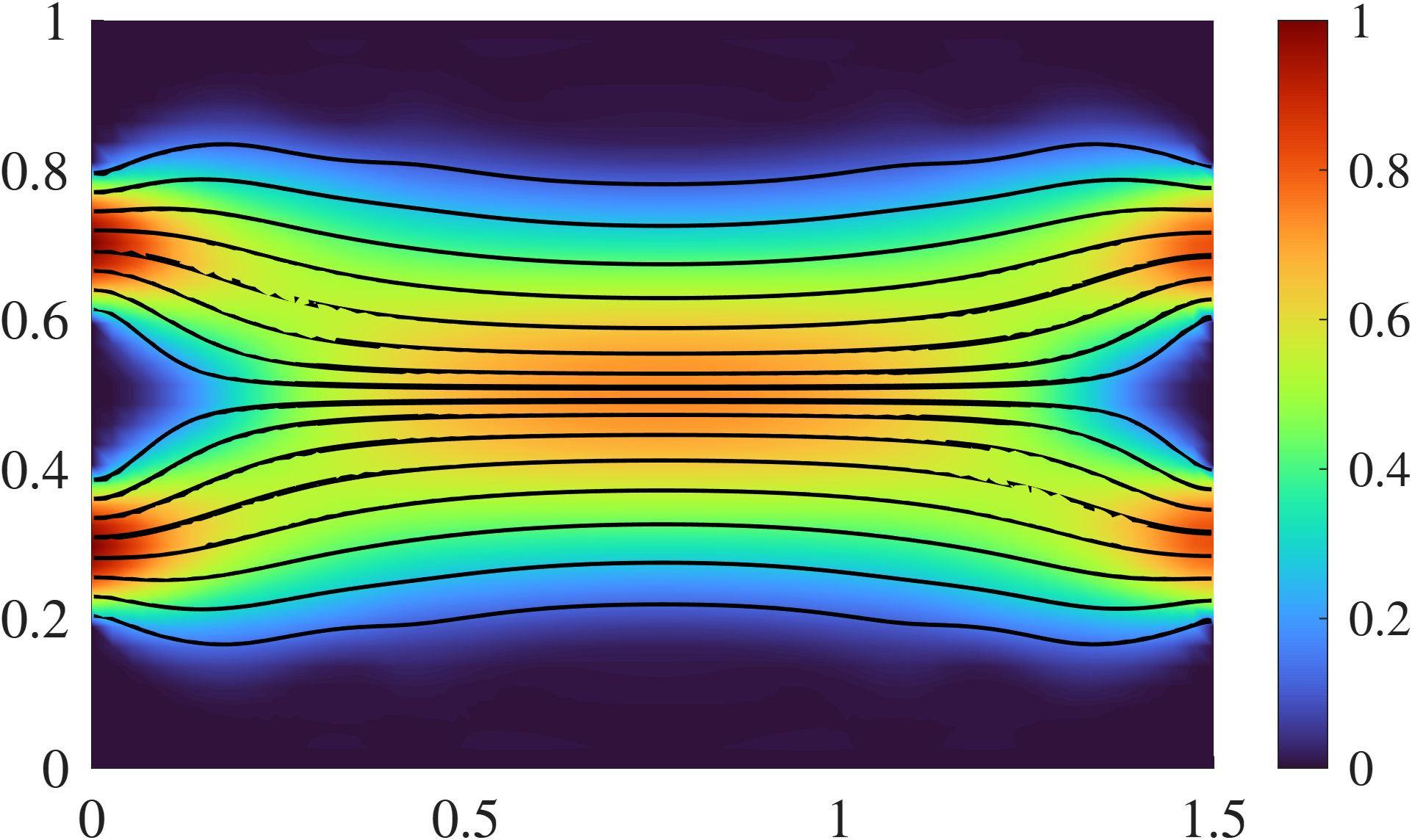}
		\caption{Double-pipe velocity }
	\end{subfigure}
	\caption{Density-based fluid TO illustrative examples.}
	\label{fig:fluidBenchmarks}
\end{figure}

Figure~\ref{fig:fluidBenchmarks} shows the prescribed boundary conditions and the optimized channel geometries with overlaid velocity magnitude fields for two illustrative problems: (1) a pipe bend and (2) a double-pipe configuration. In both cases, the optimization objective is the minimization of the total energy dissipation, similar to examples posed in \cite{Alexandersen2022}.

The problems are formulated using the SIMP material interpolation with penalization parameter $p=3$ and solved using the MMA. The computational domain is discretized into $2000$ finite elements, and a volume fraction constraint of $V_f = 0.3$ is imposed. A uniform inlet velocity of $U_{\mathrm{in}} = 1~\mathrm{m\,s^{-1}}$ is prescribed. The fluid is assumed incompressible with density $\rho = 1~\mathrm{kg\,m^{-3}}$ and dynamic viscosity $\mu = 1~\mathrm{kg/(m\;s)}$.

\begin{figure*}[!h]
	\centering
	\includegraphics[width=0.55\linewidth]{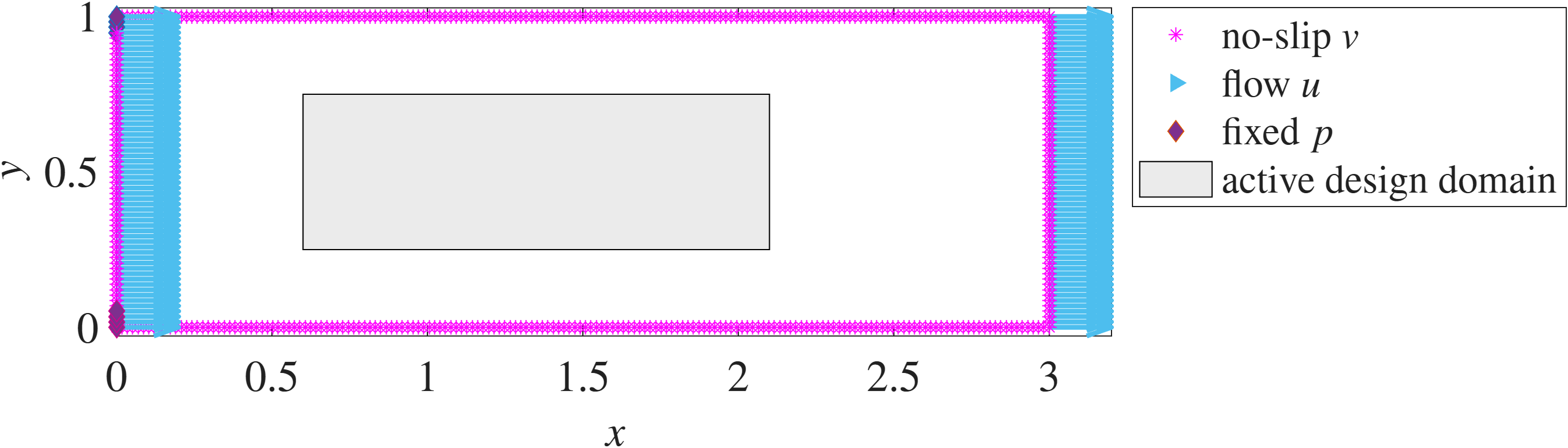}
	\caption{Wind tunnel example and boundary conditions.}
	\label{fig_windtunnel_bc}
\end{figure*}

 \begin{figure*}[t]
	\centering
    	\begin{subfigure}[b]{0.2\linewidth}
		\centering
		\includegraphics[width=0.6\linewidth]{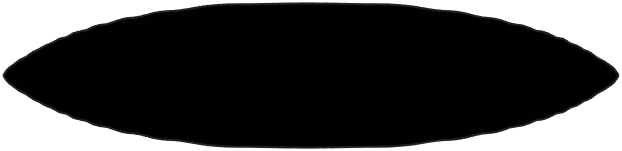}
		\caption{}
	\end{subfigure}
	\begin{subfigure}[b]{0.39\linewidth}
		\centering
		\includegraphics[width=\linewidth]{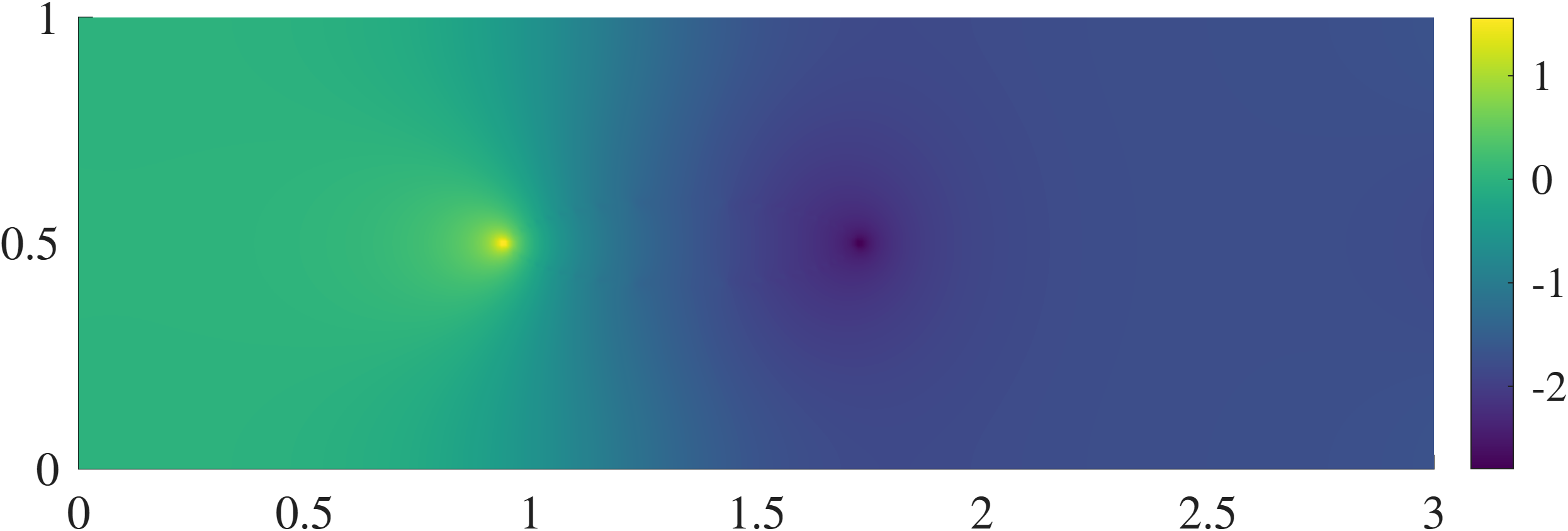}
		\caption{}
	\end{subfigure}
	\begin{subfigure}[b]{0.39\linewidth}
		\centering
		\includegraphics[width=\linewidth]{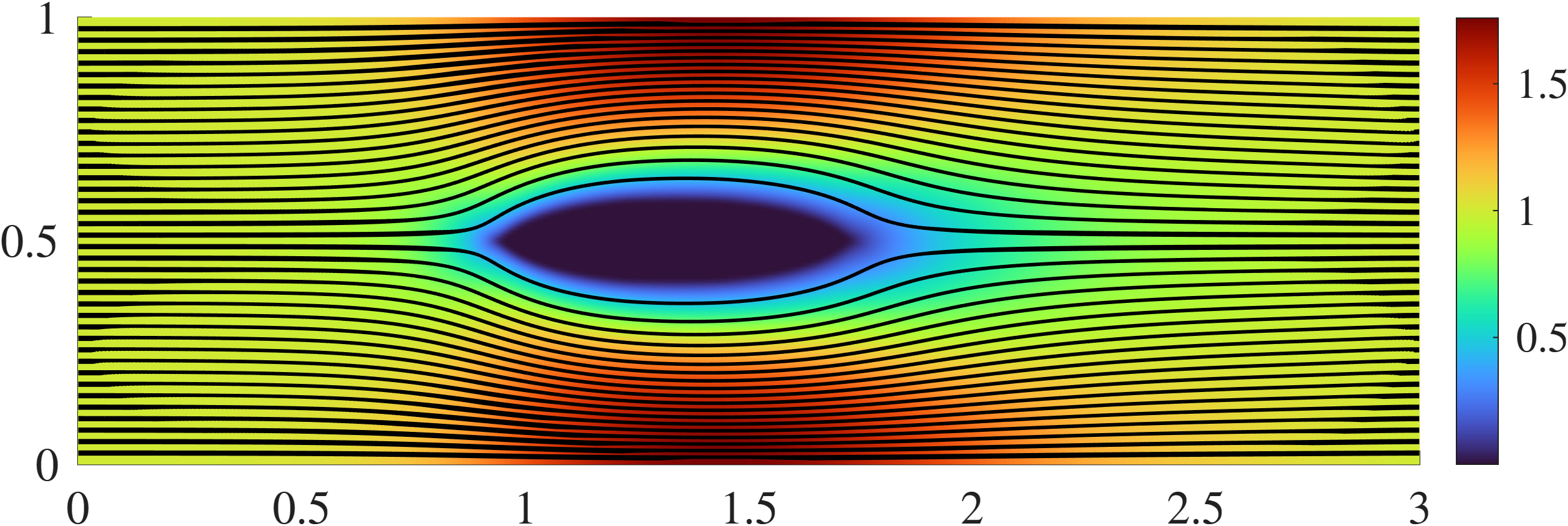}
		\caption{}
	\end{subfigure}

    \begin{subfigure}[b]{0.2\linewidth}
		\centering
		\includegraphics[width=0.6\linewidth]{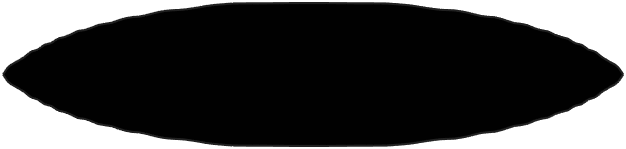}
		\caption{}
	\end{subfigure}
	\begin{subfigure}[b]{0.39\linewidth}
		\centering
		\includegraphics[width=\linewidth]{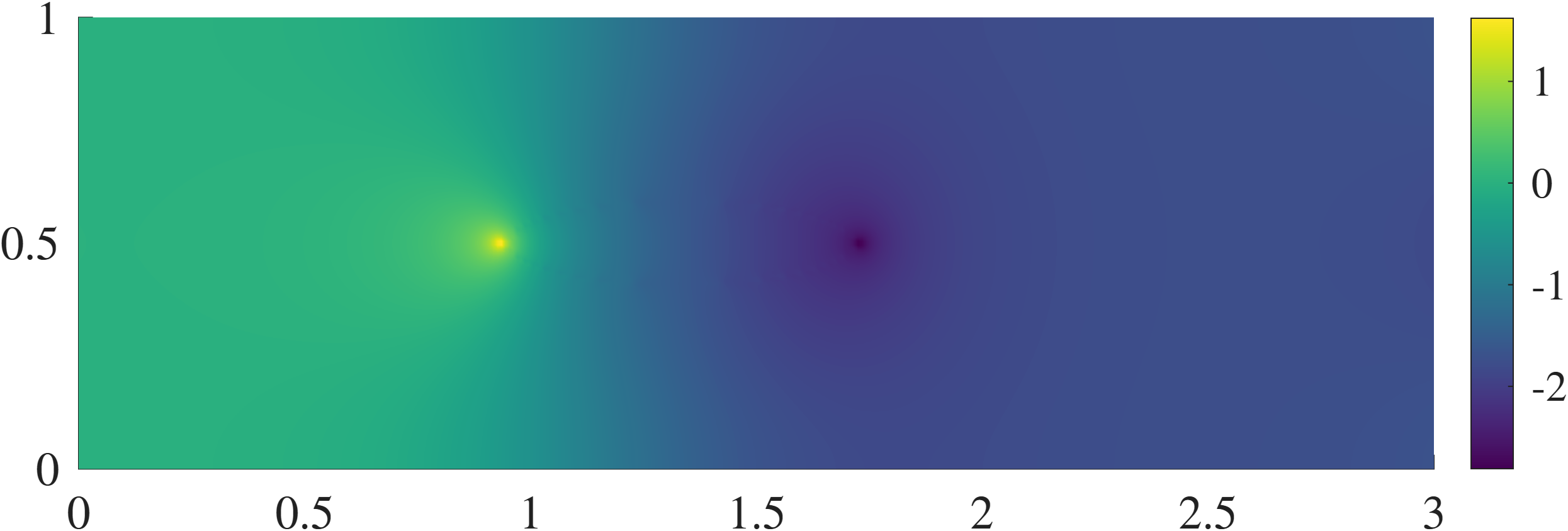}
		\caption{}
	\end{subfigure}
	\begin{subfigure}[b]{0.39\linewidth}
		\centering
		\includegraphics[width=\linewidth]{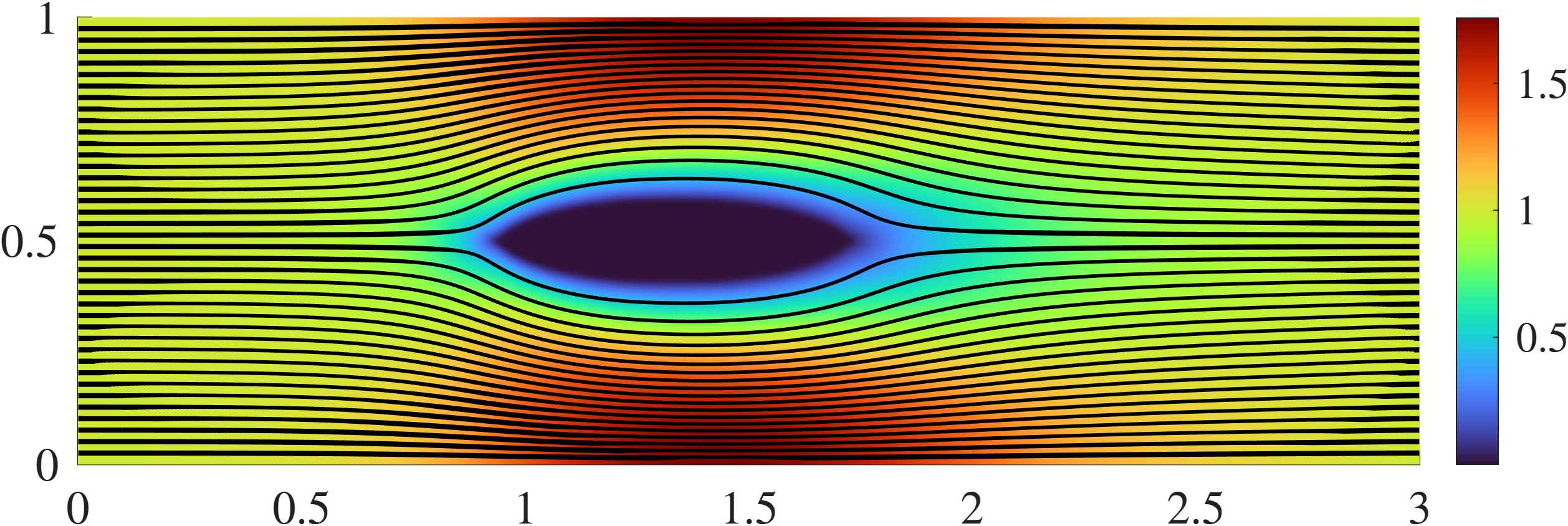}
		\caption{}
	\end{subfigure}
    
	\begin{subfigure}[b]{0.2\linewidth}
		\centering
		\includegraphics[width=0.8\linewidth]{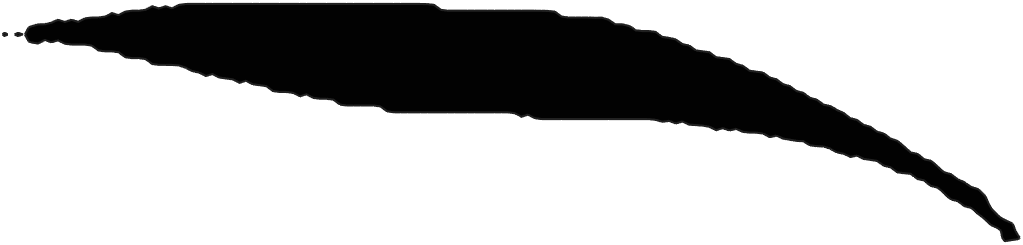}
		\caption{}
	\end{subfigure}
	\begin{subfigure}[b]{0.39\linewidth}
		\centering
		\includegraphics[width=\linewidth]{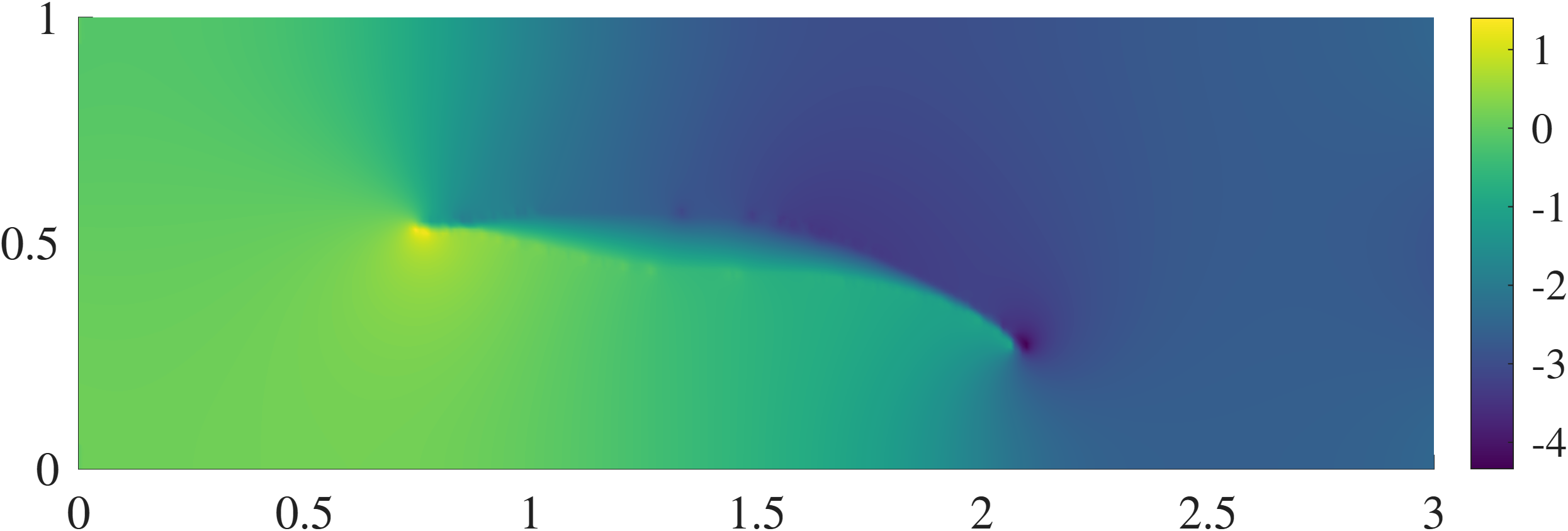}
		\caption{}
	\end{subfigure}
	\begin{subfigure}[b]{0.39\linewidth}
		\centering
		\includegraphics[width=\linewidth]{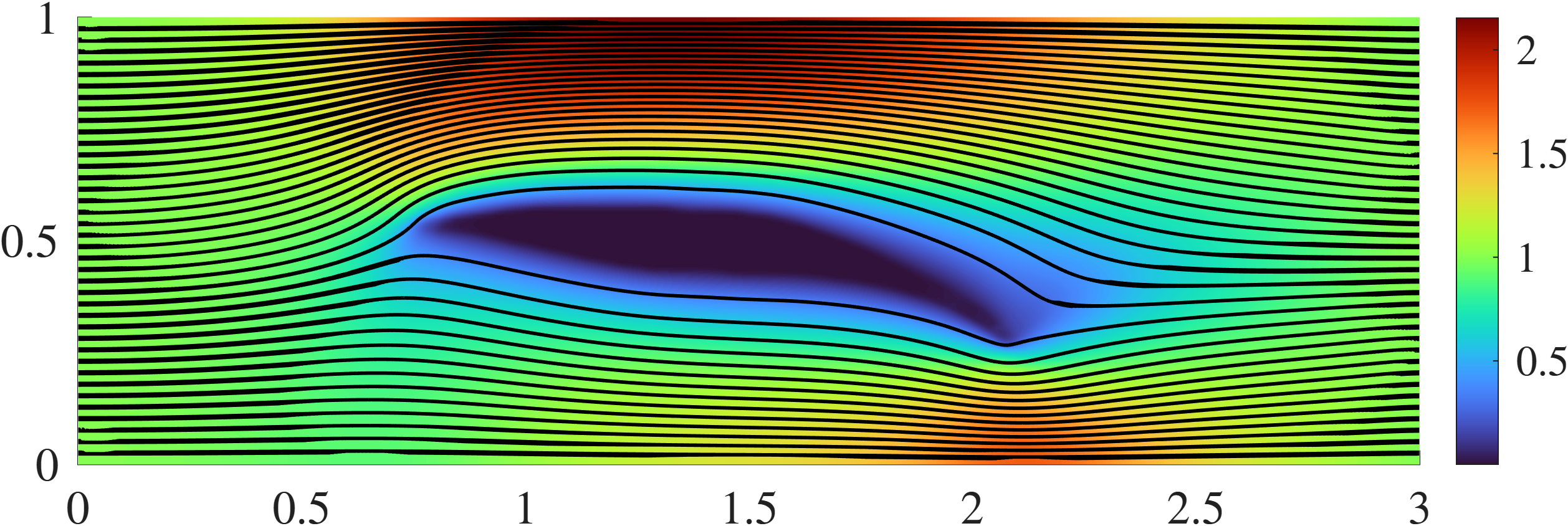}
		\caption{}
	\end{subfigure}

	\caption{Wind tunnel example for minimizing dissipated energy (first row) and drag (second row) and maximizing lift (third row).}
	\label{fig_example_fluid}
\end{figure*}

Figure~\ref{fig_windtunnel_bc} shows the wind-tunnel-style domain and boundary conditions used in this example; the domain is discretized into $40{,}000$ elements and operated at $\mathrm{Re}=10$ with inlet velocity $U_{\mathrm{in}}=1~\mathrm{m\,s^{-1}}$, assuming unit density $\rho=1~\mathrm{kg\,m^{-3}}$ and viscosity $\mu=\rho\,U_{\mathrm{ref}}\,L_c/\mathrm{Re}$ (with $U_{\mathrm{ref}}=U_{\mathrm{in}}$ and $L_c$ in $\mathrm{m}$. The design (active) subdomain is a rectangle centered at $\mathbf{c}=[1.35,\,0.5]~\mathrm{m}$ with width $w=1.5~\mathrm{m}$ and height $h=0.5~\mathrm{m}$, and the volume constraint enforces that $85\%$ of this active domain is occupied by solid (airfoil) material; the characteristic length is taken as $L_c=\sqrt{v_f\,A_{\mathrm{act}}}$ with $v_f$ the (solid) volume fraction and $A_{\mathrm{act}}=w/;h$ the active-domain area (in $\mathrm{m^2}$).

We study three objective variants under a prescribed fluid-volume constraint: (1) minimization of dissipated energy (a standard benchmark objective), (2) minimization of drag, and (3) maximization of lift (optionally with a drag constraint to avoid degenerate, overly-rotated profiles). The resulting optimized designs are visualized by an isocontour/isoline of the design field together with the corresponding pressure and velocity magnitude fields.

For this example, we can solve the fluid FEA over the entire wind-tunnel domain while optimizing the design \emph{only} within an \textit{active} subdomain. When setting up the problem, we define:
\begin{lstlisting}
%% Active Design Domain
center = [1.35,0.5];
w = 1.5;
h = 0.5;
solver = solver.createRectangularDesignDomain(center,w,h);
\end{lstlisting}

Assuming the lift and drag functionals are implemented (and can be used either as the objective or as constraints), we define:
\begin{lstlisting}
%% evaluate reference drag
beta = 1.1;
dragRef = 2.4207; % from Joe Alexandersen's paper

%% Objective and Constraints
objective = densityLift(solver);

constraints = {
    activeVolume(solver, volumeFraction)
    densityDrag(solver, beta*dragRef)
    };
\end{lstlisting}

Observe that we impose two constraints: (1) a volume constraint applied \textit{only} within the active subdomain, and (2) a drag constraint to prevent overly slender designs, as discussed in \cite{Alexandersen2022}.

Finally, we impose two manufacturing constraints: a minimum feature size filter and a physical density projection:
\begin{lstlisting}
% manufacturing constraints
rmin = 1.5;
mfgConstraints = {
    minimumFeatureSize_dist(solver, rmin)
    physicalDensity(solver)
    };
\end{lstlisting}

Figure~\ref{fig_example_fluid} compares representative solutions for the three objectives. Minimizing dissipated energy promotes well-connected flow paths that reduce both viscous losses and Brinkman-induced dissipation. Drag minimization yields a more streamlined obstruction with reduced resistance to the imposed freestream. Lift maximization produces an asymmetric (cambered) profile that generates a pressure differential across the body, consistent with an increased lift response. These examples illustrate how changing only the objective functional within the same PDE-constrained framework leads to qualitatively different, physically interpretable flow designs.

\section{Conclusion}\label{sec:conclusion}
This paper introduced \textsc{STORX} (Shape and Topology Optimization for Research and Experimentation), an open-source MATLAB-based educational framework intended to support learning, teaching, and rapid prototyping in computational design optimization. \textsc{STORX} provides a unified platform for parametric and level-set shape optimization together with topology optimization methods spanning density-based, level-set, and topological-sensitivity-driven approaches, including evolutionary and Pareto-tracing strategies. By embedding finite element analysis, sensitivity analysis, and visualization within a consistent object-oriented software structure, the framework enables users to investigate the continuum between shape and topology optimization in a transparent and reproducible workflow.

A key outcome of this work is the software architecture itself. \textsc{STORX} is organized around a clear separation of concerns and a set of abstract base-class interfaces that define the core contracts of the framework. This design allows new objective functionals, constraints (including manufacturing- and application-driven constraints), and problem variants to be incorporated as derived classes without modifying the core code base. As a result, \textsc{STORX} is positioned not only as a teaching tool but also as a practical foundation for research experimentation, where method development often requires frequent extensions and careful verification.

Through illustrative examples, we demonstrated how standard SO/TO mathematical formulations map directly to executable implementations, lowering the barrier between conceptual understanding and reproducible computational experiments. While the paper is intentionally implementation-centered and does not aim to be a comprehensive reference on finite element theory or general optimization theory, it provides the essential algorithmic and numerical components required to reproduce representative results and to serve as a starting point for further study and development. Overall, {STORX} contributes a cohesive, extensible, and transparent environment for engaging with modern shape and topology optimization methods and for accelerating hands-on research and education in computational design.

STORX has also proven effective as both an educational and research prototyping platform. The framework was integrated into a graduate-level course in the Aerospace Engineering department at the University of Kansas, where students used it for hands-on implementation of shape and topology optimization methods. Its modular architecture enabled rapid incorporation of newly introduced concepts; for example, following an invited lecture where local volume fraction constraint in TO was introduced by Jun Wu, students were able to formulate and deploy this constraint within STORX as part of a homework assignment.

Beyond structured assignments, students extended the framework in independent projects spanning antenna shape optimization, fluid-structure interaction for wing design, and TO of auxetic metamaterials. These experiences underscore STORX's accessibility and extensibility as a platform for multi-physics design optimization in both research and instruction.

We welcome bug reports, feature requests, and contributions from the community, and we encourage users to open issues and propose enhancements through the public repository. Educators interested in adopting \textsc{STORX} in graduate or advanced undergraduate courses are invited to reach out for accompanying teaching materials (e.g., syllabi, lecture modules, and assignments) and for feedback on curriculum integration.

\section*{Acknowledgements}
A.M. would like to thank the Aerospace Department at the University of Kansas for supporting this work.
A.M. also gratefully acknowledges Dr. Jun Wu, Dr. Julian Norato, Dr. Faez Ahmed, Dr. Aaditya Chandrasekhar, and Dr. Krishnan Suresh for their insightful guest lectures in the graduate-level course `Shape and Topology Optimization' at the University of Kansas, which enriched the course and contributed to the further development of STORX. The authors would like to thank Grayson Klimek for his assistance in the 3D printing process.

\section*{Declarations}
\textbf{Compliance with ethical standards} The authors declare that they have no conflict of interest.

\textbf{Replication of Results} The MATLAB code is available at \href{https://github.com/DEL-KU/storx}{https://github.com/DEL-KU/storx}. Instructors are welcome to reach out to request codes for additional modules, example files, teaching material, and homework assignments.  

\textbf{Use of Artificial Intelligence} Generative artificial intelligence tools were used to improve the clarity and readability of portions of the manuscript. All technical content, results, and interpretations were developed and verified by the authors.

\bibliography{sn-bibliography}


\begin{thebibliography}{111}
\ifx \bisbn   \undefined \def \bisbn  #1{ISBN #1}\fi
\ifx \binits  \undefined \def \binits#1{#1}\fi
\ifx \bauthor  \undefined \def \bauthor#1{#1}\fi
\ifx \batitle  \undefined \def \batitle#1{#1}\fi
\ifx \bjtitle  \undefined \def \bjtitle#1{#1}\fi
\ifx \bvolume  \undefined \def \bvolume#1{\textbf{#1}}\fi
\ifx \byear  \undefined \def \byear#1{#1}\fi
\ifx \bissue  \undefined \def \bissue#1{#1}\fi
\ifx \bfpage  \undefined \def \bfpage#1{#1}\fi
\ifx \blpage  \undefined \def \blpage #1{#1}\fi
\ifx \burl  \undefined \def \burl#1{\textsf{#1}}\fi
\ifx \doiurl  \undefined \def \doiurl#1{\url{https://doi.org/#1}}\fi
\ifx \betal  \undefined \def \betal{\textit{et al.}}\fi
\ifx \binstitute  \undefined \def \binstitute#1{#1}\fi
\ifx \binstitutionaled  \undefined \def \binstitutionaled#1{#1}\fi
\ifx \bctitle  \undefined \def \bctitle#1{#1}\fi
\ifx \beditor  \undefined \def \beditor#1{#1}\fi
\ifx \bpublisher  \undefined \def \bpublisher#1{#1}\fi
\ifx \bbtitle  \undefined \def \bbtitle#1{#1}\fi
\ifx \bedition  \undefined \def \bedition#1{#1}\fi
\ifx \bseriesno  \undefined \def \bseriesno#1{#1}\fi
\ifx \blocation  \undefined \def \blocation#1{#1}\fi
\ifx \bsertitle  \undefined \def \bsertitle#1{#1}\fi
\ifx \bsnm \undefined \def \bsnm#1{#1}\fi
\ifx \bsuffix \undefined \def \bsuffix#1{#1}\fi
\ifx \bparticle \undefined \def \bparticle#1{#1}\fi
\ifx \barticle \undefined \def \barticle#1{#1}\fi
\bibcommenthead
\ifx \bconfdate \undefined \def \bconfdate #1{#1}\fi
\ifx \botherref \undefined \def \botherref #1{#1}\fi
\ifx \url \undefined \def \url#1{\textsf{#1}}\fi
\ifx \bchapter \undefined \def \bchapter#1{#1}\fi
\ifx \bbook \undefined \def \bbook#1{#1}\fi
\ifx \bcomment \undefined \def \bcomment#1{#1}\fi
\ifx \oauthor \undefined \def \oauthor#1{#1}\fi
\ifx \citeauthoryear \undefined \def \citeauthoryear#1{#1}\fi
\ifx \endbibitem  \undefined \def \endbibitem {}\fi
\ifx \bconflocation  \undefined \def \bconflocation#1{#1}\fi
\ifx \arxivurl  \undefined \def \arxivurl#1{\textsf{#1}}\fi
\csname PreBibitemsHook\endcsname

\bibitem[\protect\citeauthoryear{Pironneau}{2005}]{pironneau2005optimal}
\begin{bchapter}
\bauthor{\bsnm{Pironneau}, \binits{O.}}:
\bctitle{Optimal shape design for elliptic systems}.
In: \bbtitle{System Modeling and Optimization: Proceedings of the 10th IFIP Conference New York City, USA, August 31--September 4, 1981},
pp. \bfpage{42}--\blpage{66}
(\byear{2005}).
\bcomment{Springer}
\end{bchapter}
\endbibitem

\bibitem[\protect\citeauthoryear{Sokolowski and Zol{\'e}sio}{1992}]{sokolowski1992introduction}
\begin{bbook}
\bauthor{\bsnm{Sokolowski}, \binits{J.}},
\bauthor{\bsnm{Zol{\'e}sio}, \binits{J.-P.}}:
\bbtitle{Introduction to Shape Optimization},
pp. \bfpage{5}--\blpage{12}.
\bpublisher{Springer},
\blocation{Berlin, Heidelberg}
(\byear{1992})
\end{bbook}
\endbibitem

\bibitem[\protect\citeauthoryear{Haftka and G{\"u}rdal}{2012}]{haftka2012elements}
\begin{bbook}
\bauthor{\bsnm{Haftka}, \binits{R.T.}},
\bauthor{\bsnm{G{\"u}rdal}, \binits{Z.}}:
\bbtitle{Elements of Structural Optimization}
vol. \bseriesno{11}.
\bpublisher{Springer},
\blocation{Berlin, Heidelberg}
(\byear{2012})
\end{bbook}
\endbibitem

\bibitem[\protect\citeauthoryear{Bends{\o}e and Kikuchi}{1988}]{bendsoe1988generating}
\begin{barticle}
\bauthor{\bsnm{Bends{\o}e}, \binits{M.P.}},
\bauthor{\bsnm{Kikuchi}, \binits{N.}}:
\batitle{Generating optimal topologies in structural design using a homogenization method}.
\bjtitle{Computer methods in applied mechanics and engineering}
\bvolume{71}(\bissue{2}),
\bfpage{197}--\blpage{224}
(\byear{1988})
\end{barticle}
\endbibitem

\bibitem[\protect\citeauthoryear{Allaire et~al.}{2005}]{allaire2005structural}
\begin{barticle}
\bauthor{\bsnm{Allaire}, \binits{G.}},
\bauthor{\bsnm{Gournay}, \binits{F.d.}},
\bauthor{\bsnm{Jouve}, \binits{F.}},
\bauthor{\bsnm{Toader}, \binits{A.-M.}}:
\batitle{Structural optimization using topological and shape sensitivity via a level set method}.
\bjtitle{Control and cybernetics}
\bvolume{34}(\bissue{1}),
\bfpage{59}--\blpage{80}
(\byear{2005})
\end{barticle}
\endbibitem

\bibitem[\protect\citeauthoryear{Novotny et~al.}{2007}]{novotny2007topological}
\begin{barticle}
\bauthor{\bsnm{Novotny}, \binits{A.A.}},
\bauthor{\bsnm{Feij{\'o}o}, \binits{R.A.}},
\bauthor{\bsnm{Taroco}, \binits{E.}},
\bauthor{\bsnm{Padra}, \binits{C.}}:
\batitle{Topological sensitivity analysis for three-dimensional linear elasticity problem}.
\bjtitle{Computer Methods in Applied Mechanics and Engineering}
\bvolume{196}(\bissue{41-44}),
\bfpage{4354}--\blpage{4364}
(\byear{2007})
\end{barticle}
\endbibitem

\bibitem[\protect\citeauthoryear{Sigmund}{2001}]{Sigmund2001}
\begin{barticle}
\bauthor{\bsnm{Sigmund}, \binits{O.}}:
\batitle{A 99 line topology optimization code written in matlab}.
\bjtitle{Structural and Multidisciplinary Optimization}
\bvolume{21}(\bissue{2}),
\bfpage{120}--\blpage{127}
(\byear{2001})
\doiurl{10.1007/s001580050176}
\end{barticle}
\endbibitem

\bibitem[\protect\citeauthoryear{Andreassen et~al.}{2011}]{Andreassen2011}
\begin{barticle}
\bauthor{\bsnm{Andreassen}, \binits{E.}},
\bauthor{\bsnm{Clausen}, \binits{A.}},
\bauthor{\bsnm{Schevenels}, \binits{M.}},
\bauthor{\bsnm{Lazarov}, \binits{B.S.}},
\bauthor{\bsnm{Sigmund}, \binits{O.}}:
\batitle{Efficient topology optimization in matlab using 88 lines of code}.
\bjtitle{Structural and Multidisciplinary Optimization}
\bvolume{43}(\bissue{1}),
\bfpage{1}--\blpage{16}
(\byear{2011})
\doiurl{10.1007/s00158-010-0594-7}
\end{barticle}
\endbibitem

\bibitem[\protect\citeauthoryear{Talischi et~al.}{2012}]{talischi2012polytop}
\begin{barticle}
\bauthor{\bsnm{Talischi}, \binits{C.}},
\bauthor{\bsnm{Paulino}, \binits{G.H.}},
\bauthor{\bsnm{Pereira}, \binits{A.}},
\bauthor{\bsnm{Menezes}, \binits{I.F.}}:
\batitle{Polytop: a matlab implementation of a general topology optimization framework using unstructured polygonal finite element meshes}.
\bjtitle{Structural and Multidisciplinary Optimization}
\bvolume{45}(\bissue{3}),
\bfpage{329}--\blpage{357}
(\byear{2012})
\end{barticle}
\endbibitem

\bibitem[\protect\citeauthoryear{Liu and Tovar}{2014}]{LiuTovar2014}
\begin{barticle}
\bauthor{\bsnm{Liu}, \binits{K.}},
\bauthor{\bsnm{Tovar}, \binits{A.}}:
\batitle{An efficient 3d topology optimization code written in matlab}.
\bjtitle{Structural and Multidisciplinary Optimization}
\bvolume{50}(\bissue{6}),
\bfpage{1175}--\blpage{1196}
(\byear{2014})
\doiurl{10.1007/s00158-014-1107-x}
\end{barticle}
\endbibitem

\bibitem[\protect\citeauthoryear{Wang et~al.}{2025}]{wang2025efficient}
\begin{barticle}
\bauthor{\bsnm{Wang}, \binits{J.}},
\bauthor{\bsnm{Aage}, \binits{N.}},
\bauthor{\bsnm{Wu}, \binits{J.}},
\bauthor{\bsnm{Sigmund}, \binits{O.}},
\bauthor{\bsnm{Westermann}, \binits{R.}}:
\batitle{Efficient large-scale 3d topology optimization with matrix-free matlab code}.
\bjtitle{Structural and Multidisciplinary Optimization}
\bvolume{68}(\bissue{9}),
\bfpage{1}--\blpage{16}
(\byear{2025})
\end{barticle}
\endbibitem

\bibitem[\protect\citeauthoryear{Aage}{2014}]{AagePETSc2014}
\begin{botherref}
\oauthor{\bsnm{Aage}, \binits{N.}}:
Topology optimization using PETSc.
\url{https://www.topopt.mek.dtu.dk/-/media/subsites/topopt/apps/dokumenter-og-filer-til-apps/petsc-1-.pdf}
(2014)
\end{botherref}
\endbibitem

\bibitem[\protect\citeauthoryear{Challis}{2010}]{challis2010discrete}
\begin{barticle}
\bauthor{\bsnm{Challis}, \binits{V.J.}}:
\batitle{A discrete level-set topology optimization code written in matlab}.
\bjtitle{Structural and multidisciplinary optimization}
\bvolume{41},
\bfpage{453}--\blpage{464}
(\byear{2010})
\end{barticle}
\endbibitem

\bibitem[\protect\citeauthoryear{Zhang et~al.}{2016}]{zhang2016new}
\begin{barticle}
\bauthor{\bsnm{Zhang}, \binits{W.}},
\bauthor{\bsnm{Yuan}, \binits{J.}},
\bauthor{\bsnm{Zhang}, \binits{J.}},
\bauthor{\bsnm{Guo}, \binits{X.}}:
\batitle{A new topology optimization approach based on moving morphable components (mmc) and the ersatz material model}.
\bjtitle{Structural and Multidisciplinary Optimization}
\bvolume{53}(\bissue{6}),
\bfpage{1243}--\blpage{1260}
(\byear{2016})
\end{barticle}
\endbibitem

\bibitem[\protect\citeauthoryear{Wei et~al.}{2018}]{Wei2018}
\begin{barticle}
\bauthor{\bsnm{Wei}, \binits{P.}},
\bauthor{\bsnm{Wang}, \binits{M.}},
\bauthor{\bsnm{Chen}, \binits{S.}},
\bauthor{\bsnm{Wang}, \binits{M.}},
\bauthor{\bsnm{Wang}, \binits{X.}}:
\batitle{An 88-line matlab code for the parameterized level set method based topology optimization}.
\bjtitle{Structural and Multidisciplinary Optimization}
\bvolume{58}(\bissue{4}),
\bfpage{1541}--\blpage{1558}
(\byear{2018})
\doiurl{10.1007/s00158-018-1904-8}
\end{barticle}
\endbibitem

\bibitem[\protect\citeauthoryear{Amir}{2021}]{amir2021efficient}
\begin{barticle}
\bauthor{\bsnm{Amir}, \binits{O.}}:
\batitle{Efficient stress-constrained topology optimization using inexact design sensitivities}.
\bjtitle{International Journal for Numerical Methods in Engineering}
\bvolume{122}(\bissue{13}),
\bfpage{3241}--\blpage{3272}
(\byear{2021})
\end{barticle}
\endbibitem

\bibitem[\protect\citeauthoryear{Deng et~al.}{2022}]{deng2022efficient}
\begin{barticle}
\bauthor{\bsnm{Deng}, \binits{H.}},
\bauthor{\bsnm{Vulimiri}, \binits{P.S.}},
\bauthor{\bsnm{To}, \binits{A.C.}}:
\batitle{An efficient 146-line 3d sensitivity analysis code of stress-based topology optimization written in matlab}.
\bjtitle{Optimization and Engineering}
\bvolume{23}(\bissue{3}),
\bfpage{1733}--\blpage{1757}
(\byear{2022})
\end{barticle}
\endbibitem

\bibitem[\protect\citeauthoryear{Giraldo-Londo{\~n}o and Paulino}{2021}]{giraldo2021polystress}
\begin{barticle}
\bauthor{\bsnm{Giraldo-Londo{\~n}o}, \binits{O.}},
\bauthor{\bsnm{Paulino}, \binits{G.H.}}:
\batitle{Polystress: a matlab implementation for local stress-constrained topology optimization using the augmented lagrangian method}.
\bjtitle{Structural and Multidisciplinary Optimization}
\bvolume{63}(\bissue{4}),
\bfpage{2065}--\blpage{2097}
(\byear{2021})
\end{barticle}
\endbibitem

\bibitem[\protect\citeauthoryear{Alexandersen}{2022}]{Alexandersen2022}
\begin{botherref}
\oauthor{\bsnm{Alexandersen}, \binits{J.}}:
A detailed introduction to density-based topology optimisation of fluid flow problems with implementation in {MATLAB}.
\url{https://arxiv.org/pdf/2207.13695}
(2022)
\end{botherref}
\endbibitem

\bibitem[\protect\citeauthoryear{{dolfin-adjoint developers}}{2023}]{DolfinAdjointStokesTO}
\begin{botherref}
\oauthor{\bsnm{{dolfin-adjoint developers}}}:
Topology optimisation of fluids in Stokes flow (tutorial).
\url{https://www.dolfin-adjoint.org/en/release/documentation/stokes-topology/stokes-topology.html}.
Accessed 2025-10-12
(2023)
\end{botherref}
\endbibitem

\bibitem[\protect\citeauthoryear{Tavakoli and Mohseni}{2014}]{tavakoli2014alternating}
\begin{barticle}
\bauthor{\bsnm{Tavakoli}, \binits{R.}},
\bauthor{\bsnm{Mohseni}, \binits{S.M.}}:
\batitle{Alternating active-phase algorithm for multimaterial topology optimization problems: a 115-line matlab implementation}.
\bjtitle{Structural and Multidisciplinary Optimization}
\bvolume{49}(\bissue{4}),
\bfpage{621}--\blpage{642}
(\byear{2014})
\end{barticle}
\endbibitem

\bibitem[\protect\citeauthoryear{Zheng et~al.}{2024}]{zheng2024efficient}
\begin{barticle}
\bauthor{\bsnm{Zheng}, \binits{R.}},
\bauthor{\bsnm{Yi}, \binits{B.}},
\bauthor{\bsnm{Peng}, \binits{X.}},
\bauthor{\bsnm{Yoon}, \binits{G.-H.}}:
\batitle{An efficient code for the multi-material topology optimization of 2d/3d continuum structures written in matlab}.
\bjtitle{Applied Sciences}
\bvolume{14}(\bissue{2}),
\bfpage{657}
(\byear{2024})
\end{barticle}
\endbibitem

\bibitem[\protect\citeauthoryear{Chandrasekhar and Suresh}{2021}]{TOuNN2021}
\begin{barticle}
\bauthor{\bsnm{Chandrasekhar}, \binits{A.}},
\bauthor{\bsnm{Suresh}, \binits{K.}}:
\batitle{Tounn: Topology optimization using neural networks}.
\bjtitle{Structural and Multidisciplinary Optimization}
\bvolume{63},
\bfpage{1135}--\blpage{1149}
(\byear{2021})
\doiurl{10.1007/s00158-020-02748-4}
\end{barticle}
\endbibitem

\bibitem[\protect\citeauthoryear{Lingaard and contributors}{2020--2025}]{DeepTopoptGitHub}
\begin{botherref}
\oauthor{\bsnm{Lingaard}, \binits{E.}},
\oauthor{\bsnm{contributors}}:
deep-topopt: deep learning-based topology optimization (PyTorch).
\url{https://github.com/elingaard/deep-topopt}
(2020--2025)
\end{botherref}
\endbibitem

\bibitem[\protect\citeauthoryear{Xie and Steven}{1997}]{xie1997basic}
\begin{botherref}
\oauthor{\bsnm{Xie}, \binits{Y.M.}},
\oauthor{\bsnm{Steven}, \binits{G.P.}}:
Basic evolutionary structural optimization.
Springer
(1997)
\end{botherref}
\endbibitem

\bibitem[\protect\citeauthoryear{Bends{\o}e and Sigmund}{2004}]{martin2004topology}
\begin{botherref}
\oauthor{\bsnm{Bends{\o}e}, \binits{M.P.}},
\oauthor{\bsnm{Sigmund}, \binits{O.}}:
Topology optimization: theory, methods, and applications.
Springer
(2004)
\end{botherref}
\endbibitem

\bibitem[\protect\citeauthoryear{Ferrari and Sigmund}{2020}]{ferrari2020new}
\begin{barticle}
\bauthor{\bsnm{Ferrari}, \binits{F.}},
\bauthor{\bsnm{Sigmund}, \binits{O.}}:
\batitle{A new generation 99 line matlab code for compliance topology optimization and its extension to 3d}.
\bjtitle{Structural and Multidisciplinary Optimization}
\bvolume{62}(\bissue{4}),
\bfpage{2211}--\blpage{2228}
(\byear{2020})
\end{barticle}
\endbibitem

\bibitem[\protect\citeauthoryear{Smith and Norato}{2020}]{SmithNorato2020}
\begin{barticle}
\bauthor{\bsnm{Smith}, \binits{H.}},
\bauthor{\bsnm{Norato}, \binits{J.A.}}:
\batitle{A {MATLAB} code for topology optimization using the geometry projection method}.
\bjtitle{Structural and Multidisciplinary Optimization}
\bvolume{62},
\bfpage{1579}--\blpage{1594}
(\byear{2020})
\doiurl{10.1007/s00158-020-02552-0}
\end{barticle}
\endbibitem

\bibitem[\protect\citeauthoryear{Coniglio et~al.}{2019}]{coniglio2019generalized}
\begin{botherref}
\oauthor{\bsnm{Coniglio}, \binits{S.}},
\oauthor{\bsnm{Morlier}, \binits{J.}},
\oauthor{\bsnm{Gogu}, \binits{C.}},
\oauthor{\bsnm{Amargier}, \binits{R.}}:
Generalized geometry projection: a unified approach for geometric feature based topology optimization.
Archives of Computational Methods in Engineering,
1--38
(2019)
\end{botherref}
\endbibitem

\bibitem[\protect\citeauthoryear{Gao et~al.}{2021}]{gao2021igatop}
\begin{barticle}
\bauthor{\bsnm{Gao}, \binits{J.}},
\bauthor{\bsnm{Wang}, \binits{L.}},
\bauthor{\bsnm{Luo}, \binits{Z.}},
\bauthor{\bsnm{Gao}, \binits{L.}}:
\batitle{Igatop: an implementation of topology optimization for structures using iga in matlab}.
\bjtitle{Structural and Multidisciplinary Optimization}
\bvolume{64}(\bissue{3}),
\bfpage{1669}--\blpage{1700}
(\byear{2021})
\end{barticle}
\endbibitem

\bibitem[\protect\citeauthoryear{Chen et~al.}{2019}]{chen2019213}
\begin{barticle}
\bauthor{\bsnm{Chen}, \binits{Q.}},
\bauthor{\bsnm{Zhang}, \binits{X.}},
\bauthor{\bsnm{Zhu}, \binits{B.}}:
\batitle{A 213-line topology optimization code for geometrically nonlinear structures}.
\bjtitle{Structural and Multidisciplinary Optimization}
\bvolume{59}(\bissue{5}),
\bfpage{1863}--\blpage{1879}
(\byear{2019})
\end{barticle}
\endbibitem

\bibitem[\protect\citeauthoryear{Giraldo-Londono and Paulino}{2021}]{giraldo2021polydyna}
\begin{barticle}
\bauthor{\bsnm{Giraldo-Londono}, \binits{O.}},
\bauthor{\bsnm{Paulino}, \binits{G.H.}}:
\batitle{Polydyna: a matlab implementation for topology optimization of structures subjected to dynamic loads}.
\bjtitle{Structural and Multidisciplinary Optimization}
\bvolume{64}(\bissue{2}),
\bfpage{957}--\blpage{990}
(\byear{2021})
\end{barticle}
\endbibitem

\bibitem[\protect\citeauthoryear{Wang et~al.}{2021}]{wang2021comprehensive}
\begin{barticle}
\bauthor{\bsnm{Wang}, \binits{C.}},
\bauthor{\bsnm{Zhao}, \binits{Z.}},
\bauthor{\bsnm{Zhou}, \binits{M.}},
\bauthor{\bsnm{Sigmund}, \binits{O.}},
\bauthor{\bsnm{Zhang}, \binits{X.S.}}:
\batitle{A comprehensive review of educational articles on structural and multidisciplinary optimization}.
\bjtitle{Structural and Multidisciplinary Optimization}
\bvolume{64}(\bissue{5}),
\bfpage{2827}--\blpage{2880}
(\byear{2021})
\end{barticle}
\endbibitem

\bibitem[\protect\citeauthoryear{Ferrari et~al.}{2021}]{ferrari2021topology}
\begin{barticle}
\bauthor{\bsnm{Ferrari}, \binits{F.}},
\bauthor{\bsnm{Sigmund}, \binits{O.}},
\bauthor{\bsnm{Guest}, \binits{J.K.}}:
\batitle{Topology optimization with linearized buckling criteria in 250 lines of matlab}.
\bjtitle{Structural and Multidisciplinary Optimization}
\bvolume{63}(\bissue{6}),
\bfpage{3045}--\blpage{3066}
(\byear{2021})
\end{barticle}
\endbibitem

\bibitem[\protect\citeauthoryear{Huang}{2023}]{huang2023matlab}
\begin{botherref}
\oauthor{\bsnm{Huang}, \binits{X.}}:
A matlab code of topology optimization by imposing the implicit floating projection constraint
(2023)
\end{botherref}
\endbibitem

\bibitem[\protect\citeauthoryear{Gao et~al.}{2019}]{gao2019concurrent}
\begin{barticle}
\bauthor{\bsnm{Gao}, \binits{J.}},
\bauthor{\bsnm{Luo}, \binits{Z.}},
\bauthor{\bsnm{Xia}, \binits{L.}},
\bauthor{\bsnm{Gao}, \binits{L.}}:
\batitle{Concurrent topology optimization of multiscale composite structures in matlab}.
\bjtitle{Structural and Multidisciplinary Optimization}
\bvolume{60}(\bissue{6}),
\bfpage{2621}--\blpage{2651}
(\byear{2019})
\end{barticle}
\endbibitem

\bibitem[\protect\citeauthoryear{Christiansen and Sigmund}{2021}]{christiansen2021compact}
\begin{barticle}
\bauthor{\bsnm{Christiansen}, \binits{R.E.}},
\bauthor{\bsnm{Sigmund}, \binits{O.}}:
\batitle{Compact 200 line matlab code for inverse design in photonics by topology optimization: tutorial}.
\bjtitle{Journal of the Optical Society of America B}
\bvolume{38}(\bissue{2}),
\bfpage{510}--\blpage{520}
(\byear{2021})
\end{barticle}
\endbibitem

\bibitem[\protect\citeauthoryear{{FEniCS Project}}{2024}]{FEniCSSIMPDemo}
\begin{botherref}
\oauthor{\bsnm{{FEniCS Project}}}:
Topology optimization using the SIMP method (demo).
\url{https://comet-fenics.readthedocs.io/en/latest/demo/topology_optimization/simp_topology_optimization.html}.
Accessed 2025-10-12
(2024)
\end{botherref}
\endbibitem

\bibitem[\protect\citeauthoryear{Ferguson and contributors}{2019--2025}]{FEniCSTopOptRepo}
\begin{botherref}
\oauthor{\bsnm{Ferguson}, \binits{Z.}},
\oauthor{\bsnm{contributors}}:
fenics-topopt: Topology optimization with FEniCS.
\url{https://github.com/zfergus/fenics-topopt}
(2019--2025)
\end{botherref}
\endbibitem

\bibitem[\protect\citeauthoryear{{TopOpt Group, DTU}}{2018}]{DTUPythonCodes}
\begin{botherref}
\oauthor{\bsnm{{TopOpt Group, DTU}}}:
Topology optimization codes written in Python (educational).
\url{https://www.topopt.mek.dtu.dk/apps-and-software/topology-optimization-codes-written-in-python}.
Accessed 2025-10-12
(2018)
\end{botherref}
\endbibitem

\bibitem[\protect\citeauthoryear{Huang and contributors}{2021--2025}]{TopOptjlDocs}
\begin{botherref}
\oauthor{\bsnm{Huang}, \binits{Y.}},
\oauthor{\bsnm{contributors}}:
TopOpt.jl Documentation.
\url{https://juliatopopt.github.io/TopOpt.jl/}
(2021--2025)
\end{botherref}
\endbibitem

\bibitem[\protect\citeauthoryear{contributors}{2021--2025}]{TopOptjlGitHub}
\begin{botherref}
\oauthor{\bsnm{contributors}, \binits{J.}}:
TopOpt.jl GitHub repository.
\url{https://github.com/JuliaTopOpt/TopOpt.jl}
(2021--2025)
\end{botherref}
\endbibitem

\bibitem[\protect\citeauthoryear{Packages}{2022}]{TopOptJLPkg}
\begin{botherref}
\oauthor{\bsnm{Packages}, \binits{J.}}:
TopOpt\_jl: Julia package (port of top88).
\url{https://juliapackages.com/p/topopt_jl}.
Accessed 2025-10-12
(2022)
\end{botherref}
\endbibitem

\bibitem[\protect\citeauthoryear{Van~Dijk et~al.}{2013}]{van2013level}
\begin{barticle}
\bauthor{\bsnm{Van~Dijk}, \binits{N.P.}},
\bauthor{\bsnm{Maute}, \binits{K.}},
\bauthor{\bsnm{Langelaar}, \binits{M.}},
\bauthor{\bsnm{Van~Keulen}, \binits{F.}}:
\batitle{Level-set methods for structural topology optimization: a review}.
\bjtitle{Structural and Multidisciplinary Optimization}
\bvolume{48},
\bfpage{437}--\blpage{472}
(\byear{2013})
\end{barticle}
\endbibitem

\bibitem[\protect\citeauthoryear{Sigmund and Maute}{2013}]{sigmund2013topology}
\begin{barticle}
\bauthor{\bsnm{Sigmund}, \binits{O.}},
\bauthor{\bsnm{Maute}, \binits{K.}}:
\batitle{Topology optimization approaches: A comparative review}.
\bjtitle{Structural and multidisciplinary optimization}
\bvolume{48}(\bissue{6}),
\bfpage{1031}--\blpage{1055}
(\byear{2013})
\end{barticle}
\endbibitem

\bibitem[\protect\citeauthoryear{Eschenauer and Olhoff}{2001}]{eschenauer2001topology}
\begin{barticle}
\bauthor{\bsnm{Eschenauer}, \binits{H.A.}},
\bauthor{\bsnm{Olhoff}, \binits{N.}}:
\batitle{Topology optimization of continuum structures: a review}.
\bjtitle{Appl. Mech. Rev.}
\bvolume{54}(\bissue{4}),
\bfpage{331}--\blpage{390}
(\byear{2001})
\end{barticle}
\endbibitem

\bibitem[\protect\citeauthoryear{Rozvany}{2009}]{rozvany2009critical}
\begin{barticle}
\bauthor{\bsnm{Rozvany}, \binits{G.I.}}:
\batitle{A critical review of established methods of structural topology optimization}.
\bjtitle{Structural and multidisciplinary optimization}
\bvolume{37}(\bissue{3}),
\bfpage{217}--\blpage{237}
(\byear{2009})
\end{barticle}
\endbibitem

\bibitem[\protect\citeauthoryear{Deaton and Grandhi}{2014}]{deaton2014survey}
\begin{barticle}
\bauthor{\bsnm{Deaton}, \binits{J.D.}},
\bauthor{\bsnm{Grandhi}, \binits{R.V.}}:
\batitle{A survey of structural and multidisciplinary continuum topology optimization: post 2000}.
\bjtitle{Structural and multidisciplinary optimization}
\bvolume{49}(\bissue{1}),
\bfpage{1}--\blpage{38}
(\byear{2014})
\end{barticle}
\endbibitem

\bibitem[\protect\citeauthoryear{Wu et~al.}{2021}]{wu2021topology}
\begin{barticle}
\bauthor{\bsnm{Wu}, \binits{J.}},
\bauthor{\bsnm{Sigmund}, \binits{O.}},
\bauthor{\bsnm{Groen}, \binits{J.P.}}:
\batitle{Topology optimization of multi-scale structures: a review}.
\bjtitle{Structural and Multidisciplinary Optimization}
\bvolume{63}(\bissue{3}),
\bfpage{1455}--\blpage{1480}
(\byear{2021})
\end{barticle}
\endbibitem

\bibitem[\protect\citeauthoryear{Bends{\o}e}{1989}]{bendsoe1989optimal}
\begin{barticle}
\bauthor{\bsnm{Bends{\o}e}, \binits{M.P.}}:
\batitle{Optimal shape design as a material distribution problem}.
\bjtitle{Structural optimization}
\bvolume{1},
\bfpage{193}--\blpage{202}
(\byear{1989})
\end{barticle}
\endbibitem

\bibitem[\protect\citeauthoryear{Sigmund}{2001}]{sigmund200199}
\begin{barticle}
\bauthor{\bsnm{Sigmund}, \binits{O.}}:
\batitle{A 99 line topology optimization code written in matlab}.
\bjtitle{Structural and multidisciplinary optimization}
\bvolume{21},
\bfpage{120}--\blpage{127}
(\byear{2001})
\end{barticle}
\endbibitem

\bibitem[\protect\citeauthoryear{Bends{\o}e and Sigmund}{1999}]{bendsoe1999material}
\begin{barticle}
\bauthor{\bsnm{Bends{\o}e}, \binits{M.P.}},
\bauthor{\bsnm{Sigmund}, \binits{O.}}:
\batitle{Material interpolation schemes in topology optimization}.
\bjtitle{Archive of applied mechanics}
\bvolume{69},
\bfpage{635}--\blpage{654}
(\byear{1999})
\end{barticle}
\endbibitem

\bibitem[\protect\citeauthoryear{Xie and Steven}{1993}]{xie_simple_1993}
\begin{barticle}
\bauthor{\bsnm{Xie}, \binits{Y.M.}},
\bauthor{\bsnm{Steven}, \binits{G.P.}}:
\batitle{A simple evolutionary procedure for structural optimization}.
\bjtitle{Computers \& Structures}
\bvolume{49}(\bissue{5}),
\bfpage{885}--\blpage{896}
(\byear{1993})
\doiurl{10.1016/0045-7949(93)90035-C} .
Accessed 2014-12-02
\end{barticle}
\endbibitem

\bibitem[\protect\citeauthoryear{Allaire et~al.}{2004}]{allaire2004topology}
\begin{barticle}
\bauthor{\bsnm{Allaire}, \binits{G.}},
\bauthor{\bsnm{Jouve}, \binits{F.}},
\bauthor{\bsnm{Maillot}, \binits{H.}}:
\batitle{Topology optimization for minimum stress design with the homogenization method}.
\bjtitle{Structural and Multidisciplinary Optimization}
\bvolume{28}(\bissue{2}),
\bfpage{87}--\blpage{98}
(\byear{2004})
\end{barticle}
\endbibitem

\bibitem[\protect\citeauthoryear{Feijoo et~al.}{2005}]{feijoo2005topological}
\begin{botherref}
\oauthor{\bsnm{Feijoo}, \binits{R.}},
\oauthor{\bsnm{Novotny}, \binits{A.}},
\oauthor{\bsnm{Taroco}, \binits{E.}},
\oauthor{\bsnm{Padra}, \binits{C.}}:
The topological-shape sensitivity method in two-dimensional linear elasticity topology design.
Applications of computational mechanics in structures and fluids
(2005)
\end{botherref}
\endbibitem

\bibitem[\protect\citeauthoryear{Burger et~al.}{2004}]{burger2004incorporating}
\begin{barticle}
\bauthor{\bsnm{Burger}, \binits{M.}},
\bauthor{\bsnm{Hackl}, \binits{B.}},
\bauthor{\bsnm{Ring}, \binits{W.}}:
\batitle{Incorporating topological derivatives into level set methods}.
\bjtitle{Journal of computational physics}
\bvolume{194}(\bissue{1}),
\bfpage{344}--\blpage{362}
(\byear{2004})
\end{barticle}
\endbibitem

\bibitem[\protect\citeauthoryear{Sigmund and Torquato}{1997}]{sigmund1997design}
\begin{barticle}
\bauthor{\bsnm{Sigmund}, \binits{O.}},
\bauthor{\bsnm{Torquato}, \binits{S.}}:
\batitle{Design of materials with extreme thermal expansion using a three-phase topology optimization method}.
\bjtitle{Journal of the Mechanics and Physics of Solids}
\bvolume{45}(\bissue{6}),
\bfpage{1037}--\blpage{1067}
(\byear{1997})
\end{barticle}
\endbibitem

\bibitem[\protect\citeauthoryear{Deaton and Grandhi}{2016}]{deaton2016stress}
\begin{barticle}
\bauthor{\bsnm{Deaton}, \binits{J.D.}},
\bauthor{\bsnm{Grandhi}, \binits{R.V.}}:
\batitle{Stress-based design of thermal structures via topology optimization}.
\bjtitle{Structural and Multidisciplinary Optimization}
\bvolume{53}(\bissue{2}),
\bfpage{253}--\blpage{270}
(\byear{2016})
\end{barticle}
\endbibitem

\bibitem[\protect\citeauthoryear{Dbouk}{2017}]{dbouk2017review}
\begin{barticle}
\bauthor{\bsnm{Dbouk}, \binits{T.}}:
\batitle{A review about the engineering design of optimal heat transfer systems using topology optimization}.
\bjtitle{Applied Thermal Engineering}
\bvolume{112},
\bfpage{841}--\blpage{854}
(\byear{2017})
\end{barticle}
\endbibitem

\bibitem[\protect\citeauthoryear{Borrvall and Petersson}{2003}]{borrvall2003topology}
\begin{barticle}
\bauthor{\bsnm{Borrvall}, \binits{T.}},
\bauthor{\bsnm{Petersson}, \binits{J.}}:
\batitle{Topology optimization of fluids in stokes flow}.
\bjtitle{International journal for numerical methods in fluids}
\bvolume{41}(\bissue{1}),
\bfpage{77}--\blpage{107}
(\byear{2003})
\end{barticle}
\endbibitem

\bibitem[\protect\citeauthoryear{Lin et~al.}{2015}]{lin2015topology}
\begin{barticle}
\bauthor{\bsnm{Lin}, \binits{S.}},
\bauthor{\bsnm{Zhao}, \binits{L.}},
\bauthor{\bsnm{Guest}, \binits{J.K.}},
\bauthor{\bsnm{Weihs}, \binits{T.P.}},
\bauthor{\bsnm{Liu}, \binits{Z.}}:
\batitle{Topology optimization of fixed-geometry fluid diodes}.
\bjtitle{Journal of Mechanical Design}
\bvolume{137}(\bissue{8}),
\bfpage{081402}
(\byear{2015})
\end{barticle}
\endbibitem

\bibitem[\protect\citeauthoryear{Alexandersen and Andreasen}{2020}]{AlexandersenReview2020}
\begin{barticle}
\bauthor{\bsnm{Alexandersen}, \binits{J.}},
\bauthor{\bsnm{Andreasen}, \binits{C.S.}}:
\batitle{A review of topology optimisation for fluid-based problems}.
\bjtitle{Fluids}
\bvolume{5}(\bissue{1}),
\bfpage{29}
(\byear{2020})
\doiurl{10.3390/fluids5010029}
\end{barticle}
\endbibitem

\bibitem[\protect\citeauthoryear{Dilgen et~al.}{2019}]{dilgen2019topology}
\begin{barticle}
\bauthor{\bsnm{Dilgen}, \binits{C.B.}},
\bauthor{\bsnm{Dilgen}, \binits{S.B.}},
\bauthor{\bsnm{Aage}, \binits{N.}},
\bauthor{\bsnm{Jensen}, \binits{J.S.}}:
\batitle{Topology optimization of acoustic mechanical interaction problems: a comparative review}.
\bjtitle{Structural and Multidisciplinary Optimization}
\bvolume{60}(\bissue{2}),
\bfpage{779}--\blpage{801}
(\byear{2019})
\end{barticle}
\endbibitem

\bibitem[\protect\citeauthoryear{D{\"u}hring et~al.}{2008}]{duhring2008acoustic}
\begin{barticle}
\bauthor{\bsnm{D{\"u}hring}, \binits{M.B.}},
\bauthor{\bsnm{Jensen}, \binits{J.S.}},
\bauthor{\bsnm{Sigmund}, \binits{O.}}:
\batitle{Acoustic design by topology optimization}.
\bjtitle{Journal of sound and vibration}
\bvolume{317}(\bissue{3-5}),
\bfpage{557}--\blpage{575}
(\byear{2008})
\end{barticle}
\endbibitem

\bibitem[\protect\citeauthoryear{Yi and Youn}{2016}]{yi2016comprehensive}
\begin{barticle}
\bauthor{\bsnm{Yi}, \binits{G.}},
\bauthor{\bsnm{Youn}, \binits{B.D.}}:
\batitle{A comprehensive survey on topology optimization of phononic crystals}.
\bjtitle{Structural and Multidisciplinary Optimization}
\bvolume{54}(\bissue{5}),
\bfpage{1315}--\blpage{1344}
(\byear{2016})
\end{barticle}
\endbibitem

\bibitem[\protect\citeauthoryear{Sigmund and S{\o}ndergaard~Jensen}{2003}]{sigmund2003systematic}
\begin{barticle}
\bauthor{\bsnm{Sigmund}, \binits{O.}},
\bauthor{\bsnm{S{\o}ndergaard~Jensen}, \binits{J.}}:
\batitle{Systematic design of phononic band--gap materials and structures by topology optimization}.
\bjtitle{Philosophical Transactions of the Royal Society of London. Series A: Mathematical, Physical and Engineering Sciences}
\bvolume{361}(\bissue{1806}),
\bfpage{1001}--\blpage{1019}
(\byear{2003})
\end{barticle}
\endbibitem

\bibitem[\protect\citeauthoryear{Lucchini et~al.}{2022}]{lucchini2022topology}
\begin{barticle}
\bauthor{\bsnm{Lucchini}, \binits{F.}},
\bauthor{\bsnm{Torchio}, \binits{R.}},
\bauthor{\bsnm{Cirimele}, \binits{V.}},
\bauthor{\bsnm{Alotto}, \binits{P.}},
\bauthor{\bsnm{Bettini}, \binits{P.}}:
\batitle{Topology optimization for electromagnetics: A survey}.
\bjtitle{IEEE Access}
\bvolume{10},
\bfpage{98593}--\blpage{98611}
(\byear{2022})
\end{barticle}
\endbibitem

\bibitem[\protect\citeauthoryear{Zhou et~al.}{2015}]{zhou2015minimum}
\begin{barticle}
\bauthor{\bsnm{Zhou}, \binits{M.}},
\bauthor{\bsnm{Lazarov}, \binits{B.S.}},
\bauthor{\bsnm{Wang}, \binits{F.}},
\bauthor{\bsnm{Sigmund}, \binits{O.}}:
\batitle{Minimum length scale in topology optimization by geometric constraints}.
\bjtitle{Computer Methods in Applied Mechanics and Engineering}
\bvolume{293},
\bfpage{266}--\blpage{282}
(\byear{2015})
\end{barticle}
\endbibitem

\bibitem[\protect\citeauthoryear{Sigmund}{2007}]{sigmund2007morphology}
\begin{barticle}
\bauthor{\bsnm{Sigmund}, \binits{O.}}:
\batitle{Morphology-based black and white filters for topology optimization}.
\bjtitle{Structural and Multidisciplinary Optimization}
\bvolume{33}(\bissue{4}),
\bfpage{401}--\blpage{424}
(\byear{2007})
\end{barticle}
\endbibitem

\bibitem[\protect\citeauthoryear{Liu et~al.}{2018}]{liu2018current}
\begin{barticle}
\bauthor{\bsnm{Liu}, \binits{J.}},
\bauthor{\bsnm{Gaynor}, \binits{A.T.}},
\bauthor{\bsnm{Chen}, \binits{S.}},
\bauthor{\bsnm{Kang}, \binits{Z.}},
\bauthor{\bsnm{Suresh}, \binits{K.}},
\bauthor{\bsnm{Takezawa}, \binits{A.}},
\bauthor{\bsnm{Li}, \binits{L.}},
\bauthor{\bsnm{Kato}, \binits{J.}},
\bauthor{\bsnm{Tang}, \binits{J.}},
\bauthor{\bsnm{Wang}, \binits{C.C.}}, \betal:
\batitle{Current and future trends in topology optimization for additive manufacturing}.
\bjtitle{Structural and multidisciplinary optimization}
\bvolume{57}(\bissue{6}),
\bfpage{2457}--\blpage{2483}
(\byear{2018})
\end{barticle}
\endbibitem

\bibitem[\protect\citeauthoryear{Mirzendehdel and Suresh}{2016}]{mirzendehdel2016support}
\begin{barticle}
\bauthor{\bsnm{Mirzendehdel}, \binits{A.M.}},
\bauthor{\bsnm{Suresh}, \binits{K.}}:
\batitle{Support structure constrained topology optimization for additive manufacturing}.
\bjtitle{Computer-Aided Design}
\bvolume{81},
\bfpage{1}--\blpage{13}
(\byear{2016})
\end{barticle}
\endbibitem

\bibitem[\protect\citeauthoryear{Mirzendehdel et~al.}{2018}]{mirzendehdel2018strength}
\begin{barticle}
\bauthor{\bsnm{Mirzendehdel}, \binits{A.M.}},
\bauthor{\bsnm{Rankouhi}, \binits{B.}},
\bauthor{\bsnm{Suresh}, \binits{K.}}:
\batitle{Strength-based topology optimization for anisotropic parts}.
\bjtitle{Additive Manufacturing}
\bvolume{19},
\bfpage{104}--\blpage{113}
(\byear{2018})
\end{barticle}
\endbibitem

\bibitem[\protect\citeauthoryear{Mirzendehdel et~al.}{2020}]{mirzendehdel2020topology}
\begin{barticle}
\bauthor{\bsnm{Mirzendehdel}, \binits{A.M.}},
\bauthor{\bsnm{Behandish}, \binits{M.}},
\bauthor{\bsnm{Nelaturi}, \binits{S.}}:
\batitle{Topology optimization with accessibility constraint for multi-axis machining}.
\bjtitle{Computer-Aided Design}
\bvolume{122},
\bfpage{102825}
(\byear{2020})
\end{barticle}
\endbibitem

\bibitem[\protect\citeauthoryear{Mirzendehdel et~al.}{2022}]{mirzendehdel2022topology}
\begin{barticle}
\bauthor{\bsnm{Mirzendehdel}, \binits{A.M.}},
\bauthor{\bsnm{Behandish}, \binits{M.}},
\bauthor{\bsnm{Nelaturi}, \binits{S.}}:
\batitle{Topology optimization for manufacturing with accessible support structures}.
\bjtitle{Computer-Aided Design}
\bvolume{142},
\bfpage{103117}
(\byear{2022})
\end{barticle}
\endbibitem

\bibitem[\protect\citeauthoryear{Suresh}{2013}]{suresh2013efficient}
\begin{barticle}
\bauthor{\bsnm{Suresh}, \binits{K.}}:
\batitle{Efficient generation of large-scale pareto-optimal topologies}.
\bjtitle{Structural and Multidisciplinary Optimization}
\bvolume{47}(\bissue{1}),
\bfpage{49}--\blpage{61}
(\byear{2013})
\end{barticle}
\endbibitem

\bibitem[\protect\citeauthoryear{Challis et~al.}{2014}]{challis2014high}
\begin{barticle}
\bauthor{\bsnm{Challis}, \binits{V.J.}},
\bauthor{\bsnm{Roberts}, \binits{A.P.}},
\bauthor{\bsnm{Grotowski}, \binits{J.F.}}:
\batitle{High resolution topology optimization using graphics processing units (gpus)}.
\bjtitle{Structural and Multidisciplinary Optimization}
\bvolume{49}(\bissue{2}),
\bfpage{315}--\blpage{325}
(\byear{2014})
\end{barticle}
\endbibitem

\bibitem[\protect\citeauthoryear{Herrero-P{\'e}rez and Castej{\'o}n}{2021}]{herrero2021multi}
\begin{barticle}
\bauthor{\bsnm{Herrero-P{\'e}rez}, \binits{D.}},
\bauthor{\bsnm{Castej{\'o}n}, \binits{P.J.M.}}:
\batitle{Multi-gpu acceleration of large-scale density-based topology optimization}.
\bjtitle{Advances in Engineering Software}
\bvolume{157},
\bfpage{103006}
(\byear{2021})
\end{barticle}
\endbibitem

\bibitem[\protect\citeauthoryear{Iyer et~al.}{2024}]{iyer2024pato}
\begin{barticle}
\bauthor{\bsnm{Iyer}, \binits{N.}},
\bauthor{\bsnm{Mirzendehdel}, \binits{A.M.}},
\bauthor{\bsnm{Raghavan}, \binits{S.}},
\bauthor{\bsnm{Jiao}, \binits{Y.}},
\bauthor{\bsnm{Ulu}, \binits{E.}},
\bauthor{\bsnm{Behandish}, \binits{M.}},
\bauthor{\bsnm{Nelaturi}, \binits{S.}},
\bauthor{\bsnm{Robinson}, \binits{D.}}:
\batitle{Pato: producibility-aware topology optimization using deep learning for metal additive manufacturing}.
\bjtitle{International Journal on Interactive Design and Manufacturing (IJIDeM)}
\bvolume{18}(\bissue{10}),
\bfpage{7459}--\blpage{7476}
(\byear{2024})
\end{barticle}
\endbibitem

\bibitem[\protect\citeauthoryear{Chandrasekhar et~al.}{2023}]{chandrasekhar2023frc}
\begin{barticle}
\bauthor{\bsnm{Chandrasekhar}, \binits{A.}},
\bauthor{\bsnm{Mirzendehdel}, \binits{A.}},
\bauthor{\bsnm{Behandish}, \binits{M.}},
\bauthor{\bsnm{Suresh}, \binits{K.}}:
\batitle{Frc-tounn: Topology optimization of continuous fiber reinforced composites using neural network}.
\bjtitle{Computer-Aided Design}
\bvolume{156},
\bfpage{103449}
(\byear{2023})
\end{barticle}
\endbibitem

\bibitem[\protect\citeauthoryear{Shin et~al.}{2023}]{shin2023topology}
\begin{barticle}
\bauthor{\bsnm{Shin}, \binits{S.}},
\bauthor{\bsnm{Shin}, \binits{D.}},
\bauthor{\bsnm{Kang}, \binits{N.}}:
\batitle{Topology optimization via machine learning and deep learning: a review}.
\bjtitle{Journal of Computational Design and Engineering}
\bvolume{10}(\bissue{4}),
\bfpage{1736}--\blpage{1766}
(\byear{2023})
\end{barticle}
\endbibitem

\bibitem[\protect\citeauthoryear{Padhy et~al.}{2025}]{padhy2025pilltop}
\begin{botherref}
\oauthor{\bsnm{Padhy}, \binits{R.K.}},
\oauthor{\bsnm{Chandrasekhar}, \binits{A.}},
\oauthor{\bsnm{Mirzendehdel}, \binits{A.M.}}:
Pilltop: Multi-material topology optimization of polypills for prescribed drug-release kinetics.
arXiv preprint arXiv:2512.09154
(2025)
\end{botherref}
\endbibitem

\bibitem[\protect\citeauthoryear{Chandrasekhar and Suresh}{2021}]{chandrasekhar2021tounn}
\begin{barticle}
\bauthor{\bsnm{Chandrasekhar}, \binits{A.}},
\bauthor{\bsnm{Suresh}, \binits{K.}}:
\batitle{Tounn: topology optimization using neural networks}.
\bjtitle{Structural and Multidisciplinary Optimization}
\bvolume{63}(\bissue{3}),
\bfpage{1135}--\blpage{1149}
(\byear{2021})
\end{barticle}
\endbibitem

\bibitem[\protect\citeauthoryear{Regenwetter et~al.}{2022}]{regenwetter2022deep}
\begin{barticle}
\bauthor{\bsnm{Regenwetter}, \binits{L.}},
\bauthor{\bsnm{Nobari}, \binits{A.H.}},
\bauthor{\bsnm{Ahmed}, \binits{F.}}:
\batitle{Deep generative models in engineering design: A review}.
\bjtitle{Journal of Mechanical Design}
\bvolume{144}(\bissue{7}),
\bfpage{071704}
(\byear{2022})
\end{barticle}
\endbibitem

\bibitem[\protect\citeauthoryear{Herrmann et~al.}{2024}]{herrmann2024neural}
\begin{barticle}
\bauthor{\bsnm{Herrmann}, \binits{L.}},
\bauthor{\bsnm{Sigmund}, \binits{O.}},
\bauthor{\bsnm{Li}, \binits{V.M.}},
\bauthor{\bsnm{Vogl}, \binits{C.}},
\bauthor{\bsnm{Kollmannsberger}, \binits{S.}}:
\batitle{On neural networks for generating better local optima in topology optimization}.
\bjtitle{Structural and Multidisciplinary Optimization}
\bvolume{67}(\bissue{11}),
\bfpage{192}
(\byear{2024})
\end{barticle}
\endbibitem

\bibitem[\protect\citeauthoryear{Woldseth et~al.}{2022}]{woldseth2022use}
\begin{barticle}
\bauthor{\bsnm{Woldseth}, \binits{R.V.}},
\bauthor{\bsnm{Aage}, \binits{N.}},
\bauthor{\bsnm{B{\ae}rentzen}, \binits{J.A.}},
\bauthor{\bsnm{Sigmund}, \binits{O.}}:
\batitle{On the use of artificial neural networks in topology optimisation}.
\bjtitle{Structural and Multidisciplinary Optimization}
\bvolume{65}(\bissue{10}),
\bfpage{294}
(\byear{2022})
\end{barticle}
\endbibitem

\bibitem[\protect\citeauthoryear{Xia et~al.}{2018}]{xia2018bi}
\begin{barticle}
\bauthor{\bsnm{Xia}, \binits{L.}},
\bauthor{\bsnm{Xia}, \binits{Q.}},
\bauthor{\bsnm{Huang}, \binits{X.}},
\bauthor{\bsnm{Xie}, \binits{Y.M.}}:
\batitle{Bi-directional evolutionary structural optimization on advanced structures and materials: a comprehensive review}.
\bjtitle{Archives of Computational Methods in Engineering}
\bvolume{25}(\bissue{2}),
\bfpage{437}--\blpage{478}
(\byear{2018})
\end{barticle}
\endbibitem

\bibitem[\protect\citeauthoryear{Suresh}{2010}]{suresh2010199}
\begin{barticle}
\bauthor{\bsnm{Suresh}, \binits{K.}}:
\batitle{A 199-line matlab code for pareto-optimal tracing in topology optimization}.
\bjtitle{Structural and Multidisciplinary Optimization}
\bvolume{42}(\bissue{5}),
\bfpage{665}--\blpage{679}
(\byear{2010})
\end{barticle}
\endbibitem

\bibitem[\protect\citeauthoryear{Guest et~al.}{2004}]{guest2004achieving}
\begin{barticle}
\bauthor{\bsnm{Guest}, \binits{J.K.}},
\bauthor{\bsnm{Pr{\'e}vost}, \binits{J.H.}},
\bauthor{\bsnm{Belytschko}, \binits{T.}}:
\batitle{Achieving minimum length scale in topology optimization using nodal design variables and projection functions}.
\bjtitle{International journal for numerical methods in engineering}
\bvolume{61}(\bissue{2}),
\bfpage{238}--\blpage{254}
(\byear{2004})
\end{barticle}
\endbibitem

\bibitem[\protect\citeauthoryear{Lazarov and Sigmund}{2011}]{lazarov2011filters}
\begin{barticle}
\bauthor{\bsnm{Lazarov}, \binits{B.S.}},
\bauthor{\bsnm{Sigmund}, \binits{O.}}:
\batitle{Filters in topology optimization based on helmholtz-type differential equations}.
\bjtitle{International journal for numerical methods in engineering}
\bvolume{86}(\bissue{6}),
\bfpage{765}--\blpage{781}
(\byear{2011})
\end{barticle}
\endbibitem

\bibitem[\protect\citeauthoryear{Wu et~al.}{2017}]{wu2017infill}
\begin{barticle}
\bauthor{\bsnm{Wu}, \binits{J.}},
\bauthor{\bsnm{Aage}, \binits{N.}},
\bauthor{\bsnm{Westermann}, \binits{R.}},
\bauthor{\bsnm{Sigmund}, \binits{O.}}:
\batitle{Infill optimization for additive manufacturing—approaching bone-like porous structures}.
\bjtitle{IEEE transactions on visualization and computer graphics}
\bvolume{24}(\bissue{2}),
\bfpage{1127}--\blpage{1140}
(\byear{2017})
\end{barticle}
\endbibitem

\bibitem[\protect\citeauthoryear{Zhang and Zhu}{2018}]{zhang2018topology}
\begin{botherref}
\oauthor{\bsnm{Zhang}, \binits{X.}},
\oauthor{\bsnm{Zhu}, \binits{B.}}:
Topology optimization of compliant mechanisms.
Springer
(2018)
\end{botherref}
\endbibitem

\bibitem[\protect\citeauthoryear{Ansola~Loyola et~al.}{2018}]{ansola2018sequential}
\begin{barticle}
\bauthor{\bsnm{Ansola~Loyola}, \binits{R.}},
\bauthor{\bsnm{Querin}, \binits{O.M.}},
\bauthor{\bsnm{Garaigordobil~Jim{\'e}nez}, \binits{A.}},
\bauthor{\bsnm{Alonso~Gordoa}, \binits{C.}}:
\batitle{A sequential element rejection and admission (sera) topology optimization code written in matlab}.
\bjtitle{Structural and Multidisciplinary Optimization}
\bvolume{58}(\bissue{3}),
\bfpage{1297}--\blpage{1310}
(\byear{2018})
\end{barticle}
\endbibitem

\bibitem[\protect\citeauthoryear{Zhuang et~al.}{2024}]{Zhuang2024}
\begin{barticle}
\bauthor{\bsnm{Zhuang}, \binits{Z.}},
\bauthor{\bsnm{Zhang}, \binits{W.}},
\bauthor{\bsnm{Zhang}, \binits{Y.}},
\bauthor{\bsnm{Li}, \binits{Z.}},
\bauthor{\bsnm{Shu}, \binits{C.}}:
\batitle{A 262-line matlab code for the level set topology optimization based on the estimated gradient field in the body-fitted mesh}.
\bjtitle{Structural and Multidisciplinary Optimization}
(\byear{2024})
\doiurl{10.1007/s00158-024-03891-y}
\end{barticle}
\endbibitem

\bibitem[\protect\citeauthoryear{Andreasen et~al.}{2020}]{andreasen2020level}
\begin{barticle}
\bauthor{\bsnm{Andreasen}, \binits{C.S.}},
\bauthor{\bsnm{Elingaard}, \binits{M.O.}},
\bauthor{\bsnm{Aage}, \binits{N.}}:
\batitle{Level set topology and shape optimization by density methods using cut elements with length scale control}.
\bjtitle{Structural and Multidisciplinary Optimization}
\bvolume{62}(\bissue{2}),
\bfpage{685}--\blpage{707}
(\byear{2020})
\end{barticle}
\endbibitem

\bibitem[\protect\citeauthoryear{Hunter et~al.}{2017}]{Hunter2007william}
\begin{botherref}
\oauthor{\bsnm{Hunter}, \binits{W.}}, et al.:
ToPy - Topology optimization with Python.
GitHub
(2017)
\end{botherref}
\endbibitem

\bibitem[\protect\citeauthoryear{Uihlein et~al.}{2025}]{uihlein2025140}
\begin{botherref}
\oauthor{\bsnm{Uihlein}, \binits{A.}},
\oauthor{\bsnm{Sigmund}, \binits{O.}},
\oauthor{\bsnm{Stingl}, \binits{M.}}:
A 140 line matlab code for topology optimization problems with probabilistic parameters.
arXiv preprint arXiv:2505.10421
(2025)
\end{botherref}
\endbibitem

\bibitem[\protect\citeauthoryear{Xia and Breitkopf}{2015}]{xia2015design}
\begin{barticle}
\bauthor{\bsnm{Xia}, \binits{L.}},
\bauthor{\bsnm{Breitkopf}, \binits{P.}}:
\batitle{Design of materials using topology optimization and energy-based homogenization approach in matlab}.
\bjtitle{Structural and multidisciplinary optimization}
\bvolume{52}(\bissue{6}),
\bfpage{1229}--\blpage{1241}
(\byear{2015})
\end{barticle}
\endbibitem

\bibitem[\protect\citeauthoryear{Huang and Xie}{2010}]{huang2010evolutionary}
\begin{botherref}
\oauthor{\bsnm{Huang}, \binits{X.}},
\oauthor{\bsnm{Xie}, \binits{M.}}:
Evolutionary topology optimization of continuum structures: methods and applications.
John Wiley \& Sons
(2010)
\end{botherref}
\endbibitem

\bibitem[\protect\citeauthoryear{Biyikli and To}{2015}]{biyikli2015proportional}
\begin{barticle}
\bauthor{\bsnm{Biyikli}, \binits{E.}},
\bauthor{\bsnm{To}, \binits{A.C.}}:
\batitle{Proportional topology optimization: a new non-sensitivity method for solving stress constrained and minimum compliance problems and its implementation in matlab}.
\bjtitle{PloS one}
\bvolume{10}(\bissue{12}),
\bfpage{0145041}
(\byear{2015})
\end{barticle}
\endbibitem

\bibitem[\protect\citeauthoryear{Liang and Cheng}{2020}]{liang2020further}
\begin{barticle}
\bauthor{\bsnm{Liang}, \binits{Y.}},
\bauthor{\bsnm{Cheng}, \binits{G.}}:
\batitle{Further elaborations on topology optimization via sequential integer programming and canonical relaxation algorithm and 128-line matlab code}.
\bjtitle{Structural and Multidisciplinary Optimization}
\bvolume{61}(\bissue{1}),
\bfpage{411}--\blpage{431}
(\byear{2020})
\end{barticle}
\endbibitem

\bibitem[\protect\citeauthoryear{R.~Najafabadi et~al.}{2021}]{r2021smart}
\begin{barticle}
\bauthor{\bsnm{R.~Najafabadi}, \binits{H.}},
\bauthor{\bsnm{G.~Goto}, \binits{T.}},
\bauthor{\bsnm{Falheiro}, \binits{M.S.}},
\bauthor{\bsnm{C.~Martins}, \binits{T.}},
\bauthor{\bsnm{Barari}, \binits{A.}},
\bauthor{\bsnm{SG~Tsuzuki}, \binits{M.}}:
\batitle{Smart topology optimization using adaptive neighborhood simulated annealing}.
\bjtitle{Applied Sciences}
\bvolume{11}(\bissue{11}),
\bfpage{5257}
(\byear{2021})
\end{barticle}
\endbibitem

\bibitem[\protect\citeauthoryear{Liu et~al.}{2021}]{liu2021ode}
\begin{barticle}
\bauthor{\bsnm{Liu}, \binits{Y.}},
\bauthor{\bsnm{Yang}, \binits{C.}},
\bauthor{\bsnm{Wei}, \binits{P.}},
\bauthor{\bsnm{Zhou}, \binits{P.}},
\bauthor{\bsnm{Du}, \binits{J.}}:
\batitle{An ode-driven level-set density method for topology optimization}.
\bjtitle{Computer Methods in Applied Mechanics and Engineering}
\bvolume{387},
\bfpage{114159}
(\byear{2021})
\end{barticle}
\endbibitem

\bibitem[\protect\citeauthoryear{Wang and Kang}{2021}]{wang2021matlab}
\begin{barticle}
\bauthor{\bsnm{Wang}, \binits{Y.}},
\bauthor{\bsnm{Kang}, \binits{Z.}}:
\batitle{Matlab implementations of velocity field level set method for topology optimization: an 80-line code for 2d and a 100-line code for 3d problems}.
\bjtitle{Structural and Multidisciplinary Optimization}
\bvolume{64}(\bissue{6}),
\bfpage{4325}--\blpage{4342}
(\byear{2021})
\end{barticle}
\endbibitem

\bibitem[\protect\citeauthoryear{Chandrasekhar et~al.}{2022}]{GMTOuNN2022}
\begin{botherref}
\oauthor{\bsnm{Chandrasekhar}, \binits{A.}},
\oauthor{\bsnm{Sridhara}, \binits{S.}},
\oauthor{\bsnm{Suresh}, \binits{K.}}:
GM-TOuNN: Graded Multiscale Topology Optimization using Neural Networks.
\url{https://arxiv.org/abs/2204.06682}
(2022)
\end{botherref}
\endbibitem

\bibitem[\protect\citeauthoryear{Lin et~al.}{2020}]{lin2020ansys}
\begin{barticle}
\bauthor{\bsnm{Lin}, \binits{H.}},
\bauthor{\bsnm{Xu}, \binits{A.}},
\bauthor{\bsnm{Misra}, \binits{A.}},
\bauthor{\bsnm{Zhao}, \binits{R.}}:
\batitle{An ansys apdl code for topology optimization of structures with multi-constraints using the beso method with dynamic evolution rate (der-beso)}.
\bjtitle{Structural and Multidisciplinary Optimization}
\bvolume{62}(\bissue{4}),
\bfpage{2229}--\blpage{2254}
(\byear{2020})
\end{barticle}
\endbibitem

\bibitem[\protect\citeauthoryear{Deng et~al.}{2022}]{deng2022self}
\begin{barticle}
\bauthor{\bsnm{Deng}, \binits{C.}},
\bauthor{\bsnm{Wang}, \binits{Y.}},
\bauthor{\bsnm{Qin}, \binits{C.}},
\bauthor{\bsnm{Fu}, \binits{Y.}},
\bauthor{\bsnm{Lu}, \binits{W.}}:
\batitle{Self-directed online machine learning for topology optimization}.
\bjtitle{Nature communications}
\bvolume{13}(\bissue{1}),
\bfpage{388}
(\byear{2022})
\end{barticle}
\endbibitem

\bibitem[\protect\citeauthoryear{Zuo and Xie}{2015}]{zuo2015simple}
\begin{barticle}
\bauthor{\bsnm{Zuo}, \binits{Z.H.}},
\bauthor{\bsnm{Xie}, \binits{Y.M.}}:
\batitle{A simple and compact python code for complex 3d topology optimization}.
\bjtitle{Advances in Engineering Software}
\bvolume{85},
\bfpage{1}--\blpage{11}
(\byear{2015})
\end{barticle}
\endbibitem

\bibitem[\protect\citeauthoryear{Ibhadode et~al.}{2024}]{ibhadode2024freeto}
\begin{barticle}
\bauthor{\bsnm{Ibhadode}, \binits{O.}},
\bauthor{\bsnm{Fu}, \binits{Y.-F.}},
\bauthor{\bsnm{Qureshi}, \binits{A.}}:
\batitle{Freeto-freeform 3d topology optimization using a structured mesh with smooth boundaries in matlab}.
\bjtitle{Advances in Engineering Software}
\bvolume{198},
\bfpage{103790}
(\byear{2024})
\end{barticle}
\endbibitem

\bibitem[\protect\citeauthoryear{Jia et~al.}{2024}]{jia2024fenitop}
\begin{barticle}
\bauthor{\bsnm{Jia}, \binits{Y.}},
\bauthor{\bsnm{Wang}, \binits{C.}},
\bauthor{\bsnm{Zhang}, \binits{X.S.}}:
\batitle{Fenitop: a simple fenicsx implementation for 2d and 3d topology optimization supporting parallel computing}.
\bjtitle{Structural and Multidisciplinary Optimization}
\bvolume{67}(\bissue{8}),
\bfpage{140}
(\byear{2024})
\end{barticle}
\endbibitem

\bibitem[\protect\citeauthoryear{Ooms et~al.}{2023}]{ooms2023compliance}
\begin{barticle}
\bauthor{\bsnm{Ooms}, \binits{T.}},
\bauthor{\bsnm{Vantyghem}, \binits{G.}},
\bauthor{\bsnm{Thienpont}, \binits{T.}},
\bauthor{\bsnm{Van~Coile}, \binits{R.}},
\bauthor{\bsnm{De~Corte}, \binits{W.}}:
\batitle{Compliance-based topology optimization of structural components subjected to thermo-mechanical loading}.
\bjtitle{Structural and Multidisciplinary Optimization}
\bvolume{66}(\bissue{6}),
\bfpage{126}
(\byear{2023})
\end{barticle}
\endbibitem

\bibitem[\protect\citeauthoryear{Ooms et~al.}{2024}]{ooms2024thermoelastic}
\begin{barticle}
\bauthor{\bsnm{Ooms}, \binits{T.}},
\bauthor{\bsnm{Vantyghem}, \binits{G.}},
\bauthor{\bsnm{Thienpont}, \binits{T.}},
\bauthor{\bsnm{Van~Coile}, \binits{R.}},
\bauthor{\bsnm{De~Corte}, \binits{W.}}:
\batitle{Thermoelastic topology optimization of structural components at elevated temperatures considering transient heat conduction}.
\bjtitle{Engineering with Computers}
\bvolume{40}(\bissue{4}),
\bfpage{2183}--\blpage{2207}
(\byear{2024})
\end{barticle}
\endbibitem

\end{thebibliography}

\appendix
\section{List of Open-Source Software for Topology Optimization} \label{app:reviewTable}
\onecolumn
\begin{center}
\footnotesize
\renewcommand{\arraystretch}{1.2}

\rowcolors{2}{lightgray}{white}

\begin{longtable}{ >{\raggedright\arraybackslash}p{2.8cm} >{\raggedright\arraybackslash}p{1.9cm} >{\raggedright\arraybackslash}p{0.8cm} >{\raggedright\arraybackslash}p{3.1cm} >{\raggedright\arraybackslash}p{2.cm} >{\raggedright\arraybackslash}p{1.5cm} >{\centering\arraybackslash}p{1cm}} \caption{Educational and tutorial papers for shape and topology optimization.} \label{tab:educational_tutorials}\\ \rowcolor{white} \hline \textbf{Reference} & \textbf{Function} & \textbf{nD} & \textbf{Domain} & \textbf{Method} & \textbf{Language} & \textbf{Year}\\ \hline \endfirsthead \caption[]{(continued)}\\ \hline \textbf{Reference} & \textbf{Function} & \textbf{nD} & \textbf{Domain} & \textbf{Method} & \textbf{Language} & \textbf{Year}\\ \hline \endhead \hline \endfoot \hline \endlastfoot

Sigmund~\cite{Sigmund2001} & \texttt{top99} & 2D & Structural compliance & SIMP + OC & MATLAB & 2001\\

Andreassen et~al.~\cite{Andreassen2011} & \texttt{top88}& 2D &
Structural compliance (vectorization/speedup) & 
SIMP + density filter & 
MATLAB & 2011\\

Liu \& Tovar~\cite{LiuTovar2014} & 
 \texttt{top3d} &3D& Structural and thermal compliance & 
SIMP + OC & 
MATLAB & 2014\\

Wei et~al.~\cite{Wei2018} & \texttt{TOPRBF}& 2D &
Structural compliance & 
Parameterized level-set (RBF) & 
MATLAB & 2018\\

Tavakoli \& Mohseni~\cite{tavakoli2014alternating}&  \texttt{multi\_top}& 2D& Structural and thermal compliance & Alternating active-phase for multi-material& MATLAB & 2013 \\

Zhang et~al.~\cite{zhang2018topology}&\texttt{Hf\_CM}& 2D & Compliant mechanisms& Level-set&MATLAB&2018\\

Chen et~al.~\cite{chen2019213}&\texttt{TOGN213} &2D& Geometric nonlinear&Density + MMA&MATLAB&2018\\

Challis~\cite{challis2010discrete} & \texttt{top\_levelset}&2D&
Structural compliance & Discrete level-set & MATLAB & 2010\\

Zhang et~al.~\cite{zhang2016new}& \texttt{MMC188} & 2D & Structural compliance& Moving Morphable Components & MATLAB & 2016\\

Talischi et~al.~\cite{talischi2012polytop}& \texttt{PolyTop} & 2D & Structural compliance& Density w/ unstructured polygon mesh & MATALB & 2012\\

Rub{\'e}n et~al.~\cite{ansola2018sequential}&\texttt{sera} & 2D & Structural compliance&Density w/ SERA & MATALB & 2018\\

Zhuang et~al.~\cite{Zhuang2024} &  \texttt{GFLS262}& 2D&
Structural compliance&Body-fitted level-set & MATLAB & 2024\\

Smith \& Norato~\cite{SmithNorato2020} & \texttt{GPTO}& 2-3D&
Structural compliance&Geometry projection & 
MATLAB & 2020\\

Alexandersen~\cite{Alexandersen2022} &\texttt{topFlow} & 2D & Fluid flow& Density & MATLAB & 2022\\

Ferrari \& Sigmund~\cite{ferrari2020new}& \texttt{top99neo} & 2-3D&Structural compliance & Density & MATLAB & 2020\\

Wang et~al.~\cite{wang2025efficient}&\texttt{TOP3D\_XL}& 3D& Structural compliance&SIMP w/ Matrix-free FEA& MATLAB & 2025\\

Andreasen~et~al.\cite{andreasen2020level}& \texttt{topcut}& 2D &  Structural compliance&hybrid density and level-set & MATLAB & 2020\\

Hunter~et~al.\cite{Hunter2007william}&\texttt{ToPY}&2D& Structural compliance &Density& Python& 2017\\

Uihlein~et~al.~\cite{uihlein2025140}&\texttt{topS140}& 2D& Structural compliance& Density w/ stochastic gradient &MATLAB & 2025\\

Suresh~\cite{suresh2010199}& \texttt{Pareto... OptimalTracing}& 2D&Structural compliance& Topological sensitivity&MATLAB&2010\\

Ferrari~et~al.~\cite{ferrari2021topology}&\texttt{topBuck250}&2D & Structural buckling &Density &MATLAB&2021\\

Amir~\cite{amir2021efficient} & \texttt{minVpnorm\_adpt}&2D&Structural stress&Density &MATLAB&2021\\

Deng~et~al.~\cite{deng2022efficient}& \texttt{stress\_minimize} & 2D & Structural stress&Density &MATLAB&2022\\

 Giraldo-Londo{\~n}o \& Paulino ~\cite{giraldo2021polystress}&
 \texttt{PolyStress}&2D&Structural stress&Density &MATLAB&2021\\

 Xia \& Breitkopf~\cite{xia2015design}&
 \texttt{topX}&2D&Microstructures&Density &MATLAB&2015\\

Giraldo-Londo{\~n}o \& Paulino~\cite{giraldo2021polydyna}& \texttt{PolyDyna} & 2D& Structural dynamic loads& Density&MATLAB&2021\\

Christiansen \& Sigmund \cite{christiansen2021compact}&\texttt{top200EM} &2D&Photonic structures& Density&MATLAB&2021\\

Xie \& Steven~\cite{xie1997basic} &\texttt{ESO2D} & 2D & Structural compliance & evolutionary & MATLAB & 1997 \\

Huang \& Xie \cite{huang2010evolutionary} & \texttt{BESO2D} & 2D & Structural compliance & Bi-directional evolutionary & MATLAB & 2010 \\

Bends{\o}e \& Sigmund~\cite{martin2004topology} & \texttt{top91,top105} & 2D & Structural compliance & Density & MATLAB & 2004 \\

Gao~et~al.~\cite{gao2021igatop} &  \texttt{IgaTop} & 2D & Structural compliance & Iso-geometric analysis & MATLAB & 2021 \\

Biyikli \& To~\cite{biyikli2015proportional}&  \texttt{PTOc, PTOs} & 2D & Structural compliance and stress & Density & MATLAB & 2015 \\

Coniglio~et~al.~\cite{coniglio2019generalized}&\texttt{GGP}& 2D & Structural compliance & Geometry projection & MATLAB & 2019 \\

Liang \& Cheng~\cite{liang2020further}&\texttt{DVTOPCRA}& 2D & Structural compliance & Density & MATLAB & 2020 \\

Najafabadi~et~al. \cite{r2021smart}&\texttt{top2f,top3f}& 2D & Structural compliance & Simulated annealing & MATLAB & 2021 \\

Liu~et~al.~\cite{liu2021ode}&\texttt{top58} \texttt{top105} \texttt{top108} \texttt{top53,top80}& 2-3D & Structural \& thermal compliance and eigen-frequency  & Hybrid density and level-set & MATLAB & 2021 \\

Wang \& Kang~\cite{wang2021matlab}&\texttt{VFLSM} \texttt{VFLSM3D}& 2-3D & Structural compliance & Level-set & MATLAB & 2021 \\

Chandrasekhar et~al.~\cite{TOuNN2021} &\texttt{TOuNN} & 2D & Structural compliance & Neural network assisted density & Python & 2021\\

Chandrasekhar et~al.~\cite{GMTOuNN2022} &\texttt{GMTOuNN} & 2D & Multiscale  & Neural network assisted density & Python & 2022\\

Gao~et~al.~\cite{gao2019concurrent}&\texttt{ConTop2D} \texttt{ConTop3D} & 2-3D & Multiscale  & Density & MATLAB & 2019\\

Huang~\cite{huang2023matlab}&\texttt{FPTO} & 2D &  Structural compliance and compliant mechanism &Density  & MATLAB & 2023\\

Lin et~al.~\cite{lin2020ansys}&\texttt{DER-BESO} & 2D &  Structural compliance &Bi-directional evolutionary  & ANSYS APDL & 2020\\

Zheng et~al.~\cite{zheng2024efficient}&\texttt{Multimaterial2d} \texttt{Multimaterial3d} & 2-3D &  Structural compliance & Multimaterial Density & MATLAB & 2024\\

Deng et~al.~\cite{deng2022self}&\texttt{solo} & 2D &  Structural compliance, heat, fluid & Density & MATLAB & 2022\\

Xia~et al. \cite{xia2018bi}&\texttt{esoL}, \texttt{esoX}& 2D &  Structural compliance \& microsturcture & Bi-directional evolutionary & MATLAB & 2016\\

Zou \& Xie \cite{zuo2015simple}&\texttt{BESO\_Basic}, \texttt{BESO\_Adv\_ML}, \texttt{BESO\_Adv\_NL}& 3D &  Structural compliance & Bi-directional evolutionary & Python & 2015\\

Ibhadod~et al. \cite{ibhadode2024freeto}& FreeTO & 3D & Structural & SIMP and SEMDOT  & MATLAB & 2024\\

Comet-FEniCS Numerical Tour~\cite{FEniCSSIMPDemo} &
\texttt{simp\_topopt} & 2D &
Structural compliance &
SIMP (alternate minimization / OC-like update) &
Python and FEniCS & 2023 \\

Bleyer~\cite{FEniCSSIMPDemo} 
& \texttt{simp\_topology\_optimization} 
& 2D 
& Structural compliance 
& SIMP (alt. minimization / penalization schedule) 
& Python and FEniCS 
& 2018 \\

Farrell~\cite{DolfinAdjointStokesTO} 
& \texttt{stokes-topology} 
& 2D 
& Fluids (Stokes) 
& Density + Brinkman penalization + adjoint + IPOPT 
& Python and \texttt{dolfin\_adjoint} 
& 2017 \\

Aage~\cite{DTUPythonCodes} 
& \texttt{DTU\_TopOpt\_Python} 
& 2D 
& Structural (teaching) 
& SIMP variants 
& Python 
& 2019 \\

Huang \& Tarek~\cite{TopOptjlDocs,TopOptjlGitHub,TopOptJLPkg} 
& \texttt{TopOpt.jl} 
& 2-3D 
& Structural (package) 
& SIMP + AD 
& Julia 
& 2021 \\

Aage et~al.~\cite{AagePETSc2014} 
& \texttt{TopOpt\_PETSc} 
& 2-3D 
& Large-scale structural compliance 
& SIMP{+}MMA 
& C{+}{+} and PETSc 
& 2014 \\

Gupta et~al.~\cite{FEniCSTopOptRepo} 
& \texttt{topo-fenics} 
& 2-3D 
& Structural compliance 
& SIMP 
& Python and FEniCS 
& 2020 \\

Elingaard~\cite{DeepTopoptGitHub} 
& \texttt{deep-topopt} 
& 2D 
& ML exemplar (compliance) 
& DL surrogate + SIMP-style compliance loss 
& Python and PyTorch 
& 2021 \\

Jia et~al.~\cite{jia2024fenitop} 
& \texttt{FEniTop} 
& 2-3D 
& Structural
& Density w/ Auto Diff.
& Python and FEniCS 
& 2024 \\

Ooms et~al.~\cite{ooms2023compliance,ooms2024thermoelastic} 
& \texttt{top\_tml\_shc} 
& 2D 
& Thermoelastic steady-state \& transient heat conduction
& Density
& MATLAB 
& 2023, 2024 \\
\hline
\end{longtable}
\end{center}
\twocolumn

\section{Boundary Representation in STORX}\label{app:brep}

\begin{figure}[H]
		\centering
		\includegraphics[width=0.75\linewidth]{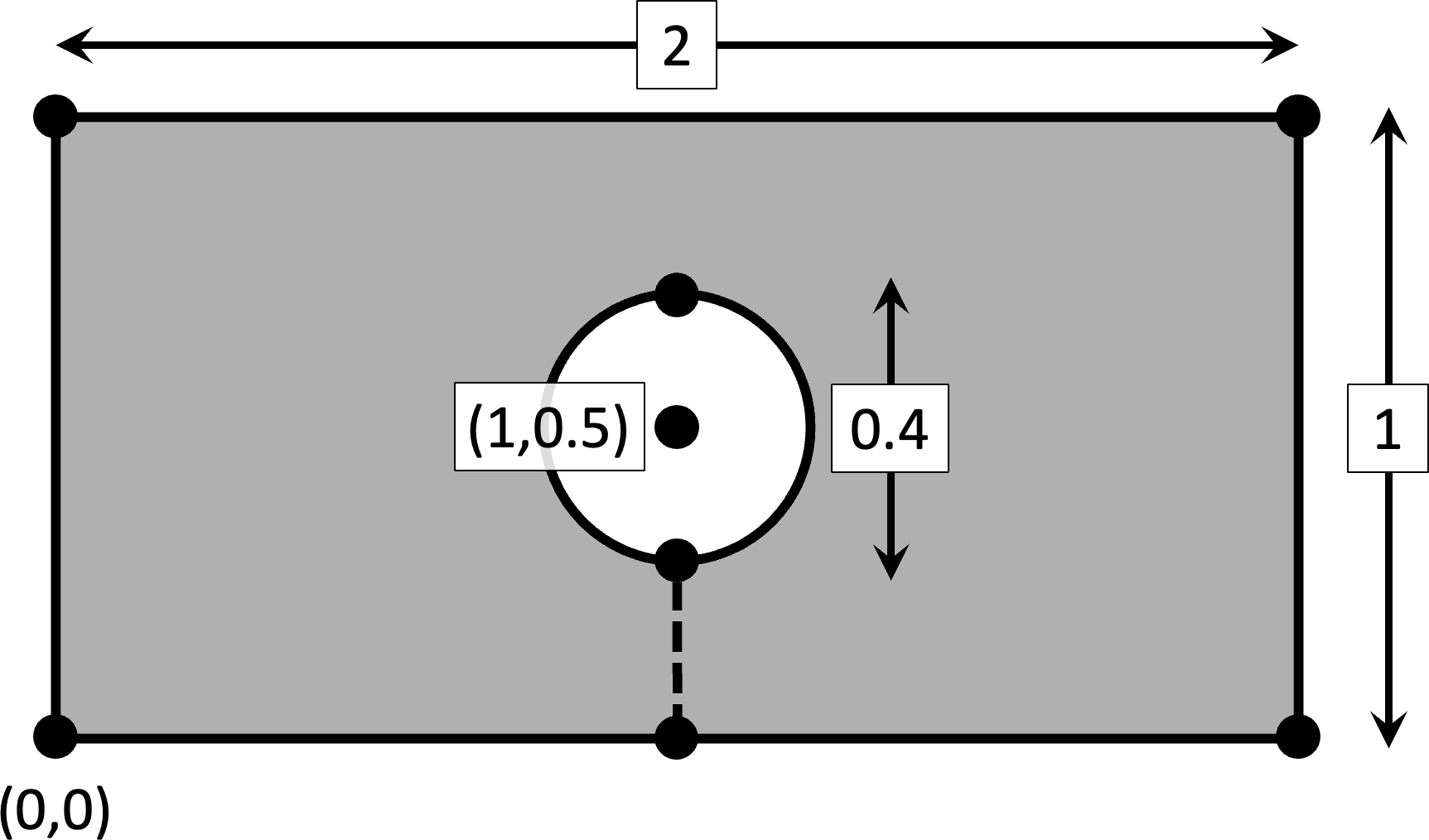}
		\caption{Rectangle model with internal hole.}
		\label{fig_brep_beamWithHole}
	\end{figure}

STORX represents 2D B-Rep models through separate vertex and edge data structures, which store geometric coordinates and topological connectivity, respectively.
Edge type and direction are specified according to the conventions in Tables \ref{tab:segTypes} and \ref{tab:segDirs}.
\vspace{-0.5cm}
\begin{table}[!h]
\captionsetup{justification=raggedright,singlelinecheck=false}
\caption{Segment types}

\begin{tabular*}{\linewidth}{@{\extracolsep{\fill}}lc}
\hline 
Segment Type & Value \\  
\hline 
Line Segment & 1 \\
Arc Segment & 2 \\
Construction Segment & $-1$ \\
\hline
\end{tabular*}\label{tab:segTypes}
\end{table}
\vspace{-1.5cm}
\begin{table}[!h]
\captionsetup{justification=raggedright,singlelinecheck=false}
\caption{Arc segment direction}
\label{table_arcDirection}

\begin{tabular*}{\linewidth}{@{\extracolsep{\fill}}lc}
\hline 
Direction & Sign \\  
\hline 
Clockwise (CW) & $+$ \\
Counterclockwise (CCW) & $-$ \\
\hline
\end{tabular*}\label{tab:segDirs}
\end{table}
\vspace{-0.5cm}
	Let us consider the rectangle with an internal hole of radius 0.2 with its center located at $(1,0.5)$ shown in Fig. \ref{fig_brep_beamWithHole}. 

\vspace{-1cm}

\begin{table}[!h]
\captionsetup{justification=raggedright,singlelinecheck=false}
\caption{Vertices}
\label{table_rectVertices_hole}

\begin{tabular*}{\linewidth}{@{\extracolsep{\fill}}ccc}
\hline 
Vertex & $x$ & $y$ \\  
\hline 
1 & 0 & 0 \\
2 & 1 & 0 \\
3 & 1 & 0.3 \\
4 & 1 & 0.7 \\
5 & 2 & 0 \\
6 & 2 & 1 \\
7 & 0 & 1 \\
8 & 1 & 0.5 \\
\hline
\end{tabular*}
\end{table}

\vspace{-0.5em}

\begin{table}[!h]
\captionsetup{justification=raggedright,singlelinecheck=false}
\caption{Edges}
\label{table_rectEdgeshole}

\begin{tabular*}{\linewidth}{@{\extracolsep{\fill}}ccccc}
\hline  
Edge & Type & $v_1$ & $v_2$ & Other \\  
\hline 
1 & Line & 1 & 2 & None \\
2 & Construction & 2 & 3 & None \\
3 & Arc & 3 & 4 & CW about 8 \\
4 & Arc & 4 & 3 & CW about 8 \\
5 & Construction & 3 & 2 & None \\
6 & Line & 2 & 5 & None \\
7 & Line & 5 & 6 & None \\
8 & Line & 6 & 7 & None \\
9 & Line & 7 & 1 & None \\
\hline
\end{tabular*}
\end{table}
    
    \vspace{-0.5cm}
It is worth emphasizing once again that every edge should be traversed such that the material always lies to the left.
   
To create internal holes while maintaining edge connectivity, we use construction line segments to connect the boundary to a point inside the domain. In this example, we connect $v_2=(1,0)$ to $v_3=(1,0.3)$. Next, we create an arc segment from $v_3$ to $v_4=(1,0.7)$ about the center $v_8=(1,0.5)$ in a clockwise (CW) direction. To complete the circle, we create another arc segment, this time going from $v_4$ to $v_3$. Finally, to get back outer boundary, we create another construction segment from $v_3$ to $v_2$ and define the rest of the shape similar to previous examples.

The two data structures for vertices and edges for rectangle with an internal hole are captured in  Table \ref{table_rectVertices_hole} and Table \ref{table_rectEdgeshole}, respectively.

\begin{figure}[!h]
		\centering
		\includegraphics[width=0.75\linewidth]{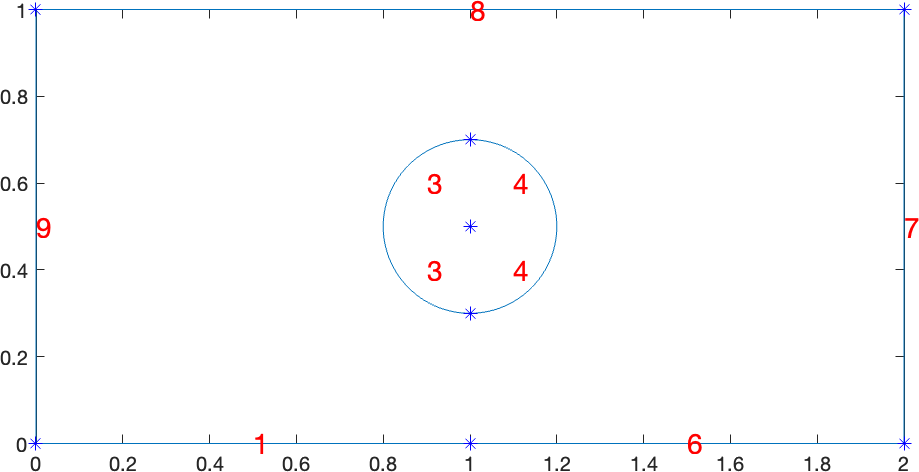}
		\caption{Rectangle with an internal hole.}
		\label{fig_brep_beamWHole}
	\end{figure}

We can define the model as:
    
\begin{lstlisting}
BeamWithHole.vertices = [0 0;1 0;1 0.3;1 0.7;2 0;2 1;0 1;1 0.5]';
BeamWithHole.segments = [1 1 2 0; -1 2 3 0; 2 3 4 8; 2 4 3 8; -1 3 2 0; 1 2 5 0; 1 5 6 0; 1 6 7 0; 1 7 1 0]';
\end{lstlisting}

To create the beam with an internal hole and visualize it, we write:
\begin{lstlisting}
brepClass = @brep2d;
geom = brep2d(BeamWithHole);
geom.plotGeometry();
\end{lstlisting}

The resulting model is illustrated in Fig. \ref{fig_brep_beamWHole}.
Observe that in Fig. \ref{fig_brep_beamWHole}, the edges number 2 and 5 are construction segments and are not visible.
\end{document}